# Homological Quantum Codes Beyond the Toric Code

Von der Fakultät für Mathematik, Informatik und Naturwissenschaften der RWTH Aachen University zur Erlangung des akademischen Grades eines Doktors der Naturwissenschaften genemigte Dissertation

vorgelegt von

M. Sc. Physics

Nikolas P. Breuckmann

aus Duisburg

Berichter: Universitätsprofessorin Dr. rer. nat. Barbara M. Terhal
Universitätsprofessor Dr. rer. nat. Hendrik Bluhm
Dr. Earl Campbell

Tag der mündlichen Prüfung: 17. November 2017

Diese Dissertation ist auf den Internetseiten der Universitätsbibliothek verfügbar.

# Contents

# List of Figures

# List of Tables

# Abstract

Computer architectures which exploit quantum mechanical effects can solve computing tasks that are otherwise impossible to perform. A quantum computer operates on a number of small quantum mechanical systems, known as quantum bits, or *qubits*. Since these systems are realized on the scale of atoms, they are very prone to errors. Errors occur when the environment interacts with the qubits, a process called *decoherence*. It is widely accepted that it will not be possible to shield qubits completely from the outside world. If one were to perform a quantum computation on the qubits directly, then after a short period of time the information present in the qubits would be lost. To counter decoherence the state of a qubit can be encoded into multiple physical ones. This is called a *quantum error correcting code*. Performing quantum error correction allows one to extend the life time of the encoded qubit arbitrarily, assuming that the rate of errors remains below a certain *threshold* value. The use of quantum codes creates an *overhead* in resources, as for every logical qubit many more physical qubits are needed. The resource overhead for fault-tolerance is problematic, since realizing qubits will be costly, and in the early stages of building quantum computers the number of physical qubits will be limited.

The currently favored coding architecture is the *toric code* and its variant the *surface code* in which the physical qubits are put on a square grid in which interactions are only between nearest neighbors. In this thesis we will explore quantum codes in which qubits interact as if they were nearest neighbors in more exotic spaces.

In the first part we will consider closed surfaces with constant negative curvature. We show how such surfaces can be constructed and enumerate all quantum codes derived from them which have less than 10.000 physical qubits. For codes that are extremal in a certain sense we perform numerical simulations to determine the value of their threshold. Furthermore, we give evidence that these codes can be used for more overhead efficient storage as compared to the surface code by orders of magnitude. We also show how to read and write the encoded qubits while keeping their connectivity low.

In the second part we consider codes in which qubits are layed-out according to a four-dimensional geometry. Such codes allow for much simpler decoding schemes compared to codes which are two-dimensional. In particular, measurements do not necessarily have to be repeated to obtain reliable information about the error and the classical hardware performing the error correction is greatly simplified. We perform numerical simulations to analyze the performance of these codes using decoders based on local updates. We also introduce a novel decoder based on techniques from machine learning and image recognition to decode four-dimensional codes.

# Zusammenfassung

Durch Ausnutzung quantenmechanischer Effekte ist es prinzipiell möglich Berechnungen durchzuführen welche für einen klassischen Computer unmöglich sind. Quantencomputer basieren, im Gegensatz zu klassischen Computern, auf sogenannten *Qubits*, wobei jedes Qubit ein Quantensystem mit zwei Zuständen ist. Durch Wechselwirkungen mit der Umgebung verlieren die Qubits ihre Koheränzeigenschaften welche notwendig für Quantenberechnungen sind.

Ein vielversprechender Ansatz um der Dekohärenz entgegenzuwirken ist es den Zustand jedes einzelnen Qubits in den globalen Zustand eines Vielteilchensystems zu kodieren. Solche Systeme werden *Quantencodes* genannt. Durch lokale Messungen in einem Quantencode ist es möglich Informationen über entstandene Fehler zu erhalten und diese Fehler zu korrigieren. Dies geschieht ohne dabei den kodierten Zustand zu beeinflussen. Solange sich die Fehlerrate unter einem gewissen Schwellwert befindet, lässt sich die Lebensdauer der kodierten Zustände und damit die Dauer einer Berechnung im Quantencomputer, beliebig verlängern. Allerdings führt der Einsatz von Quantencodes dazu, dass mehr physische Qubits benötigt werden. Diesen Mehraufwand gilt es zu minimieren, da die Herstellung von Qubits sehr aufwendig ist und in absehbarer Zeit nur eine geringe Anzahl von Qubits zur Verfügung stehen wird. Momentan konzentriert sich ein Großteil der Aufmerksamkeit auf den sogenannten *Toric Code* und den *Surface Code*. In diesen Codes werden die Qubits auf einem quadratischen Gitter plaziert und können nur mit ihren direkten Nachbarn interagieren. Diese Dissertation behandelt Quantencodes in welchen die Konnektivität der Qubits durch die Geometrie von etwas komplexeren Räumen bestimmt wird.

Der erste Teil dieser Arbeit behandelt den Fall von geschlossenen, zweidimensionalen Flächen welche negativ gekrümmt sind. Wir zeigen wie solche Flächen mittels Reflektionsgruppen konstruiert werden können und welche Eigenschaften die aus ihnen gewonnenen Quantencodes haben. Unsere Konstruktion erlaubt es uns Familien von Quantencodes mit bis zu 10.000 physischen

Qubits zu generieren. Mittels Monte-Carlo-Simulationen analysieren wir die Fehlerunterdrückungsrate und den kritischen Schwellwert unterhalb dessen die Lebensdauer der kodierten Qubits beliebig verlängert werden kann. Mittels eines empirischen Ausdrucks für die Fehlerunterdrückungsrate argumentieren wir das im Vergleich zum Surface Code mehrere Größenordnungen an physischen Qubits weniger benötigt werden um den gleichen Schutz vor Fehlern gewährleisten zu können. Darüberhinaus zeigen wir wie die kodierten Zustände manipuliert werden können, wobei die Konnektivität der physischen Qubits niedrig bleibt.

Im zweiten Teil dieser Arbeit analysieren wir Quantencodes welche aus vierdimensionalen Räumen gewonnen werden. Der Vorteil genenüber den zweidimensionalen Quantencodes ist, dass die Fehlerkorrektur hier konzeptionell viel einfacher ist. Wir analysieren und vergleichen verschiedene Fehlerkorrekturmechanismen basierend auf lokalen Korrekturen mittels Monte-Carlo-Simulationen. Darüberhinaus stellen wir einen neuen Dekoder vor, welcher auf maschinellem Lernen basiert.

# Acknowledgments

First and foremost, I want to thank Barbara who has been my supervisor for pretty much all my academic life. I would like to thank her for giving me academic and general life advice ("Science is like cooking. You stir in some pots and maybe something good comes out of one of them."). I have greatly benefited from her supervision during my time at the IQI.

I would like to thank Prof. Hendrik Bluhm, Dr. Earl Campbell, Prof. David DiVincenzo and Prof. Stefan Wessel for agreeing to be members of my defense committee. Another thanks to David for pointing out reference [2] which was very helpful in understanding tessellations.

Thanks to my parents Lucia and Peter for encouraging me to follow my interests and to my family and friends for supporting me.

I would like to thank Hélène Barton for shielding me from bureaucratic duties and for her cheerful attitude. Thanks to my office mates Susanne Richer and Ben Criger and Xiaotong Ni without whom my days would have been very dull. I want to thank the members of the "Mathematicians Anonymous" (MA) meetings: Johannes Keller, Jascha Ulrich, Christoph Ohm, Tobias Kühn, Joris Dolderer and Jonathan Schmidt-Dominé. The MA meetings lasted for over 3 years and were always enjoyable which, to a big part, is owed to Johannes' enthusiasm and Jascha's ability to find an intuition for the most obscure definitions. Johannes managed to convince me that differential geometry is an interesting subject to study and thereby unknowingly prepared me for what turned out to become part of my PhD project. Thanks to Jascha and Prof. Fabian Hassler for countless fun discussions about all kinds of things. I learned a lot of stuff from Christophe Vuillot who also has been a great traveling companion to various conferences. Thanks to Xiaotong Ni for teaching me about artificial neural networks while simultaneously introducing me to the TV series Westworld, which made the learning experience more intense. Thanks to Anirudh Krishna for the great time during his stay at the IQI and for some of the best board game experiences I have ever had. Thanks to Ananda Roy and Fabio Pedrocchi for regularly visiting my office to discuss physics, jokes and gym work-outs. Thanks to Alessandro Ciani, Stefano Bosco, Kasper Duivenvoorden and Prof. Maarten Wegewijs for providing copious amounts of coffee on a daily basis. Thanks to Manuel Rispler for fun discussions and for taking me to the CCC in Hamburg. During my PhD I was fortunate to co-supervise two excellent undergrads Friederike Metz and Jonathan Conrad. Generally, I want to thank all past and present members of the IQI. Special thanks to the team of the RWTH Computer Cluster for not deleting my account (especially during the numerical simulations in 4D).

I would like to thank Prof. Robert König for his hospitality during my stay at TU Munich and for very interesting discussions. I would also like to thank him for sharing notes explaining Dehn twists. Many thanks to the people in Palo Alto for inviting me over to California. I very much enjoyed my stay and I am looking forward to continue working with you guys. I would also like to thank Jens Eberhardt and Prof. Wolfgang Soergel for their hospitality and interesting discussions during my stay at Freiburg University.

There are many more people that I have had very illuminating discussions on quantum codes. Outside of the IQI these are in particular Earl Campbell, Nicolas Delfosse, Michael Kastoryano, Robert König, Naomi Nickerson and Leonid Pryadko. Thanks to Earl Campbell for suggesting the semi-hyperbolic construction and to Nicolas Delfosse for discussions on their distance. Thanks to Prof. Alex Lubotzky for pointing out references regarding the construction of 4D hyperbolic codes and to Sergey Bravyi for sharing his algorithm to efficiently determine distances of 2D CSS codes.

Finally I would like to thank my partner John Kwan for all his support.

# Chapter 1

# Introduction

## 1.1 Prologue

The quantum computer's internal state is affected by inescapable interactions with the environment. This is problematic since quantum algorithms involve large-scale superpositions which due to the interactions with the environment start to decay, a process called *decoherence*. In the last twenty years there has been a tremendous amount of progress on the experimental side in gaining control over quantum systems and shielding them from the debilitating effects of noise. However, to execute quantum algorithms with arbitrary long run-times we need to be able to obtain an arbitrary amount of error suppression. To counter decoherence the state of a qubit can be encoded into multiple physical ones. This is called a *quantum error correcting code*. More specifically, *active* quantum error correction is a process through which we gain information on the error that occurred through measurements on the system. This information is processed by a regular (non-quantum) computer to infer an operation which reverses the effects of the error. By repeating this process of error inference and correction one can extend the life time of the encoded qubit arbitrarily, assuming that the rate of errors remains below a certain *threshold value*.

Apart from being an important ingredient for scalable quantum computing, there is also a fundamental interest in obtaining stable quantum memories: Highly entangled states tend to decohere so that most natural systems do not explore the full Hilbert space. In this sense we can probe the laws of quantum mechanics by showing that complex entangled states can be realized and kept stable.

The first theoretical study in fault-tolerance was conducted by John von Neumann [3] who showed that a classical computer which is subject to noise can be made less noisy by *multiplexing*: The computer's internal state is copied several times and the algorithm is run on each copy. Occasionally there is a majority vote between all copies. Von Neumann showed that as long as the bits in each copy flip with a probability below a certain threshold value, the multiplexing technique can prevent the error rate from rising.

As fault-tolerance is linked to scalability, it is important that the overhead in resources it requires are modest. The threshold theorem [4] establishes that this is indeed possible in theory. It states that if we have components which fail with a probability at most $p$ we can perform arbitrary quantum computations of some desired error rate $\varepsilon$, as long as $p$ is below some threshold which depends on the fault-tolerance architecture. Importantly, there is only a polylogarithmic overhead in the size of the quantum computation we want to perform and a polynomial overhead in the inverse of the target error rate $\varepsilon$. Current quantum technologies achieve error rates which are close to the point at which fault-tolerance becomes feasible. The question remains as to which fault-tolerance architecture provides the lowest overhead in resources to achieve error suppression [5].

So far, all experimental efforts are focused on realizing a quantum code called the *surface code* in which the qubits and their interactions are put into a *planar layout*. More specifically, qubits are located on a square grid where only nearest neighbors can interact. Some physical realizations of qubits are very limited in the way they can be made to interact with each other. A planar layout is certainly helpful in those cases, making the surface code a preferred choice. However, there are several proposed physical implementations of quantum computers in which interactions do not have to be planar. For example, there are proposals of modular architectures in which each module holds a small number of qubits. The modules are interconnected by photonic links which are not restricted to be planar [6].

In this thesis we explore quantum code constructions which utilize non-planarity in several ways: We show how to construct quantum codes derived from negatively curved spaces called hyperbolic surfaces which can potentially reduce the resource overhead when compared to currently favored quantum error correction codes. Such codes are called hyperbolic codes and were first mentioned in [7] where it was shown that hyperbolic codes have a finite encoding rate. This means that when increasing the size of the code the amount of data that can be stored grows proportionally. In [8] it was shown that hyperbolic codes are in some sense optimal finite rate codes with 2D connectivity. In [9] it was argued that hyperbolic codes are optimal even when allowing for arbitrary $D$-dimensional connectivity. Although, this was later shown not to be the case [10]. We expand on this earlier work by

1. using mathematical tools that allow one to find any such code with an exhaustive searching procedure,

2. analyzing the noise threshold of several interesting examples and

3. showing how a hyperbolic surface code can outperform planar quantum codes in terms of error suppression and resource overhead for a fixed number of encoded qubits.

We perform an exhaustive search and enumerate all hyperbolic surfaces which can be used to define quantum codes with less than $10^4$ physical qubits. We also consider variations of the construction, such as planar hyperbolic codes and semi-hyperbolic codes. Finally, we discuss how to transfer data in and out of a hyperbolic code from another topological quantum code in a fault-tolerant way and show how to manipulate the data encoded within a 2D hyperbolic code.

Furthermore, we investigate higher dimensional codes which have the advantage of offering more robustness against faulty measurements. We consider various decoding schemes for such codes. One property that these decoders have in common is that they reduce the overhead in classical processing that is required. We analyze their performance by conducting numerical simulations.

Additionally, we introduce a decoding scheme based on machine learning and we perform numerical simulations to determine its performance as well. Machine learning was already considered to decode quantum codes [11, 12, 13, 14]. Our architecture differs from earlier ones in that ours is explicitly *scalable*. We show that convolutional neural networks can make use of the translational invariance present in the decoding problem of 4D quantum codes which allows us to train a neural network once and use the result for arbitrary system sizes.

## 1.2 Outline

In **Chapter 2** we will introduce preliminary material, including background on quantum error correction with topics such as error models, the error-correcting conditions and stabilizer codes.

We also review a recipe for constructing quantum codes from tessellated spaces. We will use a well-known quantum code called the *toric code* as an example for this construction. Readers already familiar with the toric code may skip this chapter and use it as a reference in following chapters.

In **Chapter 3** we show how to construct quantum codes from negatively curved spaces called *hyperbolic surfaces*. We will first introduce the necessary background on curved spaces and their tessellations. We show how to construct tessellations of closed surfaces using Coxeter groups, giving rise to *hyperbolic surface codes*. The rest of the chapter is dedicated to the examination of these codes.

In **Chapter 4** we analyze the performance of 2D hyperbolic codes by conducting numerical simulations, making different assumptions on the error model. We furthermore give an approximation for the probability of corrupting the encoded quantum information in the limit of low noise rates. Finally we show that hyperbolic codes can provide overhead savings in orders of magnitude as compared to currently favored topological error correction schemes.

In **Chapter 5** we discuss codes which are derived from four-dimensional spaces and their advantages over 2D codes. We review the well-known 4D toric code and introduce a variant of the 4D toric with open boundaries which we call the *tesseract code*. Furthermore, we discuss codes derived from four-dimensional curved spaces.

In **Chapter 6** we introduce different decoding schemes for 4D codes which are analyzed numerically. These decoding schemes have in common that they can be performed by very primitive classical hardware, offering an advantage over the more complex decoding problem for 2D codes. We consider decoders which can be highly parallelized, such as cellular automata which offer the additional advantage of being realizable using very primitive hardware. We also investigate the use of machine learning and artificial neural networks for decoding. We show how certain networks are particularly adapted to the decoding of higher-dimensional quantum codes.

Finally, in **Chapter 7** we give a conclusion and discuss interesting directions for future work.

# Chapter 2

# Preliminary material

This chapter serves as a review of concepts that will appear in later chapters. A reader already familiar with topological quantum codes may skip this chapter and use it as a reference. First we will introduce some basic concepts of quantum fault-tolerance and quantum error correction. The next topic will be tessellations. Tessellating a space simply means that it is covered by polygons without leaving gaps. Everyday examples of tessellations are tiles covering a wall or stained glass windows. The main result of this section will be Theorem 2.10 which shows that tessellations naturally give rise to quantum codes. The properties of such codes can be analyzed using tools from algebraic topology which we introduce directly afterwards. Finally we discuss how these codes can be decoded.

## 2.1 Quantum error correction

### 2.1.1 Background

#### Qubits

In classical computation the smallest unit of information is a bit which can have the states 0 or 1. The analogon in quantum information is the *qubit* which is a two-level quantum system. In a closed quantum system all states are *pure states* which are (normalized) vectors in a two-dimensional Hilbert space $\mathcal{H} = \mathbb{C}^2$. We assume that there is a distinguished basis $\{|0\rangle, |1\rangle\}$ called the *computational basis*. Any pure qubit state is hence given by $|\psi\rangle = \alpha|0\rangle + \beta|1\rangle$ with $\alpha, \beta \in \mathbb{C}$.

#### Quantum gates and circuits

The time evolution of a quantum system is described by a unitary operator which rotates the state vector. An arbitrary rotation can be approximated by a small set of fixed rotations called *quantum gates*. The sequence of applying different gates can be visualized by a *quantum circuit* which consists of $n$ wires symbolizing qubits. Time goes from left to right. The application of a gate is indicated by a symbol on the wire (or several wires if the gate acts on multiple qubits). Quantum circuits may also be thought of as tensor networks with distinguished input and output. Quantum circuits may also include single-qubit measurements in the computational basis.

### 2.1.2 Quantum error correction

Before we discuss quantum codes, let us briefly review some other ideas on how to mitigate quantum errors. There are several different methods to combat decoherence. For example, in many systems the qubits are subject to so-called $1/f$ noise which leads to a slow, systematic drift of the state. By applying a fast unitary rotation this drift is averaged out. This procedure is known as *dynamic decoupling*.

Another approach is to consider a subspace which suffers only from little or no decoherence from interactions with the environment. Consider the following simple example where the system consists of two spins: If the main contribution of the interaction with the environment results in an application of $\exp(i\phi Z)$ then one can choose a subspace which is mostly unaffected. Namely, the subspace spanned by $|01\rangle + |10\rangle$ and $|01\rangle - |10\rangle$ stays invariant under this type of noise. Hence we can encode the state of a single spin $\alpha|\bar{0}\rangle + \beta|\bar{1}\rangle$ by choosing $|\bar{0}\rangle = (|01\rangle + |10\rangle)/\sqrt{2}$ and $|\bar{1}\rangle = (|10\rangle + |01\rangle)/\sqrt{2}$. This example can be generalized to larger systems which suffer from systematic noise [15].

Another method which is appealing is the use of systems which exhibit so-called *topological order* at non-zero temperature [16]. However, it is not known whether such systems exists in less than 4 dimensions [5, 17] although they are unlikely to exist in 2D [18].

A different approach (and the focus of this thesis) is to preserve a subspace by means of *active quantum error correction*. Generally, active quantum error correction consists of three parts: First classical data about the state is gathered by performing measurements on the system. The data allow us to learn something about what error was applied to the system, while not disturbing the state that we want to preserve. This data is called the *error syndrome* or simply *syndrome*. In the second step the syndrome is classically processed to infer a unitary operation called *recovery operation* which is applied in the third step. The recovery operation reverses the effects of the noise.

But how can we do measurements without disturbing the encoded space? - The basic idea is to split the Hilbert space of the quantum memory $\mathcal{H}_M$ into a *code space* $\mathcal{C}$ and its orthogonal complement $\mathcal{C}^\perp$. Furthermore, the code space consists of two tensor factors: One factor $\mathcal{A}$ is the encoded space which we will also call the *logical subspace*; the other factor $\mathcal{B}$ gives the redundancy which makes it possible to protect against errors. This is the subspace on which we perform the syndrome measurements.

$$\mathcal{H}_M = \underbrace{\mathcal{A} \otimes \mathcal{B}}_{\mathcal{C}} \oplus \mathcal{C}^\perp \tag{2.1}$$

In Section 2.1.5 we will review *stabilizer codes* which are quantum codes which have a particular simple mathematical structure among other desirable properties.

### 2.1.3 Quantum errors

Quantum errors occur due to the presence of an environment which interacts with the quantum memory. The Hilbert space of the quantum memory $\mathcal{H}_M$ is in fact part of a larger space $\mathcal{H}_M \otimes \mathcal{H}_E$, where $\mathcal{H}_E$ is the Hilbert space of the *environment*. The basic assumption is that we can only observe and manipulate $\mathcal{H}_M$ whereas the Hilbert space of the environment is very large. This is problematic as generally there will be entanglement present between the memory and the environment and due to the size of $\mathcal{H}_E$ we can not describe the evolution of the full system.

To be able to describe the state of the memory we thus have to consider *probabilistic mixtures* of quantum states, meaning that the state of the memory is in a certain quantum state $|\phi_i\rangle$ with probability $p_i$. Such ensembles are conveniently expressed in terms of *density matrices*. They are defined as

$$\rho := \sum_i p_i |\phi_i\rangle \langle\phi_i| . \tag{2.2}$$

Note that when there is no ambiguity about what quantum state we are in, i.e. we have an ensemble of a single quantum state $|\phi_1\rangle$ with $p_1 = 1$ then $\rho$ is a projection of rank 1. These states are called *pure states*, as opposed to non-trivial ensembles of multiple states which are called *mixed states*. Density matrices corresponding to pure states are characterized by the equation $\mathrm{tr}(\rho^2) = 1$.

Quantum mechanics tells us that when we let some sate $|\phi\rangle$ evolve for some time we end up with a unitary rotation of that state $U|\phi\rangle$. Hence, in the language of density matrices the same time evolution of the pure state $\rho = |\phi\rangle\langle\phi|$ is given by $U\rho U^\dagger$.

A measurement is described by a collection of matrices $M_k$, where $k$ labels the different measurement outcomes, satisfying the completeness equation $\sum_k M_k^\dagger M_k = I$. The result $k$ occurs with probability $\text{Prob}(k) = \text{tr}\left(M_k^\dagger M_k \rho\right)$ and the state after the measurement is $M_k \rho M_k^\dagger / \text{Prob}(k)$. We will be mostly concerned with *projective measurements* where every $M_k$ is a projector. In this case we have that $M_k^\dagger M_k = M_k$ since projectors are Hermitian $M_k^\dagger = M_k$ and idempotent $M_k^2 = M_k$. The probability of measuring outcome $k$ thus simplifies to $\text{Prob}(k) = \text{tr}(M_k\rho)$ and the completeness condition states that the sum of projectors gives the identity $\sum_k M_k = I$.

Let us now consider the density operator of the full system of memory and environment $\rho$ which we assume to be a pure state. This state has the form $\rho = |\phi\rangle\langle\phi|$ for some $|\phi\rangle \in \mathcal{H}_M \otimes \mathcal{H}_E$. It turns out that there is a (unique) operation which maps the density matrix of the full system $\rho$ onto a density matrix $\rho_M$ describing the state of the memory, fulfilling the following condition: Any measurement on the full system $\rho$ of the form $M_k \otimes I_E$ is identical to the measurement of $M_k$ on $\rho_M$. By identical we mean that the outcome statistics defined by $\text{Prob}(k)$ is the same in both cases. This operation is called a *partial trace* and is defined as follows: Let $|e_i\rangle$ be a basis of $\mathcal{H}_M$. For density matrices of the form $|e_i\rangle\langle e_j| \otimes \rho_E$ we define

$$\text{tr}_E\left(|e_i\rangle\langle e_j| \otimes \rho_E\right) := |e_i\rangle\langle e_j| \cdot \text{tr}(\rho_E) \tag{2.3}$$

and for general $\rho = \sum_{i,j} |e_i\rangle\langle e_j| \otimes \rho_E^{i,j}$ by linear extension

$$\text{tr}_E(\rho) := \sum_{i,j} \text{tr}_E\left(|e_i\rangle\langle e_j| \otimes \rho_E^{i,j}\right). \tag{2.4}$$

Equipped with the formalism of density matrices and partial traces we are now prepared to tackle the problem of describing the interaction of our memory with the environment, without having to consider the full unitary time evolution on $\mathcal{H}_M \otimes \mathcal{H}_E$. The interaction of the memory with the environment will instead be described by *quantum operations* which are maps between the density operators of the memory. We will now derive the form of these quantum operations using the partial trace.

Without loss of generality we can assume that the environment is in a pure state $|\phi_0\rangle$. If $\rho_E$ were not pure we would have $\sum_{i=1}^n p_i |\phi_i\rangle\langle\phi_i|$. In this case we could extend the Hilbert space artificially by introducing a fictitious system $\mathcal{H}_R$ of dimension $n$ with basis states $|e_i\rangle$. Now, consider the pure state $\rho_{E'} = |\psi\rangle\langle\psi| \in \mathcal{H}_{E'}$ with $|\psi\rangle = \sum_{i=1}^{n} \sqrt{p_i}\,|\phi_i\rangle\,|e_i\rangle := \mathcal{H}_E \otimes \mathcal{H}_R$. Note that $\mathrm{tr}_R(\rho_{E'}) = \rho_E$. This process is known as *purification*.

We can extend $|\phi_0\rangle$ to a basis $|\phi_i\rangle$ of $\mathcal{H}_E$. Let us further assume that the system of memory and environment start out as a product state and evolve for some time, which implements some unitary $U$. We end up in the state $U(\rho_M \otimes |\phi_0\rangle\langle\phi_0|)U^\dagger$. Tracing out the environment gives

$$\mathrm{tr}_E\left(U(\rho_M \otimes |\phi_0\rangle\langle\phi_0|)U^\dagger\right) = \sum_k A_k \rho_M A_k^\dagger, \tag{2.5}$$

where $A_k$ are operators acting on $\mathcal{H}_M$ with entries

$$(A_k)_{i,j} = (\langle e_i| \otimes \langle\phi_k|)\,U\,(|e_j\rangle \otimes |\phi_0\rangle). \tag{2.6}$$

Since the dimension of $\mathcal{H}_E$ is assumed to be larger than the dimension of $\mathcal{H}_M$, the matrices $A_k$ must be linearly dependent. More specifically, assuming that the Hilbert space of the quantum memory $\mathcal{H}_M$ has dimension $N$ then there are at most $N^2$ linear independent operators $A_k$. The operators $A_k$ are called *Kraus operators*. The action of Kraus operators on a quantum state as in Equation 2.5 is referred to as a *quantum channel*. The word *channel* is borrowed from classical information science and describes the process of transmitting information from a sender to a receiver or equivalently as a transmission through time when the information is stored in a memory.

A common simplifying assumption is that errors occur independently on single qubits and according to the same probability distribution. In other words we assume that each qubit interacts with a separate environment and that all these interactions are the same. For such a channel the Kraus operators $A_k$ are identical single qubit operators. The Pauli operators form a basis of the single qubit operators, so that for any single-qubit noise channel acting on qubit $i$ of a density matrix $\rho$ we can expand $A_k = \alpha_k I + \beta_k X_i + \gamma_k Y_i + \delta_k Z_i$. $X_i$, $Y_i$ and $Z_i$ act as the respective Pauli operator on the $i$th qubit and as the identity on all other qubits.

A widely used channel to simulate noise in a quantum system is the *independent $X$-$Z$ error model*. It is analogous to the symmetric bit-flip channel in classical information theory where each bit is flipped independently according to a fixed probability. In the independent $X$-$Z$ error model each qubit in the quantum memory undergoes a Pauli-$X$ or a Pauli-$Z$ rotation, with the same probability $p$. In the terminology of quantum channels we have a channel $\mathcal{N} = \mathcal{N}_Z \circ \mathcal{N}_X(\rho)$ where the Kraus operators of $\mathcal{N}_X$ are

$$A_{0,i} = \sqrt{1-p}\, I, \qquad A_{1,i} = \sqrt{p}\, X, \tag{2.7}$$

and the Kraus operators of $\mathcal{N}_Z$ are

$$A_{0,i} = \sqrt{1-p}\, I, \qquad A_{1,i} = \sqrt{p}\, Z, \tag{2.8}$$

where $i$ labels the physical qubits of the quantum memory. The independent $X$-$Z$ error model, although not realistic, does not make any assumptions on the underlying hardware. It is assumed that if a quantum memory is well protected against the independent $X$-$Z$ error model then it will also show good performance against other independent error models with similar error probabilities.

### 2.1.4 Quantum error correction conditions

In the previous section we have seen how interactions with the environment can be modeled by quantum channels which operate on the density matrix of a quantum memory. Let us assume that the noise channel is given in its Kraus representation $\mathcal{N}[\cdot] = \sum_k A_k[\cdot]A_k^\dagger$. Not all Kraus operators $A_k$ are necessarily errors, meaning that they may have trivial action on the code space $\mathcal{C}$. The Kraus operators which do not have trivial action on the code space are errors and we will denote them by $E_k$. They give rise to the error channel $\mathcal{E}[\cdot] = \sum_k E_k[\cdot]E_k^\dagger$. Since some $A_k$ might be missing in $\mathcal{E}$ it is in general not trace preserving. A *recovery operation* $\mathcal{R}$ *can correct* $\mathcal{E}$ if for all Kraus operators $E_k$ and any density matrix of the code space $\rho_\mathcal{C}$ we have

$$\mathcal{R} \circ \mathcal{E}[\rho_\mathcal{C}] \propto \rho_\mathcal{C}. \tag{2.9}$$

We cannot expect that a recovery operation exists for arbitrary error channels. For example, an error channel that randomly permutes the basis of the code space and otherwise acts trivially can a priori not be corrected for. It turns out that there exists a mathematical condition which determines whether a recovery operation $\mathcal{R}$ can exist.

**Theorem 2.1** (Quantum Error Correction Conditions [19, 20]). *Let $|\phi_i\rangle$ be a basis of the code space $\mathcal{C}$ and consider the error channel $\mathcal{E}[\cdot] = \sum_k E_k[\cdot]E_k^\dagger$. There exists a recovery operation $\mathcal{R}$ which corrects $\mathcal{E}$ if and only if*

$$\langle\phi_i| E_k^\dagger E_l |\phi_j\rangle = c_{k,l}\delta_{ij} \quad \forall i,j,k,l \tag{2.10}$$

*where $c_{k,l}$ are the entries of a Hermitian matrix $C$ and $\delta_{ij}$ is the Kronecker delta.*

*Proof sketch.* The necessity in Theorem 2.1 is proved by applying the definitions and rearranging terms. The sufficiency is proved by explicitly constructing a recovery map $\mathcal{R}$. This is done by diagonalizing $C$ via a unitary $U = (u_{i,j})$. Defining $F_m = \sum_k u_{m,k} E_k$ gives us an alternative Kraus representation of $\mathcal{E}$ since $\sum_m F_m \rho F_m^\dagger = \sum_{k,l} \left( \sum_m u_{m,l}^* u_{m,k} \right) E_k \rho E_l^\dagger = \sum_k E_k \rho E_k^\dagger$. Substituting the above into Equation 2.10 gives $\langle \phi_i | E_k^\dagger E_l | \phi_j \rangle = \lambda_k \delta_{kl} \delta_{ij}$ where $\lambda_k$ are the eigenvalues of $C$. For all $k$ with $\lambda_k \neq 0$ we define $R_k = \frac{1}{\sqrt{\lambda_k}} \sum_i |\phi_i\rangle \langle \phi_i | F_k^\dagger$. We verify Equation 2.9 for every $|\phi_m\rangle \langle \phi_n|$ individually:

$$\begin{aligned} \mathcal{R} \circ \mathcal{E}[|\phi_m\rangle \langle \phi_n|] &= \sum_k R_k \left( \sum_l F_l |\phi_m\rangle \langle \phi_n| F_l^\dagger \right) R_k^\dagger \\ &= \sum_{k:\lambda_k \neq 0} \frac{1}{\lambda_k} \sum_i |\phi_i\rangle \langle \phi_i| F_k^\dagger \left( \sum_l F_l |\phi_m\rangle \langle \phi_n| F_l^\dagger \right) \sum_j F_k |\phi_j\rangle \langle \phi_j| \end{aligned} \tag{2.11}$$

Inserting Equation 2.10 gives:

$$\begin{aligned} \mathcal{R} \circ \mathcal{E}[|\phi_m\rangle \langle \phi_n|] &= \sum_{k:\lambda_k \neq 0} \frac{1}{\lambda_k} \sum_{i,j,l} (\lambda_k \delta_{lk})^2 \delta_{im} \delta_{nj} |\phi_i\rangle \langle \phi_j| \\ &= \left( \sum_k \lambda_k \right) |\phi_m\rangle \langle \phi_n| \\ &\propto |\phi_m\rangle \langle \phi_n| \end{aligned} \tag{2.12}$$

Note that if $\mathcal{E}$ is trace preserving we have $\sum_k \lambda_k = 1$ and we get an equality in the last line.
For more details see [15, 19, 20, 21]. ■

What is the intuitive meaning of Theorem 2.1? – The Kronecker delta in Equation 2.10 implies that if we start with two orthogonal code states $|\phi_i\rangle$ and $|\phi_j\rangle$ with $i \neq j$ then any pair of errors $E_k$ and $E_l$ will map them onto states which are orthogonal as well. For $k = l$ we see that, geometrically, a correctable error can only act as a rotation and a scaling on the code space. Correctable errors *cannot* be projections within the code space and neither can they change the normalized inner products between code states, i.e. they do not shear or distort the code space. More succinctly we may say that Theorem 2.1 states that the code space is mapped onto one of several copies of itself lying within the full Hilbert space of all physical qubits. The recovery operation $\mathcal{R}$ that we constructed maps from the copies back into the original code space.

The factor $c_{k,l}$ takes *degeneracy* into account. Degeneracy is a property of the quantum code. It means that different errors $E_k \neq E_l$ can affect the code space in the same way (but may have different effects on the rest of the Hilbert space). In terms of the intuitive picture given above this means that different errors can map onto the same copy of the code space. If the code is not degenerate then $C$ will be diagonal. In Section 2.1.5 we will examine examples of highly degenerate codes.

### 2.1.5 Stabilizer codes

We will now introduce a framework based on group theory which very naturally leads to the decomposition of the Hilbert space as in Equation 2.1. It is called the *stabilizer formalism* and was devised by Gottesman [22].

#### The Pauli group

Let $n$ be the number of physical qubits. The Hilbert space of each individual qubit is $\mathbb{C}^2$ so that the full Hilbert space of the quantum memory is $\mathcal{H}_M = \bigotimes_{i=1}^n \mathbb{C}^2 = \mathbb{C}^{2^n}$. We define the *Pauli group* $\mathcal{P}_n$ acting on $n$ qubits as

$$\mathcal{P}_n := \langle i, X_j, Z_j \mid j \in \{1,\dots,n\} \rangle = \left\{ \phi \bigotimes_{j=1}^{n} P_j \;\middle|\; \phi \in \{\pm 1, \pm i\}, P_j \in \{I, X, Y, Z\} \right\}, \tag{2.13}$$

where $X$, $Y$ and $Z$ are the Pauli matrices

$$X = \begin{bmatrix} 0 & 1 \\ 1 & 0 \end{bmatrix}, \quad Z = \begin{bmatrix} 1 & 0 \\ 0 & -1 \end{bmatrix}, \quad Y = iXZ = \begin{bmatrix} 0 & -i \\ i & 0 \end{bmatrix}. \tag{2.14}$$

The number of generators of the Pauli group $\mathcal{P}_n$ is $2n+1$ and its order is $4^{n+1}$ since any tensor factor can be either $I$, $X$, $Y$ or $Z$ and we have four values for the phase $\phi$. Each element of the Pauli group performs either a rotation by $\pi$ around the $x$-, $y$- or $z$-axis on each qubit (or leaves it invariant) and applies a global phase $\phi$. The *weight* $\mathrm{wt}(g)$ of a Pauli group element $g \in \mathcal{P}_n$ is the number of qubits on which it acts non-trivially. All three Pauli operators have eigenvalues $\pm 1$. Since the eigenvalues of the tensor product of two linear maps are given by the products of all pairs of eigenvalues of the individual maps, we have that each element in the Pauli group which is not proportional to the identity has a $2^{n-1}$-fold degenerate eigenvalue $+1$ and $2^{n-1}$-fold degenerate eigenvalue $-1$. Since every Pauli operator $X$, $Y$ and $Z$ squares to the identity $I$ we have that every element in the Pauli group squares to $\pm I$. If the phase of a Pauli group element is $\phi = \pm 1$ then it is self-inverse. Since two different Pauli operators either commute or anti-commute, we have that two elements in the Pauli group commute if and only if the number of different Pauli operators acting on the same qubits is even. Similarly, if the number of different Pauli operators acting on the same qubits is odd the two Pauli group elements anti-commute.

### The stabilizer formalism

The stabilizer formalism uses the properties of the Pauli group elements to define subspaces of the $n$-qubit Hilbert space. The central object of the stabilizer formalism is the stabilizer group:

**Definition 2.2.** A *stabilizer group* $S$ is a subgroup of the Pauli group $\mathcal{P}_n$ which is abelian and which does not contain $-I$. The elements of $S$ are called *stabilizers*.

Usually, a stabilizer group comes with a distinguished set of generators which, for reasons that will become appearant later on, are called *stabilizer checks*.

Since $-I$ and $\pm iI$ are not elements in $S$, all of the stabilizers have the eigenvalue $+1$. We call a subset of stabilizers *independent* if the group they generate becomes smaller if any of them are omitted. Any subset of stabilizers containing the identity is not independent. Hence we have that every stabilizer of an independent set has $2^{n-1}$-fold degenerate eigenvalue $+1$ and $2^{n-1}$-fold degenerate eigenvalue $-1$. Since any two stabilizers commute they must share a common $+1$-eigenspace.

**Definition 2.3.** A *stabilizer code* $\mathcal{C}$ is the common $+1$-eigenspace of all elements of a stabilizer group $S \subset \mathcal{P}_n$.

$$\mathcal{C} = \{|\psi\rangle \in \mathcal{H}_M \mid s|\psi\rangle = |\psi\rangle \quad \forall s \in S\} \tag{2.15}$$

### Properties of stabilizer codes

What is the dimension of the code space $\mathcal{C}$ stabilized by $S$? Let $\mathrm{gens}(S)$ be a set of $r$ *independent* generators of $S$. The projector onto the code space is

$$P_{\mathcal{C}} = \prod_{g \in \mathrm{gens}(S)} \frac{I+g}{2} = \frac{1}{2^r} \sum_{g \in S} g \tag{2.16}$$

The second equality holds since all elements of $S$ are self-inverse and because $S$ is abelian so that all elements of the group can be generated by taking products of all subsets of the generators. Since the generators are independent we obtain every element of $S$ exactly once. From Equation 2.16 we see that the dimension of the code space is $\dim\mathcal{C} = 2^{n-r}$: Since the trace of a Kronecker product is equal to the product of the individual traces $\mathrm{tr}(A\otimes B) = \mathrm{tr}(A)\mathrm{tr}(B)$ and since all Pauli matrices are traceless, we have that all elements of $S$, except for the identity $I$, are traceless. The identity $I$ has trace $2^n$ and we obtain $\dim\mathcal{C} = \mathrm{tr}(P_{\mathcal{C}}) = 2^{n-r}$. We say that $\mathcal{C}$ encodes $k := n-r$ *logical qubits*.

Given a stabilizer code in form of the stabilizer group $S$, the *logical operators* are those elements of the Pauli group which act non-trivially on the code space, but leave the code space as a whole invariant. This is the case for all operators which leave the stabilizer group as a whole invariant under conjugation. These operators form the normalizer of the stabilizer group in $\mathcal{P}_n$:

$$N(S) = \{g \in \mathcal{P}_n \mid gsg^\dagger \in S \quad \forall s \in S\} \tag{2.17}$$

It follows immediately from the definition that $S \subset N(S)$. Since all elements of $S$ have trivial action on the code space we call the elements of $N(S)\setminus S$ the *logical operators*. It can be shown (see [21]) that the group of non-trivial logical operators is isomorphic to the Pauli group on $k$ qubits $\mathcal{P}_k/\langle iI\rangle$ (disregarding global phases). Hence we can find operators $\overline{X}_1,\dots,\overline{X}_k$ and $\overline{Z}_1,\dots,\overline{Z}_k$ which generate the logical Pauli group.

The weight of the non-trivial logical operators is indicative of how well the code $\mathcal{C}$ can protect against random single qubit errors.

**Definition 2.4.** The *distance $d$* of a quantum stabilizer code $\mathcal{C}$ is the minimum weight of a logical operator

$$d = \min_{g\in N(S)\setminus S} \mathrm{wt}(g). \tag{2.18}$$

It is common to denote a code as an $[[n,k,d]]$-code, where $n$ is the number of physical qubits, $k$ is the number of logical qubits and $d$ is the distance.

### Error correction procedure

To prepare an encoded zero state $|\overline{0}\rangle^{\otimes k}$ for a stabilizer code we first prepare the state $|0\rangle^{\otimes n}$ on all physical qubits. Then we perform a projective measurement of all stabilizer checks $s_1,\dots,s_{n-k}$

and the logical operators $\overline{Z}_1, \ldots, \overline{Z}_k$. By the laws of quantum mechanics, the resulting state will be some eigenstate of the measured operators. Some of the eigenvalues of the stabilizer checks may be $-1$. To fix this we can perform single qubit rotations which anti-commute with such stabilizers to obtain a code state. Note that these corrections may also anti-commute with the $\overline{Z}_i$, so that the measurement values have to be updated. Similarly, the eigenvalues of some the logical operators $\overline{Z}_i$ may be $-1$ which means that the $i$th logical qubit is in state $|1\rangle$. This can be corrected for by applying the corresponding $\overline{X}_i$. This can be done as all $\overline{X}_i$ commute with all stabilizers.

We have seen in Section 2.1.3 that correcting Pauli errors suffices to correct for arbitrary linear combinations of those Pauli errors. A Pauli error which is not in $N(S)$ will, by definition, anti-commute with some of the stabilizer checks, changing their value on the encoded state from $+1$ to $-1$. The bit string indicating which values of stabilizer checks have been flipped is called the *syndrome*. From the syndrome we infer a recovery operation which hopefully reverses the effect of the error on the code state. Since generally multiple different errors give rise to the same syndrome we have to make a choice. This choice can be based on the error model. Note that the Pauli error $E$ that actually happened may be different from the inferred recovery $R$. But as long as $R \cdot E \in S$, the action on the encoded state will be trivial. If on the other hand $R \cdot E \in N(S) \setminus S$ then the encoded information has been corrupted. For independent single qubit errrors there are two canonical decoding strategies:

The first is *minimum-weight decoding* which determines the Pauli group element with the lowest weight which gives rise to the same syndrome as the actual Pauli error. Since Pauli group elements square to the identity (up to a global phase) the recovery simply consists of applying the inferred Pauli group element $R$. This is an example of degeneracy that we mentioned earlier.

The minimum-weight decoder is not optimal as it does not take multiplicities into account. There may be situations in which there are only a small number of potential minimum-weight errors but a vast number of errors with slightly larger weight. Hence, the optimal strategy compares all elements of the Pauli group which are consistent with the syndrome. This is called the *maximum-likelihood decoder*.

If we imagine the environment having to exert energy proportional to the error, then we can think of minimum-weight decoding as energy minimization, whereas maximum-likelihood decoding corresponds to free-energy minimization. Both decoding problems are NP-hard in general [23]. However, we will see that for certain important classes of stabilizer codes efficient decoders do exist.

### Relationship to linear codes

Stabilizer codes are closely related to a class of error-correcting codes known from classical coding theory via the CSS[1] construction.

**Definition 2.5.** A binary linear classical $(n,k)$-code $W$ is a subspace of dimension $k$ of $\mathbb{Z}_2^n$, were $\mathbb{Z}_2$ denotes the field with two elements.

There are two ways of representing a classical linear code: either as the image of a matrix $G$ called the *generator matrix* or as the kernel of matrix $H$ called the *parity check matrix*.

To construct a stabilizer quantum code let us assume we have two classical linear codes $W_1$ and $W_2$ with the following property: every element of $W_1$ is orthogonal to every element of $W_2$, i.e. for all $a \in W_1$ and $b \in W_2$ the inner product vanishes $\langle a,b\rangle = \sum_i a_i b_i = 0$. Note that arithmetic is modulo 2. In this case we write $W_1 \perp W_2$. Equivalently, if the codes are represented by their generating matrices $G_1$ and $G_2$, we have the condition

$$G_1^T \cdot G_2 = 0. \tag{2.19}$$

Given two such orthogonal codes we define a stabilizer group

$$S = \langle X^a := X^{a_1} \otimes \cdots \otimes X^{a_n},\ Z^b := Z^{b_1} \otimes \cdots \otimes Z^{b_n} \mid a \in W_1, b \in W_2\rangle. \tag{2.20}$$

According to Definition 2.2 the group $S$ needs to be abelian. This is true if and only if all elements $X^a$ commute with elements $Z^b$. As the single qubit Pauli-$X$ and Pauli-$Z$ anti-commute we have that $X^a$ and $Z^b$ commute if and only if their support is even which in turn is equivalent to $\langle a,b\rangle = 0$. Hence Equation 2.20 is indeed a stabilizer group.

The parameters of a CSS code are completely determined by the properties of $W_1$ and $W_2$: as the number of encoded qubits is $k = n - r$ with $r$ being the number of independent generators of $S$ we have

$$k = n - \dim(W_1) - \dim(W_2) = \dim(W_1^\perp / W_2) = \dim(W_2^\perp / W_1). \tag{2.21}$$

[1] Standing for Calderbank, Shor and Steane.

where $W_j^\perp := \{a \in \mathbb{Z}_2^n \mid \langle a,b \rangle = 0\ \forall b \in W_j\}$ is the orthogonal complement of $W_j$. The distance is the minimum weight Pauli operator which commutes with all elements in $S$ but is not an element of $S$. Hence we obtain

$$d = \min\{\mathrm{wt}(a) \mid a \in (W_1^\perp \setminus W_2) \cup (W_2^\perp \setminus W_1)\} \tag{2.22}$$

where here $\mathrm{wt}(a)$ is the number of entries $a_i$ which are equal to 1, also known as the *Hamming weight*.

CSS codes can be visualized in the form of *Tanner graphs*. The physical qubits are circular nodes in the Tanner graph. The stabilizer checks are square boxes. There is an edge between a box and a circle if the stabilizer acts on the qubit. The graph is tripartite with the partitions being the set of qubits, the set of $Z$-checks, and the set of $X$-checks. The graph is laid out such that all $Z$-checks, qubits and $X$-checks form horizontal lines (in this order). Due to Equation 2.19 we have:

**Lemma 2.6.** *Consider any pair of boxes in a Tanner graph, representing an $X$-check and a $Z$-check. In a valid Tanner graph the number of circles connected to both boxes must be even.*

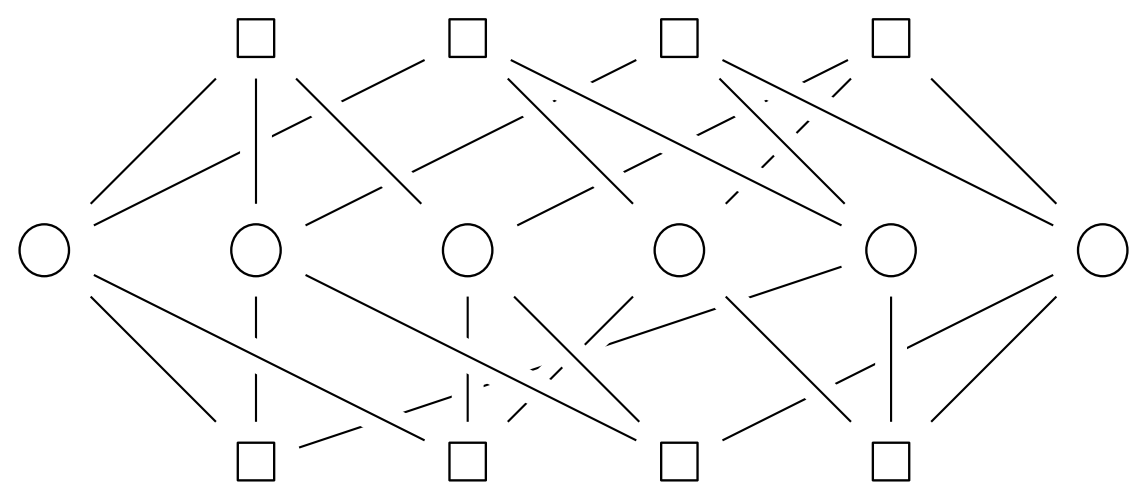

Figure 2.1: Tanner graph of a CSS code.

## 2.2 Tesselations and cellulations of manifolds

In this section we are going to discuss how to subdivide a manifold by polytopes. At the end of this section we will see how these two topics are related.

### 2.2.1 Tesselations of surfaces

Before we give the general definition of a tesselation for manifolds of arbitrary dimension (Section 2.2.3) let us first consider the two-dimensional case. Let $S$ be a two-dimensional manifold (a surface) without boundary which is *Riemannian*, which means that it is equipped with a metric and thus a distance $d_S$. A *tesselation* of $S$ consists of a covering of $S$ by a set of polygons $\{P_1, P_2, \dots\}$. Formally, each polygon is a compact, two-dimensional, Riemannian manifold which is simply connected and has a boundary consisting of a finite number of geodesics. Each polygon comes with a distance-preserving function $\phi_j : P_j \to S$, which means that for any two points $a, b \in P_j$ we have $d_S(\phi_j(a), \phi_j(b)) = d_{P_j}(a, b)$ where $d_S$ and $d_{P_j}$ are the distances in $S$ and $P_j$. Distance-preserving functions are called *isometries*. Note that isometries are necessarily injective: If two points are being mapped onto the same point then they have zero distance in the image. By definition they must have zero distance in the preimage as well and hence they coincide. Furthermore, since distances are preserved, angles are preserved as well. The images of the edges and vertices of the polygons are also called vertices and edges, respectively. The images of two distinct polygons must be either disjoint, have a single vertex in common or they must share an entire edge. Familiar examples of two-dimensional tessellations are the square tessellation and the hexagonal tessellation. Both are tessellations of the Euclidean plane $\mathbb{E}^2$.

### 2.2.2 Higher-dimensional polytopes

To generalize two-dimensional tesselations to arbitray dimensions we have to first introduce higher dimensional polytopes: A *D-dimensional (euclidean) polytope* is a compact subset $P \subset \mathbb{R}^D$ bounded by a finite number $k$ of $D-1$ - dimensional hyperplanes. Since $P$ is compact the number of hyperplanes must be strictly larger than the dimension $k > D$. Each hyperplane of $P$ is defined by an equation

$$a_i \cdot x = b_i \tag{2.23}$$

for some non-zero $a_i \in \mathbb{R}^D$ and $b_i \in \mathbb{R}$ with $i = 1, \dots, k$. If $P$ is *convex* then all of its points lie in the intersection of the half-spaces given by its bounding hyperplanes. Each half space is determined by an inequality $a_i \cdot x \leq b_i$ and the set of points belonging to the polytope $x \in P$ is simply the set of all $x \in \mathbb{R}^D$ which satisfy all those linear inequalities simultaneously. As all hyperplanes are assumed to be distinct, the inequalities must be independent. To allow for tessellations of non-Euclidean spaces we also consider all conformal maps of Euclidean polytopes.

Note that a hyperplane is given by all points which satisfy strict equality for one particular inequality. Lets call the set of these points a *facet* $\Pi$ of the polytope $P$. Since one of the inequalities is replaced by an equality, the facet is a $D-1$-dimensional convex polytope defined by all the remaining inequalities, which in general will have become dependent. The facets of $\Pi$ are themselves $D-2$-dimensional polytopes. We obtain a descending chain of facets labeled by their dimension $\Pi_{D-1},\ldots,\Pi_1,\Pi_0$ where the last two facets are the only convex 1-dimensional and 0-dimensional polytopes: The edge and the vertex. All of these sub-polytopes $\Pi_i$ are determined by a subset of $D-i$ of the inequalities in Equation 2.23 for which strict equality holds. We will call the set of all $i$-dimensional sub-polytopes *$i$-cells*. Two $i$-cells overlap if and only if the two sets of linear equations that define them together form a consistent set of equations. If $j$ is the rank of this set of linear equations, then the $i$-cells overlap on a $D-j$-cell.

An example of a higher-dimensional polytope is the 4-dimensional hypercube, also known as *tesseract*. The 8 hyperplanes bounding the tesseract are defined by the equations $x_i = \pm 1$ with $i \in \{1,\ldots,4\}$. Each hyperplane is 3-dimensional and the 3-cells at the boundary of the tesseract are cubes. There is one cube for each hyperplane, so there are 8 cubes in total. Lets consider the hyperplane defined by $x_1 = 1$. The cube contained in the hyperplane has squares at its boundary which are defined by additionally demanding that $x_i = \pm 1$ for $i \in \{2,3,4\}$, so that there are 6 squares bounding the cube. Since any two cubes overlap on a square there are $6 \cdot 8/2 = 24$ squares in a tesseract. The edges are located at the intersection of 3 hyperplanes. Relaxing any one of the equalities gives a square, so that there are 3 squares incident to an edge. Since obviously 4 edges form the boundary of a square and we know that there are 24 squares in the tesseract we get that the number of edges of a tesseract is $24 \cdot 4/3 = 32$. A similar line of reasoning shows that there are 16 vertices in a tesseract.

### 2.2.3 Higher-dimensional tesselations and cellulations

We can now generalize the definition of two-dimensional tesselations to arbitrary dimension.

**Definition 2.7.** A *tesselation of a $D$-dimensional Riemannian manifold $M$* formally consists of a set of $D$-dimensional polytopes $\{P_1,P_2,\ldots\}$ which are embedded via isometries $\phi_j : P_j \to M$. The images of $i$-cells under the mapping $\phi_j$ are also called $i$-cells. The images of two distinct polygons are either disjoint or they overlap on exactly one of their facets. The union of all images must cover $M$, i.e. $\bigcup_j \text{im}\ \phi_j = M$.

The essential property of the embedding functions $\phi_j$ is that they are injective and continuous, so that they leave the overlap between the cells invariant and keep cells from intersecting. Such mappings are called *homeomorphisms*. An injective homeomorphism can *stretch and deform* the polytopes while an isometry is *rigid*. We generalize Definition 2.7 by substituting the word "isometries" with "injective homeomorphisms" which gives us the definition of a *cellulation* of the manifold $M$. The cellulation does not respect distances but it does respect the topology of the manifold. In this case we do not need the manifold $M$ to have a metric. A cellulated manifold is also called a *cell complex*.

Note that the boundary of a $D+1$-dimensional, convex polygon itself satisfies the above definition: The set of polygons being its facets and the manifold being cellulated is the $D$-dimensional sphere $\mathbb{S}^D$. For example, the five platonic solids are two-dimensional cellulations of the sphere $\mathbb{S}^2$. Similarly, the tesseract is a cellulation of the 3-dimensional sphere $\mathbb{S}^3$ where the set of polytopes consists of 8 cubes.[2]

### 2.2.4 The Hasse diagram

The information of how the cells of a cellulation are contained in each other can be visualized by a *Hasse diagram*. Each node in the diagram corresponds to a cell of the cellulation. Two nodes are connected if and only if (a) the cell corresponding to one node is contained in the cell corresponding to the other node and (b) the difference of the dimensionalities of the cells is 1. The Hasse diagram is therefore a multipartite graph in which each partition or *level* corresponds to the set of all $i$-cells. The number of levels is therefore $D+1$. The levels are ordered vertically with the highest dimensional cells at the top (see Figure 2.2). A cell from a lower level is contained in a cell at a higher level if there exists a path connecting the two and which goes through every level at most once. For every cellulation there exists an associated cellulation called the *dual cellulation* defined as follows:

[2] In the literature some authors take the opposite view: In [2] the author interprets what we would call a $D$-dimensional cellulation to be a degenerate $D+1$-dimensional polytope. The cellulation forms a $D$-dimensional hyperplane in $\mathbb{R}^{D+1}$ and the degenerate polytope is a half-space bounded by this hyperplane.

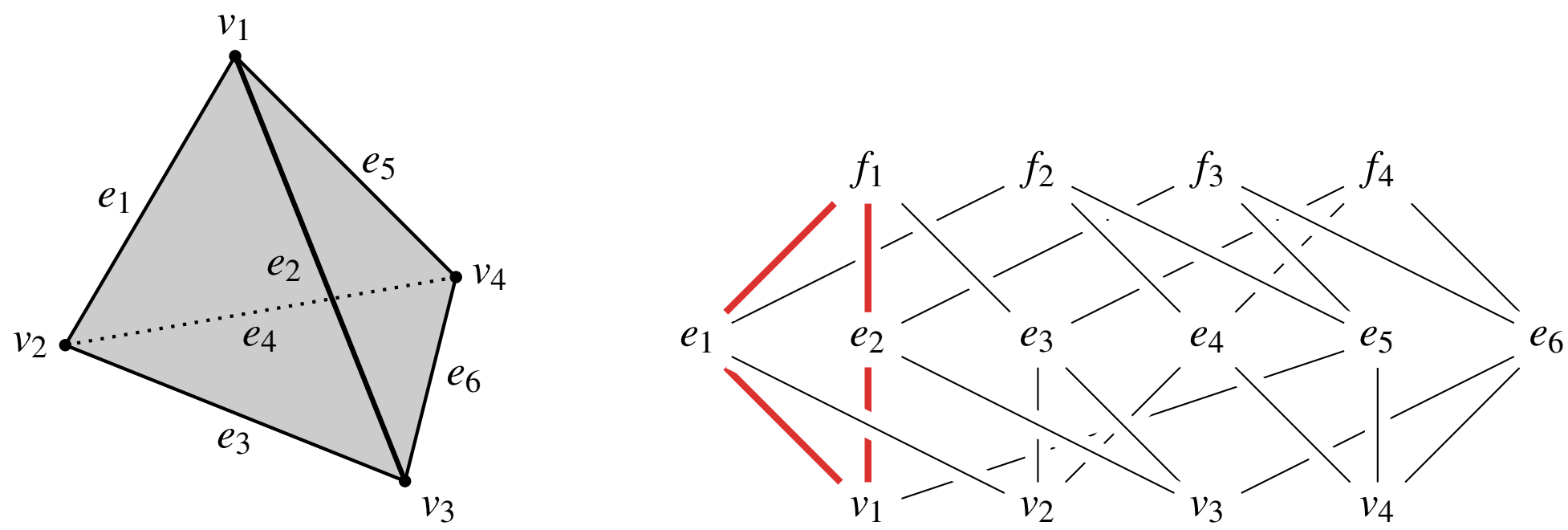

Figure 2.2: A tetrahedron (left) and its corresponding Hasse diagram (right). The faces of the tetrahedron are labeled: $f_1$ front left, $f_2$ upper back, $f_3$ front right, $f_4$ bottom. The face $f_1$ and the vertex $v_1$ overlap and consequently there exists a path between the two in the diagram. They are both incident to two edges $e_1$ and $e_2$ (highlighted). The Hasse diagram is symmetric under horizontal reflection (up to a reordering of elements in each level), since the tetrahedron is self-dual.

**Definition 2.8.** The *dual cellulation* $X^*$ of a given $D$-dimensional (primal) cellulation $X$ has the same Hasse diagram except that every level $i$ in the dual cellulation is the level $D-i$ in the primal cellulation.[3]

We will now show how a cellulation of some manifold can be used to define a quantum code. An essential ingredient is the following lemma.

**Lemma 2.9.** *Let $i \in \{1,\ldots,D-1\}$ and let $c_{i+1}$ and $c_{i-1}$ be two nodes from level $i+1$ and $i-1$ of the Hasse diagram of a cellulation. The number of cells in the ith level which are connected to both $c_{i+1}$ and $c_{i-1}$ is either zero or two.*

*Proof.* For $i \in \{1,\ldots,D-1\}$ let us pick any $i+1$-cell $c_{i+1}$ and $i-1$-cell $c_{i-1}$ in the cellulation. If $c_{i+1}$ and $c_{i-1}$ are not disjoint then by construction $c_{i-1} \subset c_{i+1}$. In this case there exists a map $\phi_j$ so that the preimages of $c_{i+1}$ and $c_{i-1}$ are an $i+1$-cell and an $i-1$-cell of some convex polytope $P_j$. Furthermore, since $c_{i-1} \subset c_{i+1}$ we have that $\phi_j^{-1}(c_{i-1})$ is a facet of a facet of $\phi_j^{-1}(c_{i+1})$. All points in $\phi_j^{-1}(c_{i-1})$ satisfy $D-(i-1)$ of the inequalities of Equation 2.23 with equality and a subset of $D-(i+1)$ of these equalities define $\phi_j^{-1}(c_{i+1})$. Hence there must be exactly two facets of $\phi_j^{-1}(c_{i+1})$ which are defined by adding on of the two additional linear equations which determine $\phi_j^{-1}(c_{i-1})$. ■

[3]For a geometric construction of the dual *tesselation* see [2].

Together with Lemma 2.6 we obtain the following theorem.

**Theorem 2.10.** *Any subgraph of a Hasse diagram of a cellulation, consisting of three consecutive levels $(i-1,i,i+1)$ for $i \in \{1\ldots,D-1\}$, defines the Tanner graph of a CSS code.*

Let us take the Hasse diagram of the tetrahedron shown in Figure 2.2 as an example. To interpret the Hasse diagram as a Tanner graph we identify the nodes labeled $e_1,\ldots,e_6$ with qubits. The top row, representing the faces, are the $Z$-checks which operate as Pauli-$Z$ on all qubits connected to them. A single face $f_i$ (a triangle) hence corresponds to an operator which acts as Pauli-$Z$ on all qubits $e_j$ (edges) in its boundary ($e_j \subset f_i$). Similarly, $X$-checks are associated with the vertices $v_i$ of the tetrahedron. They operate as Pauli-$X$ on all qubits $e_j$ (edges) which are connected to it ($v_i \subset e_j$). Only $r = 6$ of the 8 stabilizer checks are independent since multiplying either all $Z$-checks or all $X$-checks gives the identity. Hence the number of independent constraints imposed by the stabilizer checks is equal to the number of degrees of freedom in the system (the 6 physical qubits). Hence this example is a trivial code, encoding $k = n - k = 0$ qubits. It turns out that this fact is due to the *topology* of the tetrahedron. More generally, the properties of codes defined via Theorem 2.10 depend on the topology of the cellulated manifold. In Section 2.3 we will introduce tools from algebraic topology which allow us to analyze the properties of such codes in a systematic way by relating the topology of the cellulation which is encoded in the Hasse diagram to objects from linear algebra.

## 2.3 Homological quantum codes

In this section we describe quantum codes derived from cellulations of manifolds in the language of homology. The objects of homology theory will very naturally describe the properties of these codes. Hence, we are going to call them *homological codes*.

### 2.3.1 $\mathbb{Z}_2$-Homology

In $\mathbb{Z}_2$-homology, statements about the topology of a manifold are turned into statements of linear algebra over the field $\mathbb{Z}_2 = \{0,1\}$ in which all operations are carried out modulo 2.

### Chains

For a given cell complex $X$ the subsets of the set of all $i$-cells in $X$ span a $\mathbb{Z}_2$-vector space where the addition of two subsets is given by their symmetric difference (the set of elements which are contained in one set or the other but not in both). The standard basis of this space is given by all sets containing a single $i$-cell. We will identify these sets with the standard basis of $\mathbb{Z}_2^{m_i}$ where $m_i$ is the number of $i$-cells contained in $X$. We denote these vector spaces as $C_i(X)$ or simply $C_i$ and call their elements *i-chains*.

### Boundary operators

For $i \in \{0\ldots,D\}$ we define the boundary operator $\partial_i$

$$\partial_i : C_i \to C_{i-1}, \tag{2.24}$$

by its action on the basis vectors: let $c_i \in C_i$ be a single $i$-cell, then $\partial_i(c_i) \in C_{i-1}$ is the sum of all $(i-1)$-cells incident to $c_i$. Alternatively we can define the boundary operator $\partial_i$ as the incidence matrix between level $i$ and level $i-1$ of the Hasse diagram. The columns are indexed by the $i$-cells and the rows by $i-1$-cells. An entry is 1 if the cells indexing the row and column are connected by an edge and 0 otherwise.

An important property of the boundary operator is that applying it twice gives the zero-map

$$\partial_i \circ \partial_{i+1} = 0. \tag{2.25}$$

This is an immediate consequence of Lemma 2.9. Remember that every $i$-cell corresponds to a $i$-dimensional, convex polytope. We have seen in Section 2.2 that the facets of such a polytope form a tessellation of $\mathbb{S}^{i-1}$ at its boundary. The statement made by Equation 2.25 in terms of topology is that the $i$-dimensional sphere at the boundary of an $i+1$-cell has no boundary.

In addition to the boundary operator we define the dual map called the *coboundary operator*

$$\delta_i : C_i \to C_{i+1}, \tag{2.26}$$

which assigns to each $i$-cell all $i+1$-cells which are incident to it. As for the boundary operator, we will represent $\delta_i$ as a matrix in the standard bases of $C_i$ and $C_{i+1}$. Note that as a matrix $\delta_i$ is the incidence matrix between level $i$ and level $i+1$ of the Hasse diagram. Hence, the coboundary operator is the transpose of the boundary operator of one dimension higher

$$\delta_i = \partial_{i+1}^T. \tag{2.27}$$

### Boundaries and cycles

The boundary operator and coboundary operator are used to define subspaces of $C_1$. From the boundary operator one obtains the *cycle spaces* $Z_i = \ker \partial_i$ and the *boundary spaces* $B_i = \mathrm{im}\ \partial_{i+1}$. Due to Equation 2.25, we always have that the boundary spaces are subspaces of the cycle spaces $B_i \subseteq Z_i$. Intuitively, the cycle space $Z_i$ represents all $i$-chains which have no boundary. For example, if $i = 1$ then elements of $Z_i$ are collections of closed loops and if $i = 2$ then elements of $Z_i$ are collections of closed surfaces.

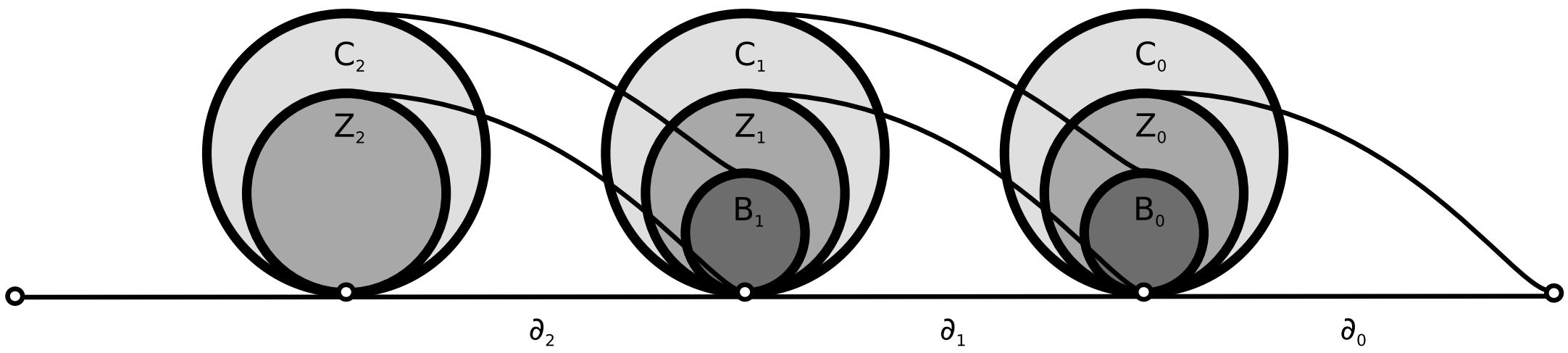

Figure 2.3: Illustration of the homology for a two-dimensional cellulation. The boundary operator $\partial_i$ maps the cycle space $Z_i$ onto $\{0\}$. The boundary space $B_i$ is the image of $C_{i+1}$ under $\partial_{i+1}$. The elements of $Z_i \setminus B_i$ span the homology group $H_i$.

### Essential cycles

Certainly, all boundaries of $i$-chains have no boundary (which is the content of Equation 2.25) but there might be other cycles which are not the boundary of a higher dimensional chain. Such $i$-cycles have to close up on themselves without enclosing any $i+1$-dimensional volume. This can only happen if the manifold that is being cellulated has non-trivial topology. Cycles which are not boundaries are called *essential cycles*. The essential cycles surround $i$-dimensional "holes" of the cellulated manifold. We consider essential cycles as being equivalent if they differ by a boundary. The spaces of equivalent essential $i$-cycles are the *homology groups* $H_i = Z_i/B_i$. The number of holes in the manifold is the number of inequivalent essential $i$-cycles which span the the $i$th homology group $H_i$. The dimension of the $i$th homology group is also called the *ith Betti number* of the manifold and it is invariant under the choice of cellulation. A proof of this fact can be found in [24].

It is common to formally define $\partial_{D+1}$ to be the zero map, so that

$$H_D = \ker \partial_D / \mathrm{im}\ \partial_{D+1} = \ker \partial_D / \{0\} = Z_D. \tag{2.28}$$

It is only possible for a $D$-chain to be boundaryless if it covers the whole cell complex. If we allow $X$ to contain cellulations of disconnected closed manifolds then $H_D$ counts the number of disconnected components.

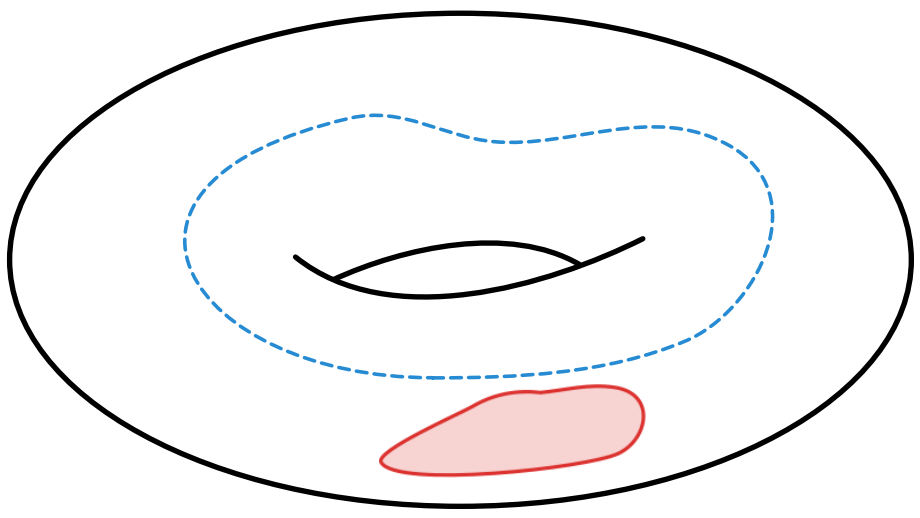

Figure 2.4: Homology of a torus. The solid red line is a 1-cycle which the boundary of the opaque region. The dashed blue line is an essential 1-cycle as it is not the boundary of any region on the torus. The sum of the solid and the red line is also an essential 1-cycle.

### Cohomology

The coboundary operator defines the *cocycle space* $Z^i = \ker \delta_i$ and the *coboundary space* $B^i = \mathrm{im}\ \delta_{i-1}$ with $B^i \subseteq Z^i$. From Equation 2.27 it follows that

$$\langle \delta_i a, b \rangle_{C_{i+1}} = \langle a, \partial_{i+1} b \rangle_{C_i}, \tag{2.29}$$

where $\langle \cdot, \cdot \rangle_{C_i}$ denotes the standard inner product in $C_i$. As arithmetic is done modulo 2 the inner product between two chains is 1 if their supports overlap on an odd number of cells and 0 if their supports overlap on an even number of cells. Together with Equation 2.25 it follows that the cocycle space is the orthogonal complement of the boundary space and the coboundary space is the orthogonal complement of the cycle space, i.e.

$$B^i = Z_i^\perp \text{ and } Z^i = B_i^\perp. \tag{2.30}$$

In other words, the coboundaries are exactly those chains which have even overlap with the cycles and the cocycles are exactly those chains which have even overlap with the boundaries.

We define the *cohomology groups* as $H^i = Z^i/B^i$. They contain the equivalence classes of essential $i$-cocycles. The 0th cohomology group is defined as $H^0 = \ker \delta_0$. We may think of $\delta_0$ as a gradient[4] which can be seen as follows: Let $c \in C_0$ be a 0-chain, then $\delta_0$ takes any two vertices $v$ and $w$ connected by an edge and assigns to that edge the sum $c_v + c_w$ which is non-zero if and only if $c_v \neq c_w$. The only assignment of values to the vertices which is in the kernel of the gradient is constant (on a connected component). Hence we have that $H^0$ counts the connected components of $X$, similar to $H_D$.

### 2.3.2 Duality

In Section 2.2 we have seen that the $i$-cells of a cell complex $X$ correspond to the $D-i$-cells of the dual cell complex $X^*$. By linear extension, this gives rise to an isomorphism

$$* : C_i(X) \to C_{D-i}(X^*) \tag{2.31}$$

of the $i$-chains of $X$ to the $D-i$-chains of $X^*$. We first note that going to the dual chains leaves the inner product (even or oddness of their overlap) invariant

$$\langle a, b \rangle_{C_i} = \langle *a, *b \rangle_{C^*_{D-i}}. \tag{2.32}$$

Directly from the definitions it is also clear that applying the coboundary operator to a chain of a cell complex is equivalent to going to the dual complex and applying the boundary operator of the complementary dimension

$$\delta_i = *^{-1} \circ \partial_{D-i} \circ * \tag{2.33}$$

[4]This is exactly true in a “continouous version” of homology theory, called de Rham cohomology.

Or in diagrammatic form:

$$\begin{array}{ccc} C_i & \xrightarrow{\delta_i} & C_{i+1} \\ {\scriptstyle *}\downarrow & & \downarrow{\scriptstyle *} \\ C^*_{D-i} & \xrightarrow[\partial_{D-i}]{} & C^*_{D-i-1} \end{array} \tag{2.34}$$

Equation 2.33 can also be understood in terms of the Hasse diagram: Going to the dual cellulation means we reflect the Hasse diagram along the horizontal axis which exchanges level $i$ with level $D-i$. Since $\delta_i$ is the incidence matrix of the Hasse diagram between the level $i$ and level $i+1$ it must be the same as the incidence matrix between level $D-i$ and level $D-i-1$ which is precisely $\partial_{D-i}$. Equation 2.33 allows us to interpret the coboundaries $B^i$ and cocycles $Z^i$ as boundaries and cycles of the dual structure (see Figure 2.5).

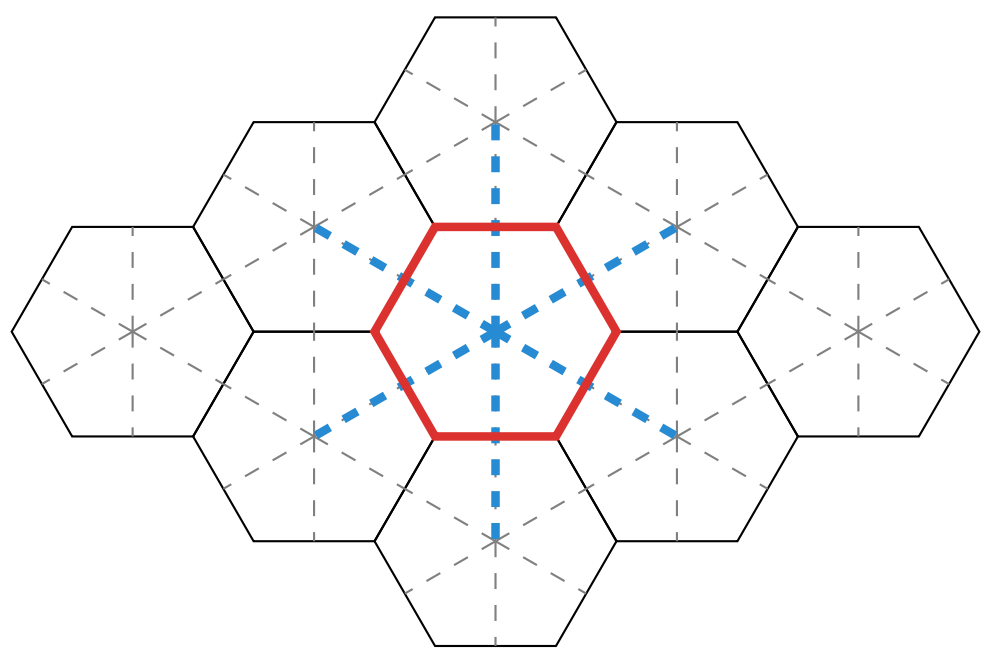

Figure 2.5: Part of a 2-dimensional tessellation by hexagons. The dual tessellation is drawn with dashed lines. The boundary of a face (red) corresponds to the coboundary of a vertex in the dual structure (blue).

### 2.3.3 Topological invariants

The homology groups $H_i$ do not depend on the cellulation of a manifold $M$. Cellulations, by definition, do not need to respect the metric of $M$. Hence the homology groups $H_i$ only contain information about the topology of the cellulated space. There are further topologial invariants that are related to the homology groups which we want to mention here.

**Definition 2.11.** The *Euler characteristic* $\chi(M)$ of a manifold $M$ is given by the alternating sum of the number of cells of a cellulation of $M$

$$\chi(M) = \sum_{i=0}^{D} (-1)^i \dim C_i. \tag{2.35}$$

From Definition 2.11 it is not obvious that the Euler characteristic is a topological invariant as the same manifold can have many different cellulations. It can be shown that the value of $\chi(M)$ does not depend on the cellulation and further that

$$\chi(M) = \sum_{i=0}^{D} (-1)^i \dim H_i. \tag{2.36}$$

*Proof.* By definition of the homology group we have

$$\dim\ker \partial_i = \dim H_i + \dim\mathrm{im}\ \partial_{i+1}. \tag{2.37}$$

The rank-nullity theorem states that for any linear map $\phi : V \to W$ we have

$$\dim V = \dim\ker \phi + \dim\mathrm{im}\ \phi. \tag{2.38}$$

Applying the rank-nullity theorem to $\partial_i$ it follows that

$$\dim C_i = \dim\ker \partial_i + \dim\mathrm{im}\ \partial_i = \dim H_i + \dim\mathrm{im}\ \partial_{i+1} + \dim\mathrm{im}\ \partial_i. \tag{2.39}$$

Now we have

$$\begin{aligned} \sum_{i=0}^{D} (-1)^i \dim C_i &= \sum_{i=0}^{D} (-1)^i \dim H_i + (-1)^D \dim\mathrm{im}\ \partial_{D+1} + (-1)^0\ \dim\mathrm{im}\ \partial_0 \\ &= \sum_{i=0}^{D} (-1)^i \dim H_i, \end{aligned} \tag{2.40}$$

where in the last equation we used the definitions $\partial_0 : C_0 \to \{0\}$ and $\partial_{D+1} : \{0\} \to C_{D+1}$ (cf. Equation 2.28). ■

Another topological invariant that we are going to use is the *genus*.

**Definition 2.12.** The *genus* $g$ of a closed, orientable surface $M$ is the maximum number of cuts along closed curves which do not cross, without rendering $M$ disconnected.

The genus is the number of "handles" of the surface. Every handle contributes two independent essential cycles (see Figure 2.6), so that $\dim H_1 = 2g$ and hence $\chi(M) = 2 - 2g$ (since $\dim H_0 = \dim H_2 = 1$). A classical result of topology is that all orientable surfaces are up to homeomorphism determined by their genus.

**Proposition 2.13.** *For $D = 2$ all closed manifolds (surfaces) which consist of a single connected component with the same* genus *are related by a homeomorphism.*

Proposition 2.13 implies that any orientable surface can be embedded in 3D Euclidean space $\mathbb{E}^3$. Such an embedding may need to stretch the surface so that the embedding is not necessarily an isometry. Note that Proposition 2.13 generally does not hold for non-orientable surfaces.

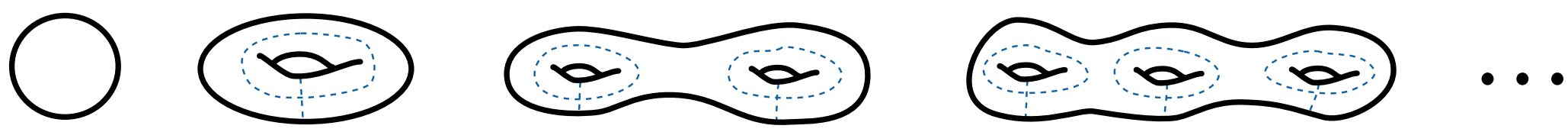

Figure 2.6: Closed orientable surfaces with genus $g \in \{0, 1, 2, 3, \dots\}$. Every "handle" contributes two independent essential 1-cycles (dashed lines) along which we can cut the surface without rendering it disconnected.

### 2.3.4 Stabilizer codes derived from $\mathbb{Z}_2$-homology

Theorem 2.10 has shown that any cellulation $X$ of a $D$-dimensional manifold gives a quantum CSS code. Using $\mathbb{Z}_2$-homology we can relate the properties of the code, namely the number of physical qubits $n$, the number of encoded qubits $k$ and the distance $d$ to properties of $X$.

We have seen in Section 2.1.5 that the $2^k$-dimensional codespace $C \subseteq \mathcal{H}$ is defined as the $+1$ eigenspace of a subgroup $S$ of the Pauli group which is abelian and which does not contain $-I$. To turn a cell complex $X$ into a stabilizer code we pick a dimension $i \in \{1, \dots, D-1\}$ and identify all $i$-cells with qubits. The boundaries of the $i+1$-cells are used to define $Z$-type check operators. For every $i+1$-cell we add a generator to $S$ which acts as $Z$ on all $i$-cells which belong to the boundary of the $i+1$-cell. Equivalently, the set of coboundaries of $i-1$-cells gives a generating set of $X$-check operators. From now on, when we refer to *the* stabilizer generators we mean this canonical set associated to the $i+1$-cells and $i-1$-cells. We can write more compactly:

$$S_Z = \{Z^c \mid c \in B_i\} \quad \text{and} \quad S_X = \{X^c \mid c \in B^i\} \tag{2.41}$$

where $Z^c = \bigotimes_{j=1}^n Z^{c_i}$ and $X^c = \bigotimes_{j=1}^n X^{c_i}$. However, most of the time we will not differentiate between the chains $c$ and the operators $Z^c$ or $X^c$.

The weight of a $Z$-check ($X$-check) corresponding to the boundary of some $i+1$-cell (coboundary of a $i-1$-cell) $c$ is thus given by the number of $i$-cells incident to $c$.

The number of *physical qubits* $n$ is simply the number of $i$-cells which span the space of $i$-chains $n = \dim C_i$. The number of *encoded qubits* $k$ can be calculated by taking the number of physical qubits and subtracting the number of restrictions that the stabilizers impose:

$$\begin{aligned} k &= \dim C_i - \dim B^i - \dim B_i \\ &= \dim C_i - \dim Z_i^{\perp} - \dim B_i \\ &= \dim C_i - (\dim C_i - \dim Z_i) - \dim B_i \\ &= \dim H_i \end{aligned} \tag{2.42}$$

There is another way of seeing this which helps the intuitive understanding of homological codes: Consider the subgroup $N(S)$ of the Pauli group which consists of those operators which commute with all elements in $S$. The action of $N(S)$ on $\mathcal{H}$ leaves the space of logical qubits as a whole invariant. The $Z$- and $X$-type elements of $N(S)$ stand in one-to-one correspondence with the cycles and cocycles by Equation 2.30. All stabilizers (the boundaries and coboundaries) have a trivial action on the logical qubits. The operators which correspond to essential cycles have a non-trivial action on the encoded qubits. The essential cycles and essential cocycles come in $\dim H_i$ pairs with odd overlap, whereas the overlap of elements from different pairs is even. We identify these pairs of essential cycles as the $X$- and $Z$-operators, each acting on one of the logical qubits.

Note that in general the generating set of $Z$- and $X$-stabilizers is not independent. Applying the rank-nullity theorem (see Equation 2.38) to $\partial_{i+1}$ and solving for $B_i = \text{im } \partial_{i+1}$ we obtain

$$\dim B_i = \dim C_{i+1} - \dim Z_{i+1} = \dim C_{i+1} - \dim H_{i+1} + \dim B_{i+1}. \tag{2.43}$$

Equation 2.43 is a recursive expression in the dimensions of the boundary spaces. It can be put into an explicit form which only depends on the number of cells of dimension $> i$ and the dimensions of the homology groups of dimension $> i$:

$$\dim B_i = \sum_{j=1}^{D-i} (-1)^{j+1} \left(\dim C_{i+j} - \dim H_{i+j}\right). \tag{2.44}$$

For the $X$-checks we obtain equivalently

$$\dim B^i = \sum_{j=1}^{i} (-1)^{j+1} \left(\dim C_{i-j} - \dim H^{i-j}\right). \tag{2.45}$$

For $D = 2$ Equation 2.44 and Equation 2.45 each give a single linear dependency due to $\dim H_0 = \dim H^2 = 1$ (assuming the code is defined on a manifold with a single connected component). For $D > 2$ there are non-trivial linear dependencies. We will discuss this in Section 5.1.

The distance $d$ of the code is given by the minimum weight of a logical operator. By the previous discussion this is the same as the minimum length of an essential $i$-cycle in the cell complex or its dual. This quantity is also known as the *combinatorial i-systole* and denoted by $\mathrm{csys}_i$ (or $\mathrm{csys}_i^*$ for the dual case).

| **CSS stabilizer code** | **Homology** |
|---|---|
| $S_Z$ | $B_i$ |
| $S_X$ | $B^i$ (or $B_i^*$) |
| $N(S_X)_Z$ | $Z_i$ |
| $N(S_Z)_X$ | $Z^i$ (or $Z_i^*$) |
| $n$ | $\dim C_i$ |
| $k$ | $\dim H_i$ |
| $d_Z$ | $\mathrm{csys}_i$ |
| $d_X$ | $\mathrm{csys}_i^*$ |
| $d$ | $\min(\mathrm{csys}_i, \mathrm{csys}_i^*)$ |

Table 2.1: Overview: Corresponding notions in the language of stabilizer codes and homology. The subscript $X$ and $Z$ restricts the set to $X$- respectively $Z$-type operators.

### 2.3.5 Error correction in homological CSS codes

For homological codes, error correction can be understood geometrically. For the remainder of this section we assume that qubits are identified with $i$-cells of a cellular complex. Every Pauli error can be identified with two chains $E_X, E_Z \in C_i$, where the support of each chain tells us where the error acts as Pauli-$X$ respectively Pauli-$Z$ on the qubits. Let us consider $E_Z$ for concreteness. The syndrome is a tuple of bits where each bit indicates whether $E_Z$ anti-commutes with a certain $X$-check (see discussion in Section 2.1.5). In the language of homology the syndrome is an $i-1$-chain. Whether $E_Z$ and an $X$-check anti-commute, can be deduced by evaluating the inner product of their corresponding $i$-chains. Since we identify $X$-checks with the coboundaries of $i-1$-cells which are the columns of $\delta_{i-1} = \partial_i^T$, we have that *the result of the $X$-check measurements is the boundary $\partial_i E_Z$*. The syndrome of the $X$-error $E_X$ is given by $\delta_i E_X$ or, equivalently, by the boundary of $*E_X$ in the dual complex.

A $Z$-error $E_Z$ is corrected by applying a $Z$-type Pauli operator with support $R_Z \in C_i$ such that the boundary of $R_Z$ coincides with the syndrome, i.e. $\partial_i R_Z = \partial_i E_Z$. After this is done, the system is back in a code state. If the sum of chains $E_Z + R_Z$ is an element of $B_i$, we applied a stabilizer and no operation was performed on the logical qubits. If, however, $E_Z + R_Z$ contains an essential cycle, i.e. $E_Z + R_Z \in Z_i \setminus B_i$, then we applied a logical operator and our encoded information is corrupted. An $X$-error is corrected in the exact same way but in the dual cellulation.

#### Minimum-weight decoder

A *minimum-weight decoder* (defined in Section 2.1.5) takes a syndrome $s_Z \in B_{i-1}$ as its input and determines a recovery chain $R_Z \in C_i$ with minimum support, such that its boundary coincides with the syndrome $s_Z$, i.e.

$$\begin{aligned} &\text{minimize} \;\; \operatorname{supp}(R_Z) \\ &\text{subject to} \;\; \partial_i R_Z = s_Z. \end{aligned} \tag{2.46}$$

When $i = 1$ this problem is called *minimum-weight perfect matching (MWM)*: In this case the boundaries are subsets of vertices with an even number of elements (see Figure 2.7). A minimum-weight recovery $R_Z$ is given by a set of paths connecting those vertices pairwise such that the total length of all paths is minimized. The decoder has succeeded if the support of the error and the recovery and the error together form a boundary $E_Z + R_Z \in B_1$.

#### Imperfect measurements

So far we assumed that the outcomes of the stabilizer measurements can be done with perfect accuracy. However, in practice such measurements have to be facilitated by a quantum circuit as shown in Figure 2.8.

Errors may happen at any point while the measurements circuits are being applied. Hence, the measurement process will be subject to noise as well. To deal with this type of noise and to obtain more reliable information on the syndrome, the measurement is repeated. We will discuss the details of this procedure in Section 4.2 when we perform numerical simulations.

### 2.3.6 The surface code

In some important experimental realizations of qubits it is desirable to only have planar interactions. This is achieved in the *surface code* which is a homological code defined from the square tessellation of a rectangular disc. Simply applying the above machinery would give us a trivial code with $k = 0$ as all 1-cycles must necessarily be boundaries. However, we can make this code non-trivial by removing $X$-checks on two opposing sides at the boundary. A collection of Pauli-$Z$ operators forming a string can now terminate where the $X$-checks have been removed so that it commutes with all remaining stabilizer checks (see Figure 2.9).

In terms of homology this is described by *relative homology* in which we consider a cellulated space $A$ and a subspace $B \subset A$. We assume that the subspace $B$ respects the cellulation, i.e. there are no cells with interior only partially support in $B$. A chain $c$ is a *cycle relative to B* if its boundary is either zero or supported within $B$ only. In calculations this is simply done by removing all columns and rows of the (co)boundary operators which correspond to cells in $B$. All previous definitions for homology carry over to relative homology, except that we write the homology groups as $H_i(A,B)$. In Figure 2.9 the space $B$ is the 1D submanifold consisting of the two gray, dashed lines. For more information on relative homology see [25]. For more information on the surface code as well as a variant which does not need relative homology see [1, 5].

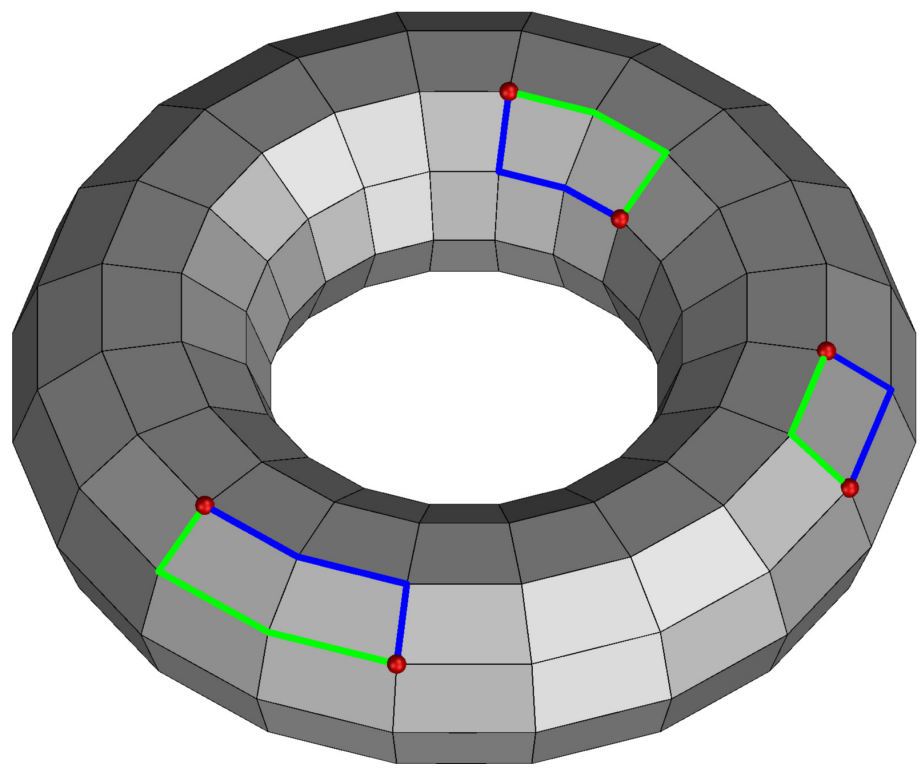

Figure 2.7: The toric code is a homological code derived from a torus tessellated by squares. The $Z$-error $E_Z$ (blue) anti-commutes with $X$-checks at its boundary points, giving rise to the syndrome $s_Z$ (red). The minimum-weight decoder finds a minimum-weight chain $R_Z$ (green) which has boundary $s_Z$, so that $E_Z + R_Z$ is a cycle. In this illustration $E_Z + R_Z \in B_1$ which means that the recovery was successful.

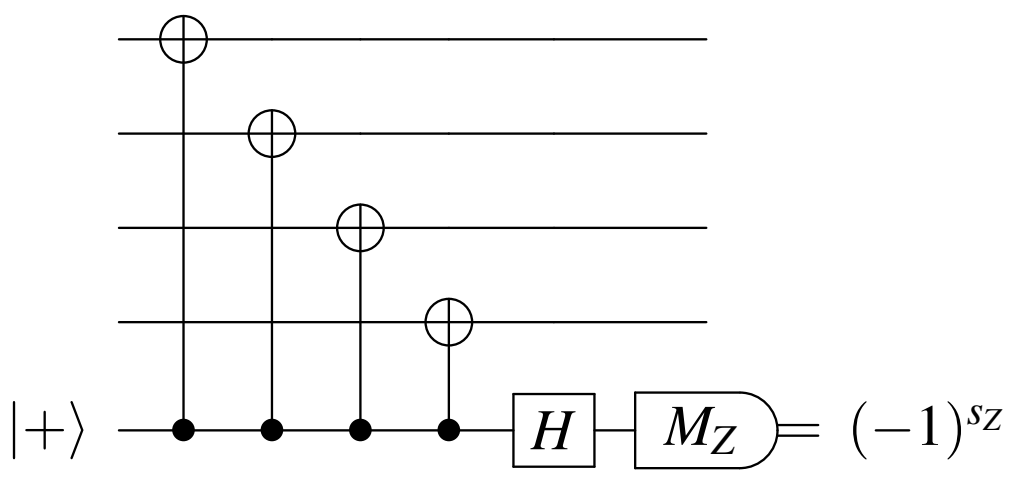

(a) Weight-4 $X$-check measurement circuit facilitating a projective measurement of $XXXX$.

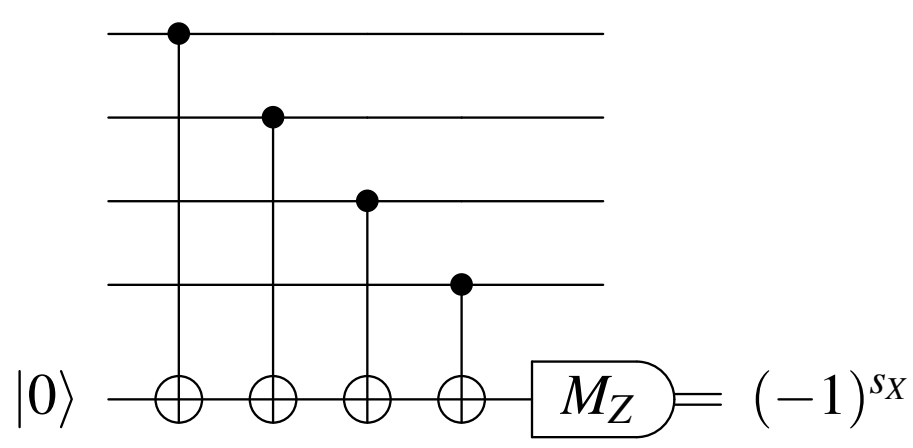

(b) Weight-4 $Z$-check measurement circuit facilitating a projective measurement of $ZZZZ$.

Figure 2.8: Circuits to facilitate stabilizer measurements. The first four wires correspond to the four physical qubits of the code. The last wire is an additional ancilla qubit.

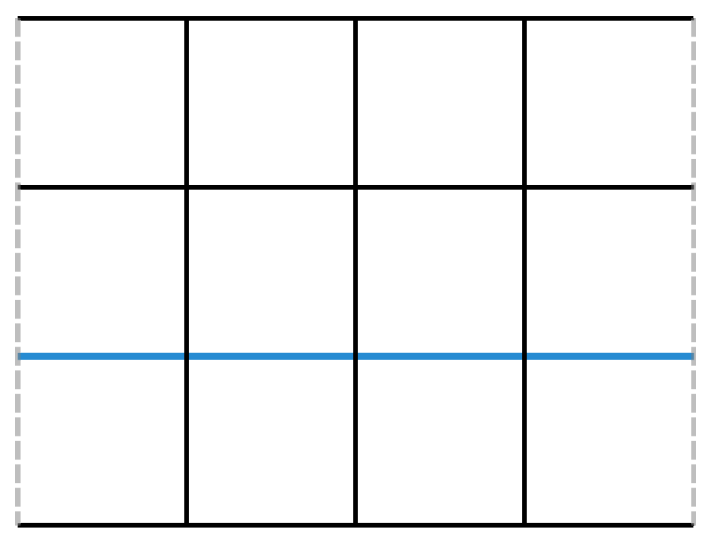

Figure 2.9: The surface code is defined on a $L \times (L-1)$-square lattice. All edges (qubits) and vertices ($X$-checks) in the lines defined by $x = 0$ and $x = L$ are removed. A (relative) essential cycle corresponding to a logical $Z$-operator is highlighted in blue. The surface code is a $[[L^2 + (L-1)^2, 1, L]]$-code.

# Chapter 3

# Two-dimensional hyperbolic codes

In this chapter we will discuss the construction and properties of quantum codes that are derived from two-dimensional, hyperbolic tessellations. We will first introduce the necessary background on curved spaces and their tessellations. We show how tessellations can be described in terms of group theory and more concretely by *reflection groups* describing the symmetries of the tessellations. The group theoretic description will turn out to be a powerful tool to enumerate all possible hyperbolic surfaces admitting such tessellations and hence all possible quantum codes derived from them. We discuss the properties of these codes that can be derived from the hyperbolic geometry. We define families of such quantum codes that we call *extremal* hyperbolic surface codes. Members of this family offer the best protection for the smallest amount of resources. We discuss two variations of the construction of hyperbolic codes. One construction makes the interactions planar, which is a desirable property for implementation. However, it is shown that this construction does not preserve some nice properties that hyperbolic codes have. The second construction which we call *semi-hyperbolic* allows for interpolating between properties of hyperbolic codes and the toric code. We show that against uncorrelated errors the extremal hyperbolic surface codes exhibit a phase transition. In the memory phase errors are polynomially suppressed in system size. Finally, we discuss how to transfer data in and out of a hyperbolic code from another topological quantum code in a fault-tolerant way. We also show how to manipulate the data encoded within a 2D hyperbolic code fault-tolerantly and we give arguments why these operations can be performed while keeping the connectivity between qubits low.

## 3.1 Regular tessellations of curved surfaces

### 3.1.1 Curvature

Before we can discuss tessellations of curved surfaces, we first need to introduce curvature itself. The following expression for the curvature of a surface which is endowed with a distance function is due to Bertrand and Puiseux [26]. The expression is elementary in that it does not rely on concepts of differential geometry.

**Definition 3.1.** Consider a circle on the surface with mid-point $p$, which is given by the set of all points of distance $r$ away from $p$. Let $C_p(r)$ be the circumference of this circle. The *curvature* $\kappa_p$ *at point* $p$ is

$$\kappa_p = \lim_{r\to 0^+} 3 \cdot \frac{2\pi r - C_p(r)}{\pi r^3} \tag{3.1}$$

If $\kappa_p$ is constant on the whole surface we simply write $\kappa$.

The curvature of the Euclidean plane $\mathbb{E}^2$ is clearly $\kappa = 0$ since all circles of radius $r$ have circumference $2\pi r$. For a non-trivial example, consider the sphere of radius $R$. A circle with radius $r$ on the sphere has circumference $C(r) = 2\pi R \sin(r/R)$. This can be seen in Figure 3.1. Using the series expansion of the sine function we obtain

$$\begin{aligned}
\kappa &= \lim_{r\to 0^+} 3 \cdot \frac{2\pi r - 2\pi R \sin(r/R)}{\pi r^3} \\
&= \lim_{r\to 0^+} 6 \cdot \frac{r - R\left(r/R - (r/R)^3/3! + O(r^7)\right)}{r^3} \\
&= \lim_{r\to 0^+} \frac{r^3/R^2 + O(r^7)}{r^3} \\
&= 1/R^2.
\end{aligned} \tag{3.2}$$

We immediately see that any sphere is positively curved and the curvature vanishes in the limit of infinite radius $R \to \infty$. The unit sphere $\mathbb{S}^2$ has curvature $\kappa = +1$.

An example of a surface with *negative curvature*, where the circumference of a circle with radius $r$ is *larger* than $2\pi r$, is the surface of a saddle. The infinite plane which at every point looks like a saddle and has $\kappa = -1$ is the *hyperbolic plane* $\mathbb{H}^2$. It can be realized in the interior of a disc.

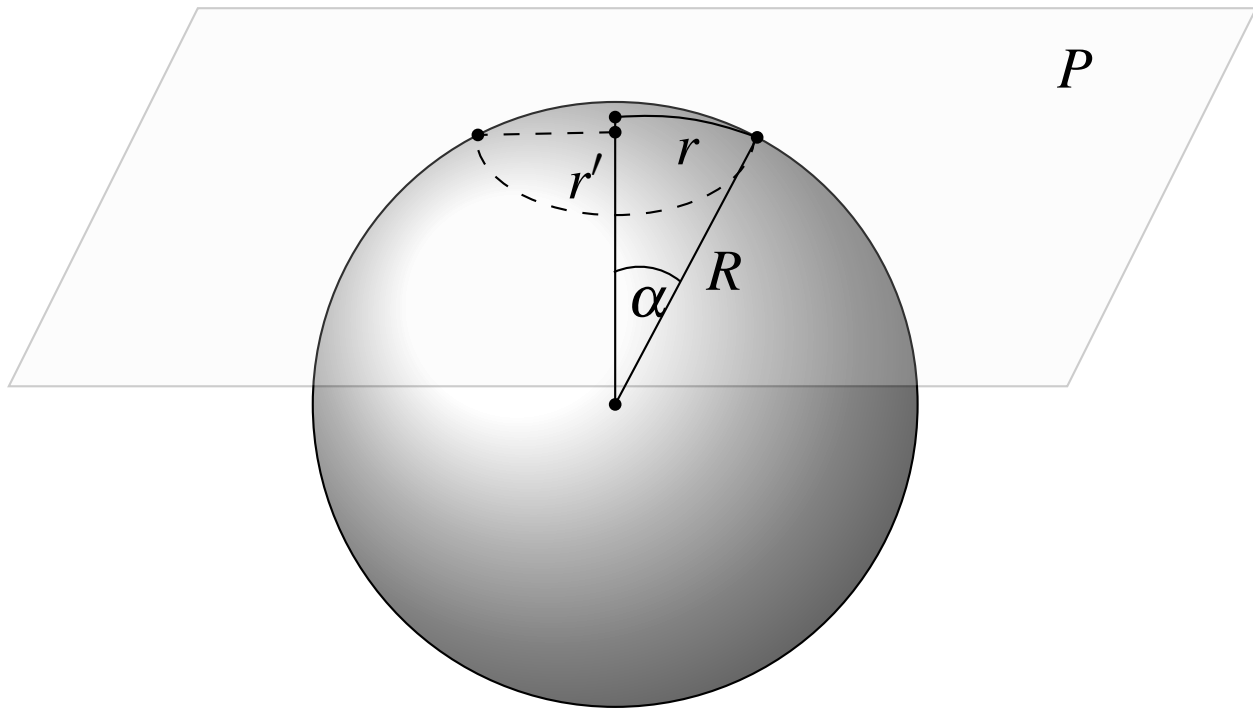

Figure 3.1: A circle on a plane $P$ which coincides with a circle on a sphere of radius $R$. The circumference of the circle on the sphere of radius $r$ is the same as the circumference of a circle on the plane $P$ of radius $r' = R\sin(\alpha)$, where $\alpha = r/R$.

**Definition 3.2** (Poincaré Disc Model)**.** The hyperbolic plane $\mathbb{H}^2$ can be realized as the set of points inside the open unit disc $D^\circ = \{x \in \mathbb{R}^2 \mid ||x||^2 < 1\}$ with the distance function

$$d(x,y) = \cosh^{-1}\left(1 + \frac{2||x-y||^2}{(1-||x||^2)(1-||y||^2)}\right) \tag{3.3}$$

and metric

$$\mathrm{d}s^2 = 4\frac{\mathrm{d}x_1^2 + \mathrm{d}x_2^2}{1-||x||^2}. \tag{3.4}$$

For examples of the Poincaré disc model, see Figure 3.2.

### 3.1.2 The Gauß-Bonnet Theorem

What does Equation 3.1 imply for tessellations of closed surfaces? – The key to answer this question will turn out be the following important theorem which establishes a connection between the curvature $\kappa_p$, the Euler characteristic (see Definition 2.11) and an effect called *angular defect*.

**Theorem 3.3** (Gauß-Bonnet)**.** *Let $M$ be a surface with curvature $\kappa_p$, boundary $\partial M$ consisting of $r$ geodesic lines and Euler characteristic $\chi(M)$. Furthermore, let $\alpha_i$ be the interior angles at which the boundary lines of $M$ meet. It holds that*

$$\int_M \kappa_p \,\mathrm{d}A + \sum_{i=1}^{r}(\pi - \alpha_i) = 2\pi\chi(M). \tag{3.5}$$

*If $M$ is closed, $\partial M = \emptyset$ and Equation 3.5 simplifies to*

$$\int_M \kappa_p \, \mathrm{d}A = 2\pi\chi(M). \tag{3.6}$$

We will now apply Theorem 3.3 to a single polygon $P$ belonging to a tessellation of a surface with constant curvature. This is possible since $P$ is a submanifold of the tesselated manifold. Let $r$ be the number of sides of $P$. $P$ consists of $r$ vertices (0-cells), $r$ edges (1-cells) and a single face (2-cell). Hence we have $\chi(P) = |V| - |E| + |F| = 1$ (see Definition 2.11 in Section 2.3.1). Applying Theorem 3.3 we obtain the following corollary.

**Corollary 3.4.** *For a polygon with $r$ sides and internal angles $\alpha_i$ on a surface with constant curvature $\kappa$ we have*

$$\sum_{i=1}^{r} \alpha_i = \kappa \cdot area(P) + (r-2)\pi. \tag{3.7}$$

*In particular, for a triangle with internal angles $\alpha$, $\beta$, $\gamma$ we have*

$$\alpha + \beta + \gamma \begin{cases} > \pi & \text{if } \kappa > 0 \\ = \pi & \text{if } \kappa = 0 \\ < \pi & \text{if } \kappa < 0. \end{cases} \tag{3.8}$$

The deviation from the behavior of angles compared to the Euclidean case ($\kappa = 0$) is what is called an *angular defect*. Equation 3.7 implies that for curved surfaces the area of a polygon is fixed by its internal angles.

### 3.1.3 Regular tessellations

**Definition 3.5.** We will call a tessellation of a surface *regular* if:

(a) all polygons $P_i$ are identical,

(b) all polygons $P_i$ are equilateral and equiangular,

(c) the same number of polygons meet at each vertex in the tessellation.

If $r$ is the number of sides of each polygon and $s$ is the number of polygons meeting at a vertex then we will denote the tessellation by the *Schläfli symbol* $\{r,s\}$.

For a tessellation of the Euclidean plane $\mathbb{E}^2$ not all values of $r$ and $s$ are allowed by Corollary 3.4. We can determine the allowed values by subdividing the polygons into triangles. Take a single polygon and draw a line from any vertex to the mid-point. Draw a second line from one of the edges which are adjacent to the vertex perpendicular to the mid-point. The two lines and the half-edge form a triangle with interior angles $\pi/2$, $\pi/r$ and $\pi/s$. Since $\mathbb{E}^2$ has no curvature $\kappa = 0$, by Equation 3.8 we must have $\pi/2 + \pi/r + \pi/s = \pi$. Hence we obtain the condition $1/r + 1/s = 1/2$ for tessellations of $\mathbb{E}^2$. The only pairs of integers satisfying this equation are $(4,4)$, $(3,6)$ and $(6,3)$, giving the square, triangular and hexagonal tessellation.

Similarly, $\{r,s\}$-tessellations of the sphere $\mathbb{S}^2$ have to obey the constraint $1/r + 1/s > 1/2$. The five pairs of integers satisfying this restriction are $(3,3)$, $(3,4)$, $(4,3)$, $(3,5)$ and $(5,3)$. The tessellations of the sphere correspond to the five Platonic solids. In the same order as above: Tetrahedron, octahedron, cube, icosahedron and dodecahedron (see Figure 3.2).

For the hyperbolic plane $\mathbb{H}^2$ the restriction is $1/r + 1/s < 1/2$ which is satisfied by *infinitely* many pairs of integers. Hence, there are infinitely many hyperbolic tessellations. A list of all regular tessellations of $\mathbb{S}^2$ and $\mathbb{E}^2$ as well as some small regular tessellations of $\mathbb{H}^2$ can be found in Figure 3.2.

### 3.1.4 Symmetries of surfaces and tessellations

In the previous section we have seen that all regular tessellations can be realized on the sphere $\mathbb{S}^2$, the Euclidean plane $\mathbb{E}^2$ and the hyperbolic plane $\mathbb{H}^2$. In the next section we will see that *any* closed surface of constant curvature can be derived from one of these three surfaces via a *compactification procedure*. For this reason we are going to call $\mathbb{S}^2$, $\mathbb{E}^2$ and $\mathbb{H}^2$ *covering spaces*. The essential ingredient of the compactification procedure is the action of symmetry groups on the covering spaces. In the following discussion we will, for concreteness, take the hyperbolic plane $\mathbb{H}^2$. However, all statements will be equally valid for $\mathbb{S}^2$ and $\mathbb{E}^2$.

#### Group actions and tessellations

The set of all distance-preserving mappings (isometries) from $\mathbb{H}^2$ onto itself forms a group under composition. This group is called the *group of isometries* $\text{Isom}(\mathbb{H}^2)$. Its elements are either reflections, rotations, translations or combinations of translations and reflections (called *glide reflections*).

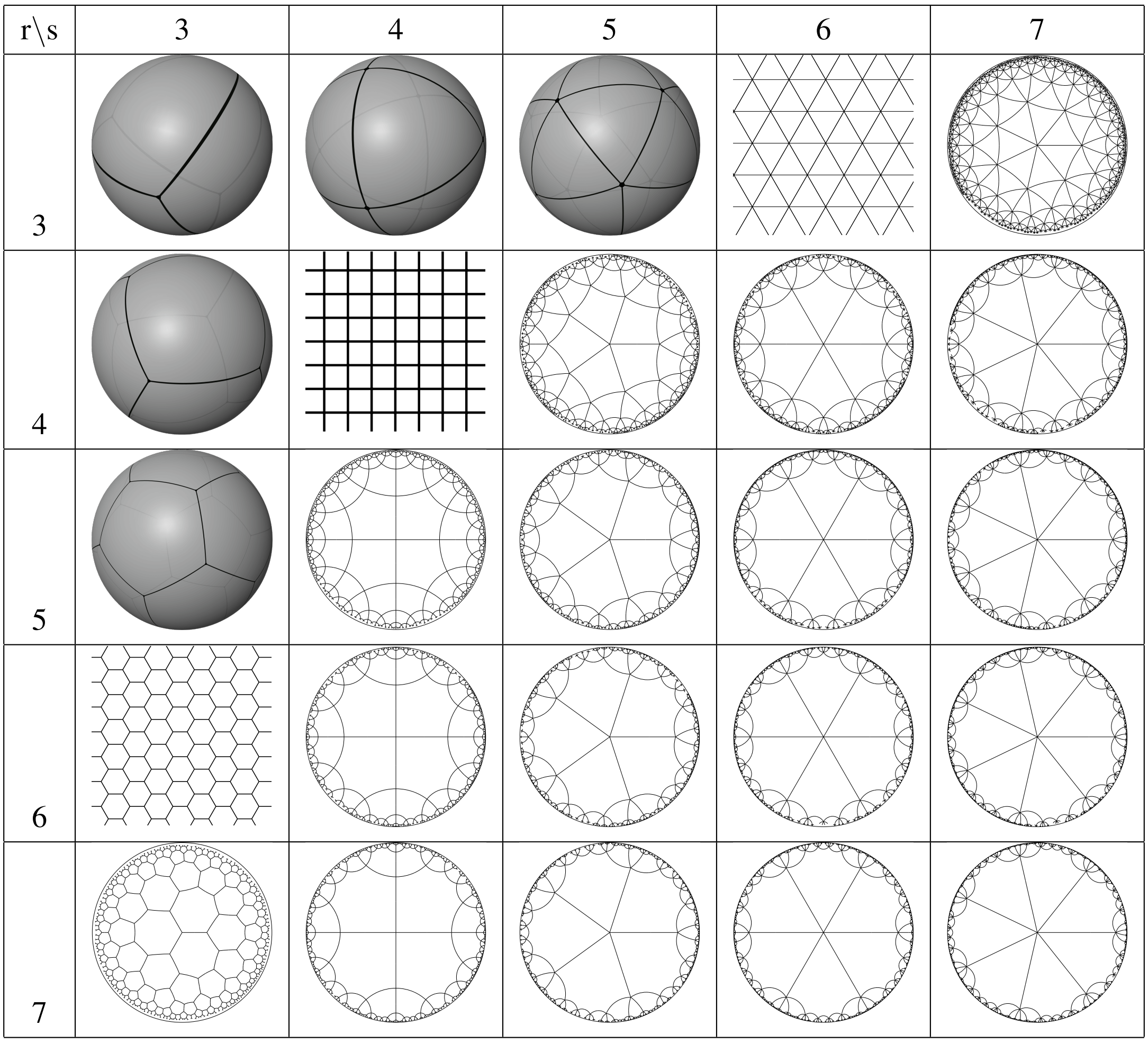

Figure 3.2: All regular $\{r,s\}$-tessellations with $r,s \leq 7$. All regular tessellations which are not shown in this figure ($r,s > 7$) are hyperbolic. The dual of a $\{r,s\}$-tessellation is given by $\{s,r\}$. Hence, taking the dual means reflecting this table along the diagonal. All tessellations on the diagonal are self-dual. The five tessellations of the sphere can be turned into the Platonic solids by flattening the faces.

Let $G$ be a subgroup of $\mathrm{Isom}(\mathbb{H}^2)$. For a point $P \in \mathbb{H}^2$, we call the set of all points that $P$ is mapped to by elements of $G$ the *orbit* of $P$ under $G$. A *fundamental domain* of a group $G$ is a part of the plane such that no two points in its interior are mapped onto each other and all points in the plane can be reached by applying an element in $G$ to a point in the fundamental domain.

The action of $G$ on the points of $\mathbb{H}^2$ induces an action on a fundamental domain $F$ of $G$. We denote the application of a $g \in G$ on $F$ by $F * g$. The identity element $e \in G$ acts as the identity map $F * e = F$ and the action is compatible with the group multiplication. By the latter we mean that the application of $gh$ on $F$ is the same as first applying $g$ to $F$ and then $h$ to the result, or in short $F * (gh) = (F * g) * h$.

### Symmetries of regular tessellations: Coxeter groups

We will now discuss *Wythoff's kaleidoscopic construction* which allows us to relate a regular tessellation $\{r,s\}$ to a group of isometries $G = G_{r,s}$.

A regular $r$-gon has $2r$ symmetries generated by the reflections through its symmetry axes. The group is called the *Dihedral group* $D_r$. This can be seen in Figure 3.3a. The symmetry axes induce a triangulation of the $r$-gon into $2r$ (right) triangles. The triangulation of the faces induces a triangulation of the whole tessellation into triangles with internal angles $\pi/2, \pi/r$ and $\pi/s$ (see Figure 3.4).

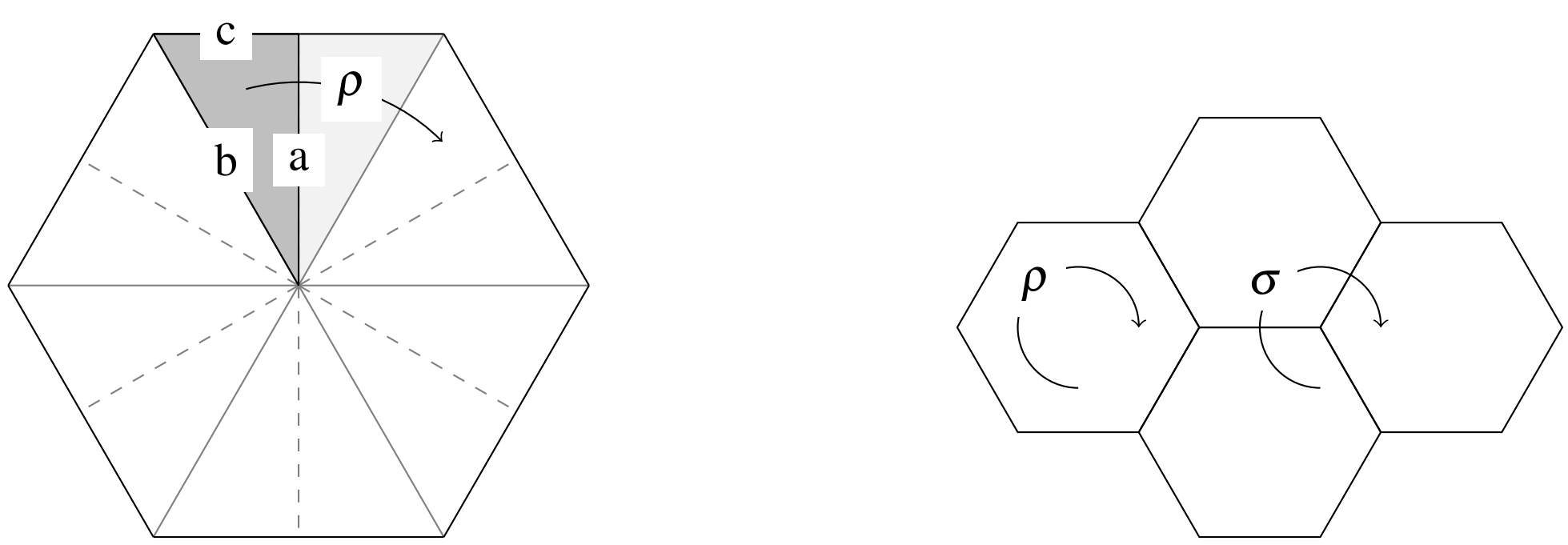

(a) Action of the symmetry group on a single face.

(b) Rotations acting on the tessellation

Figure 3.3: Group acting on the $\{6,3\}$-tessellation.

Let $G_{r,s}$ be generated by the reflections on the sides $a, b, c$ of a single triangle. We assume that $a$, $b$ and $c$ are arranged in clockwise order such that $a$ is opposite to the angle $\pi/s$, $b$ is opposite to the angle $\pi/2$ and $c$ is opposite to the angle $\pi/r$. Applying a reflection twice is the same as doing nothing, so that $a^2 = b^2 = c^2 = e$, where $e$ is the identity element of the group. Additionally, two consecutive reflections on two lines that intersect at an angle $\pi/k$ for some positive integer $k$ correspond to a rotation around the intersection point by an angle $2\pi/k$. Thus, $k$ rotations give the identity. These are all the relations that $a$, $b$ and $c$ fulfill. The multiplication of any two elements of $G_{r,s}$ is completely determined by them. Hence, we can write compactly

$$G_{r,s} = \langle a,b,c \mid a^2 = b^2 = c^2 = (ab)^r = (bc)^s = (ac)^2 = e\rangle. \tag{3.9}$$

This is called the *presentation* of the group $G_{r,s}$. One way to think about group presentations is as a set of strings or words consisting of $\{a,b,c,a^{-1},b^{-1},c^{-1}\}$. Group multiplication simply corresponds to a concatenation of words. All words which differ by a subword that is equivalent to $e$ by the rules of group multiplication (such as $g^{-1}g = e$) or any of the given relations are considered to be equal. Abstract group presentations can be directly used in computer algebra systems such as GAP or MAGMA. The groups $G_{r,s}$ are called *Coxeter groups*.

There is an important subgroup of $G_{r,s}$ which we will denote by $G^+_{r,s}$. It consists of all orientation-preserving maps (rotations and translations). The group $G^+_{r,s}$ is generated by rotations $\rho = ab$ and $\sigma = bc$ (see Figure 3.3b), so that

$$G^+_{r,s} = \langle \rho, \sigma \mid \rho^r = \sigma^s = (\rho\sigma)^2 = e\rangle. \tag{3.10}$$

The generators $\rho$ and $\sigma$ act as a clockwise $2\pi/r$ and $2\pi/s$ rotations, respectively. Note that the $G^+_{r,s}$-orbit of a triangle only covers “half” the plane as we cannot map between triangles that share an edge. We fix this by taking two triangles related by an $a$-reflection as the fundamental domain of $G^+_{r,s}$.

The important observation to make is that (by construction) each element of the orbit of the group acting on a fundamental domain is uniquely labeled by a group element. By fixing a triangle $F_0$ that is the fundamental domain of $G^+_{r,s}$, every other triangle $F$ can be labeled by the group element $g \in G^+_{r,s}$ that maps $F_0$ onto it, i.e. $F_0 * g = F$. This allows us to forget about the group action ($F_0$ was arbitrary in the first place) and only talk about the group $G^+_{r,s}$.

Let $\langle\rho\rangle = \{e,\rho,\rho^2,\ldots,\rho^{r-1}\}$ be the cyclic subgroup of $G^+_{r,s}$ that is generated by the rotation $\rho$. The faces of the $\{r,s\}$-tessellation (the $r$-gons) are in one-to-one correspondence to the triangles *up*

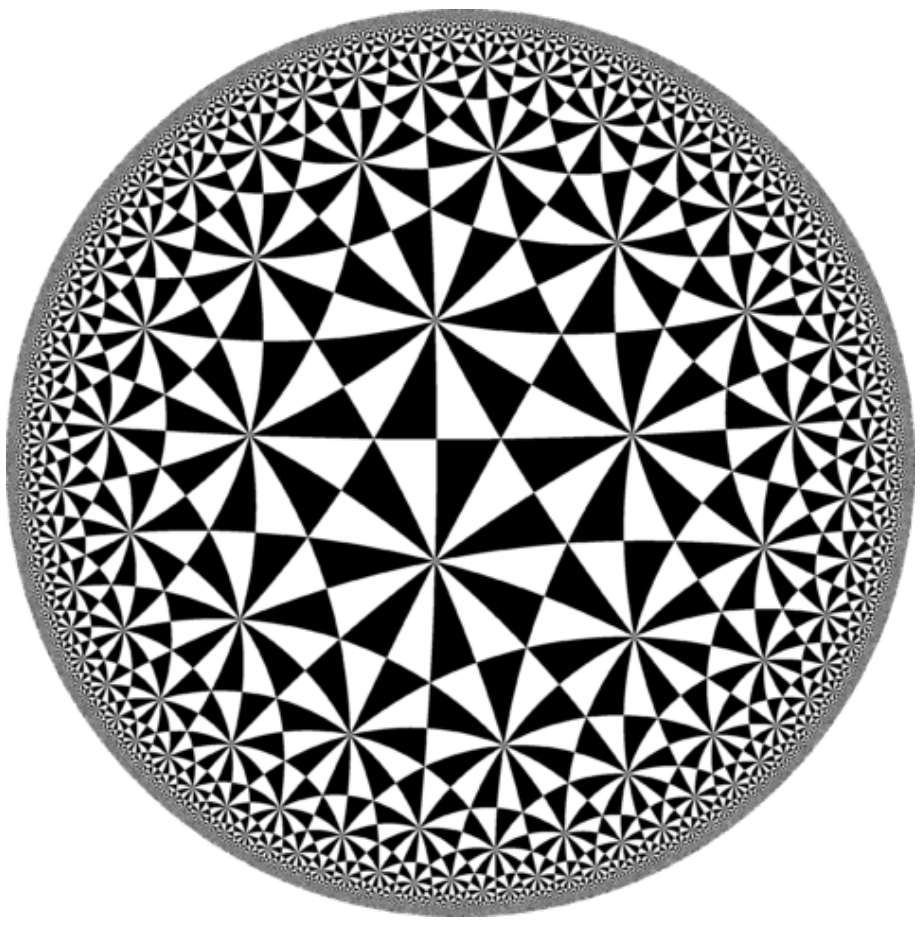

Figure 3.4: The $\{7,3\}$-tessellated hyperbolic plane $\mathbb{H}^2$ in the Poincaré disc model. The fundamental domains of $G_{7,3}$ are colored in black and white. Fundamental domains of the same color are related by an element of $G^+_{7,3}$.

*to a rotation by* $\rho$. In group theoretic language these are the *left cosets* of the subgroup $\langle\rho\rangle$ denoted by $g\langle\rho\rangle = \{g, g\rho, g\rho^2, \ldots, g\rho^{r-1}\}$ for a $g \in G^+_{r,s}$. Similarly, the vertices and edges of the tessellation can be uniquely labeled by left cosets of the cyclic subgroups $\langle\sigma\rangle$ and $\langle\rho\sigma\rangle$, respectively.

### 3.1.5 Compactification

In this section we show how isometries can be used to define closed, two-dimensional manifolds (or surfaces) which have the same curvature $\kappa_p$ at every point $p$. Just as in the previous section, we will only consider the case with negative curvature. However, all statements will be true for Euclidean and positively curved spaces as well.

Let $\Gamma$ be a subgroup of the group of isometries $\mathrm{Isom}(\mathbb{H}^2)$ acting on $\mathbb{H}^2$. We can construct a new space $\mathbb{H}^2/\Gamma$ where we identify all points that differ by the application of an element of $\Gamma$. The surface $\mathbb{H}^2/\Gamma$ is called a *quotient surface*. Formally, each point of $\mathbb{H}^2/\Gamma$ is a set of the form $\{p * g \mid g \in \Gamma\}$, i.e. the orbit of some point $p$. In this sense $\mathbb{H}^2$ is a *covering* of the quotient surface $\mathbb{H}^2/\Gamma$ and $\Gamma$ is called the *covering group*.

To avoid degenerate cases, we demand that the action of $\Gamma$ does not have any fixed points. Additionally, we assume that there exists an $\varepsilon > 0$ such that for all points $p \in \mathbb{H}^2$ the discs of radius $\varepsilon$ around every point in the orbit $p*\Gamma$ do not overlap. If $\Gamma$ meets these requirements one says that it acts *fixed-point free* and *discontinuous*. In this case there exists a fundamental domain of $\Gamma$ around every point of $\mathbb{H}^2$.

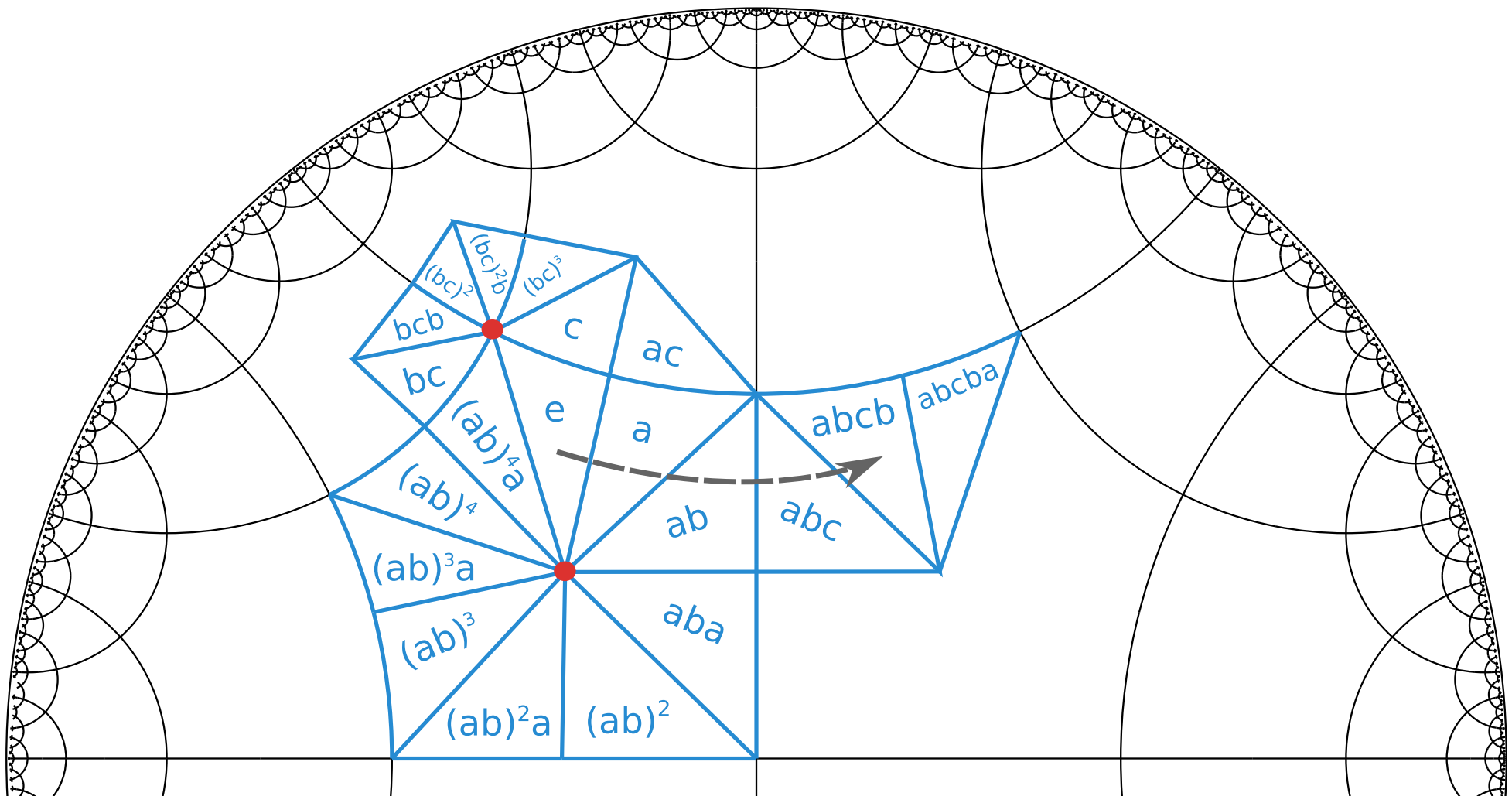

Figure 3.5: The $\{5,4\}$-tessellation of $\mathbb{H}^2$. Some fundamental triangles are identified with elements of $G_{5,4}$ (cf. Figure 3.3). The subgroup $\langle a,b\rangle$ contains all fundamental triangles belonging to a face. Its action leaves a single point in the center of the face invariant (red). The subgroup $\langle b,c\rangle$ covers all fundamental triangles belonging to a vertex while leaving it invariant (red). The group element *abcb* has no fixed-points. It is a translation along a geodesic (dashed gray arrow). The group element *abcba* has no fixed-points either, but it does not preserve orientation: It is a glide-reflection.

The shortest distance of a translation in $\Gamma$ is called the *injectivity radius* $R_{inj}$. The injectivity radius plays an important role in code construction of Section 2.3 as it provides a lower-bound on the length of the shortest essential cycle on $\mathbb{H}^2/\Gamma$. This can be seen as follows: if a tessellation is defined on a quotient surface then all cycles contained within a disc with radius $R_{inj}$ must be boundaries. This is true because the geometry within the disc is the same as the geometry of $\mathbb{H}^2$ and all cycles in $\mathbb{H}^2$ are boundaries of the set of faces that they surround. The injectivity radius therefore provides a lower bound on the code distance $d$.

Which quotient surfaces $\mathbb{H}^2/\Gamma$ admit a regular tessellation? – One condition on $\Gamma$ is that it needs to respect the tessellation structure. This means that *it must be a subgroup of the tessellation group* $G_{r,s}$. The faces, edges and vertices of the tessellation on the quotient surface $\mathbb{H}^2/\Gamma$ are similarly labeled by the action of $G_{r,s}$ on an arbitrarily chosen face. The action of the covering group $H$ should not affect the labels. This is the case if for all $h \in \Gamma$ and $g \in G_{r,s}$ we have $ghg^{-1} \in \Gamma$. We say that $\Gamma$ is a *normal subgroup* of $G_{r,s}$. If $\Gamma$ does not contain any glide-reflections ($\Gamma$ is a subgroup of $G^+_{r,s}$), its quotient surface $\mathbb{H}^2/\Gamma$ is orientable.

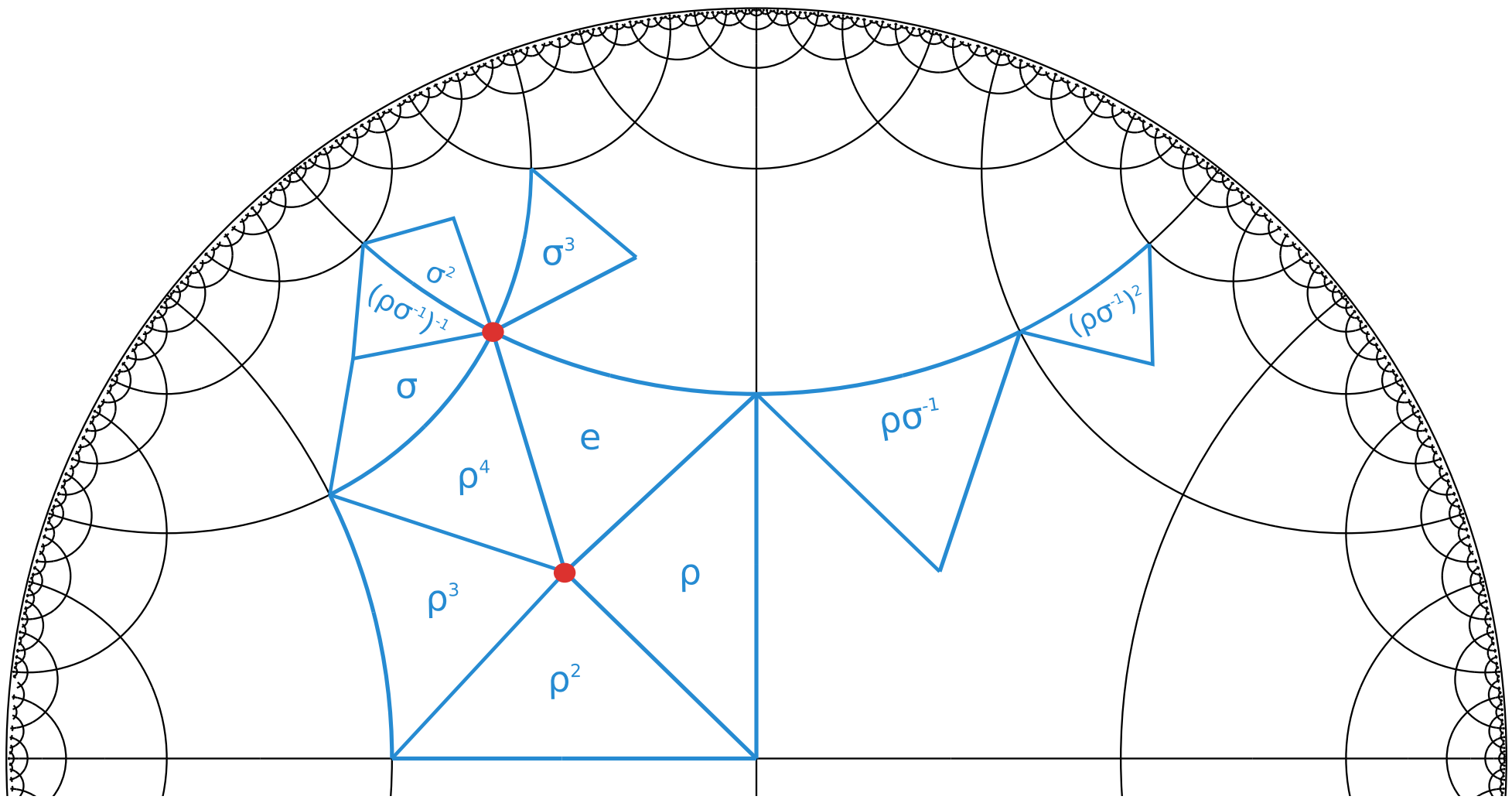

Figure 3.6: The $\{5,4\}$-tessellation of $\mathbb{H}^2$. Some fundamental triangles are labeled by elements of $G^+_{5,4}$. The subgroup $\langle\rho\rangle$ contains all fundamental triangles belonging to a face. Its action leaves a single point in the center of the face invariant (red). The subgroup $\langle\sigma\rangle$ covers all fundamental triangles belonging to a vertex while leaving it invariant (red). The group element $\rho\sigma^{-1}$ translates the fundamental triangle along a hyperbolic geodesic (cf. Figure 3.5).

Let us first consider the case where $\Gamma$ is a subgroup of $G^+_{r,s}$. In the tessellation of $\mathbb{H}^2/\Gamma$, each face can be labeled by a set of the form

$$g\langle\rho\rangle\Gamma = \{g\rho^n h \mid n \in \{1,\dots,r\}, h \in \Gamma\}, \tag{3.11}$$

for a $g \in G^+_{r,s}$. Similarly, we have a labeling of the edges and vertices of the quotient surface $\mathbb{H}^2/\Gamma$ using cosets of $\langle\rho\sigma\rangle\Gamma$ and $\langle\sigma\rangle\Gamma$ (cf. Figure 3.6). Pairs of faces, edges and vertices are incident if and only if their associated cosets share a common element. This means that the Hasse diagram of the tessellation is encoded in the group $G^+_{r,s}/\Gamma$. From this information alone, we can construct a quantum code as described in Section 2.3. Note that the tessellation is finite if and only if its symmetry group $G_{r,s}/\Gamma$ is finite.

In the more general case when $\Gamma$ is a subgroup of $G_{r,s}$ the faces, edges and vertices of the tessellation are identified with cosets with respect to $\langle a,b\rangle\Gamma$, $\langle a,c\rangle\Gamma$ and $\langle b,c\rangle\Gamma$ (cf. Figure 3.5).

The discussion above is summarized by the following proposition.

**Proposition 3.6.** *The problem of finding closed surfaces which admit a $\{r,s\}$-tessellation reduces to finding normal subgroups of $G_{r,s}$ which have no fixed-points and which give a finite quotient group.*

The procedure of compactification to obtain closed surfaces that we discussed in this section is quite general, as shown by the following theorem.

**Theorem 3.7** (Killing-Hopf)**.** *All closed surfaces of constant curvature can, up to an overall scaling, be expressed as quotient surfaces $\mathbb{S}^2/\Gamma$, $\mathbb{E}^2/\Gamma$ or $\mathbb{H}^2/\Gamma$.*

*Proof.* See [27]. ∎

**Topology of compactified spaces**

According to Theorem 3.3 the curvature of a surface does not only determine what regular tessellations are possible; it also determines the topology of the surface. Applying Theorem 3.3 to any closed surface with constant curvature $\kappa$ gives

$$\kappa \cdot \text{area}(M) = 2\pi\chi(M) = 2\pi(2 - \dim H_1) \tag{3.12}$$

where we have used Equation 2.36 and $\dim H_0 = \dim H_2 = 1$. Remarkably, Equation 3.12 allows us to compute the number of essential 1-cycles for any surface with constant curvature.

The unit sphere $\mathbb{S}^2$ is already a closed surface with $\kappa = +1$ and $\text{area}(\mathbb{S}^2) = 4\pi$. From Equation 3.12 follows that $\dim H_1 = 0$: Every cycle on the surface of a sphere is contractible. The only non-trivial quotient surface the sphere admits is called the *projective plane* $\mathbb{P}^2$, which is non-orientable. It is obtained by factoring out a reflection across a plane which intersects the sphere at a great circle. Clearly, the area of the projective plane is half the area of the sphere: $\text{area}(\mathbb{P}^2) = 2\pi$.

With Equation 3.12 it follows that $\mathbb{P}^2$ has a single essential cycle $\dim H_1 = 1$. Homological quantum codes derived from tessellations of the projective plane were analyzed in [28].[1]

Any quotient surface of the Euclidean plane $\mathbb{E}^2$ has vanishing curvature $\kappa = 0$. Hence, any such surface must contain two essential cycles, i.e. $\dim H_1 = 2$. If the covering group $\Gamma$ contains translations only, the surface $\mathbb{E}^2/\Gamma$ is a torus. If $\Gamma$ also contains a glide-reflection, the quotient surface $\mathbb{E}^2/\Gamma$ is a Klein bottle. These are the only two closed surfaces with vanishing curvature.

Quotient surfaces derived from the hyperbolic plane $\mathbb{H}^2$ have the number of their essential cycles $\dim H_1$ scaling with their area

$$\dim H_1 = \frac{\operatorname{area}(\mathbb{H}^2/\Gamma)}{2\pi} + 2. \tag{3.13}$$

## 3.2 Quantum codes from hyperbolic tessellations

We have seen in Section 3.1.5 that the number of essential cycles $\dim H_1$ grows with the area of a hyperbolic surface. We will now show how tessellated hyperbolic surfaces give us quantum codes with a *finite encoding rate*, meaning that $\lim_{n\to\infty} k/n > 0$.

By Corollary 3.4 we have that the area of a hyperbolic $r$-gon is $(2-r)\pi + \sum_{i=1}^{r} \alpha_i$. The internal angles in the regular tessellation are all equal to $\alpha_i = 2\pi/s$. If $|F|$ is the number of faces then it holds that

$$\operatorname{area}(\mathbb{H}^2/\Gamma) = |F|\pi \left(r - 2 - 2\frac{r}{s}\right) \tag{3.14}$$

In an $\{r,s\}$-tessellation each face has $r$ edges at its boundary and hence the total number of edges in the tessellation $|E| = r|F|/2$. Together with Equation 3.13 and Equation 3.14 we obtain:

$$\dim H_1 = |E| \left(1 - \frac{2}{r} - \frac{2}{s}\right) + 2 \tag{3.15}$$

We derive a quantum code from a tessellated hyperbolic surface by identifying the edges with physical qubits. From Equation 3.14 directly follows that a quantum code derived from the $\{r,s\}$-tessellation of a closed hyperbolic surface will have an *encoding rate*

$$\frac{k}{n} = 1 - \frac{2}{r} - \frac{2}{s} + \frac{2}{n}. \tag{3.16}$$

[1] It turns out that the Shor's 9-qubit code is a homological code derived from a tessellation of $\mathbb{P}^2$. The fact that it encodes a single qubit can be derived from the fact that the area of $\mathbb{P}^2$ is $2\pi$.

Each stabilizer $X$-check corresponds to a vertex of the tessellation, acting on all edges incident to the vertex. In a $\{r,s\}$-tessellation the weight of any $X$-check is therefore $s$. Similarly, each stabilizer $Z$-check corresponds to a face of the tessellation, acting on all incident edges. The weight of every $Z$-check is therefore $r$. Note that every edge, regardless of $r$ and $s$, is incident to two faces and two vertices, so that every qubit is acted upon by 4 stabilizer checks (qubit degree is 4).

The asymptotic rate ($n \to \infty$) only depends on the tessellation and it is higher for larger values of $r$ and $s$. In the previous section we established that there are infinitely many hyperbolic tessellation subject to the constraint $1/r + 1/s < 1/2$. The tessellation giving the smallest combined weight $r+s$ of the stabilizer generators is the $\{5,4\}$-tessellation (see Figure 3.2). Compare the above result to the rate of codes obtained from tessellation of quotient surfaces of the Euclidean plane. As discussed in the previous section they only encode two logical qubits, regardless of the number of physical qubits.

## 3.3 Bounds on parameters of quantum codes in 2D

### 3.3.1 Trade-offs

In [29] it was shown that for any $[[n,k,d]]$ 2D stabilizer code the number of encoded (logical) qubits $k$ and the number of physical qubits $n$ and the distance $d$ obey the trade-off relation $kd^2 \leq cn$ for some constant $c$. This result assumes that physical qubits are laid out according to an Euclidean geometry. This rate-distance trade-off bound is achieved by the toric code and the surface code in which one can, for example, encode $k$ qubits into $k$ separate surface code sheets, each with $d^2$ qubits and distance $d$, leading to a total number of $n = kd^2$ qubits. Clearly, the encoding rate $k/n$ approaches zero when one tries to encode better qubits with growing distance $d$.

In [30] Fetaya showed that two-dimensional homological codes, based on tessellations of arbitrary two-dimensional surfaces, obey the square-root bound on the distance, i.e. $d \leq O(\sqrt{n})$. In [7] it was shown that with a family of homological codes based on 4-dimensional manifolds with non-zero curvature, one can go beyond this square-root bound and encode a qubit with distance scaling as $O(\sqrt{n \log n})$. In addition, [7] showed that there exist hyperbolic surface codes which have a constant rate and logarithmic scaling distance.

A result by Delfosse [8] shows that the parameters of all 2D homological codes are subject to a trade-off.

**Theorem 3.8.** *The parameters* $[[n,k,d]]$ *of a homological quantum code derived from a tessellation of a 2D surface are subject to the inequality*

$$kd^2 \leq cn(\log k)^2, \tag{3.17}$$

*where* $c > 0$ *is some constant independent of* $n$*,* $k$ *and* $d$*.*

This result shows that from tessellated 2D surfaces we cannot derive codes with a non-vanishing asymptotic encoding rate $\lim_{n\to\infty} k/n > 0$ and superlogarithmic distance. In particular, hyperbolic codes can have at most logarithmic distance.

More intuition on why hyperbolic codes have a logarithmic distance follows from the analysis in [31] where the author considers the following procedure: Start with a single face in the hyperbolic plane. Then add all faces which share an edge with the original face. Continue adding faces which share an edge with the outermost faces. This process is iterated and each time a new layer is added. Note that whenever a new layer is added the length of the shortest path from the innermost to the outermost layer grows by one. It is shown in [31] that the total number of edges grows approximately with some constant factor in each iteration. For a regular $\{r,s\}$-tessellation this factor is lower bounded by $\sqrt{rs}$. For codes derived from hyperbolic tessellations this has implications on the scaling of the number of physical qubits versus the distance. Since there are no essential cycles of weight $< d$ for a distance $d$ code we know that within such a radius all loops must be contractible. Thus within this radius the surface looks like the tessellated hyperbolic plane. With the previous considerations we see that $n \in O\left((rs)^{d/2}\right)$ and thus $d \leq c\log(n)$. By describing the growth process with recurrence relations it is shown in [31] that we can bound

$$d \leq \frac{r}{2}\log_{\sqrt{rs}}(2n). \tag{3.18}$$

### 3.3.2 Lower bound on distance

It has been proven in [32] that for a $\{r,s\}$-tessellation of $\mathbb{H}$ for any given $q \in \mathbb{N}$ there exists a hyperbolic surface $\mathbb{H}^2/\Gamma$ tessellated by $\{r,s\}$ with combinatorial systole larger than $q$ and $n \leq C^q s/2$, where $n$ is the number of edges and

$$C \leq \frac{1}{2}(5rs)^{16rs}. \tag{3.19}$$

This means that there exists a family of quantum codes where the distance is lower bounded by a function growing logarithmically in the number of physical qubits (see also [7]). Since the logarithm is base $C$ and the value of $C$ is very large, this bound is not relevant for any practical purposes.

## 3.4 Computing the distance

In this section we describe an algorithm formulated by Bravyi [33] which allows one to efficiently compute the distance of any CSS code which can be embedded on a surface. In the context of homology this algorithm was presented in [34]. This algorithm can be extended to count the number of minimum weight logical operators $N_d^Z$ and $N_d^X$. These numbers are, in addition to the distance, a further property of CSS quantum codes which is indicative of their performance.

### 3.4.1 Distance algorithm

Since hyperbolic codes are CSS codes, the minimum weight logical operator acts as either Pauli-$X$ or Pauli-$Z$ on all of the qubits in its support. We focus on calculating $d_Z$, i.e. the minimum weight of any logical operator acting as Pauli-$Z$. The procedure for obtaining the minimal weight of a logical $X$ operator $d_X$ is identical.

Let $G = (V, E)$ be the graph consisting of all vertices and edges of the tessellation. Take a $\overline{X}$ operator, say $\overline{X}_1$ and let $E(\overline{X}_1) \subseteq E$ be its qubit support. Take two copies of the graph, $G$ and $G'$. Using these copies, we define a new graph $\tilde{G} = (\tilde{V}, \tilde{E})$ with $\tilde{V} = V \cup V'$ and the following edge set $\tilde{E}$. Omit in $\tilde{E}$ each edge $e = (u, v) \in E(\overline{X}_1)$ and $e' = (u', v') \in E(\overline{X}_1)$. For these omitted edges we instead include two new *cross-over* edges $(u, v') \in \tilde{E}$ and $(u', v) \in \tilde{E}$. For all other edges $e \in E \setminus E(\overline{X}_1)$, $e' \in E \setminus E(\overline{X}_1)$, include $e$ and $e'$ in $\tilde{E}$.

Consider the shortest graph distance $d(v, v')$ in $\tilde{G}$. Since $v \in G$ and $v' \in G'$, any path $\mathcal{P}$ from $v$ to $v'$ has to cross over an odd number of times from $G$ to $G'$ or vice versa. The path $\mathcal{P}$ can thus be mapped to a loop in the graph $G$ which has an odd overlap with the support of $\overline{X}_1$: We start at vertex $v$ and we replace each cross-over edge $(u', v)$ or $(u, v')$ that we encounter on the path $\mathcal{P}$ by the original edge $(u, v)$. We obtain a path $\mathcal{P}$ that will stay in the graph $G$ and which comes back to $v$ itself. Since the number of cross-over edges used is odd, this closed $\overline{Z}$ loop in $G$ will anti-commute with $\overline{X}_1$. This ensures that $\overline{Z}$ is indeed a logical operator. Note that this anti-commutation property is independent of which representative of $\overline{X}_1$.

Thus in order to determine the minimum weight of a $\overline{Z}$ operator which anti-commutes with $\overline{X}_1$, we iterate over the points $v$ such that $(u,v) \in E(\overline{X}_1)$, and for each choice of $v$ one determines the shortest graph distance $d(v,v')$. The shortest graph distance is calculated using Dijkstra's algorithm which is efficient in the number of vertices of the graph. One then takes the minimum over all these graph distances. Of course the found $\overline{Z}$ may also anti-commute with other $\overline{X}_i$.

In order to determine $d_Z$ we iterate over the elements of the logical operator basis $\overline{X}_1,\ldots,\overline{X}_k$. Assuming a list of $\overline{X}_1,\ldots,\overline{X}_k$, the procedure is $O(kn^2 \log n)$ where $n = |E|$.

### 3.4.2 Counting minimum weight logical operators

To determine the number $N_d^Z$ of logical $Z$ operators of minimum weight $d_Z$ we can simply apply the same procedure and keep a list of the minimum weight logical operators that we have found. In particular, given a fixed $\overline{X}_i$ and a fixed vertex $v$, one can run Dijkstra's algorithm to determine all paths between vertices $v$ and $v'$ of given minimal distance. One then collects these paths into a list, avoiding adding the same path twice. This process is iterated over $v$ and over all $\overline{X}_i$.

## 3.5 Small hyperbolic codes

In this section we will consider explicit examples of hyperbolic codes. In 3.5.1 we discuss hyperbolic codes with less than 10.000 physical qubits. We first show how to perform a complete search using a computer algebra system and then discuss the properties of the codes that we found. In 3.5.2 we discuss how small examples may be represented in Euclidean space. It turns out that some small hyperbolic codes correspond to self-intersecting polyhedra called star-polyhedra.

### 3.5.1 2D Hyperbolic codes with less than 10.000 qubits

#### Exhaustive search for tessellations using a CAS

Finding $\{r,s\}$-tessellations of closed surfaces reduces to finding normal subgroups of the symmetry group of the $\{r,s\}$-tessellation of $\mathbb{H}^2$, according to Proposition 3.6 in Section 3.1.5. The normal subgroups $\Gamma$ of $G_{r,s}$ can be found with the help of *computer algebra systems (CAS)* such as GAP or MAGMA. Given a maximal size $N$ of the quotient group the CAS can enumerate all normal subgroups $\Gamma$ such that $|G_{r,s}/\Gamma| < N$. We then check for each group $\Gamma$ whether it acts fixed-point free on $\mathbb{H}^2$ as follows: The only elements of $G_{r,s}$ which have fixed-points are the reflections $a$, $b$, $c$ and the rotations $\rho^i = (ab)^i$ and $\sigma^i = (bc)^i$ as well as all conjugates of these elements. Since $\Gamma$ is normal, we do not have to check all conjugations but only $a$,$b$, $c$, $\rho^i$ and $\sigma^i$. Hence, $\Gamma$ acts fixed-point free if it does not contain any of these elements. This is equivalent to the statement that these elements have non-trivial cosets with respect to $\Gamma$. This can be checked in the CAS by finding a representation of $G_{r,s}/\Gamma$ in a symmetric group $\phi : G_{r,s}/\Gamma \to S_m$ for some $m < N$. This is always possible since such a representation exists for any finite group. A function finding such a representation is implemented in most CAS. Once we have obtained the representation $\phi$, we have to test whether $\phi(g\Gamma)$ for $g \in \{a,b,c,\rho,\sigma\}$ is the trivial permutation, which can be done efficiently.

Each quotient group $G_{r,s}/\Gamma$ is the symmetry group of a quotient surface $\mathbb{H}^2/G_{r,s}$ tessellated by an $\{r,s\}$-tessellation. The cells of a tessellation are obtained by computing the cosets of $\langle a,b\rangle$, $\langle a,c\rangle$ and $\langle b,c\rangle$ corresponding to faces, edges and vertices, respectively. Since the group $\langle a,c\rangle = \{e,a,c,ac,ca\}$ contains 4 elements, the number of edges of a tessellation (and thus the number of qubits in the derived code) is set by the quotient group $n = |E| = |G_{r,s}/\Gamma|/4$.

Figure 3.7 shows the results of an exhaustive search for all hyperbolic codes with the number of physical qubits less than $10^4$. We chose the regular tessellation $\{5,4\}$ as it gives a code with the lowest stabilizer weights 5 and 4. We therefore consider it the most promising candidate of the hyperbolic codes. Furthermore, we consider the self-dual tessellations with $r,s \in \{5,7,13\}$. The reason for picking prime numbers is the fact that the search for normal subgroups was slowed down due to the existence of many subgroups which have fixed-points. The explanation for the existence of these subgroups is the following: the rotation groups $\langle\rho\rangle$ and $\langle\sigma\rangle$ and all of their subgroups have fixed-points. The rotation groups are isomorphic to cyclic groups $\mathbb{Z}_r$ and $\mathbb{Z}_s$ and their subgroups are all cyclic groups $\mathbb{Z}_k$ with $k$ dividing $r$ or $s$, respectively. Hence, if $r$ and $s$ are non-prime there are many subgroups which have fixed-points, slowing down the search significantly.

### Discussion of small hyperbolic codes

In each plot in Figure 3.7 a hyperbolic code is indicated by a marker, where the $x$-axis indicates its number of physical qubits and the $y$-axis indicates its distance $d$. We see that there exist hyperbolic codes with different $n$ for fixed $d$. Remember that the number of logical qubits is proportional to $n$ and does not need to be plotted. We are particularly interested in the following codes.

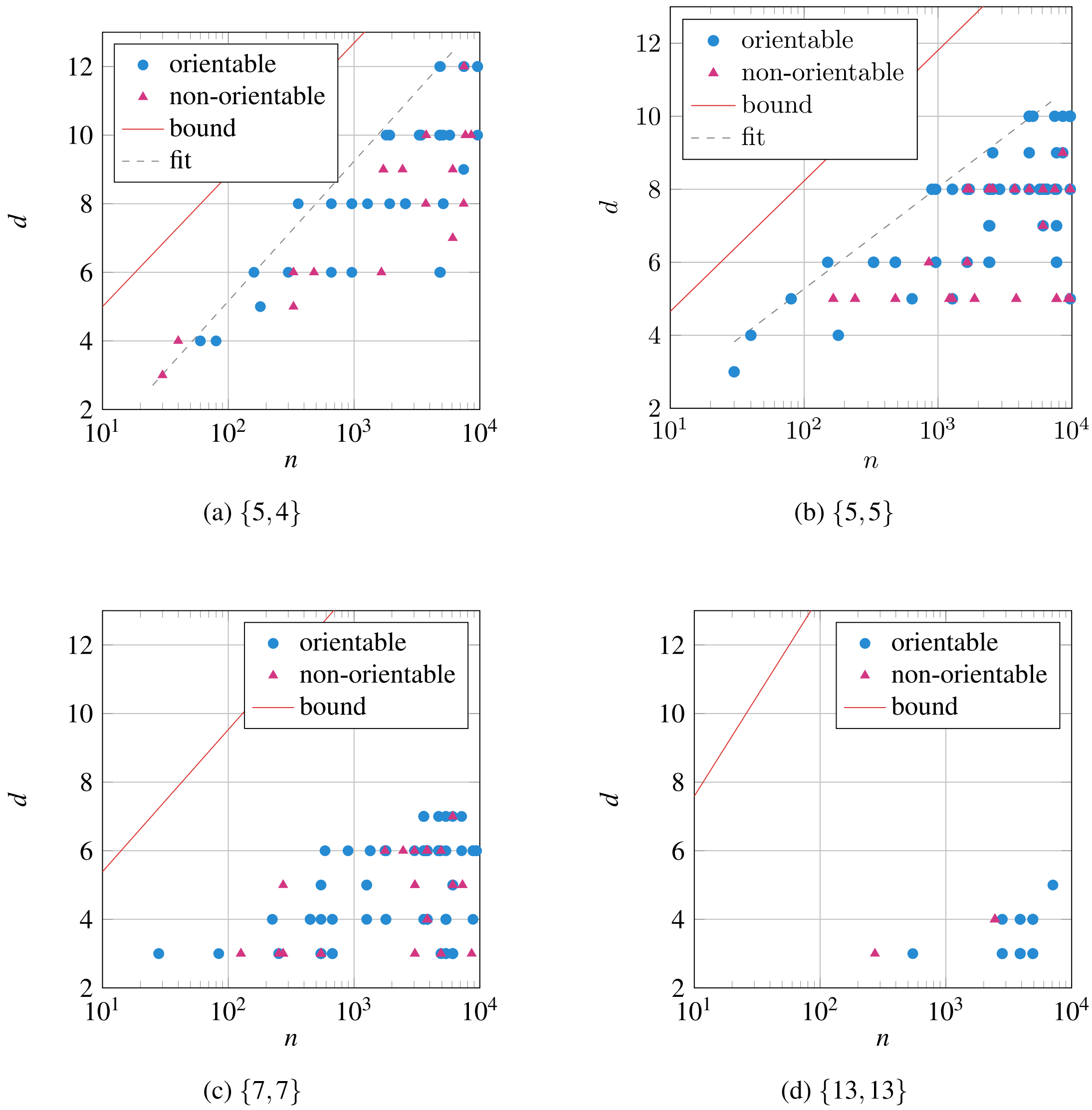

Figure 3.7: Hyperbolic codes with $n < 10^4$. They are obtained by enumerating all closed hyperbolic surfaces which admit an $\{r,s\}$-tessellation with less than $10^4$ edges. Each plot shows the number of qubits $n$ versus the distance of the codes on a log-linear scale. The red line shows the bound given in Equation 3.18.

**Definition 3.9.** A $\{r,s\}$-hyperbolic code is *extremal* if there exists no $\{r,s\}$-hyperbolic code with less physical qubits and larger or equal distance.

In Figure 3.7 the extremal hyperbolic codes are the upper leftmost ones. We can make a few observations based on the codes that we consider here. Note that these observations hold for the codes considered here and may not generalize to arbitrarily large codes. The first observation is that the extremal codes approximately lie on a line on the log-linear plots. This means that the extremal codes have their distance scaling as $\Theta(\log(n))$. There does not necessarily exist an extremal code for every distance $d \geq 3$. For example, the $\{5,4\}$-code does not have an extremal code of distance $d = 7$.

The parameters of the extremal codes are shown in Table 3.1. The table includes the number of physical qubits $n$, the number of encoded qubits $k$ and the distance $d$. The lowest weight of a purely $Z$-type ($X$-type) operator $d_Z$ ($d_X$) are the shortest essential loops (in the dual tessellation) (see Table 2.1). For small examples we give the number of those shortest loops $N_d^Z$ and $N_d^X$. The last column includes the group elements which generate the compactification group $\Gamma$. If they are purely translations, they can be expressed in terms of rotations around the center of faces $\rho$ and around vertices $\sigma$. In this case $\Gamma$ is a subgroup of the group of orientation-preserving symmetries of the tessellation $G^+_{r,s}$. If the surface is not orientable, the group $\Gamma$ contains glide-reflections which are translations followed by a subsequent reflection. These we write in terms of the basic reflections $a$, $b$ and $c$ which generate $G_{r,s}$ (see Section 3.1.5). The list of generators of larger codes was too long to fit into the table. Note that some of those group elements of the $\{5,4\}$-tessellation include the translation $\rho\sigma^{-1}$, which we have already seen in Figure 3.6.

Extremal codes with fixed parameters $[[n,k,d]]$ are not necessarily unique. For example, there exist two extremal $\{5,4\}$-codes with parameters $[[1710,173,9]]$. However, they differ in the number of lowest weight logical operators $N_d^Z$ and $N_d^X$ as well in their $X$-distance $d_X$. This difference influences the performance of the codes as we are going to see in the next chapter. It is advantageous in this case to take the code with less lowest weight representatives of logical operators, i.e. with the smallest $N_d^Z$ and $N_d^X$.

### 3.5.2 Representation of hyperbolic surfaces

A hyperbolic surface can generally not be embedded in 3D Euclidean space $\mathbb{E}^3$ without stretching it or making it self-intersect. In this section we present two different representations of hyperbolic surfaces. The first assumes that variable coupling lengths are allowed and that qubits can be located on both sides of a disk (a bilayer). It works for all codes based on orientable surfaces. The second representation relates to star-polyhedra. It preserves distances but only few examples of codes are known which have such a representation.

| Type | $n$ | $k$ | $d$ | $d_Z$ ($N_d^Z$) | $d_X$ ($N_d^X$) | translations, glide-reflections generating $\Gamma$ |
|---|---|---|---|---|---|---|
| $\{5,4\}$ | 30 | 5 | 3 | 3 (10) | 4 (75) | $abcba(cb)^2abcb$, $(bac)^6$, $(bacba)^4$ |
| $k/n \to 1/10$ | 40 | 6 | 4 | 4 (10) | 4 (20) | – |
| | 160 | 18 | 6 | 8 (500) | 6 (320) | $\sigma\rho^2(\rho\sigma^{-1})^{-2}\rho^{-1}\sigma^{-2}\rho^{-2}\sigma\rho^{-1}$ |
| | 360 | 38 | 8 | 8 (90) | 8 (5670) | $(\sigma\rho^2\sigma)^2(\rho^{-1}\sigma^{-2}\rho^{-1})^2$ |
| | 1710 | 173 | 9 | 9 (380) | 9 (28880) | – |
| | 1710 | 173 | 9 | 9 (190) | 10 (5814) | – |
| | 1800 | 182 | 10 | 10 (180) | 10 (32320) | $(\rho\sigma^{-1})^{-10}$, $\sigma\rho^2\sigma^2\rho^{-1}\sigma(\rho^2\sigma^{-1})^2(\rho\sigma^{-1})^2\sigma^{-1}\rho^{-2}\sigma\rho^{-1}$ |
| | 4860 | 488 | 12 | 12 | 12 | – |
| $\{5,5\}$ | 15 | 7 | 3 | 3 (20) | 3 (20) | $(bcba)^3$, $(bc(bca)^2)^2$, $(bc(ab)^2ca)^2$ |
| $k/n \to 1/5$ | 40 | 10 | 4 | 4 (40) | 4 (40) | $\sigma\rho^2\sigma\rho^{-1}\sigma^{-2}\rho^{-1}$ |
| | 80 | 18 | 5 | 5 (160) | 5 (160) | $\sigma(\rho\sigma^{-1})^2\sigma^{-1}\rho^{-2}\sigma\rho^{-1}$ |
| | 150 | 32 | 6 | 6 (500) | 6 (500) | $\sigma\rho^2\sigma^2\rho\sigma^{-1}\rho^{-2}\sigma^{-2}\rho^{-1}$, $\sigma(\rho\sigma^{-1})^3(\rho^{-1}\sigma)^2\rho^{-1}$ |
| | 900 | 182 | 8 | 8 (4725) | 8 (4725) | $\sigma(\rho^2\sigma^{-2})^2\rho^{-1}\sigma\rho^{-2}\sigma^2\rho^{-1}$ |
| | 2560 | 514 | 9 | 9 (10240) | 9 (10240) | – |
| | 4800 | 960 | 10 | 10 | 10 | – |
| $\{7,7\}$ | 28 | 14 | 3 | 3 (56) | 3 (56) | $\rho^3\sigma^{-2}\rho\sigma^{-1}$ |
| $k/n \to 3/7$ | 224 | 98 | 4 | 4 (336) | 4 (336) | – |
| | 273 | 119 | 5 | 5 (1092) | 5 (1092) | – |
| | 588 | 254 | 6 | 6 (7252) | 6 (7252) | – |
| | 3584 | 1538 | 7 | 7 (21504) | 7 (21504) | – |
| $\{13,13\}$ | 273 | 191 | 3 | 3 (182) | 3 (182) | – |
| $k/n \to 9/13$ | 273 | 191 | 3 | 3 (364) | 3 (364) | – |
| | 2457 | 1703 | 4 | 4 (9828) | 5 (29484) | – |
| | 7098 | 4916 | 5 | 5 | 5 | – |

Table 3.1: Extremal codes with $n < 10^4$. The last column shows the translations and glide-reflections $t_1,\dots,t_m$ which together with all of their conjugates generate $\Gamma$, i.e. $\Gamma = \langle gt_1g^{-1},\dots,gt_mg^{-1} \mid g \in G_{r,s}^{(+)}\rangle$. For orientable surfaces we use the generators $\rho$ and $\sigma$. A dash indicates that the expression for the $t_i$ is too long to fit into this table.

### Bi-layer

All codes derived from *orientable* surfaces can be realized on a bi-layer structure with holes. This follows from the fact that all orientable surfaces are completely determined by their genus (number of handles, see Proposition 2.13 in Section 2.3.3). Hence, any such surface can be deformed into a surface similar to the one in Figure 3.8 on the left. This deformation will not preserve lengths on the surface. The deformation will not make any elements of the tessellation overlap so that in the resulting layout the interactions will still be planar on the surface. Finally, we can flatten the surface to obtain a bi-layer with holes on which interactions are not crossing.

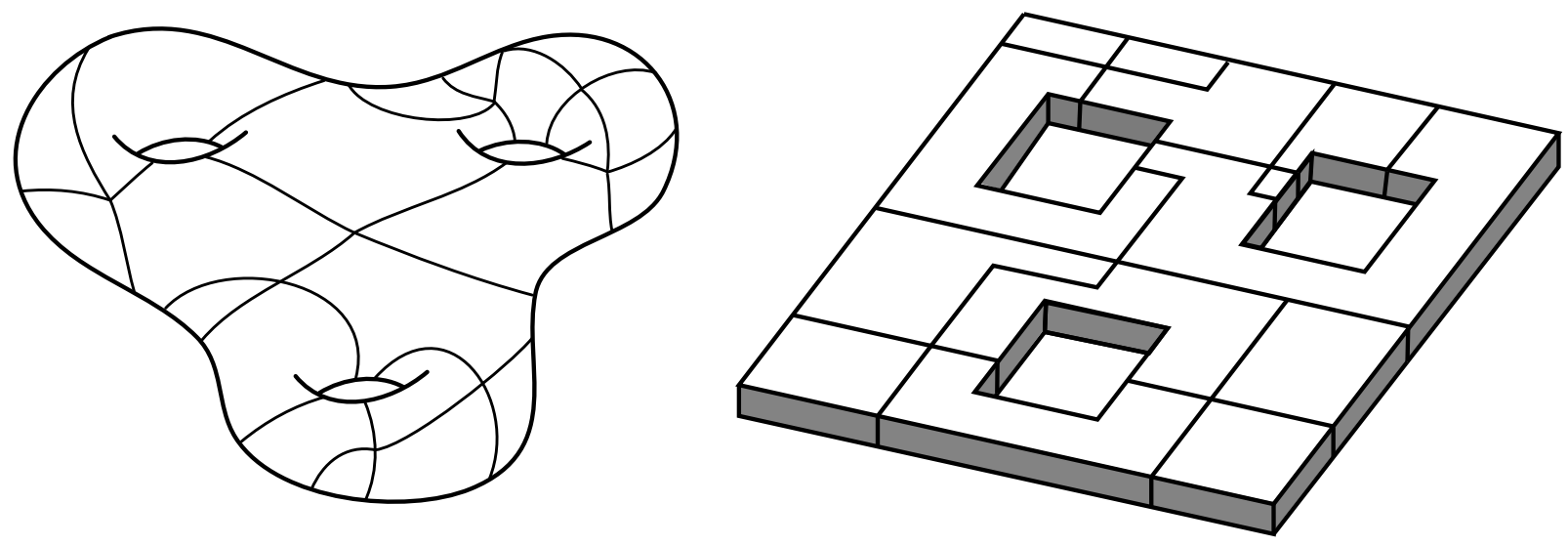

Figure 3.8: Illustration of how to turn an orientable tessellated genus-3 hyperbolic surface into a bi-layer structure. We first find an embedding of the surface into $\mathbb{E}^3$ (left). Such an embedding will not preserve distances on the surface and thus deform the tessellation. The embedding can be flattened until it corresponds to a bi-layer with $g$ holes (right).

### Star-polyhedra

Some hyperbolic surfaces correspond to certain non-convex polyhedra with intersecting faces. These polyhedra are called *star-polyhedra* as the faces meet in sharp cones and their faces can be pentagrams, giving them a star-like appearance. As star-polyhedra are self-intersecting they can be topologically non-trivial. The existence of a representation as a star-polyhedron is rather exceptional as we are going to argue later.

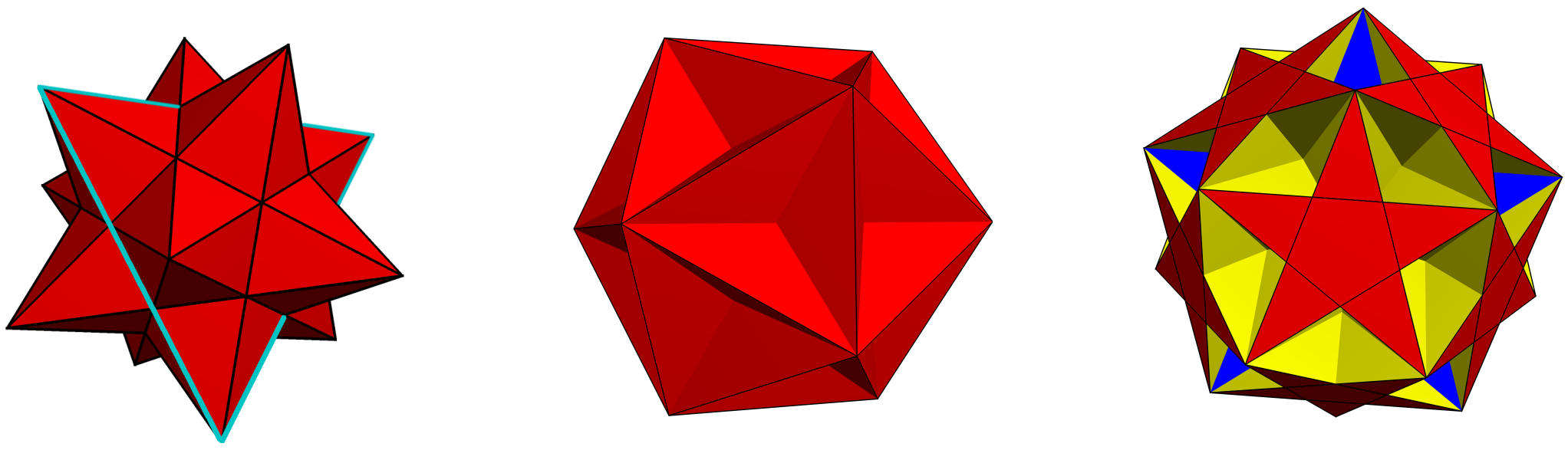

(a) small stellated dodecahedron

(b) great stellated dodecahedron

(c) ditriagonal dodecadodecahedron

Figure 3.9: Representations of two hyperbolic surfaces as star-polyhedra. (a) and (b) show two representations of the same $\{5,5\}$-tessellated genus-4 surface. (c) shows the representation of a $\{5,6\}$-tessellated genus-9 surface. In (a) an essential cycle of length 3 is highlighted.

There exist 3 hyperbolic surfaces which can be represented as star-polyhedra:

1. A $\{5,4\}$-tessellated surface with $\dim H_1 = 8$, tessellated by 24 faces, 60 edges and 30 vertices, corresponds to the *dodecadodecahedron* (see Figure 3.10). The shortest essential cycle (cocycle) has length 6 (4).

2. A $\{5,5\}$-tessellated surface with $\dim H_1 = 8$, tessellated by 12 faces, 30 edges and 12 vertices, corresponds to two star-polyhedra: The *small stellated dodecahedron* (see Figure 3.9a) and the *great stellated dodecahedron* (see Figure 3.9b).The shortest essential cycle (cocycle) has length 3 (3).

3. A $\{5,6\}$-tessellated surface with $\dim H_1 = 18$, tessellated by 24 faces, 60 edges and 20 vertices, corresponds to the *ditriagonal dodecadodecahedron* (see Figure 3.9c). The shortest essential cycle (cocycle) has length 4 (3).

The three examples were found by searching in the Great Stella Library [35].

A necessary condition for such a symmetric representation to exist is that symmetry group of the surface, which is a quotient group $G_{r,s}/\Gamma$ by Proposition 3.6, shares a subgroup with the

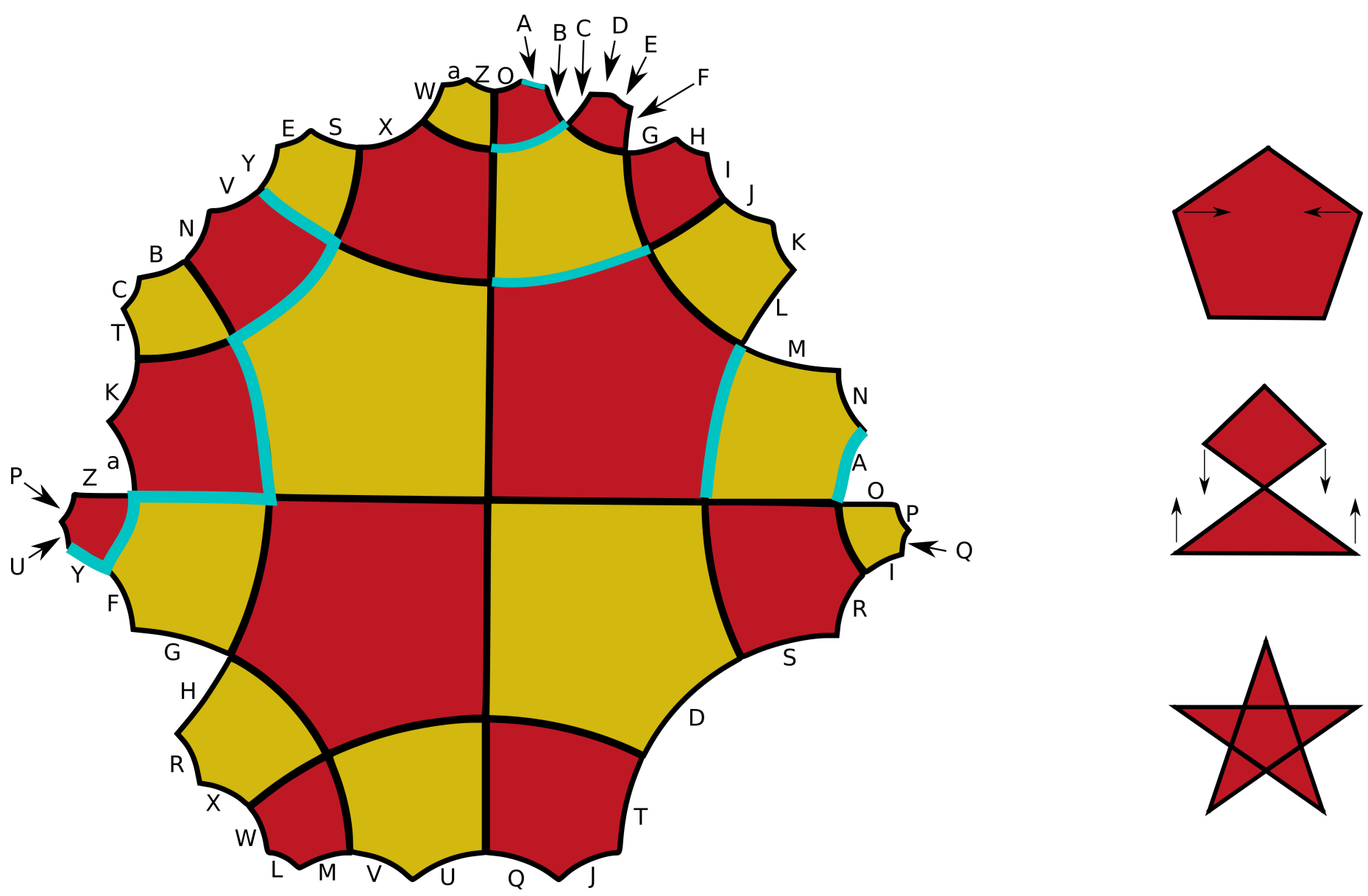

(a) $\{5,4\}$-Tessellation of a genus 4 surface. Edges at the boundary with the same label are identified. The $X$-checks are given by vertices and the $Z$-checks by faces. A logical operator $\overline{Z}$ of length 4 (an essential cycle) and a logical operator $\overline{X}$ of weight 6 (an essential cocycle) are highlighted.

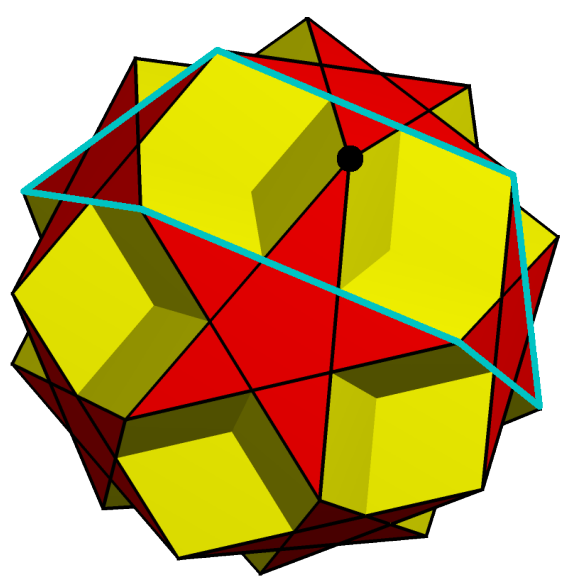

(b) Weight-6 logical $\overline{Z}$ (essential cycle)

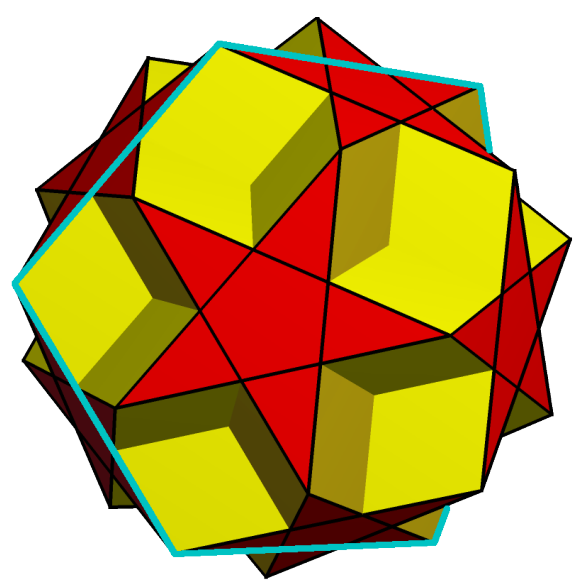

(c) Weight-6 logical $\overline{Z}$ (essential cycle)

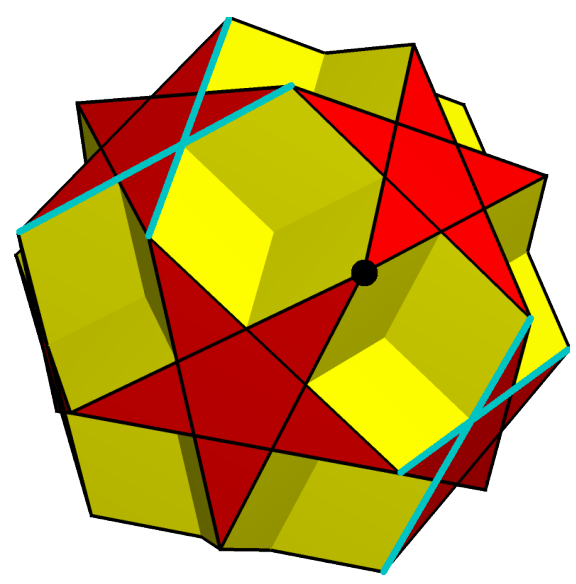

(d) Weight-4 logical $\overline{X}$ (essential cocycle)

Figure 3.10: The smallest extremal code based on a $\{5,4\}$-tessellation of an orientable surface has parameters $[[60,8,4]]$ and is related to a star-polyhedron called *dodecadodecahedron*. This can be seen as follows: Half of the faces (red) are deformed into pentagrams with self-intersecting edges, see (a). Arranging the vertices on the surface of a sphere and allowing for self-intersecting faces, gives the dodecadodecahedron, see (b)-(d). A vertex is highlighted in (b) and (d) by a dot in order to show that one can label some of the minimum-weight logical operators by the 30 vertices (cf. Table 3.1).

group of isometries of the sphere $\text{Isom}(\mathbb{S}^2) = O(3)$. The only discrete symmetries of the sphere are the symmetries of its 5 regular tessellations [27].[2] What are the symmetries of the star-polyhedra that correspond to hyperbolic surfaces? – It turns out that all four examples presented here have the symmetry group of the icosahedron $G_{3,5}$ as a subgroup. Let us show this explicitly for the dodecadodecahedron shown in Figure 3.10d. The three planes of reflective symmetry are all necessarily going through the middle of the dodecadodecahedron, cutting it in half. The three planes are (i) going through the middle of the two pentagrams meeting at the marked vertex, (ii) going through the marked vertex, cutting the pentagrams in the upper left and lower right in half (iii) going through the point where the middle and upper right pentagram meet, cutting through the pentagrams in the upper right and lower left. Plane (i) and (ii) meet at an angle $\pi/2$, plane (i) and (iii) meet at an angle $\pi/3$ and plane (i) and (iii) meet at an angle $\pi/5$, giving exactly the relations of the group $G_{3,5}$. Similarly, we can find planes of symmetry of the other examples meeting at the same angles.

## 3.6 Planar hyperbolic codes

One can try to construct a hyperbolic code which does not correspond to a closed surface, but instead to a topological disk. Such codes, if they have good distance and rate, would have a larger practical appeal, as planar architectures are easier to physically implement. We are going to discuss two variations on how to achieve this goal: First, in 3.6.1 we show how from a given hyperbolic surface one can obtain a topological disk with holes. The logical operators surround or connect these holes. Such a planar topological code was first introduced by Freedman and Meyer in [28]. In 3.6.2 we discuss a second variation which is a tessellated planar surface with some of the qubits and stabilizer checks at its boundary removed. The logical operators run between these boundaries. This encoding was introduced by Bravyi and Kitaev in [1].

### 3.6.1 Hole encoding

In [28] it was first described how a topological disk with holes can encode multiple qubits. This type of encoding is used in the surface code architecture described in [36] where qubits are encoded in both rough or smooth double-holes punctured in the surface. For this encoding, there is a simple scheme with which one can perform a CNOT gate, first described in [37].

[2] There are two infinite families of discrete subgroups of $O(3)$ which we do not consider: The dihedral groups $D_m$ and cyclic groups $\mathbb{Z}_m$.

One could imagine obtaining such a punctured surface by cutting open the hyperbolic surface. One would start with an orientable surface with $g$ handles and no boundary, nor punctures, encoding $k = \dim H_1 = 2g$ qubits. If one cuts open one handle, one has two punctures and $g-1$ handles in the remaining surface. Cutting open another disjoint handle similarly generates two punctures and cutting open the last handle leads to a surface with $2g-1$ punctures and one outer boundary (the last puncture). For a tiled surface one can make a *smooth cut* or a *rough cut*. For a smooth cut, one follows an essential cycle $\gamma$ (corresponding to a logical $\overline{Z}$) on the graph around the handle and all edges $e \in E_\gamma$ on this loop $\gamma$ are replaced by double pairs of edges, i.e. $E_\gamma \to E_\gamma^A \cup E_\gamma^B$ so that the two $Z$-checks (faces) adjoining these edges either act on edges in $E_\gamma^A$ or $E_\gamma^B$ (but not both). At each vertex on this path the number of $X$-check operators is doubled such that one $X$-check operator acts on the edges in $E_\gamma^A$ and the other on the edges in $E_\gamma^B$. Note that in this procedure, one adds as many qubits as one adds $X$-check operators as the number of edges and vertices around a loop are the same. For the original closed surface, there is one linear dependency between all $Z$-checks and one linear dependency for the $X$-checks by Equation 2.44 and Equation 2.45. For the final punctured surface with smooth holes, there is only a linear dependency between all $X$-checks: This directly implies that the punctured surface will encode $2g-1$ logical qubits. Another way of seeing that a surface with a smooth outer boundary and $2g-1$ smooth punctures or holes encodes $2g-1$ logical qubits is by enumerating the logical operators: For each hole the logical $\overline{Z}$ is a cycle around the hole, while the logical $\overline{X}$ is a cocycle to the outer boundary (see e.g. [5]). However, one does not need to use all these logical qubits.

In the smooth double-hole encoding described in [5, 36], one uses a pair of holes to encode one qubit where the logical $\overline{Z}$ is the cycle around any of the two holes and the logical $\overline{X}$ is the $X$-distance between the two holes. In this way, the surface with $2g-1$ holes can encode $g$ "double-hole qubits" where the last double-hole qubit is formed by the last hole plus the boundary 'hole'.

However, does this procedure preserve the distance of the original code and is it even possible to execute this procedure on the graph obtained from a $\{r,s\}$-tessellation of the closed hyperbolic surface? – The answer is in fact no. If we have a family of $\{r,s\}$-codes with increasing $k \sim n$ and distance lower bounded as $c_{r,s} \log n$, then logical $\overline{Z}$ operators, which act on at least on $c_{r,s} \log n$ qubits, must overlap on many qubits. If there are $\Omega(k)$ such non-intersecting loops, then there must be at least $\Omega(k\log n) = \Omega(n\log n)$ qubits in total, which is not the case. Hence, one cannot find such a set of non-overlapping loops along which to cut as logical operators must share a lot of support.

### 3.6.2 Processing the boundary

Another method of cutting the hyperbolic surface to create smooth and rough boundaries would allow one to create a coding region as in Figure 3.11. One can encode multiple qubits into such a surface code [1] by using alternating rough and smooth boundary of sufficiently large length. The boundary of the encoding region is divided into $2k$ regions and encodes $k-1$ logical qubits. The logical $\overline{Z}$ operators can start and terminate at rough boundary regions while the logical $\overline{X}$ operators start and terminate at smooth boundary regions. One can consider the asymptotic scaling of the parameters $n,k$ and $d$ of the code represented by such an encoding region, where we do not assume anything about the distribution of qubits inside the region, meaning that for such a class of codes with increasing $n$ qubits can get closer together or further apart, if we represent the qubits on the Euclidean plane. One can simply argue that for such a family of homological surface codes with boundaries the following bound should hold

$$kd \leq cn, \tag{3.20}$$

with a constant $c$. This bound shows that for codes based on tessellating a surface with open boundaries, it is not possible to have a constant encoding rate $k/n$ *and* a distance increasing as $\log n$. We argue this bound for the encoding region in Figure 3.11, but similar arguments should hold for the tessellation of a surface with holes. The simple argument goes as follows. In order to encode $k-1$ qubits, one divides the set of boundary edges $E_{\text{bound}}$ into $2k$ regions. Clearly, the total number of qubits $n \geq E_{\text{bound}}$. The logical operators run from rough to a rough or smooth to smooth boundary, hence their distance is upper-bounded by the length of a region $c|E_{\text{bound}}|/k \leq cn/k$ for some constant $c$. This is true as one can always let the logical operator run along (or close to) the boundary edges. This results in Equation 3.20. If the encoding region is a disk with many holes, we can argue similarly. We enumerate the number of edges (qubits) around the non-trivial disjoint holes which should be less than the total number of qubits $n$. The number of logical qubits $k$ scales linearly with the number of holes and the distance $d$ is at most the number of edges around each hole, resulting in Equation 3.20.

Thus we may ask whether using hyperbolic geometry gives any advantage over Euclidean geometry when considering tessellated surfaces with a boundary. The previously proven bound $kd^2 \leq cn$ is still worse than the simple bound $kd \leq cn$. It is worth noting that encoding via the boundaries as in Figure 3.11 has worse scaling than an optimal surface code encoding in the case of a Euclidean metric on the underlying qubits. In this case, the perimeter of the polygon scales at most as $\sqrt{n}$ where $n$ is the total number of qubits. This implies that the distance of the code $d \leq c\sqrt{n}/k$, which is worse than the $kd^2 \leq cn$ Euclidean bound from [38]. If we use hyperbolic geometry and imagine the encoding region as a partially-tiled hyperbolic plane, see the explicit construction below, then the number of qubits at the boundary scales like the total number of qubits, which seems promising. However, the minimum weight logical operators which run from boundary to boundary, will run along shortest paths, geodesics, which go through the interior of the region. We show a small explicit example of such a code in the next section, but the underlying construction is unlikely to give good asymptotic scaling behavior.

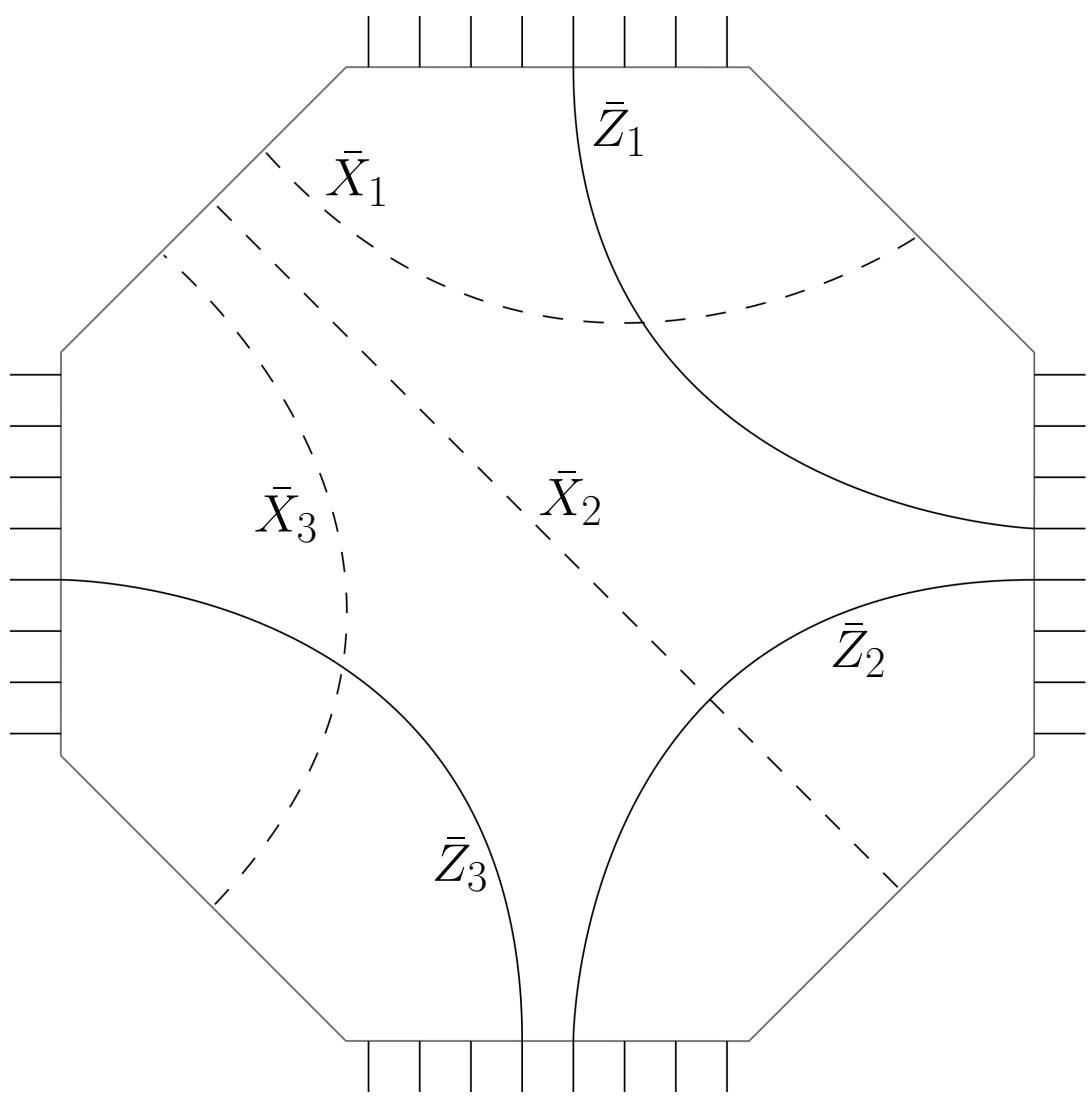

Figure 3.11: A polygon with $2k$ sides, alternatingly 'rough' and 'smooth' (see definitions in Section 3.6.3), can encode $k-1$ logical qubits [1]. Shown is $k = 4$ and a choice for the logical $\overline{X}_i, \overline{Z}_i$ operators.

### 3.6.3 Small Planar Code Example

We generate a planar graph with a finite number of vertices by taking a regular tessellation $\{r,s\}$, partially tessellating the hyperbolic plane $\mathbb{H}^2$ and then modifying the boundary so that the planar graph encodes logical qubits, as in Figure 3.11. An example of the procedure to create such a graph encoding multiple qubits consists of the following steps; we illustrate the idea in Figure 3.12 for the $\{5,5\}$-tessellation:

1. For an $\{r,s\}$-tessellation, start with a single $r$-gon (call it level-1 $r$-gon) and reflect in the edges of this $r$-gon to generate level-2 $r$-gons. Repeat this for the level-2 $r$-gons etc. so that one obtains a planar graph $G=(V,E)$ where the faces are the level-1 to level-m $r$-gons for some $m$. In the figure, we have started with 4 level-1 $r$-gons and generated only the level-2 $r$-gons and have stopped there. With every face $f$ we associate a $Z$-check acting on the qubits on the boundary of this face, and with every vertex $v$ of this graph one associates a star $X$-check acting on the qubits on edges adjacent to the vertex. At the boundaries the weight of the $X$-checks will be two or $s$ (for even $s$) or $s-1$ (for odd $s$), in the interior the $X$-checks have weight $s$. The code associated with this *starting graph G* encodes no qubits. There is one linear dependency between all $X$-checks and so the number of linearly independent $X$-checks is $V-1$. There is no linear dependency between the $Z$-checks of which there are $F$. As this is a planar graph (with Euler characteristic $\chi=1$), one has $E=V-1+F$ where $E$ is the number of edges, equal to the number of qubits $n$, hence no encoded qubits.

All boundaries in this graph are so-called smooth boundaries at which a string of $X$-errors can start and end. For this graph which encodes no qubits, such a string can be annihilated by star $X$-checks, hence there are no logical operators. Thus we need to modify this graph in order to create so-called rough boundaries. At rough boundaries a string of $Z$-errors can start. If one alternates rough and smooth boundaries as in Figure 3.11, these strings can no longer be annihilated by $X$- or $Z$-checks, but have to run from boundary to boundary. What is the procedure for creating several rough boundaries at which $Z$-strings can end?

2. First, for the given starting graph $G$, one counts the number of boundary edges in $E_{\text{bound}}$ as $|E_{\text{bound}}|$: These are defined to be the edges that the level-$k$ $r$-gons would be reflected in to generate level $k+1$ $r$-gons. In order to create a code where all encoded qubits have about the same distance, one wants to divide the set $E_{\text{bound}}$ in $2k$ equally-sized sets or regions with $k>2$ so that each subset of size $\lfloor |E_{\text{bound}}|/(2k) \rfloor$ corresponds to a rough or smooth boundary. Such a code can encode $k-1$

qubits [1]: The logical $\overline{Z}_i$, $i = 1, \ldots, k-1$ will run from the $i$th to the $(i+1)$th rough boundary, while the logical $\overline{X}_i$ will run from the smooth boundary in between the $i$th and $(i+1)$th rough boundary to the $(i+1)$th smooth boundary (see Figure 3.11). If $k$ is chosen too large, then each smooth/rough region would be too short and the shortest path between two regions would not go through the bulk. Hence one should choose $k$ such that the shortest path along the perimeter is about the same length as the shortest path through the bulk: in that case one uses the available qubit-space optimally. This choice is illustrated in Figure 3.12 where one has 60 edges along the boundary which we divide up into $10 = 2k$ regions of 6 edges each, thus encoding 4 logical qubits.

The creation of a rough boundary in a region of edges $E_{\text{bound}}^{\text{region}} \subseteq E_{\text{bound}}$ consists of the following 3 steps (variants are possible). First, one removes all $X$-check operators of weight-2 which have 2 edges among $E_{\text{bound}}^{\text{region}}$ (if the weight-2 $X$-check has an edge in $E_{\text{bound}}^{\text{region}}$ one also removes it). One could in principle also remove checks with weight more than 2 at the boundary but this makes the tessellation a bit smaller, so in this construction we prefer to remove only the weight-2 checks. Then all qubits on which only a single $Z$-check acts are removed from the tessellation and these are thus modified (in the toric code they are the weight-3 at the boundary). The removal of the $X$-checks makes it possible for a $Z$-string to start on an edge attached to a removed $X$-check. We need to make sure that such a string can only run from rough boundary to another rough boundary. Since we did not remove some $X$-checks in the rough region (some weight-5 checks in Figure 3.12), in general of weight $s$), one needs to add weight-2 $ZZ$ checks to the stabilizer, as these operators commute with all the current checks. These weight-2 checks are indicated (in red) in Figure 3.12. For the construction in Figure 3.12, one can verify that the minimum weight logical $\overline{Z}$ is of weight-4, while the minimum-weight logical $\overline{X}$ is of weight-5: we draw examples of these logical operators in the figure. The total number of physical qubits is $n = 65$. The most efficient use of the surface code (in which the tessellation is chopped off at the boundaries, see [39]) has parameters $[[d^2, 1, d]]$ (instead of $[[d^2 + (d-1)^2, 1, d]]$). Hence to encode 4 qubits with the surface code with distance 4 requires 64 physical qubits and distance 5 would require 100 qubits, showing that this simple hyperbolic construction can lead to a somewhat more efficient coding than the surface code.

In order to construct a planar graph encoding $k-1$ logical qubits, one can also generate the planar starting graph $G$ encoding no qubits by repeatedly rotating an elementary $r$-gon by $2\pi/s$ around its vertices. By this rotation, one generates a new generation of $r$-gons, the level-2 $r$-gons etc. In this construction, one removes all $X$-checks in the region where the rough boundary has to be formed. This means that no weight-2 $ZZ$ checks need to be added. Only vertices on which a single plaquette $Z$-operator acts are to be further removed.

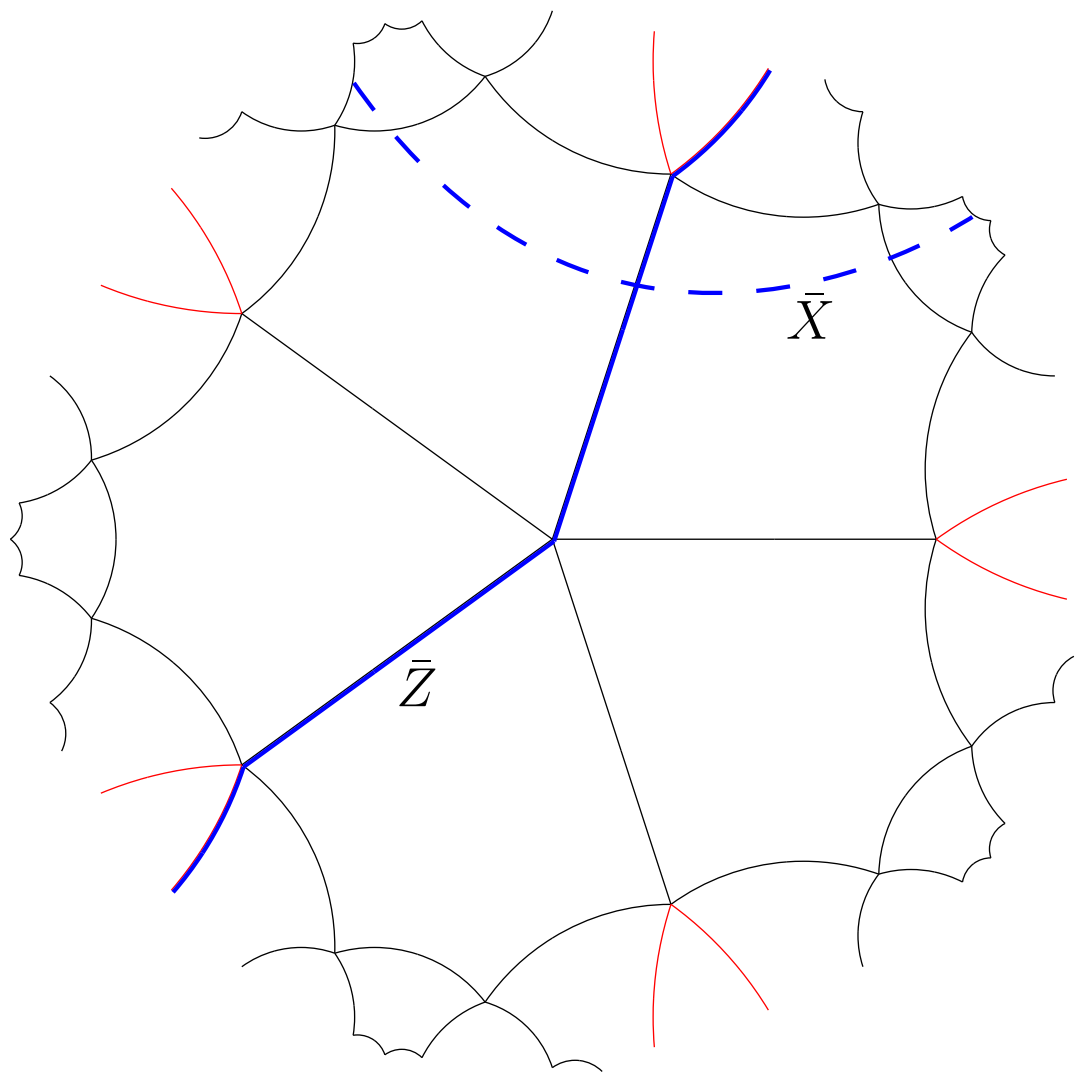

Figure 3.12: A [[65,4,4]] code based on the $\{5,5\}$-tessellation. The distance of the logical $\overline{X}$ is in fact 5 while it is 4 for the logical $\overline{Z}$. The number of boundary edges in $E_{\text{bound}}$ of the starting graph was 60 and was divided into 10 regions each with 6 edges. Shown in red are the additional weight-2 $Z$-checks.

## 3.7 Proof of threshold

In this section we will show that the hyperbolic codes have a *decoding threshold* against independent $X$- and $Z$-errors. This means that as long as the probability $p$ with which a bit-flip $X$ or phase-flip $Z$ occurs is below a certain critical value $p_c > 0$, a correct recovery operation is found with probability approaching 1 in the large code limit $n \to \infty$.

The threshold certainly depends on the algorithm used to infer the recovery operation. In the proof that we are going to present we assume that the code is decoded using a *minimum weight decoder*, which, given a syndrome, returns a recovery operation $R$ which is consistent with the syndrome and which has minimal weight among all recovery operations for the given syndrome.

Minimum-weight decoding is a good strategy when single-qubit errors are distributed uniformly and independently, as lower weight errors have higher probability. However, a minimum-weight decoder does not take degeneracies into account, i.e. multiplicities of errors with the same syndrome. But even though minimum-weight decoding is sub-optimal, it is sufficient to have a decoding threshold.

### 3.7.1 Assuming perfect measurements

Let us for now assume that we can perform the measurements of the $Z$-checks and $X$-checks perfectly. This means that for any error which acts as Pauli-$Z$ on some 1-chain $E_Z$ and as Pauli-$X$ on some 1-chain $E_X$ we can exactly determine the boundaries $\partial_1 E_Z$ and $\partial_1 E_X$ (cf. Section 2.3.5). Later we are going to relax this assumption and show that we can still have a threshold even when the measurements can contain errors.

The strategy to prove that a threshold exists is the following: We derive an upper bound $\hat{P}$ on the probability $\overline{P}$ that a logical error occurs after minimum-weight decoding. This upper bound $\hat{P}$ will depend on the physical error rate $p$ and the number of qubits $n$. Furthermore, $\hat{P}$ will be a simple expression which allows us to show that for any $p$ below a certain value $\hat{p}_c > 0$ we have $\hat{P} \to 0$ for $n \to \infty$. Since $\hat{P} > \overline{P}$ we have that $\hat{p}_c$ is a non-zero lower bound on the critical error rate $p_c > \hat{p}_c > 0$. This argument was first given in [40] for the toric code and later generalized in [41] to quantum codes with logarithmic scaling distance.

We will now derive the upper bound $\hat{P}$ on the failure probability $\overline{P}$. To avoid mentioning $X$- and $Z$-errors in every step we will only consider $Z$-errors and failure due to the appearance of $Z$-logicals. The exact same arguments apply to $X$-errors when going to the dual tessellation. Note that the overall failure rate and threshold are simply the maximum and minimum of the two cases.

Assume that $E$ is an error and $R$ is a recovery determined by the minimum-weight decoder. We will not differentiate between $\mathbb{Z}_2$-chains and sets. The sum of the error chain and the recovery chain $E+R$, interpreted as a set, is the symmetric difference $(E \setminus R) \cup (R \setminus E)$. The decoder has failed when $E+R$ contains an essential cycle. Since all essential cycles have length $\geq d$, we can bound $\overline{P}$ by the probability that $E+R$ contains a cycle $\gamma \in Z_1$ with $|\gamma| \geq d$:

$$\overline{P} = \text{Prob}\,(E+R \text{ contains an essential cycle}) \leq \sum_{\gamma \in Z_1 :\, |\gamma| \geq d} \text{Prob}(\gamma \subset E+R) \tag{3.21}$$

**Lemma 3.10.** *Assume that $\gamma \in Z_1$ is a fixed 1-cycle, $E$ is a random 1-chain where each edge is chosen uniformly and independently and $R$ is a 1-chain with the same boundary as $E$ and minimum weight.*

*For any such $E$ and $R$ we have that $\gamma \subset E+R$ implies $|\gamma \cap E| \geq |\gamma|/2$ and hence*

$$\text{Prob}\,(\gamma \subset E+R) \leq \text{Prob}\left(|\gamma \cap E| \geq \frac{|\gamma|}{2}\right). \tag{3.22}$$

*Proof.* To see that this is the case we show that $\gamma \subset E+R$ implies $|\gamma \cap E| \geq |\gamma|/2$. So let us assume that $\gamma \subset E+R$. We define $E_\gamma = (E \setminus R) \cap \gamma$ and $R_\gamma = (R \setminus E) \cap \gamma$. Note that $\gamma = E_\gamma + R_\gamma$. If we had $|E_\gamma| < |\gamma|/2$ then it would follow that $|R_\gamma| \geq |\gamma|/2$. But this would mean that we could find a recovery chain with smaller weight than $R$ by removing $R_\gamma$ from $R$ and adding $E_\gamma$ to $R$. This violates the assumption that $R$ has minimum weight and thus we must have $|\gamma \cap E| \geq |\gamma|/2$. ■

Substituting Equation 3.22 into Equation 3.21 gives

$$\overline{P} \leq \sum_{\gamma \in Z_1 \colon |\gamma| \geq d} \text{Prob}\left(|\gamma \cap E| \geq \frac{|\gamma|}{2}\right). \tag{3.23}$$

Assuming that every edge belongs to $E$ with probability $p$, we can rewrite:

$$\text{Prob}\left(|\gamma \cap E| \geq \frac{|\gamma|}{2}\right) = \sum_{i=\lceil |\gamma|/2 \rceil}^{|\gamma|} \binom{|\gamma|}{i} p^i (1-p)^{|\gamma|-i}. \tag{3.24}$$

Furthermore, the set of closed loops is contained in the set of all self-avoiding paths starting at any vertex in the tessellation. The number of self-avoiding paths of weight $w$ starting at a particular vertex in a $\{r,s\}$-tessellation can be upper bounded by $s(s-1)^{w-1}$. Note that $s|V|$ gives twice the number of edges, which is $2n$. Together with Equation 3.23 and Equation 3.24 we obtain

$$\begin{aligned} \overline{P} &\leq \sum_{\gamma \in Z_1 \colon |\gamma| \geq d} \sum_{i=\lceil |\gamma|/2 \rceil}^{|\gamma|} \binom{|\gamma|}{i} p^i (1-p)^{|\gamma|-i} \\ &\leq \sum_{w=d}^{n} |V|\, s(s-1)^{w-1} \sum_{i=\lceil w/2 \rceil}^{w} \binom{w}{i} p^i (1-p)^{w-i} \\ &\leq 2n \sum_{w=d}^{n} (s-1)^w\, 2^w p^{w/2}, \end{aligned} \tag{3.25}$$

where in the last step we have used $p^i(1-p)^{w-i} \leq p^{w/2}$ for the terms in the inner sum and the upper bound $\sum_{i=\lceil w/2 \rceil}^{w} \binom{w}{i} \leq 2^w$. By define $\alpha = 2(s-1)\sqrt{p}$ and perform the geometric sum we finally obtain the upper bound $\hat{P}$:

$$\overline{P} \leq 2n \sum_{w=d}^{n} \alpha^w = 2n \frac{\alpha^d - \alpha^{n+1}}{1-\alpha} \leq 2n \frac{\alpha^d}{1-\alpha} =: \hat{P} \tag{3.26}$$

For hyperbolic codes we have $d \geq c_{r,s} \log(n)$ with some constant $c_{r,s} > 0$ depending on the Schläfli symbol of the tessellation $\{r,s\}$ (see Section 3.3.2). Hence, for $\hat{P}$ to converge to 0 in the limit $n \to \infty$, it is sufficient that

$$p < \frac{\exp(-2/c_{r,s})}{4(s-1)^2}. \tag{3.27}$$

The same follows for $X$-errors with $4(r-1)^2$ in the denominator. Note that when we have both $Z$- and $X$-errors, we consider the application of a logical $\overline{Z}$ *or* $\overline{X}$ a failure. To take this into account we take the lower of the two bounds and obtain the following theorem.

**Theorem 3.11.** *The threshold $p_c$ of an $\{r,s\}$-hyperbolic code against independent $X$ and $Z$ errors under minimum-weight decoding is lower bounded by*

$$\hat{p}_c = \frac{\exp(-2/c_{r,s})}{4(\max(r,s)-1)^2}. \tag{3.28}$$

Note that the derivation of Theorem 3.11 did not rely on the fact that we consider hyperbolic codes. In fact, Theorem 3.11 applies to all codes with distance $d$ lower bounded by a logarithmically growing function $c\log(n)$ and weight of $X$-checks ($Z$-checks) at most $s$ ($r$). For codes where $d$ grows polynomially, as for example for the toric code, we only need to ensure in Equation 3.26 that $\alpha < 1$, so that the bound becomes $\hat{p}_c = [4(\max(r,s)-1)^2]^{-1}$.

From Equation 3.26 also follows this related result.

**Corollary 3.12.** *Consider a $\{r,s\}$-hyperbolic code with $d > c_{r,s}\log(n)$, decoded by the minimum-weight decoder. For error rate below the threshold bound $p < \hat{p}_c$ the logical error rate $\overline{P}$ is* polynomially *suppressed in the size of the code $n$.*

This has to be compared to the toric code for which the error rate $\overline{P}$ goes to zero exponentially fast if $p < p_c$.

### 3.7.2 Including noisy measurements

Theorem 3.11 can be adapted to the case when measurements are subject to noise. A simple model of noisy measurements is the *phenomenological noise model*, where it is assumed that the result of the measurement of the check operators is flipped with some probability $q$. Measurements are repeated and the minimum-weight matching is performed in a 3D space, as described in Section 2.3.5 and Section 4.2.1.

The argument to include syndrome noise is simple: First we observe that a necessary condition for the decoding to fail is still given by $E+R$ containing an essential cycle. Hence Equation 3.21 remains unchanged when taking noisy measurements into account. In fact, the only change that needs to be made is in Equation 3.25 where we bound the number of self-avoiding paths of weight $w$ paths by $s(s-1)^{w-1}$. Having added a third direction, we have to replace the connectivity $s$ with $s+2$, obtaining the bound $(s+2)(s+1)^{w-1}$. Furthermore, the number of starting-points of the self-avoiding paths has to be multiplied by the number of time steps $T$. We obtain

$$\overline{P} \leq T|V|(s+2)\frac{\alpha'}{1-\alpha'} \tag{3.29}$$

with $\alpha' = 2(s+1)\sqrt{p}$. We assume that the number of repeated measurements is at most proportional to the number of physical qubits $T \in O(n)$. In Section 4.2.2 we give numerical evidence that it suffices to choose $T = d \in O(\log(n))$ for hyperbolic codes. Together with $s|V| = 2|E| = 2n$, we get that $T|V|(s+2) \in O(n^2)$. To suppress the additional factor of $n$ we adjust Equation 3.28 and obtain:

**Theorem 3.13.** *The threshold $p_c$ of an $\{r,s\}$-hyperbolic code in the phenomenological error model with $p=q$ and decoded by a minimum-weight decoder is lower bounded by*

$$\hat{p}_c = \frac{\exp(-4/c_{r,s})}{4(\max(r,s)+1)^2}. \tag{3.30}$$

*assuming that the number of repeated measurements $T \in O(n)$.*

### 3.7.3 Threshold bounds of extremal code families

In Section 3.5.1 we have seen that the distances of extremal codes with $n < 10^4$ followed $d = c^{\text{fit}}_{r,s}\log(n)$, where $c^{\text{fit}}_{r,s}$ was estimated by a least-square fit (see Figure 3.7). Assuming that $c^{\text{fit}}_{r,s}\log(n)$ is also a good approximation for the distance of codes with $n \geq 10^4$, we obtain the bounds given in Table 3.2.

| Type | $c^{\text{fit}}_{r,s}$ | $\hat{p}_c$ perfect measurements | $\hat{p}_c$ noisy measurements |
|---|---|---|---|
| $\{5,4\}$ | 1.77 | 0.51% | 0.073% |
| $\{5,5\}$ | 1.21 | 0.30% | 0.025% |

Table 3.2: Empirical lower bounds on the thresholds of extremal code families assuming perfect and noisy measurements of the check operators. The values of $c^{\text{fit}}_{r,s}$ were obtained by a least-square fit of extremal codes with $n < 10^4$ (see Figure 3.7).

## 3.8 Semi-hyperbolic codes

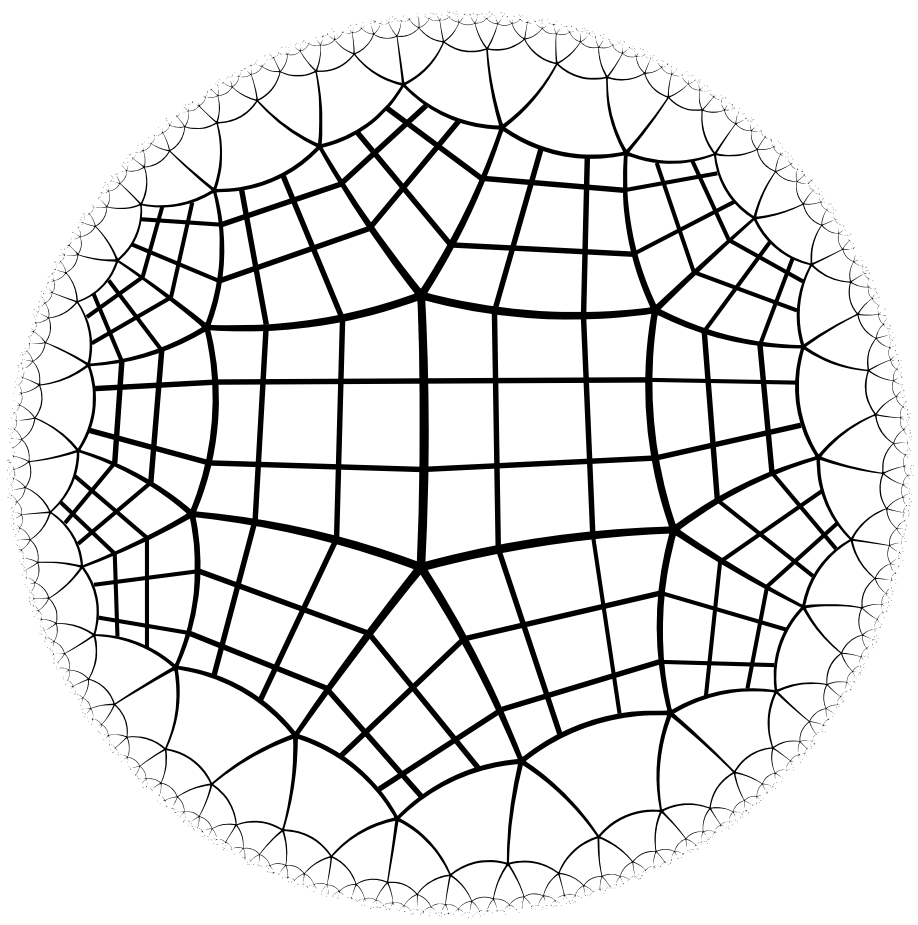

Figure 3.13: The $\{4,5\}$-tessellation with some faces replaced by a $3 \times 3$ square grid.

Consider a regular tessellation of a closed, hyperbolic surface with Schläfli-symbol $\{4,q\}$ ($q \geq 5$). In a $\{4,q\}$-tessellation it holds that $|E| = q|V|/2 = 2|F|$. Associated with this tessellation is a $\{4,q\}$-hyperbolic surface code with parameters $[[n_h, k_h, d_h]]$. We define a new tessellation with the same topology by taking every face and tessellating it with an $l \times l$ square-grid (see Figure 3.13). Essentially, we replace each square by a $\{4,4\}$-tessellation of a 2D flat space, weakening the negative curvature.

We call this refined tessellation *semi-hyperbolic* having vertices $V_{sh}$, edges $E_{sh}$ and faces $F_{sh}$

with

$$|V_{sh}| = |F|l^2 = q|V|l^2/4,$$
$$|E_{sh}| = |F| \times 2l^2 = |E|l^2, \quad (3.31)$$
$$|F_{sh}| = |F|l^2.$$

From Equation 3.31 it immediately follows that $n_{sh} = n_h l^2$. The number of encoded qubits in the hyperbolic surface code is determined by the topology of the surface which is unchanged, hence $k_{sh} = k_h$. For semi-hyperbolic codes, all $Z$-checks have weight 4 while there are two types of $X$-checks, namely the ones of weight $q$ of the original code and the new checks of weight 4 of which there are $|V|(ql^2/4-1)$ (see Figure 3.14). One can efficiently compute the distance of logical $Z$ operators and logical $X$ operators for CSS surface codes (see Section 3.4.1). Our results are listed in Table 3.3. They support the conjecture that the $\overline{Z}$-distance of the semi-hyperbolic code is $d_{sh}(\overline{Z}) = d_h l$. This would be true if the shortest non-trivial loops go over the subdivided squares *through* the vertices of the original hyperbolic tessellation. We have not been able to prove this however. Table 3.3 shows that the scaling of the $\overline{X}$-distance is clearly also growing with $l$ although the $l$-dependence is not as simple as the conjectured $l$-dependence of the $\overline{Z}$-distance.

With increasing $l$ the ratio of total curvature over the surface area vanishes so one expects that for fixed $n_h$ and increasing $l$ a semi-hyperbolic code family has similar behavior to the toric code in terms of noise threshold. We confirm this in Figure 4.3 in Section 4.1.2.

## 3.9 Fault-tolerant implementation of gates

In this section we present two schemes to realize logical operations on the qubits encoded in a hyperbolic code.[3] These schemes do not increase the connectivity between qubits. They rely on manipulating the topology of the tessellation that underlies the hyperbolic code. First, in 3.9.1 we discuss how to read out a logical qubit encoded in a hyperbolic code. To do this we make use of one-bit teleportation of quantum states. The same technique allows us to take an encoded state of another code and transfer it into the hyperbolic code. Second, in 3.9.2 we discuss a technique to manipulate the encoded qubits of a single hyperbolic code. The tessellation is *twisted* along an essential cycle of the hyperbolic surface which induces an action on the encoded qubits. It allows for the implementation of entangling gates and the permutation of the encoded qubits within a single hyperbolic code.

[3]This section was developed in discussion and collaboration with Christophe Vuillot who wrote the text.

| $n_h$ | $l$ | $n$ | $k$ | $d_Z$ | $d_X$ |
|---|---|---|---|---|---|
| 60 | 1 | 60 | 8 | 4 | 6 |
| 60 | 2 | 240 | 8 | 8 | 10 |
| 60 | 3 | 540 | 8 | 12 | 14 |
| 60 | 4 | 960 | 8 | 16 | 18 |
| 60 | 5 | 1500 | 8 | 20 | 22 |
| 60 | 10 | 6000 | 8 | 40 | 42 |
| 160 | 1 | 160 | 18 | 6 | 8 |
| 160 | 2 | 640 | 18 | 12 | 14 |
| 160 | 3 | 1440 | 18 | 18 | 20 |
| 160 | 4 | 2560 | 18 | 24 | 26 |
| 160 | 5 | 4000 | 18 | 30 | 32 |

| $n_h$ | $l$ | $n$ | $k$ | $d_Z$ | $d_X$ |
|---|---|---|---|---|---|
| 360 | 1 | 360 | 38 | 8 | 8 |
| 360 | 2 | 1440 | 38 | 16 | 16 |
| 360 | 3 | 3240 | 38 | 24 | 24 |
| 360 | 4 | 5760 | 38 | 32 | 32 |
| 360 | 5 | 9000 | 38 | 40 | 40 |
| 1800 | 1 | 1800 | 182 | 10 | 10 |

Table 3.3: Hyperbolic and semi-hyperbolic surface codes based on the $\{4,5\}$-tessellation. We give the minimum weights $d_Z$ and $d_X$ of any logical operator of $X$-type and $Z$-type, the number of qubits $n_h$ of the purely hyperbolic code, the total number of qubits $n$ of the (semi)-hyperbolic code, and the parameter $l$ used for the $l \times l$-tessellation of every square face.

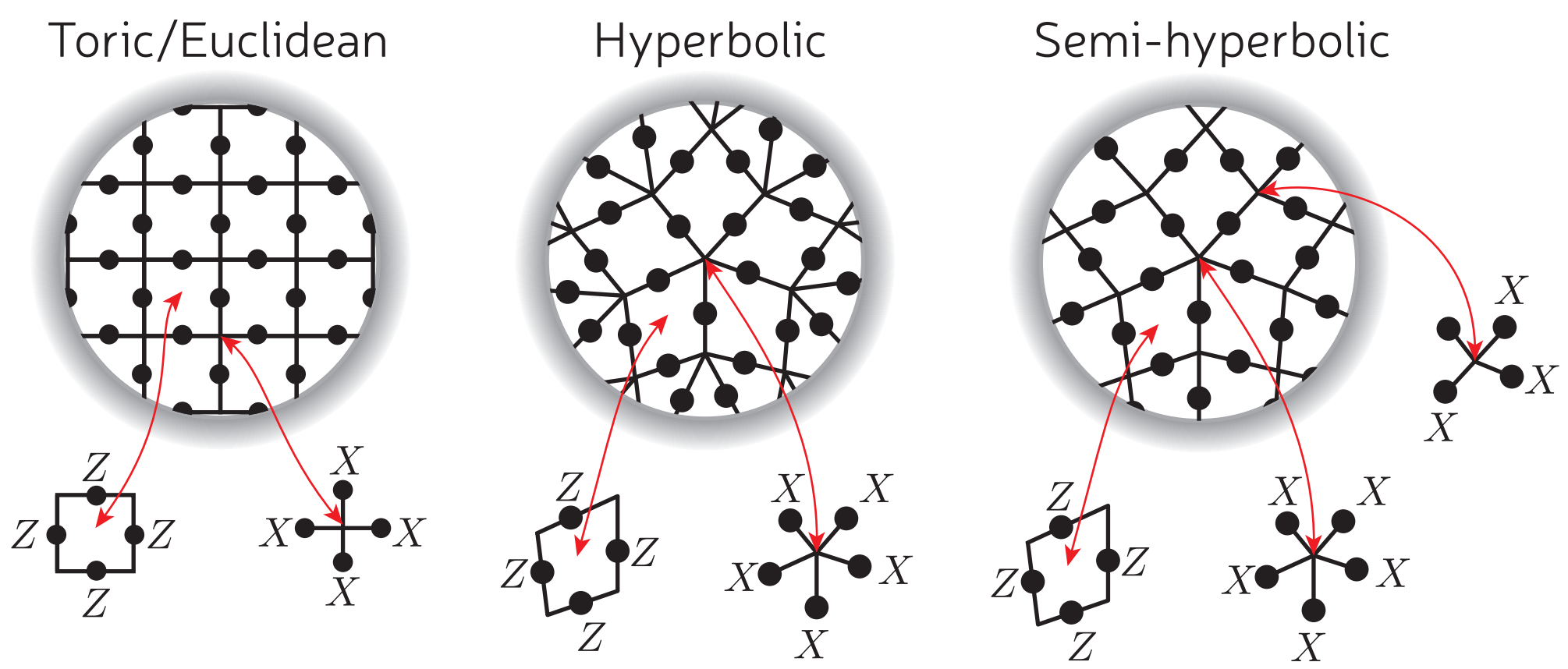

Figure 3.14: Local regions in a $\{4,4\}$-tessellation (left), $\{4,5\}$-hyperbolic tessellation (middle) and a semi-hyperbolic tessellation based on the $\{4,5\}$-tessellation (right).

### 3.9.1 Lattice Code Surgery

We will now discuss how to transfer encoded qubits from and into a hyperbolic code. The ability to perform these operations is crucial, as we are not aware of a way to fault-tolerantly implement a universal gate set, or even the full Clifford group within hyperbolic codes. We thus envision an architecture similar to that of a classical processor: At every instant, only a subset of the qubits is undergoing computation while the rest are held in storage. The storage medium here is a hyperbolic code while the computational space is thought of as a few blocks of 2D surface or color codes with magic state distillation capabilities.

In order to read or write qubits to storage, one requires the following operations on individual qubits without affecting the protection of other logical qubits:

- Measure qubit in storage in $Z$ or $X$ basis (and thus also reset individual qubit). This step can be accomplished by performing a joint $ZZ$ (resp. $XX$) measurement on the stored qubit and a qubit in the computational space which is initialized to $|0\rangle$ (resp. $|+\rangle$).

- Retrieve an encoded qubit from storage into computational space or write a qubit to storage. In order to extract logical qubits from storage to the computational space, one can implement one of the standard one-bit teleportation circuits using again $XX$ and $ZZ$ measurements (see Figure 3.15).

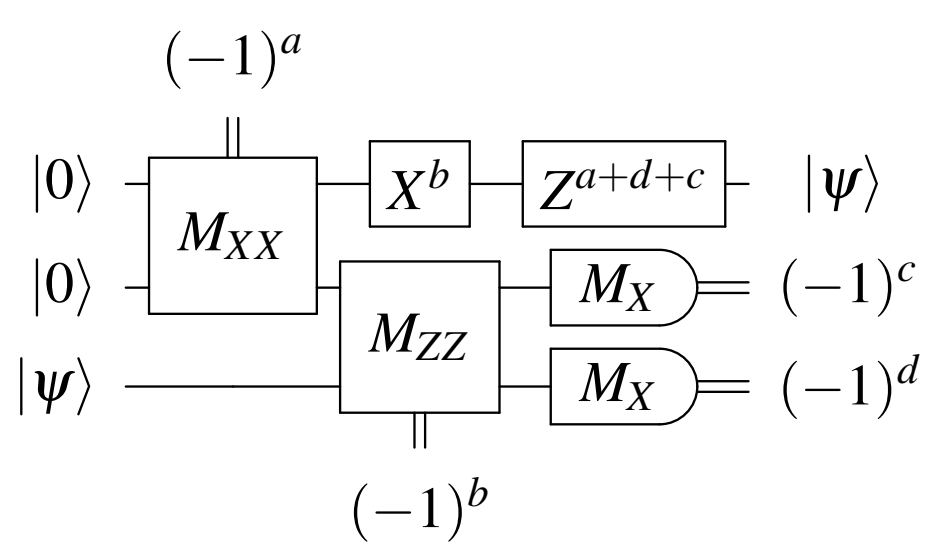

Figure 3.15: One possible circuit to realize one-bit teleportation via measurements. It uses one ancillary qubit and two weight-two joint measurements. The boxes containing $M_{XX}$ and $M_{ZZ}$ indicate a joint measurement of the two qubits involved.

The required logical $XX$ or $ZZ$ measurements between a stored qubit and a qubit in the computational space can be performed using a technique called *lattice code surgery* which was first suggested in [42].

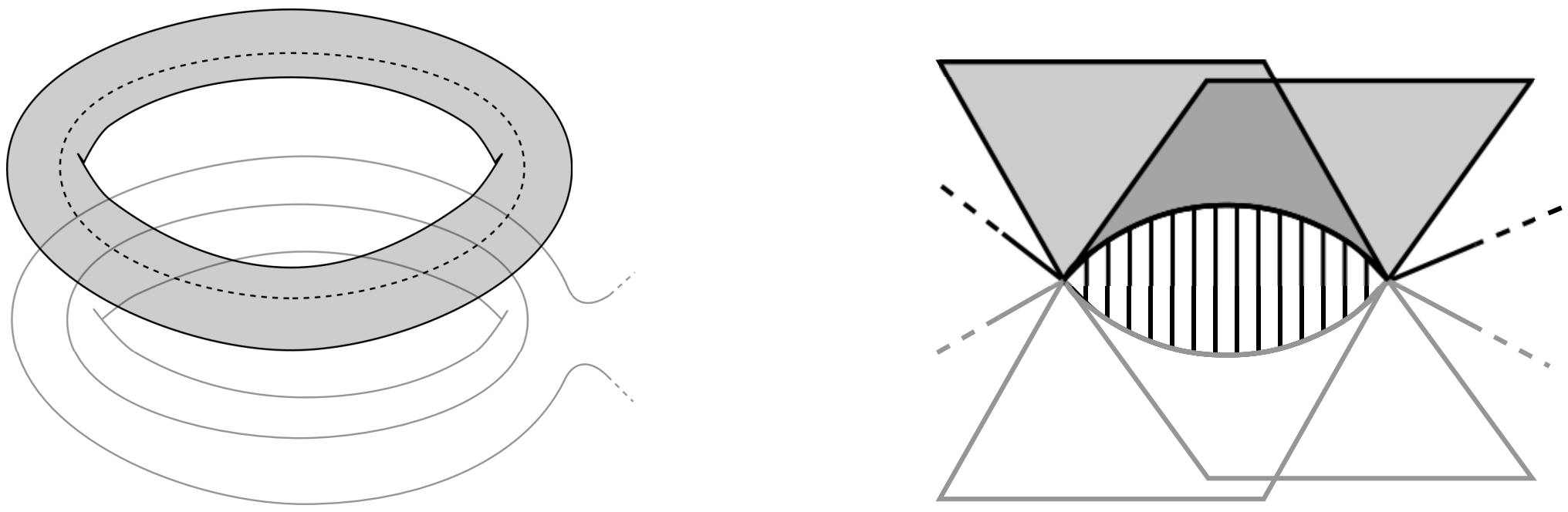

(a) (b)

Figure 3.16: (a) Positioning of the ancillary toric code (grey, on top) with respect to the storage transfer zone (white, on bottom) with $Z$ logical operators facing each other to realize a $ZZ$ measurement. (b) Local configuration of the merged lattices after measuring qubits in the support of the logical $Z$ operator in facing pairs. The paired qubits lie on the two curved edges and between them is a 2-edge face (striped) glued perpendicular to both surfaces. Note that the merger leads to $X$-checks of weight 8 (by adding a layer of qubits in between the torii one can reduce this to weight 5).

Figure 3.16a represents a possible configuration for a logical $ZZ$ measurement. Two handles are located on top of each other with two matching logical $Z$ operators facing each other. To perform the $ZZ$ measurement one measures pairs of facing qubits in the support of these two logical $Z$-operators in the $Z$ basis. Their product gives the outcome of the joint $ZZ$ measurement and the two handles are merged. The new measurements of pairs of qubits does not commute with the local $X$-checks of the separate surfaces, so these get replaced by products of these $X$-checks which do commute. The result is a merged cell complex which is no longer the cellulation of a two-dimensional manifold as it contains edges adjacent to three faces, see Figure 3.16b. One can observe that the two logical $Z$ operators become equivalent under the application of the new two-edged faces (which are elements of the new stabilizer group). Furthermore, the two corresponding logical $X$ operators have to be merged in order to commute with those two-edged faces. Error correction can be carried out in this merged phase by using the previous unchanged $Z$-checks for $X$ errors and the new merged $X$-checks for $Z$ errors. Once the outcome of the logical $ZZ$-measurement is obtained, one splits back the two handles by measuring the previous, separated $X$-checks.

This results in correlated $Z$ errors that have to be corrected in a correlated fashion, the same way as in the standard surface code lattice surgery. It is important to note that the fact that the code contains some other encoded qubits which can have a representative logical operator supported on the modified region is not a concern.

The correction, restricted to the storage space consists of applying Pauli-$Z$ operators to a subset of the qubits forming the measured $Z$ logical operator, $Z_{\text{meas}}$. Let $S_{\text{corr}} \subset \text{Supp}(Z_{\text{meas}})$ denote this subset. Equivalently, the complement of this subset, $S'_{\text{corr}} = \text{Supp}(Z_{\text{meas}}) \setminus S_{\text{corr}}$ can be used for the correction. Take some logical operator of another logical qubit of the code, $X_{\text{other}}$. Since $X_{\text{other}}$ commutes with $Z_{\text{meas}}$, it has to overlap with $\text{Supp}(Z_{\text{meas}})$ on an even number of qubits. That implies that the two choices of correction, $S_{\text{corr}}$ or $S'_{\text{corr}}$, both have the same effect on it. Both either flip its sign or both leave it invariant. Moreover, using the $X$-checks lying on $\text{Supp}(Z_{\text{meas}})$, one can move around the loop where $X_{\text{other}}$ intersects $Z_{\text{meas}}$. So there is another representative for $X_{\text{other}}$ that is unmodified by the correction. The correction just enforces that all other representative are equivalent to an unmodified one.

### 3.9.2 Dehn twists

Needing ancillary qubits and connecting these to the storage qubits is a concern for the overall connectivity and overhead. We are going to define two measures of connectivity: There is an *instantanteous qubit degree* which is the number of other qubits that a qubit has to interact with (for doing parity check measurements) at a certain point in time. We would like this degree to be a small constant throughout our schemes. Besides this notion there is a *cumulative qubit degree* which measures the total number of different qubits that a qubit has to interact with over time. For hardware with fixed connections this cumulative qubit degree should ideally be a small constant as well. For hardware which allows for switching (e.g. switches in a photonic network) the cumulative qubit degree could be allowed to grow.

If we were to decide to only have one (logical) ancillary qubit linking every storage qubit to the computational qubits then we will blow up the cumulative qubit degree of this ancillary qubit (cumulative degree scaling with the number of logical qubits $k$). On the other hand if we use one ancillary qubit for each storage qubit we give up the overhead advantage given by the (semi)-hyperbolic code. This is why we need a technique to move qubits around in storage, allowing us to read or write qubits from storage only at certain locations. Such a storage medium will not be a random access memory since the retrieval of encoded qubits depends on where they are stored in the memory. For the movement technique we propose the use of Dehn twists which is a code deformation technique using the topological nature of the code to implement operations [43, 44]. In a nutshell, Dehn twists allow us to perform CNOTs between the two qubits of one handle as well as exchanging pairs of qubits between handles. This then allows us to have designated zones for transfer from and to the computation space and move storage qubits to these zones when needed.

Our movement proposal in the form of Dehn twists leads to a growing cumulative qubit degree of some of the physical qubits in the code (cumulative degree scaling with distance $d \sim \log n \sim \log k$). In Section 3.9.2 we suggest a way in which one can modify this method leaving the cumulative degree of qubits constant at the expense of using additional space (qubits).

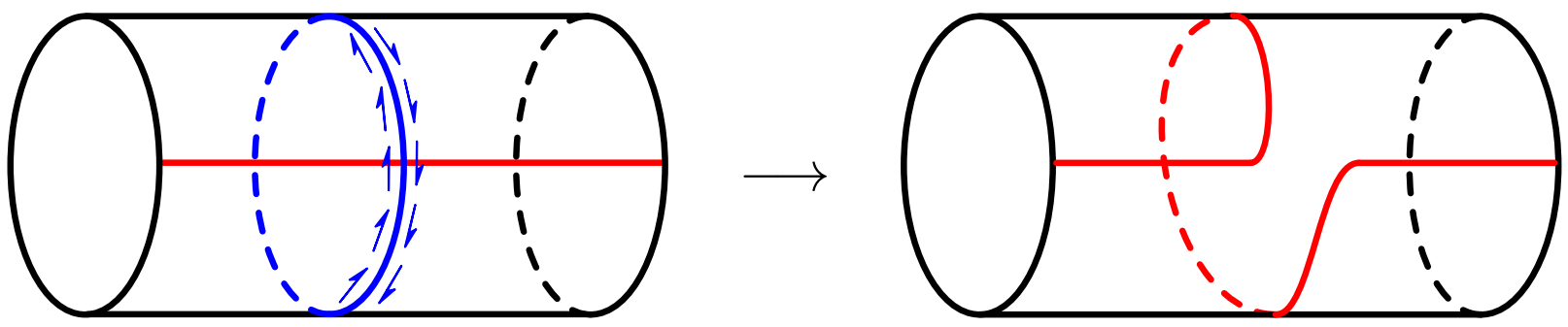

Figure 3.17: The action of a Dehn twist along the arrowed (blue) loop. It adds this loop to the (red) path crossing it.

A Dehn twist is a homeomorphic deformation of a surface, that is considered here to be closed and orientable having $g$ handles (genus $g$). Dehn twists on a closed surface $S$ are known to generate the full mapping class group $\mathrm{MCG}(S)$ of the surface [44]. The idea is to twist the surface along an essential cycle as shown in Figure 3.17. This has the effect of adding this essential cycle to any other cycle that crosses it.

We are interested in the effect of Dehn twists on the first homology group $H_1$ as elements in $H_1$ correspond to the logical $Z$ operators in our code. This space can be equipped with a standard symplectic form counting the number of crossings modulo two between loops. Acting on this space Dehn twists generate the symplectic group $\mathrm{Sp}(2g, \mathbb{Z}_2)$ as they preserve the number of crossing modulo two between loops. A possible generating set of size $3g-1$ for the full group is given in Figure 3.18.

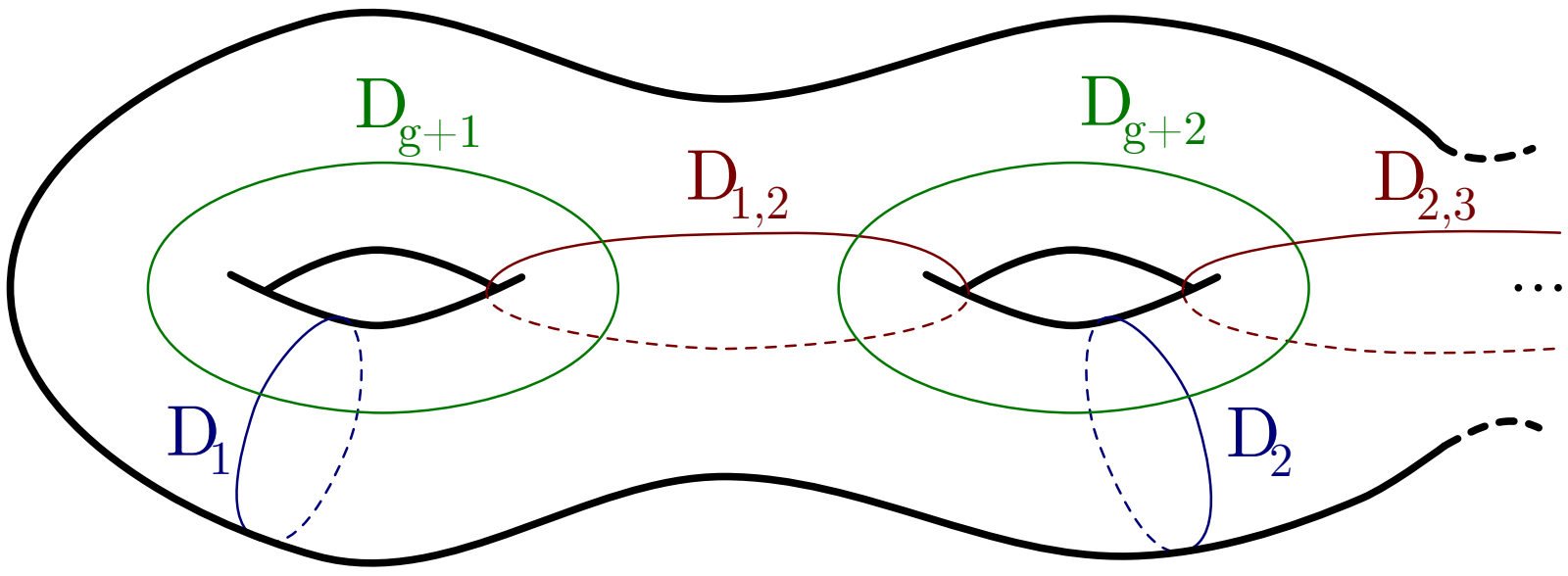

Figure 3.18: A generating set of loops for Dehn twists on a surface with $g$ handles. Each handle hosts two qubits, and at the $k$th handle we label the qubits $q_{2k-1}$ and $q_{2k}$. We choose the convention that $\overline{X}_{q_{2k-1}}$ is supported on the loop (on the dual tessellation) labelled $k$ and so $\overline{X}_{q_{2k}}$ is supported on the loop $k+g$ (on the dual tessellation). That implies that $\overline{Z}_{q_{2k-1}}$ is supported on the loop $k+g$ and $\overline{Z}_{q_{2k}}$ on the loop $k$.

It turns out that this kind of continuous deformation has a direct analog for the tiled surfaces of homological codes, so in particular hyperbolic codes. A simple example to explain the procedure is that of the toric code. Using $d$ parallel CNOTs it is possible to "dislocate" the faces by one unit along a loop as shown in Figure 3.19. Repeating the step $d$ times with CNOTs which stretch between qubits over a longer and longer range, will bring it back to its initial configuration. Tracking what happens to a $Z$ or $X$ logical operator which crosses this loop, one can easily see that the procedure acts as a CNOT on the logical operators. The control qubit $\overline{X}_{\text{control}}$ intersects the loop around which the Dehn twist is done on one vertical qubit. The successive steps apply CNOTs with this qubit as control and qubits of the $\overline{X}_{\text{target}}$ parallel to the loop as target. This gradually propagates $\overline{X}_{\text{control}}$ to $\overline{X}_{\text{target}}$. Symmetrically, $\overline{Z}_{\text{target}}$ intersects the loop on one horizontal qubit and the CNOTs propagate it to $\overline{Z}_{\text{control}}$ running around the loop. For a $\{4,5\}$-tessellation we have to modify this circuit as shown in Figure 3.20.

The question is then what useful operations on the logical qubit space these Dehn twists give us access to. It is easy to verify the action of the generating set of Dehn twists, see Figure 3.21. For our purpose, we can see that nine Dehn twists can be used to swap the qubits of two handles using the circuits in Figure 3.21 to construct SWAP operations from 3 CNOTs. By considering a larger generating set the number of Dehn twists can be reduced to seven. This can be checked by a computer.

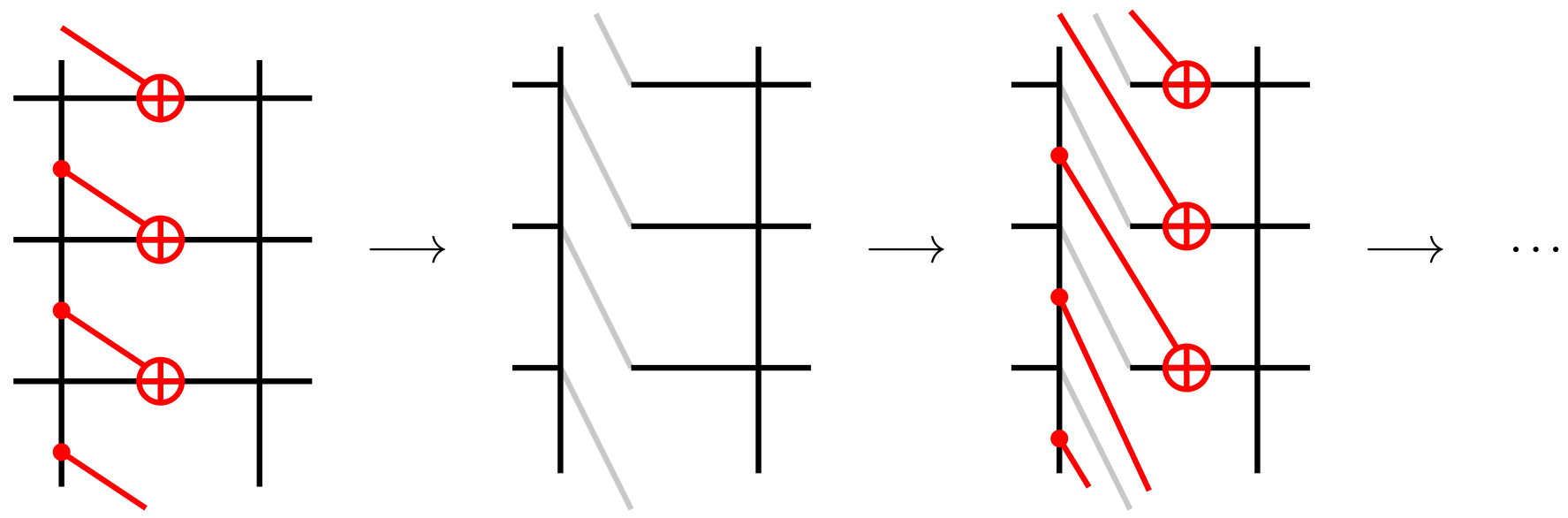

Figure 3.19: The first two steps of a Dehn twist on a toric code. The qubits are placed on the vertical and horizontal edges, each face is a $Z$-check and each vertex is a $X$-check. The subsequent steps are similar but take into account that the middle row of qubits is gradually displaced "downwards".

**Reducing Connectivity**

In each step of the Dehn twist the instantaneous qubit degree is $O(1)$. The cumulative degree of the qubits on the loop along which one does a Dehn twist is $O(d)$, with $d$ being the length of this loop. In the case of hyperbolic codes, this is logarithmic in the total number of physical qubits which is an improvement over losing all overhead or having cumulative qubit degree scaling with $k$ by employing read/write ancilla qubits. The cumulative qubit degree is increased if one performs Dehn twist along overlapping loops.

The temporal overhead of a Dehn twist, if one applies one round of error correction ($O(d)$ steps in time) between each step is $O(d^2)$. The cumulative qubit degree can be reduced using the following variation.

We can use an extended region to spread the effect of the twist and lower the connectivity requirements as well as the temporal overhead. As shown in Figure 3.22, one can choose $d$ parallel loops, and apply in parallel one step of the Dehn twist on each of the loops. This effectively realizes a Dehn twist in one go in an extended region. The connectivity required for this extended Dehn twist is constant and doing one round of error correction after this gives a total temporal overhead of $O(d)$.

There is a concern when trying to adapt this to the $\{4,5\}$-tessellation. There is no a priori guarantee that it is possible to find $d$ parallel loops. But one can make use of semi-hyperbolic modifications to help with this, basically creating more space for the twist region. Starting from a chosen $\overline{Z}$-loop, one can add parallel loops by adding qubits in the faces to one side of the loop as shown in Figure 3.23. This does not completely guarantee that one will create enough parallel columns as the tessellation is expanding to the right. Because of this, the parallel loops will grow in size demanding more twisting to complete the full operation. That said, if one step on the extended region is not enough one can repeat the step on the extended region. So if one step on the extended region twists the tessellation for a fraction of the distance, then only a constant number of steps will be needed and the cumulative degree of qubits will remain constant. Also, the total time overhead will be $O(d)$.

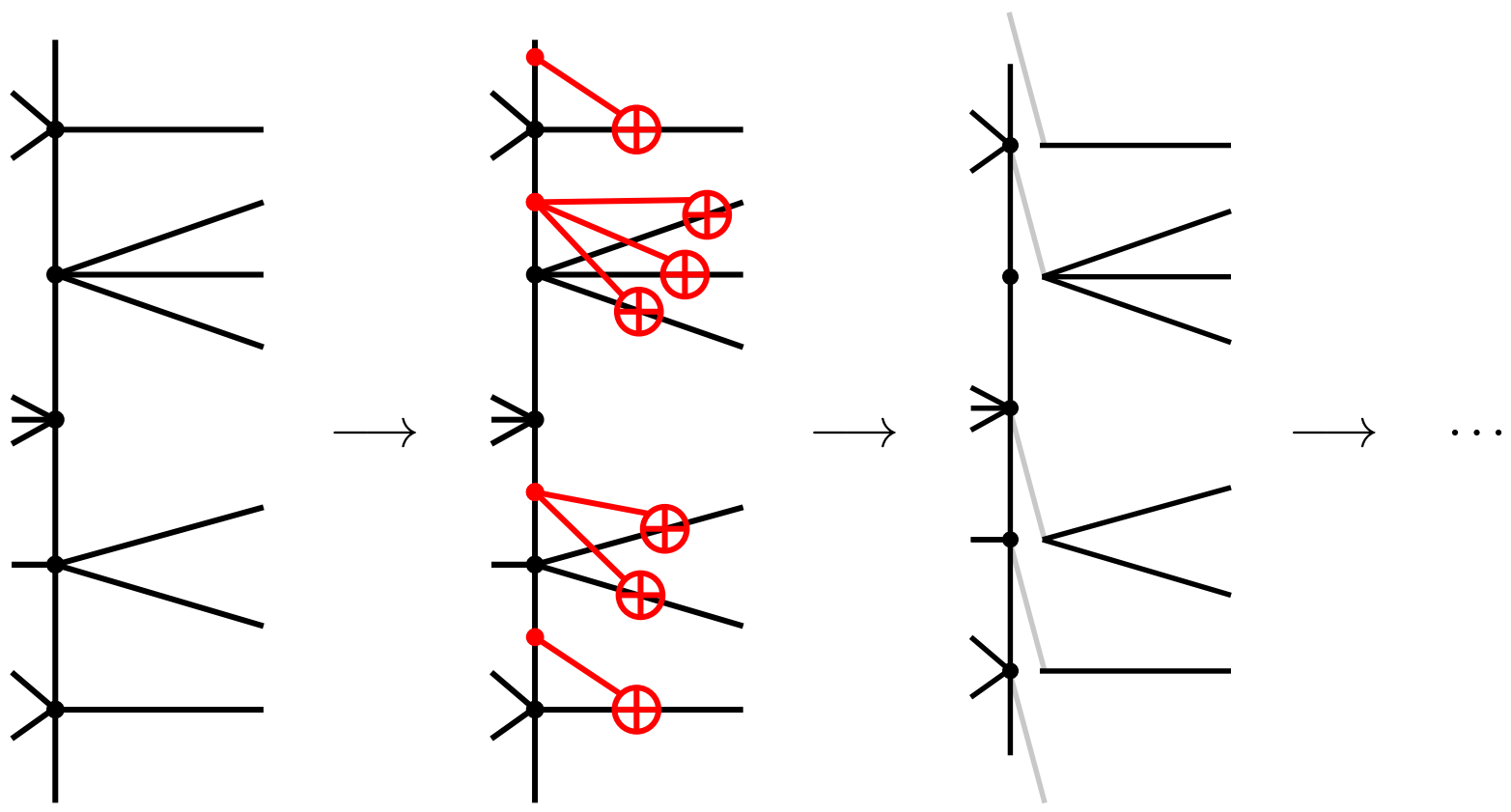

Figure 3.20: For the $\{4,5\}$-tessellation the Dehn twist procedure has to be slightly generalized. One chooses a non-trivial $\overline{Z}$-loop. The edges sticking out to one side of this loop form the support for $\overline{X}$ of the other qubit of the handle. Instead of having always exactly one edge sticking out to the right (see Figure 3.19), there can now be between zero and three edges. The modification is then to just adapt the number of target qubits for the CNOTs to this number. At intermediate steps of the Dehn twist one can observe that the $X$-checks have weight varying between 2 and 8.

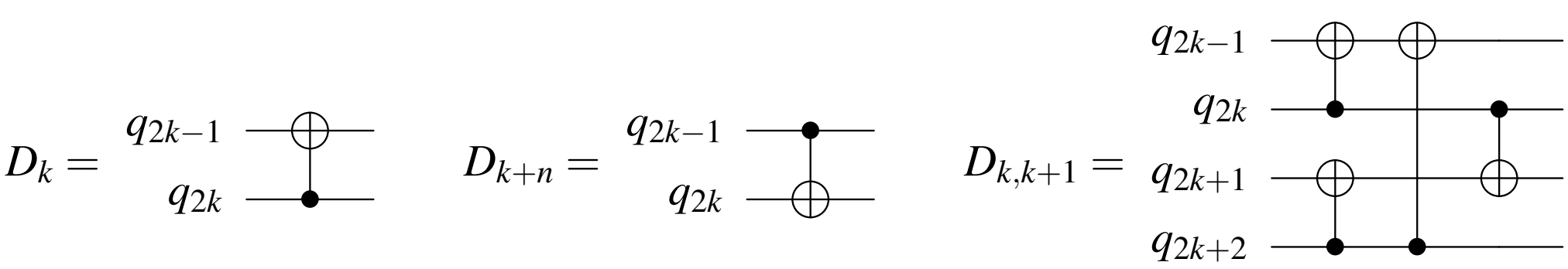

Figure 3.21: The circuits realized by the three type of generators for the Dehn twist transformations. The labelling of the Dehn twists and the qubits is the one detailed in Figure 3.18.

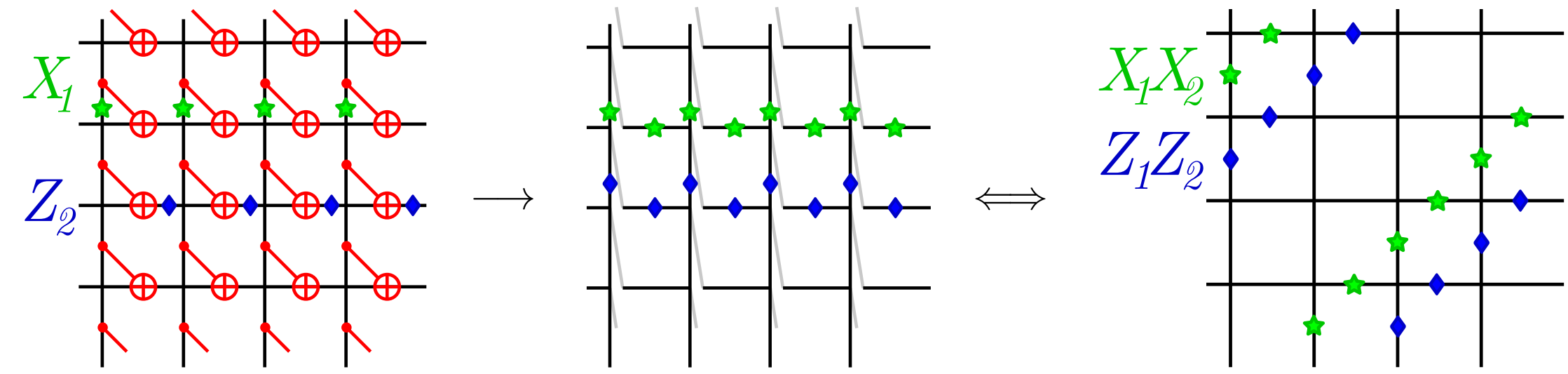

Figure 3.22: Extended Dehn twist on a distance 4 toric code. One does the 4 Dehn twist steps in parallel on $d$ parallel rows. Green stars indicate the logical $X_1$ operator and how it transforms to $X_1X_2$. Blue lozenges indicate the logical $Z_2$ operator and how it transforms to $Z_1Z_2$.

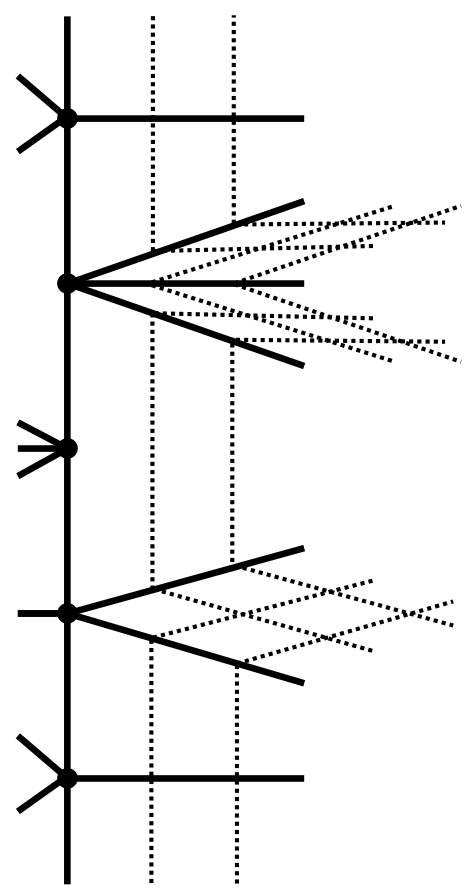

Figure 3.23: Given some initial loop in the $\{4,5\}$-tessellation, it is possible to add parallel loops to it. The dotted lines are added qubit edges that make a more fine-grained tessellation in the direction mostly "perpendicular" to the original loop. This can be somewhat problematic when the original loop takes "sharp" turns as in the middle of this example (where there is no qubit edge sticking out to the right). In this face one potentially adds a way for a logical $Z$ operator to cut a corner and that might decrease the distance by one. One should verify such properties in specific examples of interest.

# Chapter 4

# Performance of 2D hyperbolic codes

## 4.1 Threshold estimation assuming perfect measurements

### 4.1.1 Threshold of hyperbolic codes

In this section we present the results of simulated error corrections on extremal hyperbolic codes assuming that the measurements of the stabilizer checks can be done perfectly. Our error model is that of independent $X$ and $Z$ errors in which a qubit can undergo independently an $X$ error with probability $p$ and a $Z$ error with probability $p$ at each time-step. After these errors happen one applies the minimum weight matching (MWM) decoder which tries to infer what error occurred (see Section 2.3.5). The decoder succeeds if the product of the real and the inferred error is in the image of the boundary operator. We do this independently for $X$ errors as well by performing the same simulation on the dual tessellation. We gather statistics by repeating this procedure $N$ times. The probability of a logical error $\overline{P}$ on *any* of the encoded qubits is estimated by taking the ratio between failed corrections and number of trials.

Note that in previous simulations of the toric code the logical error rate against one type of error (only $X$ or $Z$) is plotted. This is due to the fact that the $\{4,4\}$-tessellation is self-dual and in the error model $X$ and $Z$ errors follow the same distribution. However, the $\{5,4\}$-code is defined on a tessellation which is not self-dual. This makes it necessary to consider both types of errors.

The results of a simulation with $N = 4 \times 10^4$ trials are shown in Figure 4.1. For increasing probability of a physical error $p$ the probability for a logical error occurring on any of the qubits $\overline{P}$ approaches $1-(1/2)^{2k}$. This is because the decoder can do no worse than random guessing and the probability to guess correctly for all $2k$ logical operators $\overline{X}_i$, $\overline{Z}_i$ is $(1/2)^{2k}$. For the toric code this is 0.75 for a single type of Pauli error.

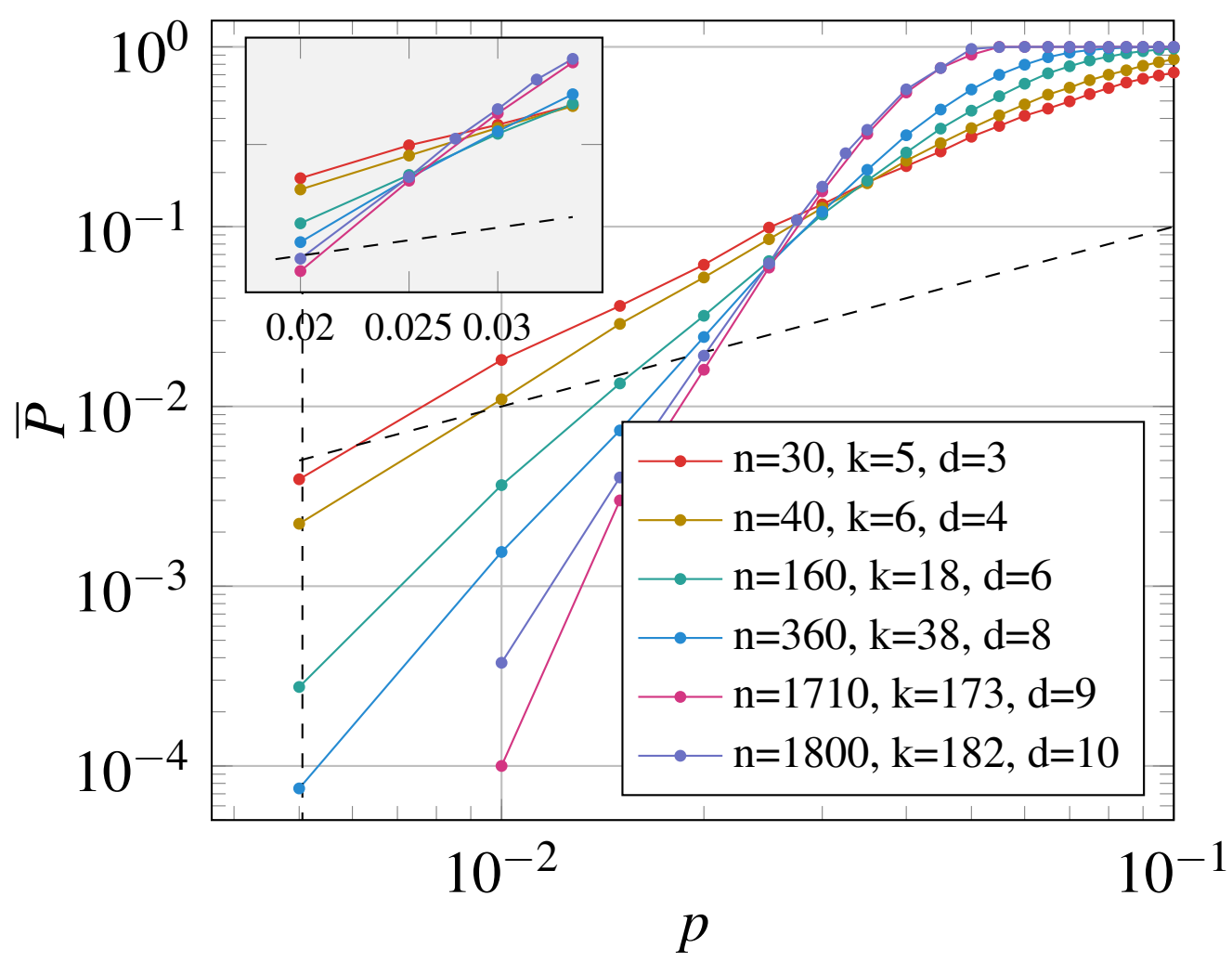

Figure 4.1: Logical error rate $\overline{P}$ against physical error rate $p$ for $\{5,4\}$-codes which are extremal. Every data point was obtained from $4 \times 10^4$ runs. The diagonal dashed line indicates $p = \overline{P}$. The vertical dashed line indicates the lower bound $\hat{p}_c = 0.51\%$ on the threshold due to Theorem 3.11 (see Table 3.2). The four largest codes seem to cross for $p$ between 2% and 3%. The code with $n = 1710$ presented here is the extremal code with the lower number of minimum weight logical operators (see Table 3.1).

In a numerical simulation of the noise threshold of the toric code (see e.g. [45]) the curves for codes of different sizes all intersect in one point, namely the threshold $p_c$, pointing to the fact that the noise threshold corresponds to a phase-transition. In Figure 4.1 we see that we do not obtain a perfect crossing of the lines. This is likely due to finite-size effects, as the crossings between codes of increasing size seem to become closer. However, the amount of data is not enough to warrant a finite-size scaling ansatz. Regarding finite-size scaling we can add the following remark: The relevant parameter for the scaling is the code distance $d$ as it corresponds to the shortest essential (co)cycle in the tessellated surface. A disk of radius $d/2$ would be indistinguishable from a disk in the infinite $\{5,4\}$-tessellated hyperbolic plane $\mathbb{H}^2$.[1] Hence, boundary effects can only appear at length scales of $O(d)$. Assuming a power law we would need data on codes with $d$ growing exponentially. However, this would imply that we need to perform simulations on system sizes growing doubly exponentially since the system size $n$ already grows exponentially with $d$ which is not feasible.

Our data suggests that for the extremal $\{5,4\}$-codes there exists a threshold between 1% and 3% (see Figure 4.1). This is compatible with the semi-analytical lower bound $\hat{p}_c = 0.5\%$ of Section 3.7.3 (see Table 3.2). For the toric code the threshold against this error model is $p_c = 10.3\%$ as determined in [45]. For hyperbolic surface codes, we expect the existence of another threshold where *all* logical qubits are potentially corrupted. In [46] the authors analyze the two distinct percolation thresholds for a prototype model of a hyperbolic tessellation. Here we only consider the threshold where an arbitrary qubit becomes corrupted. Note that the code with distance 9 performs better than the code with distance 10. This may be explained by the fact that the distance 9 code has a very small number of low-weight logical operators (see Table 3.1). Note that such an effect does not occur for the toric code as $N_d^Z = N_d^X = 2d = 2L$, i.e. the number of lowest-weight logical representatives is fixed by the distance.

The distance of the code depends on the parameters $r$ and $s$ of the tessellation. For increasing $r,s$ (and thus a code with better rate) the lower bound on the distance becomes smaller and one would expect that the threshold also goes down (at least the lower bound on the threshold given in Theorem 3.11 becomes smaller). In our simulations of various extremal $\{r,s\}$-codes we see that the error suppression capabilities go down for increasing stabilizer weight $r$ and $s$ as expected (see Figure 4.2).

### 4.1.2 Threshold of semi-hyperbolic codes

We perform the same simulation as in the previous section for semi-hyperbolic code families. For the semi-hyperbolic codes we have several choices in how to define code families. First, we can fix a purely hyperbolic $\{4,5\}$-tessellation of a hyperbolic surface and only change the $l \times l$ square tessellation of each face. As this procedure leaves the topology of the surface invariant all of these codes will encode the same number of qubits. Alternatively, we may define a family of semi-hyperbolic codes where we increase not only $l$ but also the size of the underlying hyperbolic lattice $n_h$. For example, we can choose $l$ to be proportional to $d_h$. This gives a family of codes where the encoding rate $k/n$ is polylogarithmically converging to 0.

[1] See the discussion on the injectivity radius $R_{inj}$ in Section 3.1.5.

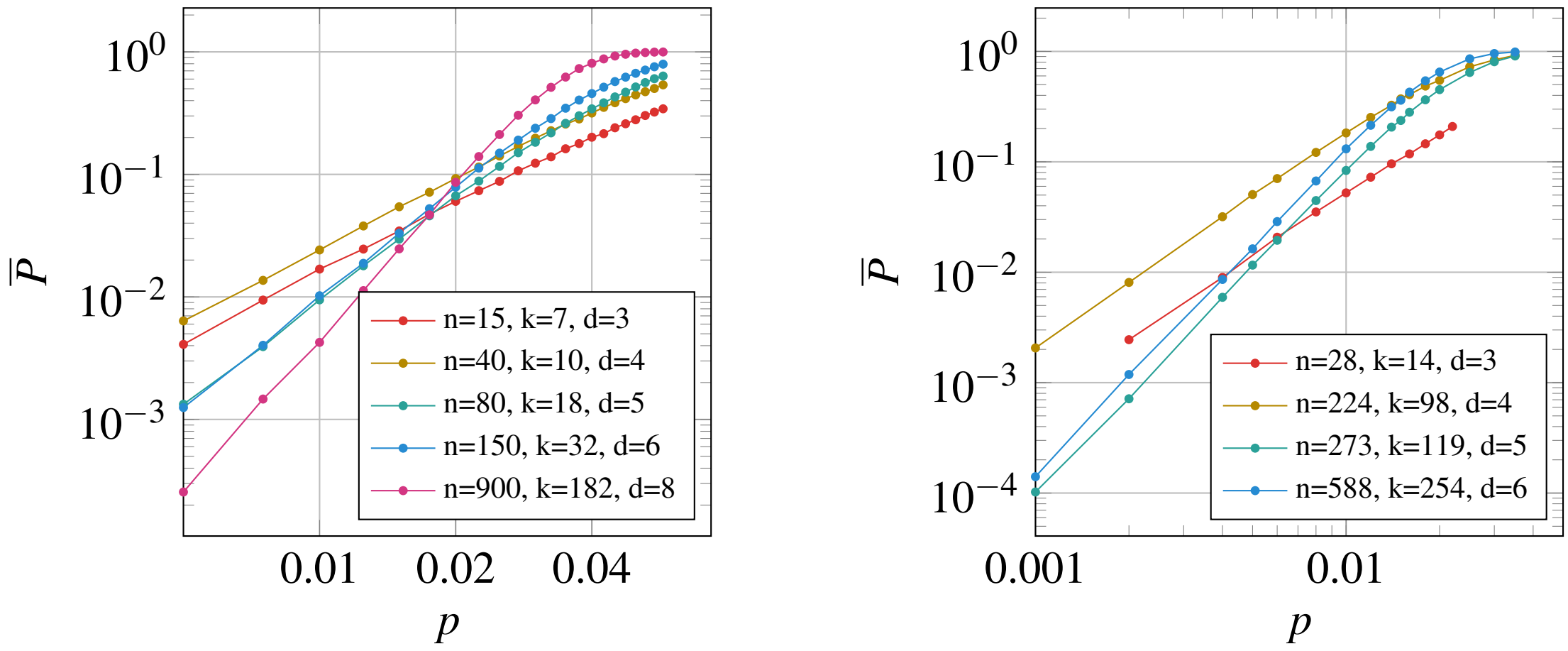

(a) $\{5,5\}$ (b) $\{7,7\}$

Figure 4.2: Results of numerical simulations of the decoding procedure for hyperbolic codes with higher encoding rate. For the $\{5,5\}$-code all lines except for $n = 40$ cross around 1.75%. The $\{7,7\}$-codes seem to be suffering from more severe finite-size effects.

We present results for semi-hyperbolic tessellations for the first family where the underlying hyperbolic tessellation is fixed in Figure 4.3. The threshold is the same as that of the toric code. Intuitively, this result can be understood by observing that this family of codes deviates from a toric code only around a constant number of vertices (cf. Figure 3.14). In Figure 4.4 we see that for the second family where we increase the size of the underlying code as well as the Euclidean square tessellation, the lines cross at about 7.9%. This code has a better threshold than the purely hyperbolic code family. We thus see that semi-hyperbolic codes allow for a trade-off between optimizing encoding rate and logical error probability.

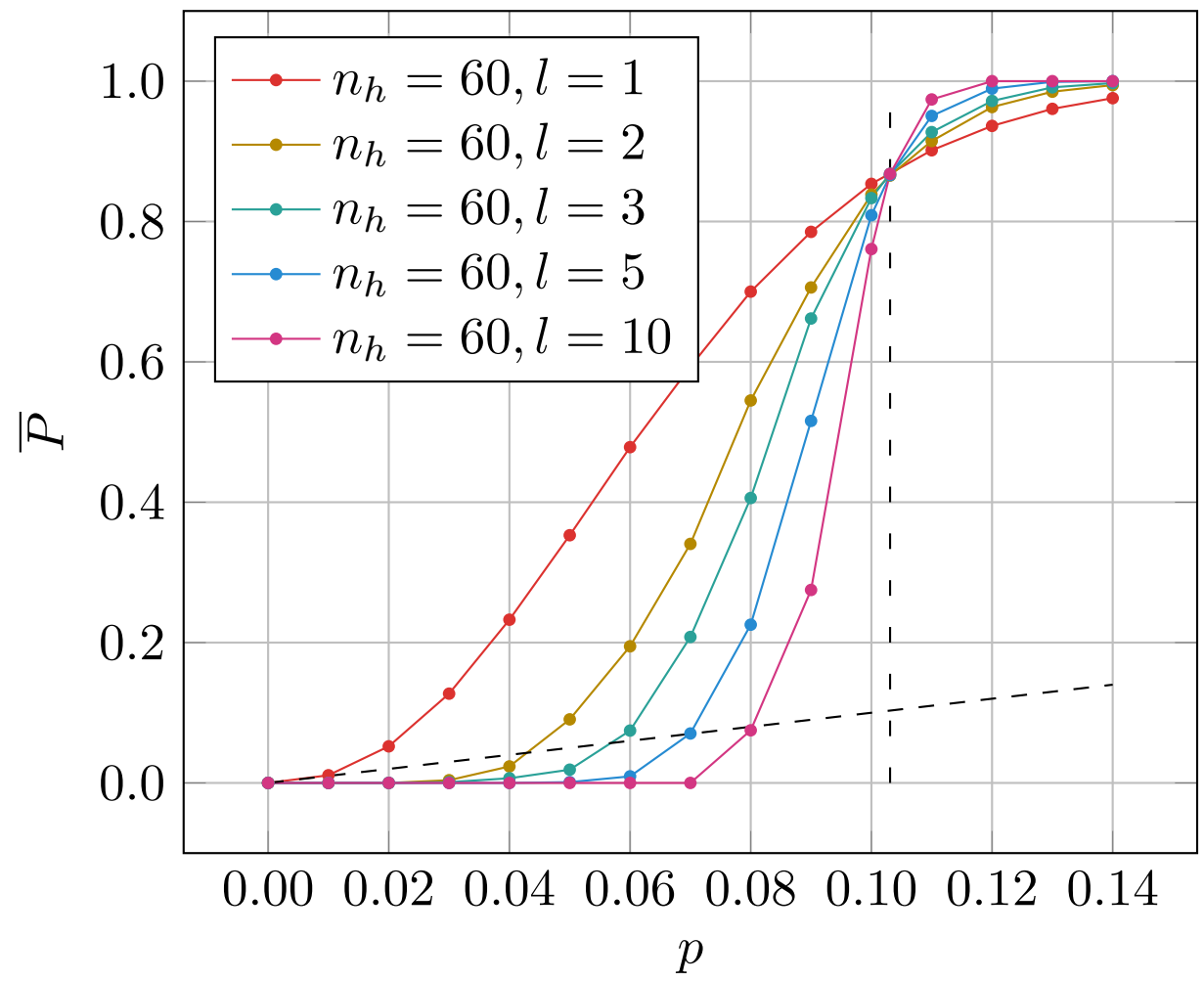

Figure 4.3: Threshold of a $\{4,5\}$-semi-hyperbolic code family, with $k = 8$ logical qubits and $l = 1,2,3,5,10$ (see Table 3.3). The stabilizer check measurements are assumed to be perfect. The case $l = 1$ is identical to the original hyperbolic code. The vertical, dashed line marks the threshold of the toric code at 10.3% and the diagonal dashed line marks $p = \overline{P}$.

## 4.2 Threshold estimation including measurement errors

### 4.2.1 Setup

We assume the same error model for the qubits where prior to each quantum error correction step (QEC step) a qubit undergoes an $X$ error with probability $p$ and a $Z$ error with probability $p$. The QEC step itself consists of an instantaneous measurement of all parity checks of the code. We model the noise on the stabilizer check measurements by assuming that each measurement result is obtained perfectly and then independently flipped with some probability $q$. In our numerical studies we restrict ourselves to $q = p$ to reduce the number of parameters.

One repeats the QEC step $T$ times and the errors are inferred based on this record using a minimum-weight matching algorithm (MWM). This modified decoding procedure was first described for the toric code in [40]. Let $G = (V,E)$ be the graph of vertices and edges in the tessellated surface. One makes $T+1$ copies of the graph $G$: Each vertex $v$ in copy $G_i$ is connected to the same vertex $v$ in copy $G_{i+1}$ via an edge, obtaining a new graph $G_{\text{time}}$ (see Figure 4.5). Each copy represents one QEC cycle in which a qubit error can take place and the entire faulty syndrome is measured. As before, we repeat the process for $Z$ errors and $X$ errors since the hyperbolic tessellations are in general not self-dual and the minimal distance of a logical $X$ is in general different from the minimal distance of a logical $Z$, see Table 3.3.

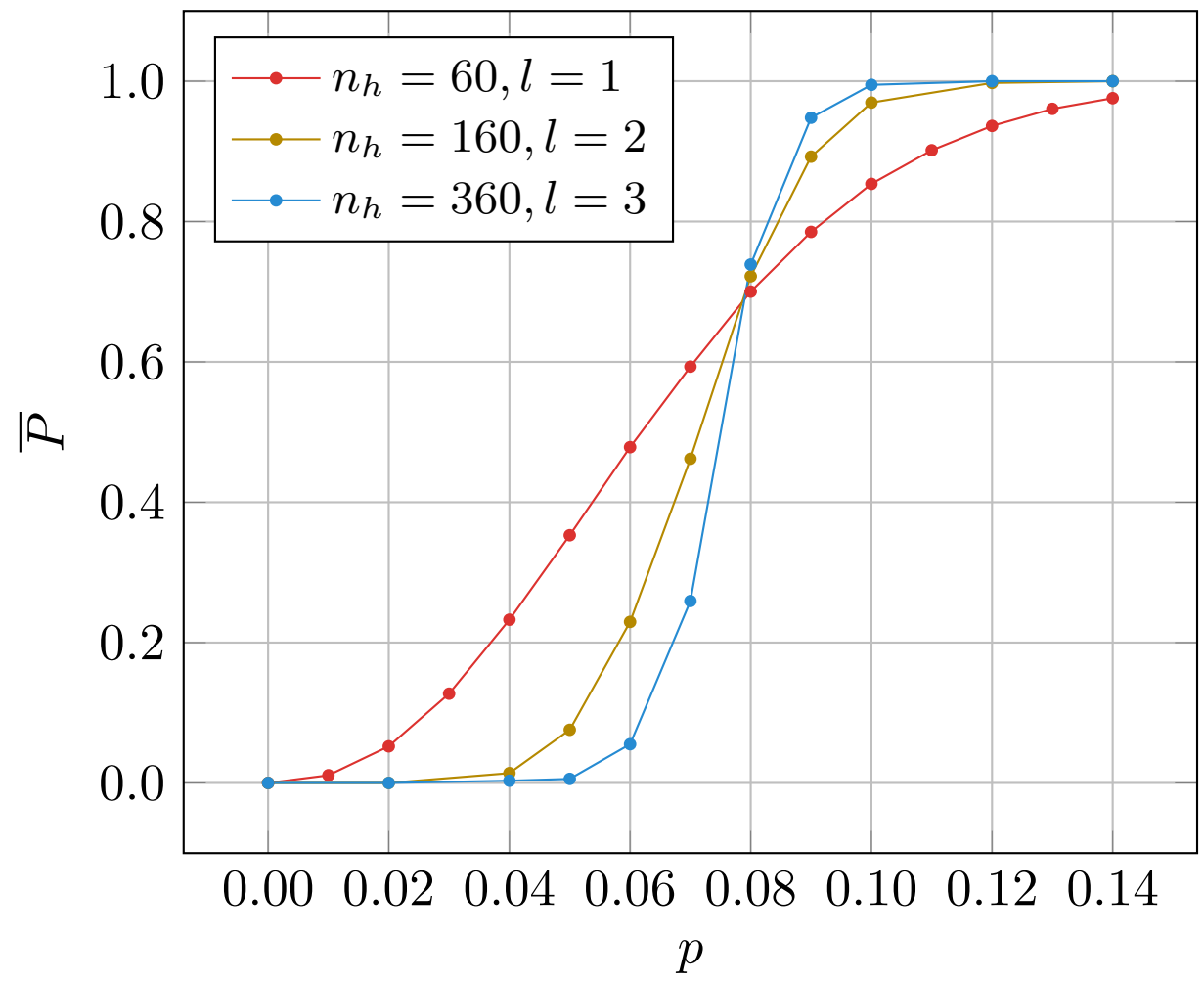

Figure 4.4: Three $\{4,5\}$-semi-hyperbolic codes $[[60,8,4]]$, $[[640,18,12]]$ and $[[3240,38,24]]$ obtained by taking $l$ proportional to the distance of the underlying hyperbolic code. The logical error probabilities cross around 7.9%.

The decoding algorithm for $Z$ errors proceeds as follows:

1. *Mark vertices:* Assume at time $t = 0$ a fictitious round of perfect QEC (no measurement or qubit error and thus all syndromes are 0).

   - For each QEC cycle at time $1 \leq t \leq T$, mark vertices where the syndrome is different from the previous time $t-1$.
   - Add a round $t = T+1$ without syndrome error and mark a vertices where the syndrome is different from the syndrome at $T$.

   The last round ensures that the total number of marked vertices is even: This step plays the role of ideal decoder and allows one to capture the logical error probability after $T$ rounds of QEC. One can thus perform minimum weight matching on the set of marked vertices:

2. *Perform MWM on marked vertices:* For each pair of marked vertices, compute the minimum distance between them using the graph distance. Feed the set of marked vertices along with the minimal distance paths between them to the MWM algorithm. The algorithm will output pairs of marked vertices such that the sum total of the weight of paths between pairs is minimized. The inferred error is the shortest path between each pair. This path consists of vertical edges (parity check measurement errors) and horizontal edges (qubit errors), see Figure 4.5.

3. *Deduce residual errors and determine whether a logical error has occurred:* We infer the errors that remain at time $T+1$ by projecting the inferred error to the last time step, obtaining a set of only horizontal edges. A horizontal edge is an element of this projected set if it was included an odd number of times in the matching. We take the real error that has occurred and project it similarly onto the $T+1$ time-slice. The product of the real and inferred $Z$ error is a closed $Z$-loop and we check whether it is a logical operator by checking whether it anti-commutes with *any* of the logical $X$ operators. If it anti-commutes, we declare it a logical failure.

For a fixed given $T$ this decoding procedure is applied to stochastically-generated $Z$ errors and repeated $N$ times so that $\overline{P}$ is estimated as $N_{\text{fail}}/N$. In addition to $\overline{P}$ one can define an *effective error probability per QEC round* $P_{\text{round}}$, which is simply defined by the equation

$$(1-P_{\text{round}})^T = 1-\overline{P}. \tag{4.1}$$

The quantity $P_{\text{round}}$ can be thought of as the average probability of a logical error occurring at any time step.

### 4.2.2 Optimal number of QEC rounds $T$

When parity check measurements are noisy, the decoding uses a record of $T$ QEC cycles. In principle correlating the syndrome record over more rounds of measurements can only improve the efficiency of the decoder per round thus lowering $P_{\text{round}}$. In [40, 45] it was shown that for a toric code with distance $d$, subject to the previously described noise model, the benefit of taking more than $T=d$ rounds of syndrome measurement is negligible. We study the variation of $P_{round}$ with $T$ in Figure 4.6. It can be seen that the improvement between successive rounds steadily decreases: After $T = d$ rounds it becomes negligible. Hence we have used $T = d$ in all further simulations.

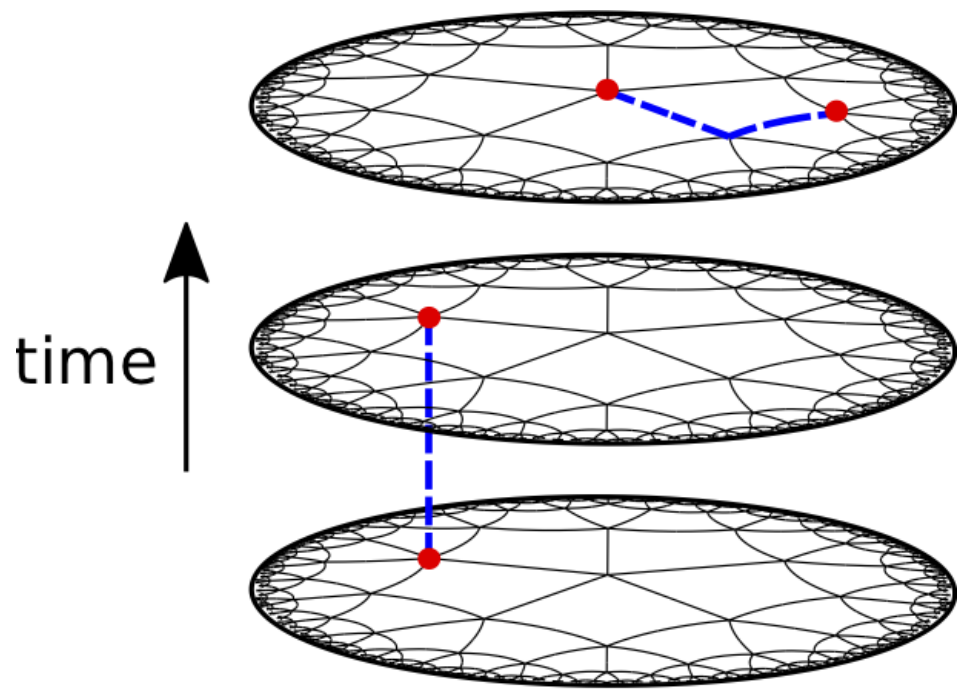

Figure 4.5: Minimum weight matching for noisy syndromes in a hyperbolic space. For ease of illustration we are showing the infinite lattice instead of a finite, compactified one and we omitted the vertical edges. There are three QEC cycles. Marked vertices are indicated by red dots. The result of the MWM is indicated by the blue, dashed lines.

Note that for the toric code the optimal number of rounds was $T \in O(\sqrt{n})$, so that the optimal number of rounds for the $\{5,4\}$-code only produces logarithmic overhead in time, rather than a polynomial overhead.

### 4.2.3 Results of the Monte Carlo simulation

We study the logical error probability $\overline{P}$ for the $\{5,4\}$-code using the decoding method and the noise model described above. The curves of the three largest codes cross at around 1.3% which can be seen in Figure 4.7. This is consistent with the lower bound of 0.7% from Table 3.2. The crossover point is somewhat lower as the 2.5% when parity check measurements are noiseless. This result may be surprising if our intuition is informed by the toric code. If syndromes can be extracted ideally, the threshold of the toric code is at 10.3% [45]. Changing to the phenomenological error model the threshold drops considerably, to around 3%, about a factor of 3 [45].

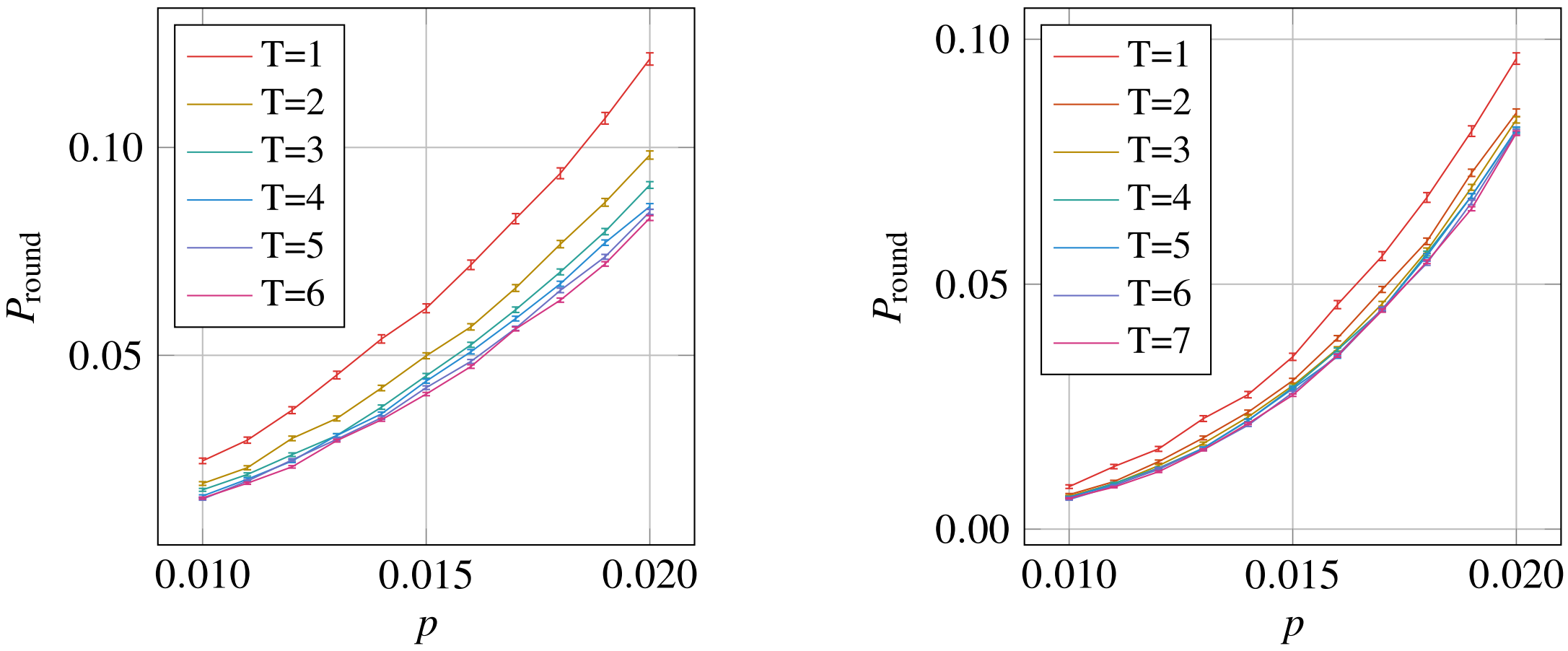

(a) Data for [[60,8,4]]

(b) Data for [[160,18,6]]

Figure 4.6: Variation of the logical error probability per round $P_{\text{round}}$ with physical error probability for the $\{5,4\}$-code.

## 4.3 Approximation of $\overline{P}$ in the low error probability limit

We focus on getting an expression for the logical error probability $\overline{P}$ when the physical error probability $p$ is low compared to the noise threshold, assuming a minimum-weight decoding method. This approach has been used for the surface code in [47].

### 4.3.1 Perfect measurements

We first consider the case of noiseless parity checks. The logical error probability for a logical $Z$ error is given by summing the probabilities of any $Z$-error to occur, times the probability of the decoder to fail on this error. In order for a minimum-weight decoder to fail, the weight of the error $E$ must be at least $|E| \geq \lceil d/2 \rceil$. This gives:

$$\overline{P}^{q=0} = \sum_{E:\, |E| \geq \lceil d/2 \rceil} \text{Prob}(\text{MWM fails on } E)\; p^{|E|}(1-p)^{n-|E|} \tag{4.2}$$

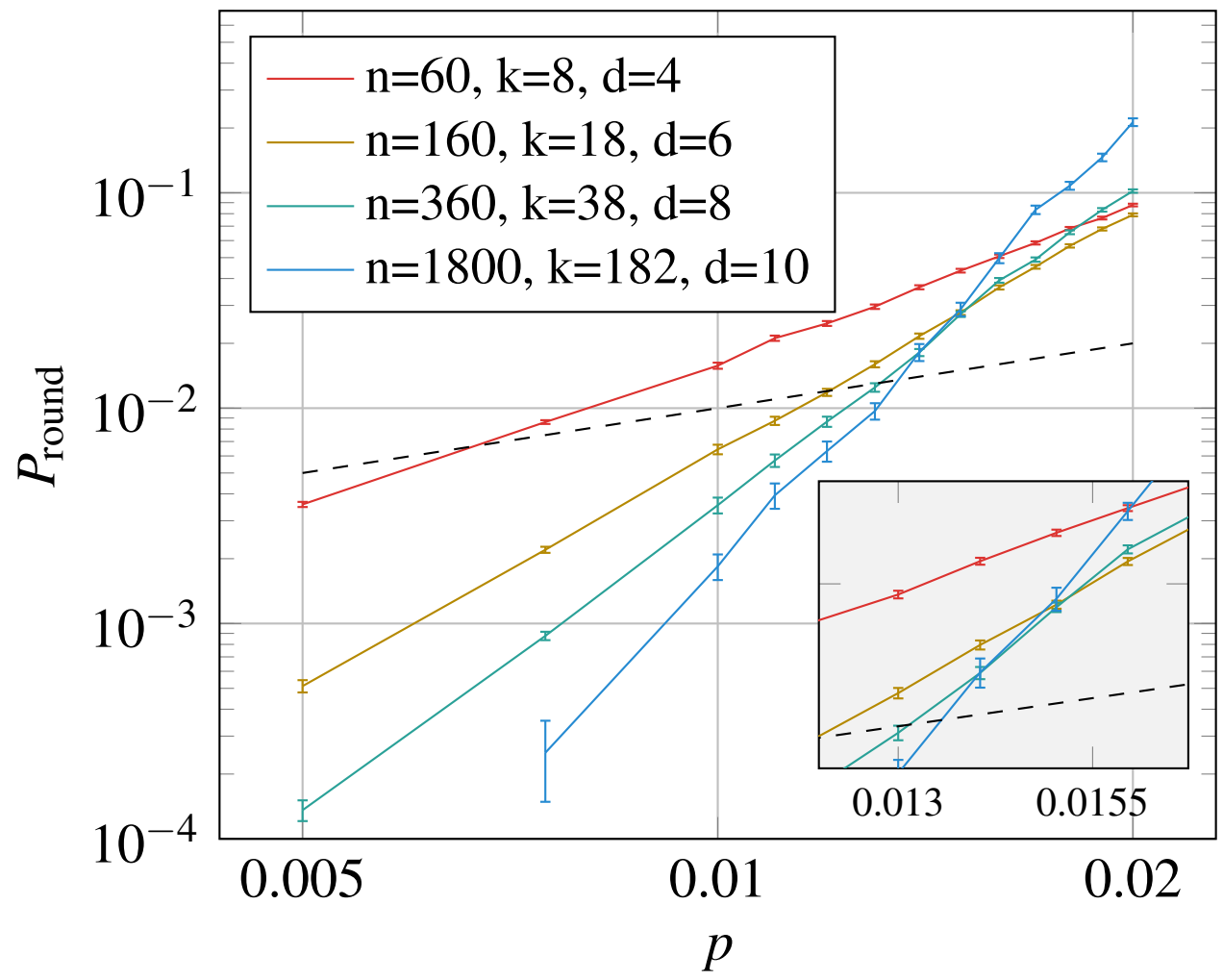

Figure 4.7: $P_{\text{round}}$ vs qubit and measurement error rate $p$. The plot above shows $P_{\text{round}}$ for $p$ in the range 0.5% to 2%. The diagonal dashed line marks $P_{\text{round}} = p$. The three largest codes seem to cross between 1.3% and 1.55%.

We are interested in the small $p$ regime of $\overline{P}$ which is a polynomial in $p$. We will thus retain only the lowest order $p^{\lceil d/2 \rceil}$ term, i.e. $\overline{P} \approx P_0^{q=0}$ where

$$P_0^{q=0} := \sum_{E:\, |E| \geq \lceil d/2 \rceil} \text{Prob}(\text{MWM fails on } E)\; p^{|E|}. \tag{4.3}$$

The errors of weight $\lceil d/2 \rceil$ on which the minimum-weight decoder fails are exactly those where all the support of the error is in the support of a weight-$d$ logical operator. There are $\binom{d}{d/2}$ of such errors. If $d$ is odd then the MWM-decoder will fail with probability 1. If $d$ is even, then there are two decodings that either lead to a successful decoding or a failure. Assuming that the decoder will choose randomly among these, the probability of failure is $1/2$ in this situation. Since the logical operators in the hyperbolic surface code will overlap on qubits, we can upper bound $P_0^{q=0}$ as

$$P_0^{q=0} \leq N_d^Z \left( \frac{3}{4} - \frac{1}{4}(-1)^d \right) \binom{d}{\lceil d/2 \rceil} p^{\lceil d/2 \rceil} \tag{4.4}$$

where $N_d$ is the number of logical operators of weight $d$. The right hand side of Equation 4.4 can be used to approximate the error probability when syndrome measurements are ideal.

### 4.3.2 Noisy measurements

For noisy parity check measurements, taking again the low $p$ limit, we can apply the same reasoning and only consider the lowest weight error configurations that can possibly lead to a logical failure. These lowest weight errors must then lie within a single time slice, so that one has

$$P_0^{q=p} \leq T N_d^Z \left(\frac{3}{4} - \frac{1}{4}(-1)^d\right) \binom{d}{\lceil d/2 \rceil} p^{\lceil d/2 \rceil}. \tag{4.5}$$

For hyperbolic codes there is no known closed expression for $N_d^Z$. However, one can compute $N_d^Z$ efficiently, see Section 3.4.2.

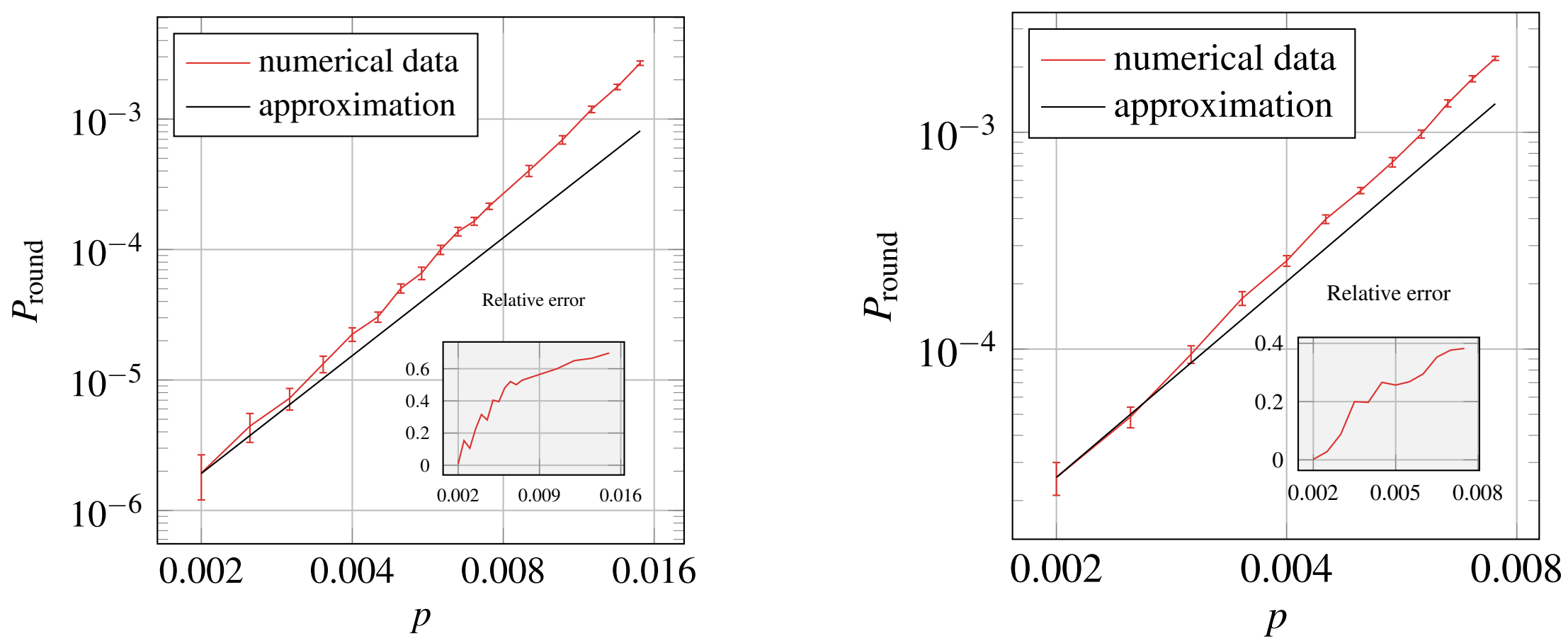

(a) Toric code with $L = 6$.

(b) $\{5,4\}$-hyperbolic code with $n = 160$.

Figure 4.8: Comparing numerical estimates for $P_{\text{round}}$ (red) with the heuristic approximation in Equation 4.5 (black). The relative error is the absolute difference between the numerical value and the approximation divided by the numerical value.

The approximation in Equation 4.5 agrees well with data obtained from numerical simulations. This can be seen in Figure 4.8 where we compute the per-round approximation in Equation 4.5 versus the numerical per-round logical error probability $P_{\text{round}}$.

## 4.4 Overhead comparison

In this section we will show that hyperbolic codes can offer an advantage over the toric code and the surface code in terms of qubit overhead. By overhead we mean the number of physical qubits needed to guarantee a certain rate of error suppression on the encoded qubits.

### 4.4.1 Perfect measurements

To compare the toric code to the hyperbolic surface codes we consider the overhead to produce logical qubits and protect them against decoherence. To do this we mark up to which physical error rate a code can protect all of its qubits with probability at least 0.999. The results are shown in Figure 4.9.

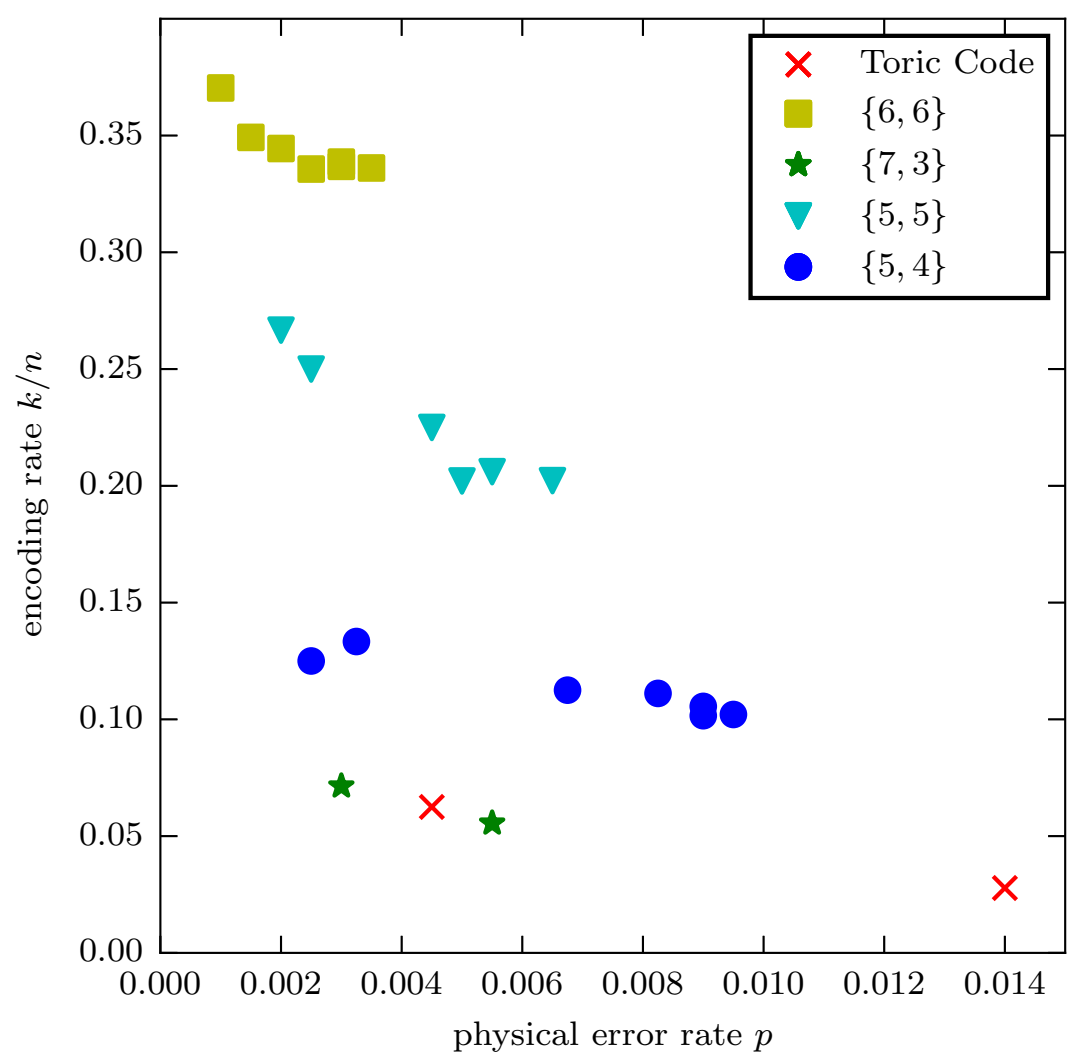

Figure 4.9: Encoding rate of a code which protects qubits with probability $> 0.999$. The number of physical qubits varies between 60 and 960. Data points are labeled by the tessellation. The two instances of the toric code are $L = 4$ and $L = 6$.

We see that the toric codes protect against errors up to a higher physical error rate. However, their overhead is quite big compared to the hyperbolic surface codes.

For each family of codes the encoding rate decreases with $n$ (see Equation 3.16). Since we are below threshold and $\overline{P}$ goes to zero for increasing $n$ the capability of the codes to protect against errors increases. For each code family we obtain a slope which converges to the asymptotic rate of the code. Codes with lower rate offer better protection against errors. The $\{5,4\}$-code which appears to offer the best error protection of all hyperbolic surface codes.

An efficient use of the surface code in which the lattice is chopped off at the boundaries (see [39]) has parameters $[[n = d^2, k = 1, d]]$. Hence to encode 8 qubits with the surface code with distance 4 requires 128 physical qubits, showing that the hyperbolic construction can lead to more efficient coding than the surface code. Note that the surface code has a higher threshold, making it more favorable for noisy qubits. For more coherent qubits hyperbolic surface codes might offer an advantage.

### 4.4.2 Noisy measurements

Let us now consider the case where the measurements of the check operators are noisy. We assume the same error model as in Section 4.2. A simple comparison between (semi)-hyperbolic codes and copies of the toric code can be done by fixing the number of logical qubits $k$, the distance $d$ and compare the number of physical qubits $n$. The parameters of the toric code are $[[2d^2, 2, d]]$. To have the same number of encoded qubits we take $k/2$ copies of the toric code, each with distance $d$, so $N_{\text{toric}} = kd^2$. For the (semi-) hyperbolic codes that we have studied one has $N_{\text{hyper}} = kd^2/(c_{4,5}\log(10k))$ assuming the asymptotic rate $k/n \to 1/10$ and a distance $d = c_{4,5}\log n$ for the $\{4,5\}$-hyperbolic code.

In order to get more insight into the possible savings one can numerically compute the maximum error probability $p_{\max}(\overline{P}_{\text{target}})$ such that $\overline{P} \leq \overline{P}_{\text{target}}$ for a surface code. Here $\overline{P}$ is the logical error probability after $T = d$ QEC rounds where $d$ is the code distance.

We have executed this numerical analysis for $\overline{P}_{\text{target}} = 10^{-5}$, resulting in the values

$$\begin{aligned} [[60, 8, 4]]&: \ p_{\max}(10^{-5}) \approx 1.5 \times 10^{-4} \\ [[160,18,6]]&: \ p_{\max}(10^{-5}) \approx 9.5 \times 10^{-4} \\ [[360,38,8]]&: \ p_{\max}(10^{-5}) \approx 1.5 \times 10^{-3} \end{aligned} \tag{4.6}$$

In order to compare the performance of the hyperbolic codes with the toric code we will focus on the largest of the three codes which has 38 logical qubits. In order to encode 38 logical qubits using the toric code, one needs 19 torii each with $2L^2$ physical qubits. If we choose all torii with $L = 3$ one has a total of 342 qubits and at $p = 1.5 \times 10^{-3}$ numerical data show that $\overline{P} = (6.8 \pm 0.7) \times 10^{-3}$ after 3 QEC rounds. Remember, that $\overline{P}$ is the probability for *any* logical qubit to be corrupted. For $L = 4$ one has 608 qubits in total and at $p = 1.5 \times 10^{-3}$ numerical runs give the estimate $\overline{P} = (9.3 \pm 0.6) \times 10^{-3}$ after 4 QEC rounds. Given that all 38 logical qubits encoded in the hyperbolic code with 360 physical qubits have a logical error probability of $10^{-5}$ after 8 QEC rounds, it clearly outperforms the toric code.

There is a version of the toric code that we will call the *rotated toric code* which has a better scaling between distance and number of physical qubits. Taking the set $[0,L]^2 \subset \mathbb{R}^2$ and identifying all points $(x,0) \sim (x,L)$ and $(0,y) \sim (L,y)$ for any $x,y \in [0,L]$ gives a torus. Instead of tessellating it with a square grid in the canonical way we choose the vertices of the tessellation to be located at integer points $(x,y) \in \{0,...,L-1\}^2$ for even $x$ and $(x,y-1/2) \in \{0,...,L-1\} \times \{1/2,...,L-1/2\}$ for odd $x$. Edges run diagonally from $(x,y)$ to $(x,y+1/2)$ and to $(x,y-1/2)$. This procedure gives a square grid on the torus, rotated by 45 degrees. Note that the shortest non-trivial loop following the edges around the torus has length $L$ while the total number of edges, and hence the number of qubits in the derived code, is $L^2$ as compared to $2L^2$ for the regular toric code. The number of encoded qubits is still 2 as there are two independent, non-trivial loops.

Using 19 rotated toric codes we can either use $L = 4$ or $L = 6$ amounting to 304 and 684 physical qubits resp. For $L = 4$ the logical error probability at $p = 1.5 \times 10^{-3}$ is numerically estimated to be $\overline{P} = (2.3 \pm 0.1) \times 10^{-2}$ after 4 QEC rounds. For $L = 6$ the logical error probability at $p = 1.5 \times 10^{-3}$ is numerically estimated to be $\overline{P} = (7.0 \pm 0.2) \times 10^{-4}$ after 6 QEC rounds.

**Overhead of semi-hyperbolic surface codes**

In order to further estimate the scaling of the logical error probability we wrote down an approximate model for the logical error probability in Section 4.3 which we used in Section 4.4.2 as the basis for further comparison. Equation 4.4 can be used to analyze the semi-hyperbolic code family in the regime where the physical error rate $p$ is low. To compare the overhead in physical qubits we fix $P_{\text{round}}$ and determine the maximum physical error probability $p_{\max}(P_{\text{round}})$. This value for $P_{\text{round}}$ was chosen such that the corresponding $p_{\max}$ is in a regime where the approximation formula is valid for all lattices considered here. In Figure 4.10 we plot the encoding rate $k/n$ against $p_{\max}$ for different code families with $P_{\text{round}} = 10^{-8}$. We ran Monte Carlo simulations for higher values of $p$ to ensure that the approximation did not deviate by more than 10% from the numerical value. Once this is established, we assume that the approximation will only become better with lower $p$.

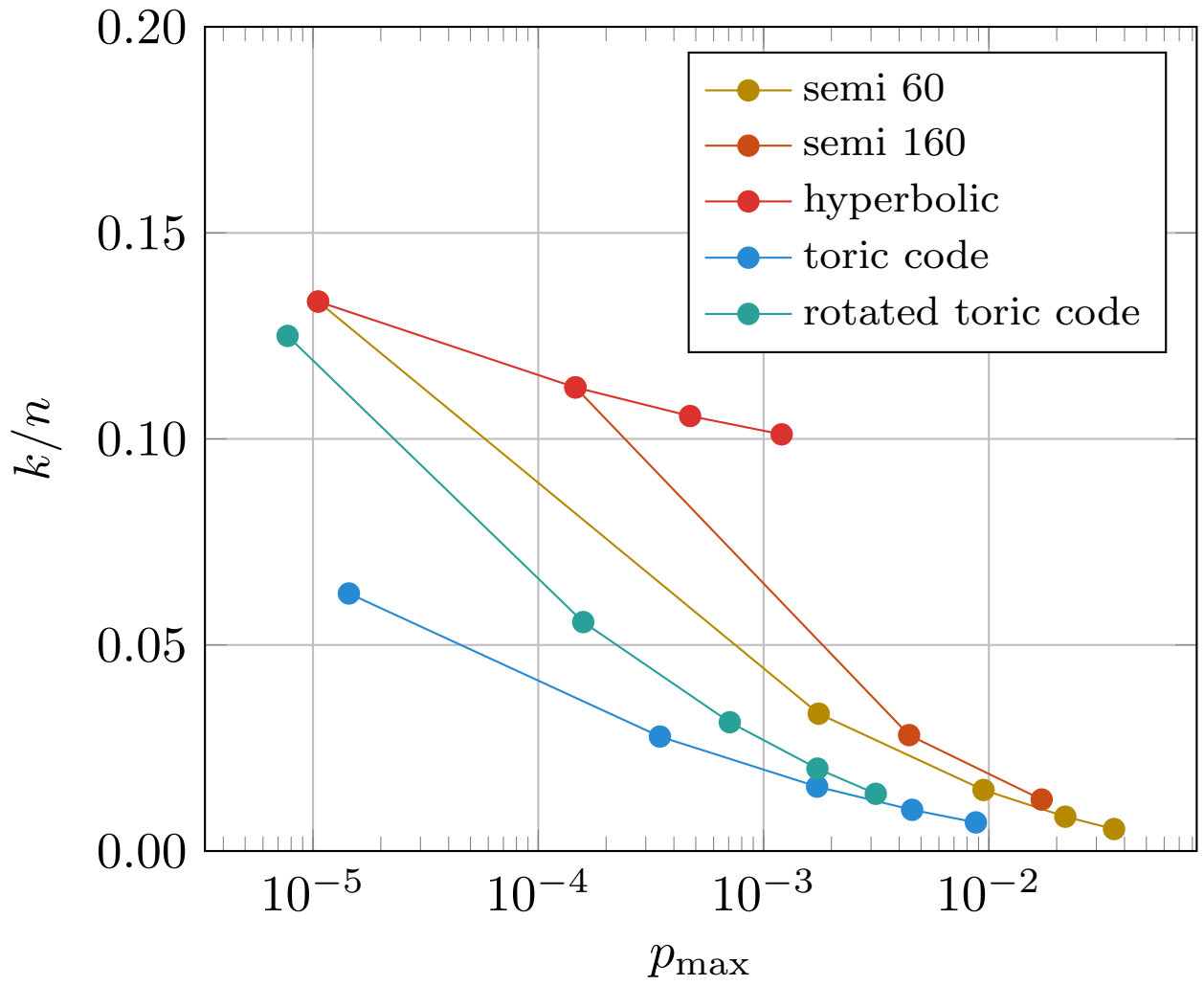

Figure 4.10: Overhead for different code families. The value of $p_{\max}(10^{-8})$ for various codes. The semi-hyperbolic codes in this figure are derived from a $\{4,5\}$-code with $n = 60$ and $n = 160$ and $l = 2,3,4$ etc. The hyperbolic codes are derived from $\{4,5\}$-tessellations with $n = 60,160,360,1800$. The toric codes considered have distance $L = 4,6,8,10,12$.

For a fixed encoding rate, we see that the semi-hyperbolic codes can offer better protection against errors than the toric code. For example, in Figure 4.10 we see that for $p_{\max} = 1.7 \times 10^{-3}$ we can choose between copies of the $L = 8$ toric code with $k/n = 0.0156$, the rotated toric code with $L = 10$ and $k/n = 0.02$, the semi-hyperbolic code with $l = 2$ from a $\{4,5\}$-tessellation with 60 edges with $k/n = 0.03$ or a hyperbolic code with $k/n = 0.1$.

# Chapter 5

# Homological codes from 4D tessellations

## 5.1 Advantages of higher-dimensional codes

Until now we have only considered codes based on tessellations of 2D manifolds. However, the recipe described in Section 2.3, turning tessellated manifolds into quantum codes, works for any dimension. Keeping the dimension low has clear advantages, such as being able to embed the code into everyday 3D space. However, codes defined from higher-dimensional manifolds do have advantages which may outweigh those of 2D codes, in particular for quantum computing architectures which allow for non-local and non-planar connectivity. For example, there are proposed modular architectures in which each module holds a small number of qubits. The modules are interconnected by photonic links which are not restricted to be planar [6, 48, 49, 50, 51]. Another example for an architecture which is not bound to planar interactions is linear optical quantum computers where qubits are realized directly using photons [52, 53, 54]. The photons interact by routing them through wave guides that are not restricted in their connectivity, allowing for non-planar interactions between qubits. A disadvantage of higher dimensional codes is that they generally have a higher qubit degree.

### 5.1.1 Check measurements in 4D homological codes

When we define a quantum code by applying the recipe of Section 2.3 to a two-dimensional tessellation, we have to identify qubits with the edges and stabilizer checks with vertices and faces.

Geometrically, errors correspond to 1D objects (collections of edges on which an error occurred). Check measurements will detect the boundary of this error chain which is zero-dimensional. Equation 2.44 tells us that there is only a single linear dependency for each $Z$-checks and $X$-checks due to the fact that $\dim H_2 = \dim H^0 = 1$ as there is only a single connected component.

To see what happens in higher dimensions, let us first study the 3D version of the toric code. Consider a 3D torus $T^3$ which can be realized by taking a box and identifying opposite sides. It can be tessellated by a $L \times L \times L$ cubic grid. We identify qubits with the faces of the tessellation ($i = 2$), so that the stabilizers are associated with the cubes and edges. In the 3D toric code there are non-trivial linear dependencies between the stabilizer checks associated with the edges. These linear dependencies stem from the fact that

$$\partial_1 \partial_2 = 0. \tag{5.1}$$

Since the support of each $X$-check is given by a row of $\partial_2$ we can interpret Equation 5.1 in terms of matrices, expressing that the rows of $\partial_1$ are the linear dependencies between the $X$-checks. Since the rows of $\partial_1$ span the image of $\delta_0$ (see Equation 2.27), the linear dependencies are given by the 0-coboundaries (see Figure 5.1). Hence, these linear dependencies are *local*.

Geometrically, Equation 5.1 simply states that the record of the violated $X$-check measurements which detect Pauli-$Z$ errors (the syndrome $s_Z$) is a collection of closed loops.

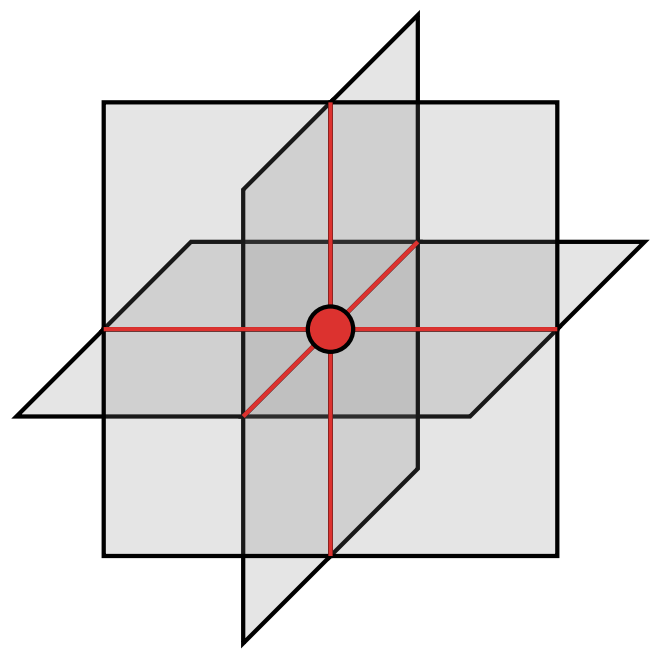

Figure 5.1: The local dependency of edge stabilizers in the 3D toric code. Taking the product of all edge-stabilizers (red) incident to a common vertex (red dot) gives the identity.

The $Z$-checks of the 3D toric code do not have such a linear dependency. They correspond to the rows of $\delta_2$ ($\partial_1^*$ in the dual) and since $\delta_3$ ($\partial_0^*$) is the zero map, the equation $\delta_3 \delta_2 = \partial_0^* \partial_1^* = 0$ is trivial. This problem can be solved by considering tessellations of four-dimensional spaces, so that both $Z$-checks and $X$-checks are subject to local constraints. We may think of the local linear dependencies as two classical linear codes which encode the syndrome.

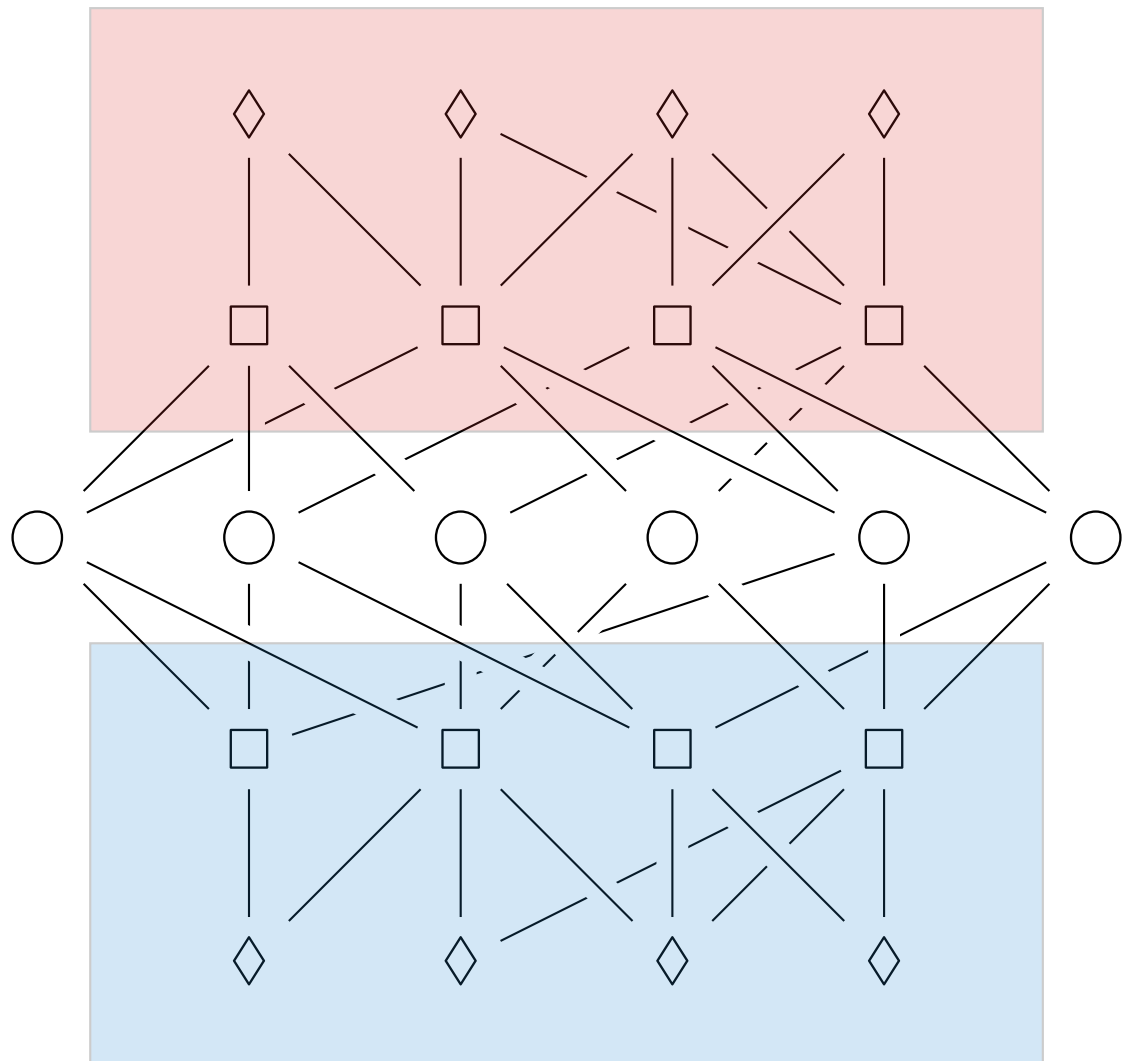

Figure 5.2: Illustration of a Hasse Diagram of a 4D tessellation. An actual 4D code will have more nodes of higher degree. The $i$th level of the diagram consists of all $i$-cells. Cells are connected by an edge if they are incident in the tessellation (see discussion in Section 2.2.4). The Hasse diagram above defines a Tanner graph describing a CSS code just as we had already seen in Section 2.2.4. In 4D there are additionally two linear codes acting on the $Z$-checks (red) and the $X$-checks (blue) of the CSS code. The syndrome checks are indicated by diamonds. They correspond to the vertices and 4-cells of the tessellation.

### 5.1.2 Single-shot fault-tolerance

Since the syndrome of a 4D code has a local redundancy, it has some built-in robustness against syndrome errors. In comparison, it is known that the 2D toric code does not have a decoding threshold in the presence of syndrome noise when only single rounds of check measurements are allowed: The parity measurements have to be repeated and the record of the syndrome measurements is decoded. This essentially implements a repetition code in time. Note that the number of repeated measurements is increased with the size of the code and thus the reliability of the syndrome record is increased as well. This is not the case for the encoding of the syndrome as the code distance clearly does not depend on the size of the system. It is therefore not obvious whether single-shot error correction can have a threshold.

In [55] the author shows that a threshold with single-shot measurements does exist for 4D homological codes. It is assumed that the decoder first performs minimum weight decoding on the syndrome (via the classical code) and then performs the standard minimum weight decoding. Note that the classical code will in general not return the actual syndrome, so that after the decoding procedure there will be residual errors left. Before we discuss details of the decoding we will introduce three constructions of 4D codes. All three are generalizations of codes that we have already seen: The toric code, the surface code and hyperbolic codes.

## 5.2 4D Toric Code

The *4D toric code* is a homological quantum code defined on a four-dimensional *hypercubic tessellation* of a 4D torus $T^4$. The set of vertices is given by

$$V = \{(x,y,z,w) \mid x,y,z,w \in \{0,\dots,L-1\}\}. \tag{5.2}$$

Two vertices are connected by an edge if and only if their coordinates differ in one position by 1 (mod $L$). In other words, edges can be identified with sets containing two vertices

$$\{(x,y,z,w),(x,y,z,w)+\vec{e}_i\}, \tag{5.3}$$

where $\vec{e}_i$ is the $i$th standard base vector and we identify $L$ with 0. The faces of the lattice are squares such as

$$(x,y,z,w),(x+1,y,z,w),(x+1,y+1,z,w),(x,y+1,z,w). \tag{5.4}$$

The faces are identified with qubits. The number of faces in a tessellation of side-length $L$ is $\binom{4}{2}L^4 = 6L^4$. This follows from the fact that every face is uniquely determined by a vertex $v$ and two coordinates $i,j \in \{x,y,z,w\}$, so that the face with base-point $v$ lies in the $i$-$j$-plane. More generally, the number of $k$-dimensional objects in the lattice (1-dimensional objects would be edges or 3-dimensional objects would be cubes) is given by $\binom{4}{k}L^4$. There are stabilizer checks for every edge and cube in the lattice. A stabilizer check associated with a particular edge acts as Pauli-$X$ on all faces which are incident to it (Figure 5.3 (left)), whereas a stabilizer check associated with a particular cube acts as Pauli-$Z$ on all faces incident to this particular cube (Figure 5.3 (right)). All stabilizer checks act on 6 qubits. Every qubit is acted upon by 4 $X$-checks as each face has 4 edges incident to it. The same holds for $Z$-checks as the hypercubic tessellation is self-dual. Each qubit is therefore acted upon by 8 check operators (the qubit degree is 8).

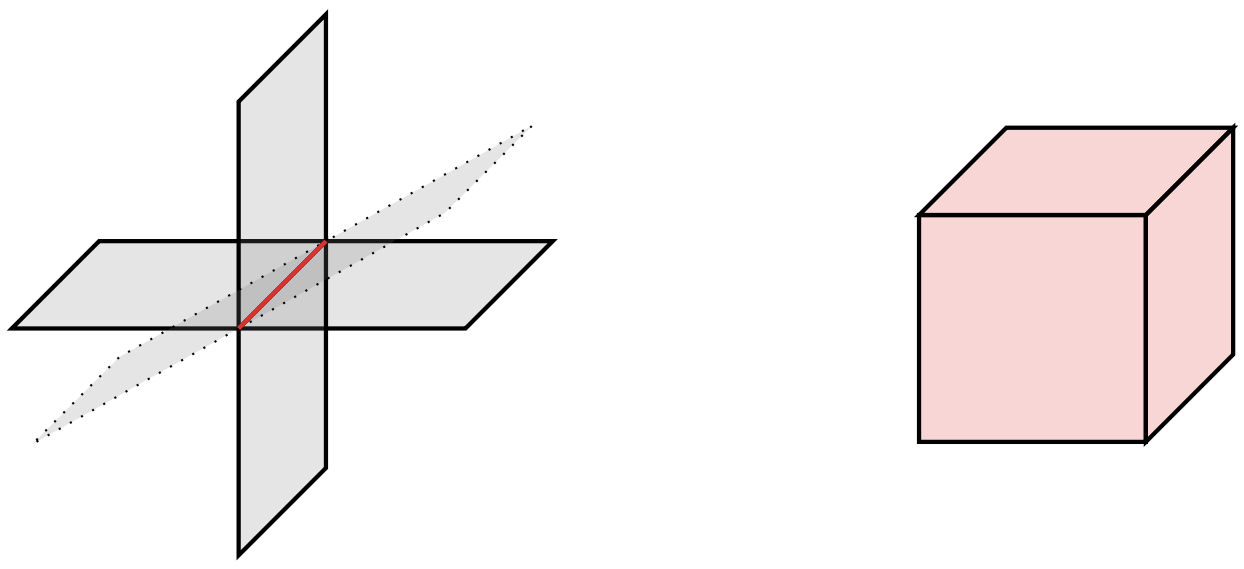

Figure 5.3: The stabilizer checks of the 4D toric code correspond to edges which act as Pauli-$X$ on all qubits incident to an edge (left) and cubes which act as Pauli-$Z$ on all qubits incident to a cube (right).

Since the topology of the lattice is non-trivial, we can consider a sheet extending over the whole $x$-$y$-plane. Due to the periodicity of the lattice this sheet has no boundary, but is not itself the boundary of any 3D volume and thus it is not the product of any $Z$-stabilizers. Therefore it must be a logical operator (an essential 2-cycle). There is one such operator for every plane in the tessellation. Each plane is labeled by two coordinates ($x$-$y$-plane, $x$-$z$-plane, ...) so that the 4D toric code encodes $\binom{4}{2} = 6$ logical qubits. A sheet extending through the whole lattice consists of at least $L^2$ faces which means that the 4D toric code has distance growing quadratically with the number of physical qubits. The parameters of the code are $[[n = 6L^4, k = 6, d = L^2]]$.

Let us consider the classical code which encodes the $Z$ syndrome $s_Z$. There is a bit for every $X$-check (edge) in the lattice, so the size of the code is $4L^4$. Moreover, there is one local dependency for every vertex. However, the local dependencies themselves are not independent. Taking the product over all vertices and all edge-checks incident to this vertex gives the identity since every edge is incident to two vertices. Hence, the number of independent checks in the classical linear code encoding the syndrome information is $L^4 - 1$. The encoded syndrome information therefore contains $4L^4 - (L^4 - 1) = 3L^4 + 1$ bits. The distance of the classical code is 4 since adding the boundary of a face (corresponding to the syndrome of a single error on a face) takes us from one valid syndrome to another valid syndrome.

## 5.3 Tesseract Code

The surface code can be generalized to a 4D code as well. Although the resulting code is four-dimensional and hence not planar, it can be advantageous to consider the tesseract code rather than the 4D toric code. One advantage is the fact that the 1st homology group $H_1$ of the tesseract code is trivial, so that any cycle is a boundary and thus a valid syndrome. This is not true for the 4D toric code where essential 1-cycles exist. As we are interested in single-shot decoding where the syndrome has to be fixed up, this is problematic as the syndrome should only be fixed to be a valid syndrome (a boundary).

Just as for the 2D surface code we start with a tessellation of a box. This time the box is a four-dimensional $L \times L \times (L-1) \times (L-1)$-box which is tessellated by hypercubes in the same fashion as the 4D toric code. We use relative homology to define a quantum code. The cells we are going to remove all cells which are contained in the 3D hyperplanes at the boundary of the box which are defined by $x = 0$, $x = L$, $y = 0$ and $y = L$ (cf. Figure 2.9). The constraints define a single connected component at the boundary that we will denote $B$. After removing the vertices in $B$ we are left with $|V| = L^2(L-1)^2$ vertices. We can count the number of all other cells by associating each with a base vertex. We have to be careful to not overcount those cells which would "stick out" of the box. For example, there are $2L^3(L-1)$ edges which point in either the $z$- or $w$-direction, but only $2L^2(L-1)^2$ edges pointing in the $x$- or $y$-direction due to removing edges at the boundary. This gives in total $|E| = 2L^2(L-1)^2 + 2L^3(L-1)$ edges. Similarly, for faces we have to consider the different planes that they can lie in. In the $x$-$y$-planes there are $(L-1)^4$. The $z$-$w$-planes are unaffected so that there are $L^4$ faces lying in this direction. The other 4 directions ($x$-$z$, $x$-$w$,...) contain $L^2(L-1)^2$ faces. This makes a total of $|F| = L^4 + 4L^2(L-1)^2 + (L-1)^4 = 6L^4 - 12L^3 + 10L^2 - 4L + 1$.

A logical $Z$ operator is given by a sheet which stretches through the $x$-$y$-plane. As all edge-checks at those boundaries have been removed the sheet will commute with all remaining $X$-stabilizers. The minimal weight of such an operator is $L^2$. It is hard to visualize, but there exists another cocycle giving rise to a $X$-logical which anti-commutes with this logical $Z$ operator.[1] It turns out that these are the only logical operators so that there is only a single logical qubit encoded in the tesseract code. To prove this we need some more advanced machinery of algebraic topology. We have deferred the proof to Appendix B. To summarize: The tesseract code is a $[[n = 6L^4 - 12L^3 + 10L^2 - 4L + 1, k = 1, d = L^2]]$-code which exhibits redundancy in the stabilizer checks.

## 5.4 4D Hyperbolic codes

More subtle than for the toric and surface code is the generalization of hyperbolic codes to 4D. The curvature at a point in space can be defined using a generalization of the Bertrand-Puiseux Theorem.

**Definition 5.1.** The curvature at point $p$ in a $D$-dimensional manifold $M$ is given by

$$\kappa_p = \lim_{r \to 0^+} (6D + 12) \frac{\text{vol}_{\mathbb{E}^D}(B(r)) - \text{vol}_M(B_p(r))}{r^{D+2} \, \text{vol}_{\mathbb{E}^D}(B(r))} \tag{5.5}$$

where $\text{vol}_{\mathbb{E}^D}(B(r))$ is the $D$-dimensional volume of a ball in Euclidean space and $\text{vol}_M(B_p(r))$ the $D$-dimensional volume of a ball of radius $r$ centered at $p$ in $M$.

We see that when $M$ is negatively curved at $p$, we find in some sense *more space* around the point $p$ than in a space with zero curvature. This is illustrated in Figure 5.4 for the 3D case.

In [10] it is shown that tessellations of 4D hyperbolic spaces give rise to quantum codes with constant rate. This also holds for the 2D hyperbolic codes discussed earlier and the proof follows along the same lines, using the higher-dimensional version of the Gauß-Bonnet Theorem (Theorem 3.3). However, for 4D hyperbolic codes the distance $d$ scales like $n^{\varepsilon}$, where $0 < \varepsilon < 0.3$. The argument can be summarized as follows: it is known that the injectivity radius[2] $R_{inj}$ of a hyperbolic manifold in any dimension is at least logarithmic in its volume [56]. A theorem by Anderson [57] states that the volume of an essential $i$-cycle is at least the volume of a ball of radius $R_{inj}$. If $i = 1$, then a 1-dimensional hyperbolic ball of radius $R$ is simply an interval $[-R, R]$ and hence has 1-dimensional volume (length) $2R$. This is translates into logarithmic distance for the 2D hyperbolic codes. However, for $i \geq 2$ we have that the $i$-dimensional volume of a ball with radius $R$ in hyperbolic space is a function in $O(e^{(i-1)R})$. Hence, for a ball of radius $R_{inj}$ has volume scaling polynomially with the total volume of the manifold. Applying the machinery of homological codes gives us a code with polynomial distance $d \in O(n^{\varepsilon})$.

[1] Alternatively, we can also exploit that, up to redefining $B$, the tessellation is self-dual.

[2] See Section 3.1.5.

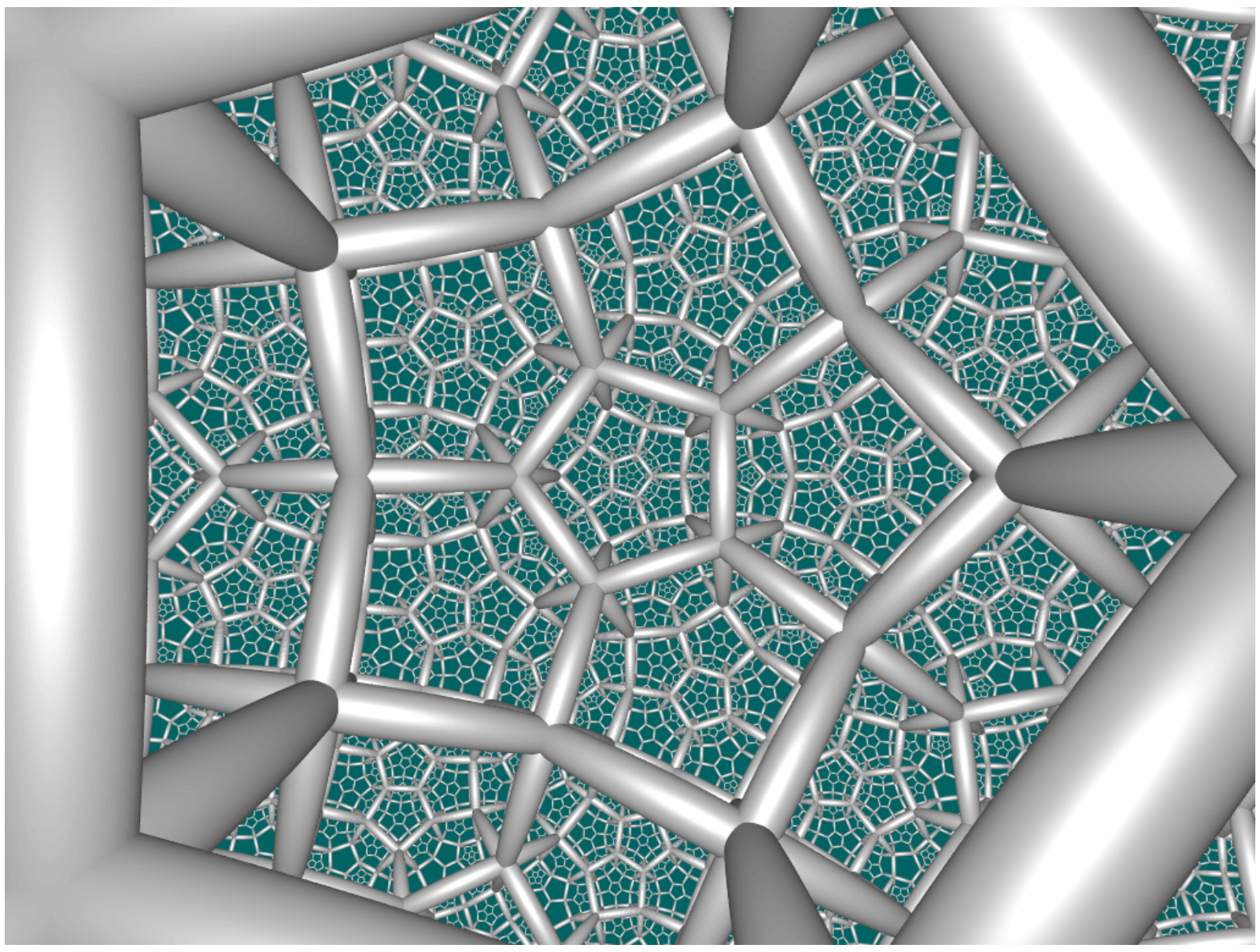

Figure 5.4: A regular tessellation of $\mathbb{H}^3$. The 3-cells are dodecahedra. Four dodecahedra are arranged around every edge, which would not be possible in a Euclidean space. Note that space is branching out in a tree-like fashion. This image was created by Roice Nelson and is distributed under copyleft CC BY-SA 3.0.

The authors of [10] do not give a construction for these codes. However, we can use the same tools that we developed in Section 3.1 for the 2D case. Unfortunately, performing an exhaustive search for codes with up to $4 \times 10^4$ physical qubits only returned a single example. This example is a code based on a tessellation of a closed four-dimensional hyperbolic manifold by 136 copies of a 4D polytope called the *120-cell*. The faces of this tessellation are pentagons and the 3-cells are dodecahedra. The code encodes $k = 197$ logical qubits into $n = 16320$ physical qubits. The stabilizer checks are both weight 12 and each qubit is acted upon by 20 stabilizer checks. A single example is unfortunately not enough to determine a decoding threshold. However, we will see in the next section that this example shows quite good performance when compared to the 4D toric code. There exists at least one more example of a hyperbolic tessellation which is obtained by identifying the opposing dodecahedra at the boundary of the 120-cell. The result is known as the *Davis manifold* (see [56]).

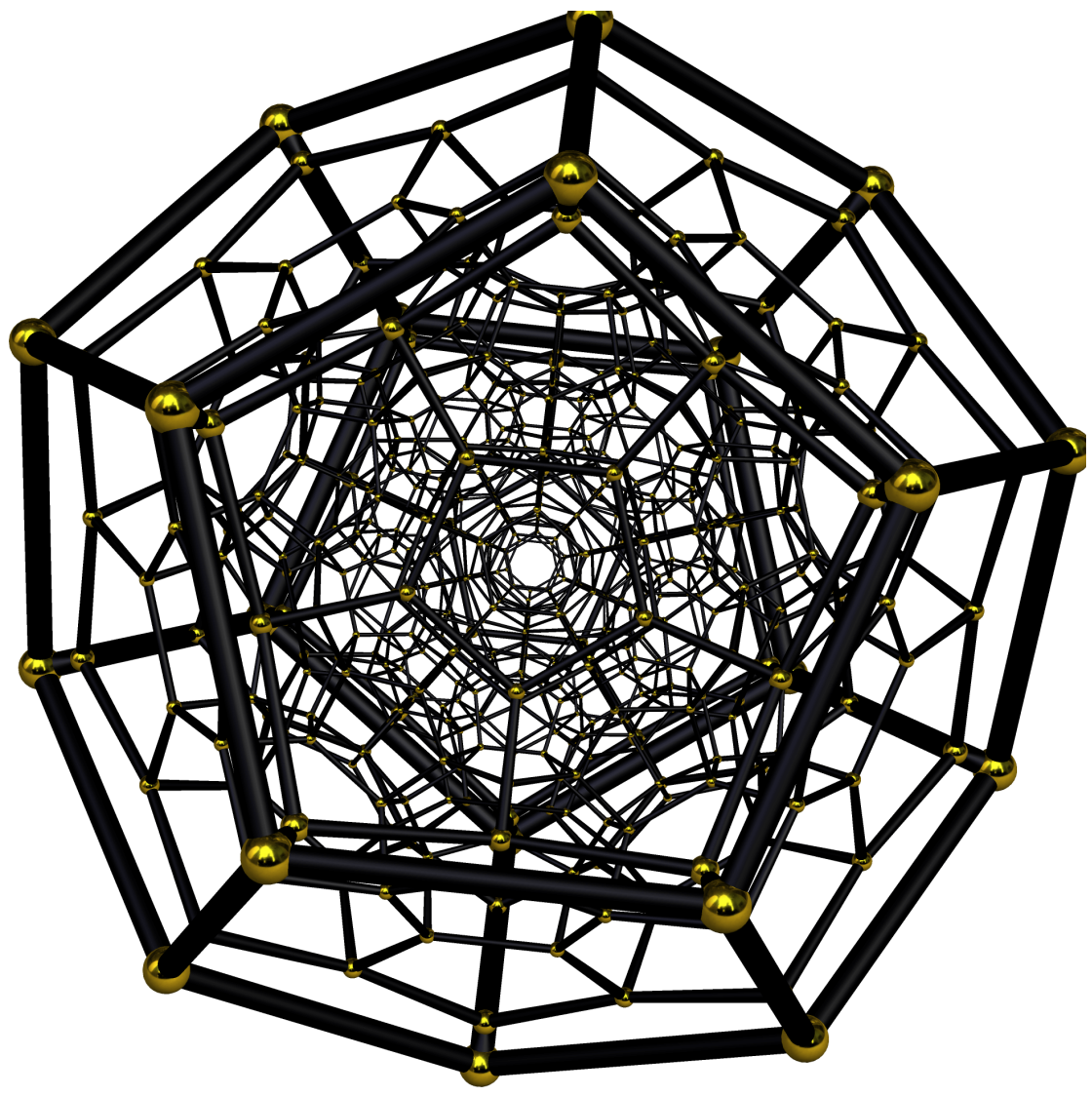

Figure 5.5: A projection of a 4D polytope called 120-cell. Its name is due to the fact that it consists of 120 dodecahedra which tessellate the 3D sphere $\mathbb{S}^3$. The reflective symmetry group subdivides this polygon into 14400 fundamental 4-simplices, analogous to the fundamental triangles of Figure 3.5.

# Chapter 6

# Decoding 4D homological codes

## 6.1 The decoding problem of 4D homological codes

### 6.1.1 Minimum-weight decoding

A minimum-weight decoder of a 4D code has to find a minimum-area surface among all surfaces that have the given syndrome as its boundary. It is known that the problem of finding such a minimum-area surface can be solved efficiently when the code is three-dimensional [58], but it is open whether it can be solved efficiently for a 4D tessellation. In [59, 60] the authors reason, based on a duality argument, that the threshold of the 4D toric code against the independent $X$-$Z$ error model under minimum-weight decoding is $p_c \approx 11.003\%$. This is the same threshold as for the 2D toric code.

In Figure 6.2 we show the results of a numerical simulation of performing error correction on the tesseract code. The global decoder was implemented by finding the solution of the optimization problem using the optimization software GUROBI. As this implementation is not efficient, we can only consider small system sizes. For the case with perfect measurements (see Figure 6.2(a)) we see that the lines cross at around $11\%$ which is to be expected if this is the threshold of the 4D toric code. Both codes only differ at the boundary which does not affect the position of the threshold.

For noisy measurements we assume the same error model as in Chapter 4. The syndrome undergoes independent bit-flips with the same probability $p$ as the qubit errors. In contrast to the 2D case we can now use the single-shot property of the tesseract code: Instead of performing several measurements we can decode the syndrome. The noisy syndrome will consist of several open and closed strings. As a valid syndrome has to be a cycle, i.e. a set of strings that do not end at a vertex, we perform minimum weight perfect matching to obtain the closest valid syndrome (this is not necessarily optimal as we will discuss in the following section). Note that this strategy would not work on the 4D toric code as we might obtain non-trivial elements in $H_1$, i.e. loops which do not correspond to boundaries which go through the whole space. For the tesseract code this can not happen since we have $\dim H_1 = \dim H^3 = 0$ which we prove in Appendix B.

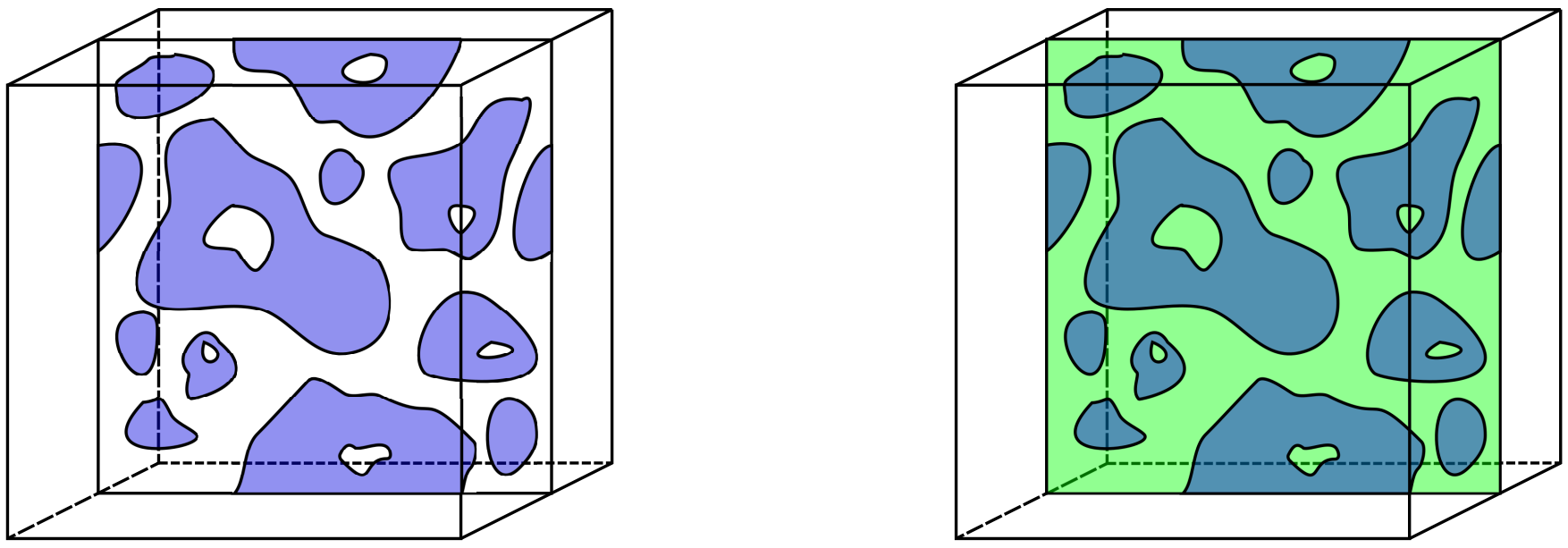

Figure 6.1: Example of failed error correction for the 4D toric code. (a) Shown is a 2D slice in a 3D slice of the full 4D space. Opposite boundaries are identified. In blue is a 'critical' high-weight error $E$ which may lead to failure. (b) Failed recovery $R$ (in green) of the error $E$ (blue). Together they form an essential 2-cycle, i.e. a logical operator.

As mentioned in the previous chapter, there will in general be residual errors left. An indicator of the decoder's performance is given by the probability distribution of those residual errors and more importantly, whether those residual errors are correctable. Hence, we set up our simulation as follows: After each QEC cycle we run the global decoder on the residual errors. If we can restore back to a code state without having applied a logical operator we call no logical failure and continue with the next time step. Otherwise we call failure and save the number of time steps until failure. In the end we obtain a value for the average number of time steps until the decoder fails depending on the physical error rate $p$ and the system size $L$. Repeating this process gives us a numerical estimate of the *average memory time* $T$. When the decoder performs well, the residual errors will be correctable and the average memory time is high.

The results of the simulation can be found in Figure 6.2(b). We see that there is an odd-even effect as codes with odd $L$ perform generally better than those with even $L$. From this data alone it is hard to say whether there is a threshold although $T$ grows with system size when $p$ is sufficiently small. We did not investigate this further as the global decoder is not efficient and hence not practical.

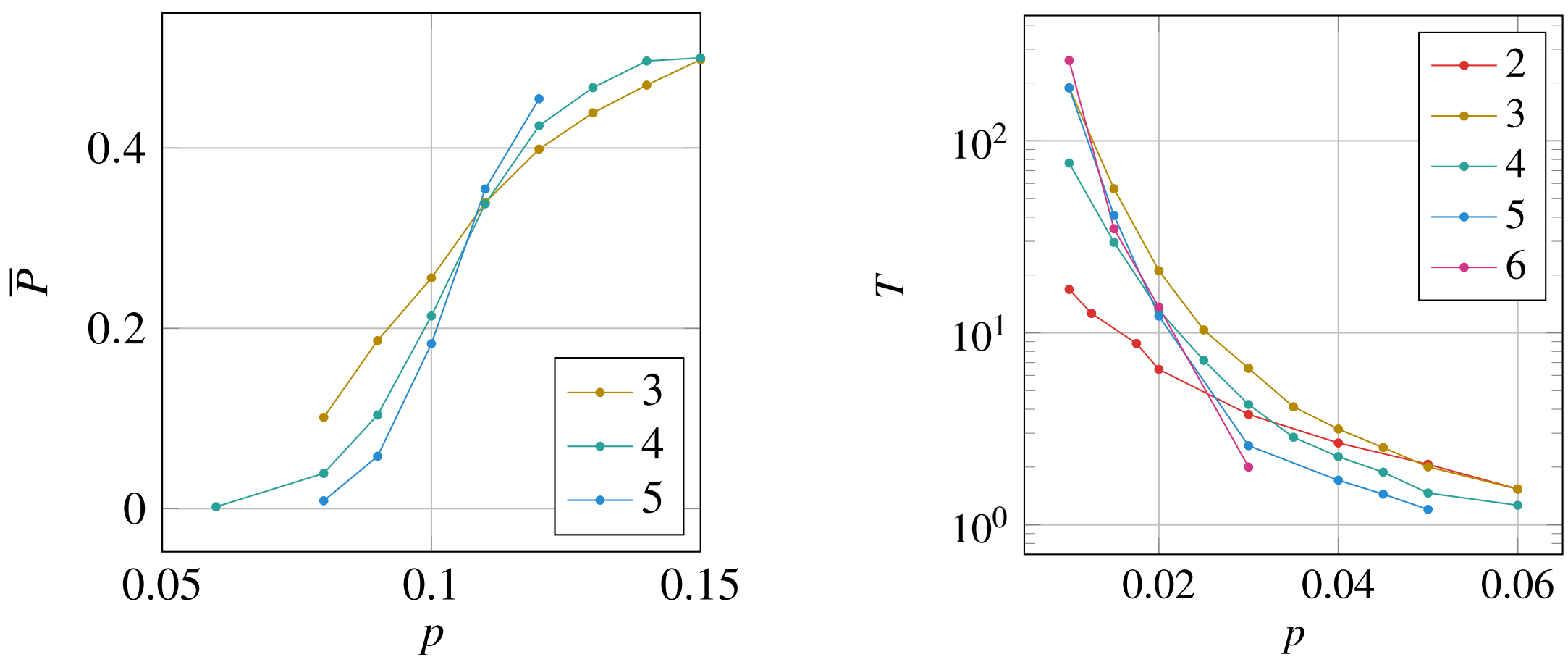

(a) perfect measurements

(b) noisy measurements

Figure 6.2: Numerical simulation for the global decoder. The logical failure probability $\overline{P}$ asymptotically goes to $1/2$ as the tesseract code encodes a single logical qubit.

There are several decoders with less optimal error-correction performance than the minimum weight decoder which are computationally efficient [40, 61, 62, 63]. The common way these decoders operate is to shorten the syndrome loops by applying a Pauli-$X$ or Pauli-$Z$ on nearby faces. And as it turns out we can make a virtue out of necessity: A rough calculation in [5] shows that assuming a quantum error correction cycle of 5MHz for an $L = 100$ surface code the classical decoder needs to process 100Gbit/s. A further disadvantage is the fact that an algorithm such as minimum-weight perfect matching has to run on a reasonably complex classical computer. This classical computer either is close to the qubits, e.g. inside a fridge, and produces thermal noise which increases the error rate, or further away which leads to an even bigger increase in the time-delay. In summary: It would be advantageous to have dedicated hardware which can be implemented using fast, low-power electrical components. This would decrease the effective rate of errors which can occur during an error correction cycle. The decoders that we are going to discuss do not have to perform more complex operations with growing system size which suggests that they are inherently more scalable than decoders like minimum-weight perfect matching.

In the next sections we are going to discuss these decoders and numerically analyze their performance. At the end of this chapter we will discuss the application of machine learning to solve the 4D decoding problem.

### 6.1.2 Energy-Barrier Limited Decoding

For all decoders that we will discuss in the following sections, it can be observed that they are unable to shrink and remove certain high-weight syndromes (assuming for simplicity that syndromes are noiseless). This implies that these decoders cannot necessarily correct a state with errors back into the code space. An example of such syndrome can be seen in Figure 6.3 where we imagine looking at a 2D plane of qubit faces. The qubit faces are flipped along a homologically non-trivial strip.

In such cases when the decoder get stuck and the syndrome does not change, we call logical failure. Note that this failure mode is related to the energy barrier [5] for the code: The errors that produce the non-local syndrome that the decoder cannot handle are precisely the errors which set the height of the energy barrier, namely $2L$ for the 4D toric code. The strip of errors can grow, without anti-commuting with yet more edge check operators, to become a logical operator which covers the whole 2D surface. Hence, once the error probability is high enough that errors are generated which locally have minimal energy (anti-commute with the minimal number of check operators), the decoder starts to fail. This is not an issue for a decoding procedure which has access to the full syndrome.

## 6.2 Hastings decoder

### 6.2.1 Decoding in local neighborhoods

In [61] Hastings proposed a local decoder for 4D homological codes based on a hyperbolic tessellation [10]. Using the fact that in negatively curved spaces the size of the boundary of a surface scales proportional to its interior, he proved that with the application of the decoder on some $O(1)$ ball (or neighborhood) the weight of an error (*i.e.* the number of qubits on which it acts non-trivially) is reduced by a constant fraction on average. The decoder is thus effective at removing errors in this geometry. We will consider its action here on the 4D toric code.

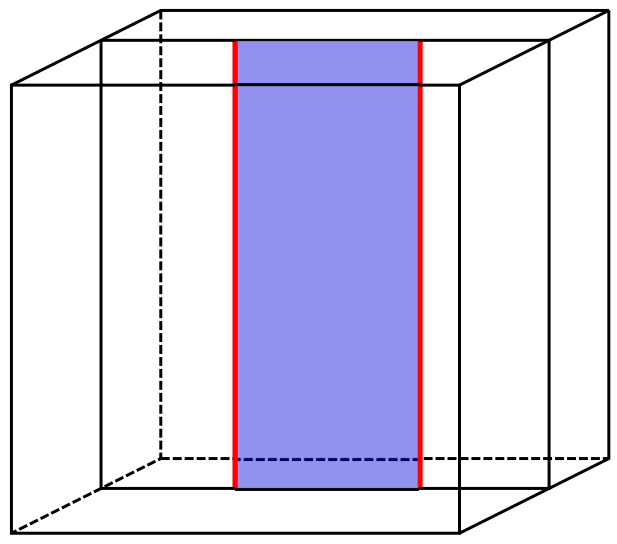

Figure 6.3: A 2D slice in a 3D slice in the full hypercubic tessellation where opposite sides are identified. A sheet of errors (blue) gives rise to a syndrome (red) on which the Hasting decoder, the cellular automaton decoders and the neural network decoder can get stuck.

We first describe the local decoding procedure assuming a perfect measurement of the check operators. We will later see that the same method works when the parity check measurements are subject to noise.

Imagine that the tessellation is split up in non-overlapping hypercubic boxes, each box $N$ defining a subset of the tessellation (i.e. a collection of vertices, edges, faces, cubes and hypercubes which form a connected region of the tessellation). For the hypercubic tessellation, the optimal choice is to take each box of side-length $l$ and placing the boxes in a grid (see Figure 6.5). This means that we can fit

$$\lfloor L/(l+1) \rfloor^4 \tag{6.1}$$

boxes of side-length $l$ into a hypercubic tessellation on a torus of size $L^4$. Note that at the boundary of the box some of the qubits on which an edge check operator acts do not need to be included inside the box, and check operators from different boxes may act on the same qubits (both of which are outside the box).

Since this arrangement does not cover the whole space and hence does not include all qubits, we will change this partition a few times in the decoding process. Consider a single box $N$ of fixed $O(1)$ size, not scaling with $L$. Denote the set of edge check operators which have non-trivial syndrome and which are contained in $N$ by $s|_N$. It may consist of closed loops which are completely confined within $N$ and for example open strings which pass through the boundary of $N$, see Figure 6.4(a). In particular given $s|_N$ we define the set of (intersection) vertices $V \in N$ as vertices in $N$ which are the boundary to an *odd* number of elements in $s|_N$. $|V|$ is always even.

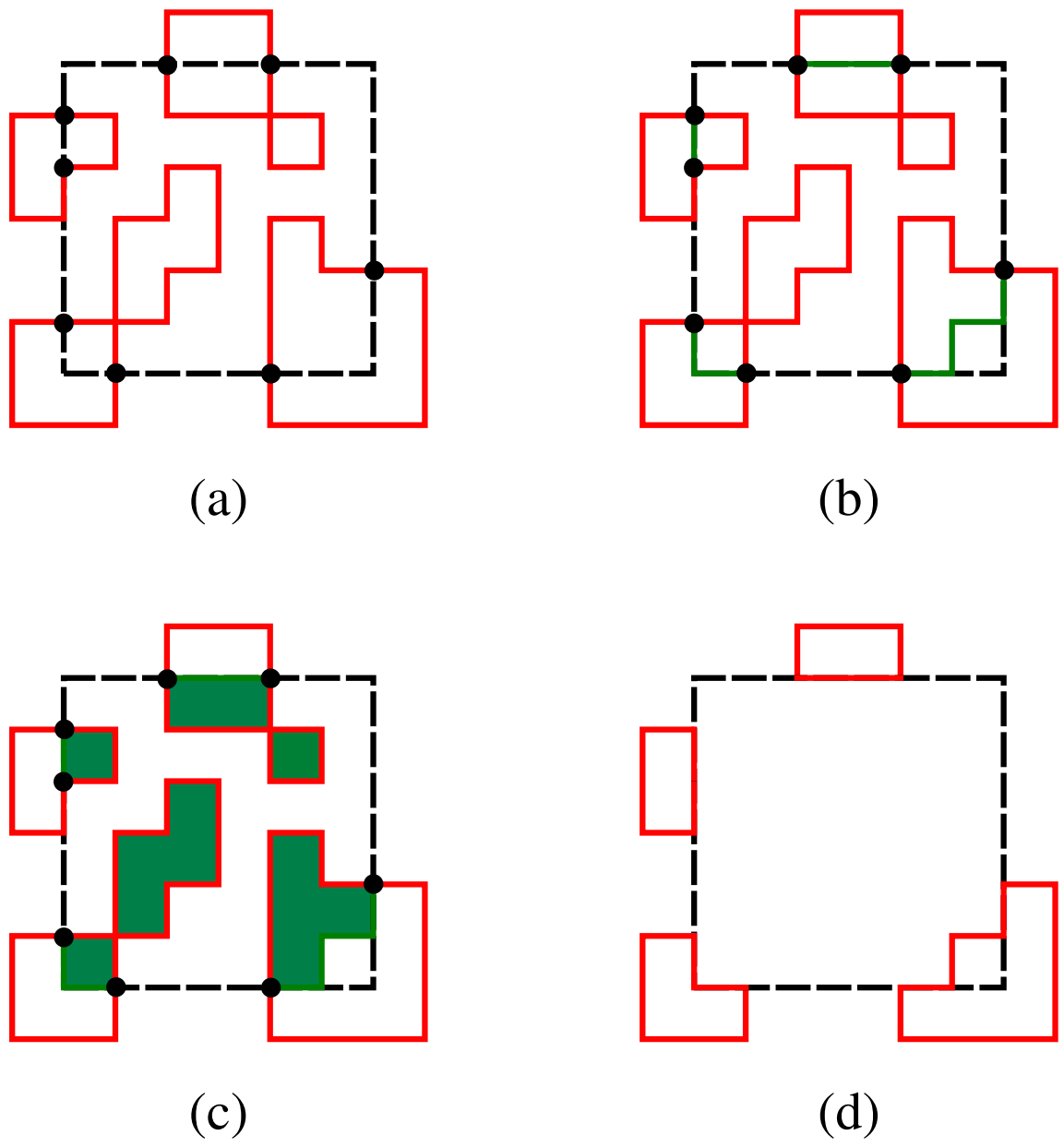

Figure 6.4: Schematic picture of local error correction procedure for a box with $O(1)$ qubits. (a) Syndrome $s$ (red) and a box $N$ (with black boundary). The black dots are the set of intersection vertices $V$ which are the vertices where $s$ intersects the boundary of the box $N$. (b) Find a set of strings $s'$ (shown in dark green) of minimal length which connects the intersection vertices $V$ (c) Find a collection of sheets $R$ with minimal area which has the closed loops $s|_N + s'$ in the interior of the neighborhood as its boundary. (d) Residual syndrome after the application of the correction $R$.

The first step of the decoder is to determine the shortest distance matching (MWM) between the vertices $V$, see Figure 6.4(b). The matching is done using the 4D Euclidean distance

$$\mathrm{dist}(x,y) = \sqrt{\sum_{i=1}^{4}(x_i - y_i)^2}. \tag{6.2}$$

We choose the path of edges connecting pairs of matched vertices to be the one which deviates least from the direct Euclidean path length. If this should be the case for more than one path of edges, we pick one uniform at random among this set. This step will always keep the length of the non-trivial syndrome in $N$ the same or shorten it. Note that using the Euclidean distance is different from taking a taxi-cab norm on the graph where one just counts the path length in terms of edges between a pair of vertices. Choosing the taxi-cab distance would not be effective in shrinking error regions as depicted in Figure 6.6.

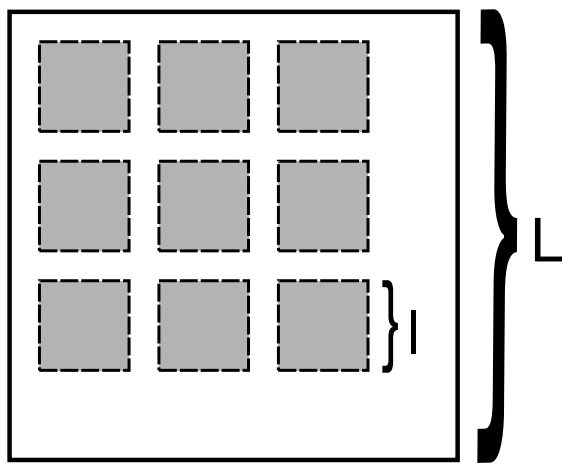

Figure 6.5: Layout of the neighborhoods in the 4D hypercubic tessellation of side-length $L$. Each neighborhood is a box with side-length $l$. In each decoding round the location of the grid of boxes is chosen at random.

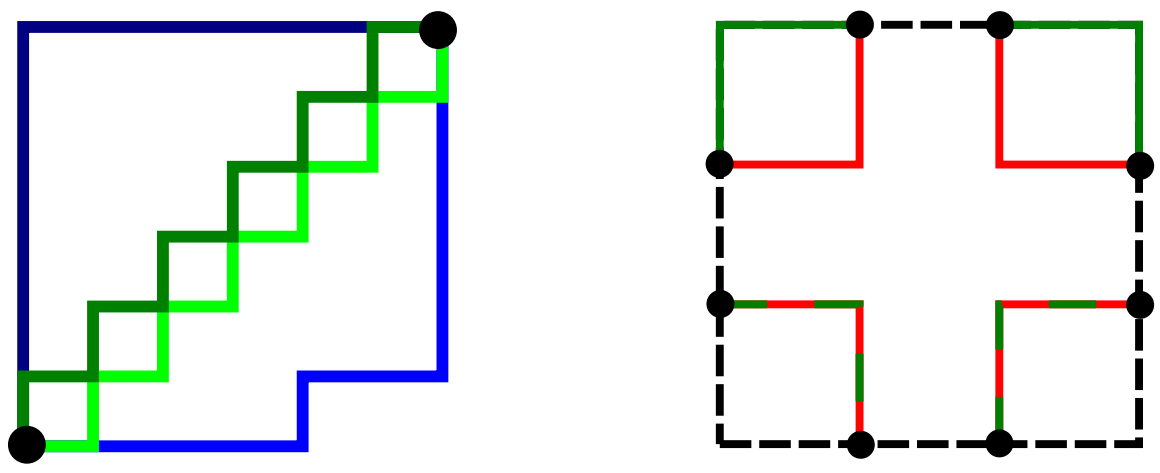

(a) (b)

Figure 6.6: Paths between pairs of matched vertices. (a) According to the taxi-cab metric all four lines connecting the two dots have the same length equal to 12. But among all paths of length 12, the green ones approximate the direct path best. (b) Syndrome (red) entering box with side length $l = 3$ at four corners. The shortest distance matching is either along the boundary (upper corners) or through the interior (lower corners). Hence only on the faces in the upper corners a correction is applied.

Let $s'$ be the resulting set of edges and note that $s_{\text{correct}} = s' + s|_N$ has the property that all vertices in $N$ touch an even number of elements of $s_{\text{correct}}$, hence $s_{\text{correct}}$ consists of only closed loops. Note that where $s'$ and $s|_N$ coincide they will cancel and they enclose no area.

In the next step we determine the smallest sheet $R$, i.e. a subset of faces in $N$, which has $s_{\text{correct}}$ as its boundary, see Figure 6.4(c). This sheet $R$ is found by solving an integer program. This is not computationally-efficient in general, but the box is of $O(1)$ size and integer programming is preferred over brute force enumeration over all possible sheets. $R$ is thus the proposed $Z$ correction for the box. If we would $Z$-flip all qubits corresponding to $R$, we are left with a residual syndrome as in Figure 6.4(d).

The decoder applies this procedure in parallel on each box and each procedure in a box clearly takes $O(1)$ computation time, hence the decoder is local. The parallel action results in a total correction $R_{\text{total}}$. Before applying $R_{\text{total}}$ or recording it as the final correction, we can repeat this decoding procedure with a different box-partition. This is useful since a partition leaves some qubits outside every box, hence no correction can take place on them.

Thus we allow the decoder to re-apply the procedure $m$ rounds, with each round being a different partition (but keeping boxes of the same size). Implicitly, it means that we allow the classical decoder to have high computational speed. After every round the total syndrome is updated given the current recovery and the next round is applied to the left-over syndrome.

When the syndrome is noisy, the measured syndrome is a collection of open strings instead of closed loops. It is possible to determine the 0-dimensional boundary of this collection and perform global MWM on this set of boundary vertices, but this would result in a *non-local* single-shot decoder. Instead we can apply the decoding procedure described above: the vertex set $V$ now simply includes vertices in the interior of $N$, see Figure 6.7(a) and Figure 6.7(b). One is left with a collection of closed loops as in Figure 6.7(c) for which we can again proceed as before.

The decoding procedure is single-shot in the sense that we only use the data obtained from a single round of syndrome measurements and the redundancy in the syndrome is used to repair the syndrome record.

### 6.2.2 Numerical simulation

In our simulation we choose the boxes of side length $l = 3$ which is the smallest non-trivial size (for $l = 2$ there is not much shrinking of syndrome loops that the decoder can do). The number of faces of a box of side-length $l$ equals $6l^4 + 12l^3 + 6l^2$. Hence, for $l = 3$ each box includes 864 qubits.

The sizes which are numerically tractable and support more than one box at a time are the tessellations with $L \in \{8, 9, 10, 11\}$ which by Equation 6.1 all support 16 boxes. During a single time step/QEC cycle the position of the grid of the boxes (as depicted in Figure 6.5) is varied randomly five times and each time the decoding process is run inside each neighborhood. We chose the number of rounds to be $m = 5$ as any increase did not improve the decoder's performance. Unfortunately, due to limited computational resources, we were not able to test the performance of the decoder on larger tessellations with an increasing number of boxes. This would have been useful to validate that the number of rounds per QEC cycle can be taken to be independent of $L$. The average memory time $T$ is estimated in a similar way as for the global decoder earlier: By checking after every QEC cycle whether the residual errors can be corrected and stopping the simulations once this is not the case anymore. Ideally, this step would be implemented by using the global decoder which finds a minimal surface. However, for the system sizes we consider here, this is computationally not feasible. Thus we let the local decoding procedure run indefinitely with perfect syndrome measurement instead.

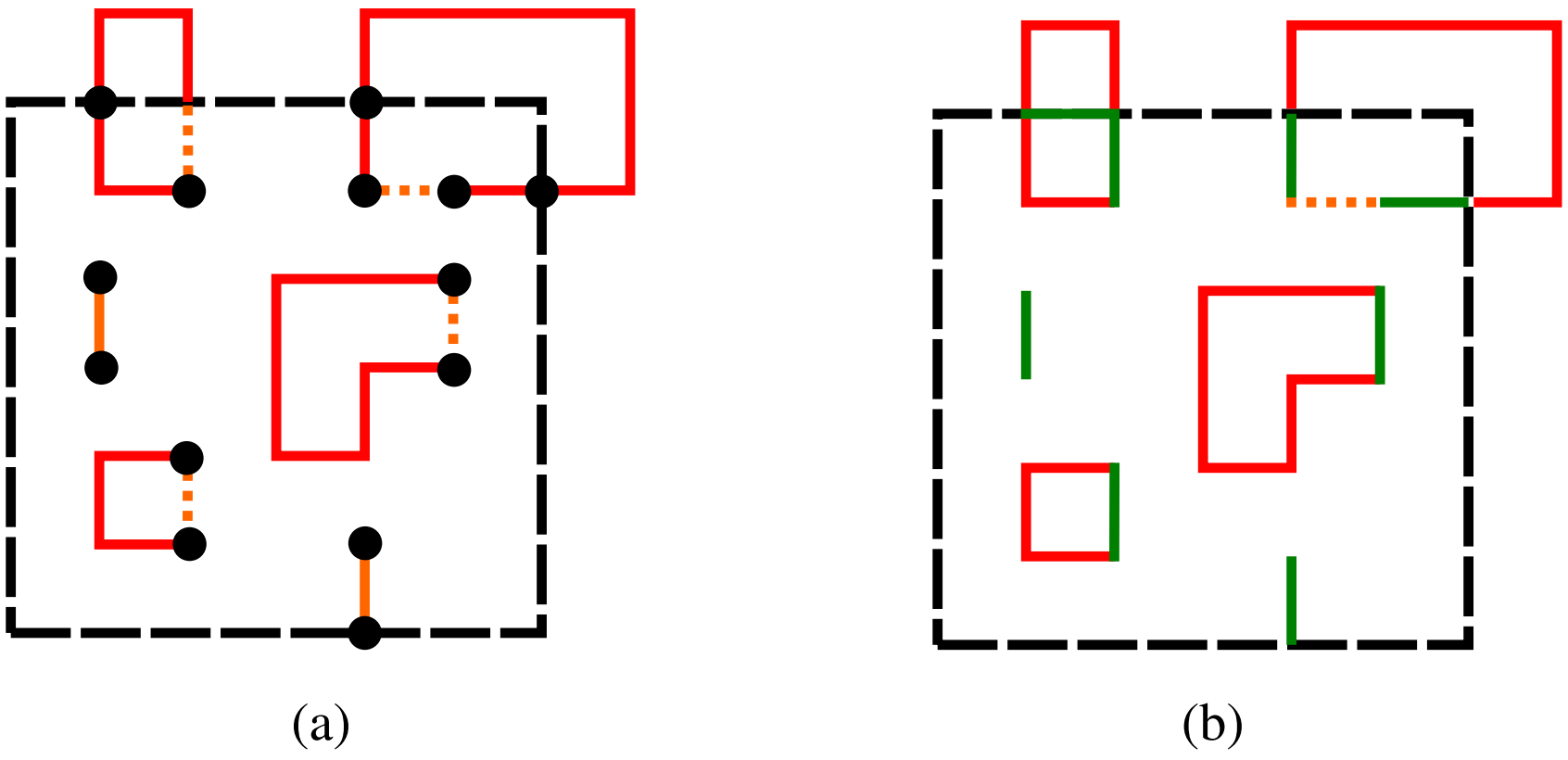

Figure 6.7: Schematic picture of local error correction procedure in the presence of erroneous syndrome measurements. (a) Local decoding neighborhood in the presence of error syndromes (red) and syndrome errors (orange). The measured syndrome consists of solid lines while dashed lines indicate edges where the syndrome error overlaps with the real syndrome. (b) After performing a matching (green lines) we are left with a collection of closed loops. We are in the same situation as in Figure 6.4(b) and can continue with the correction. Note that edges where the measured syndrome and the matching overlap are removed.

To determine the value of the critical physical error $p_c$ we assume a scaling behavior in the variable $x = (p - p_c)L^{1/\nu}$. For $p$ close to $p_c$ the memory time is well approximated by a quadratic polynomial in the scaling variable. Hence, we do a global fit of the function

$$T(p,L) = T_c + Ax + Bx^2 \tag{6.3}$$

to our data with fitting parameters $T_c$, $p_c$, $\nu$, $A$ and $B$, similar as it has been done for the 2D toric code in [45].

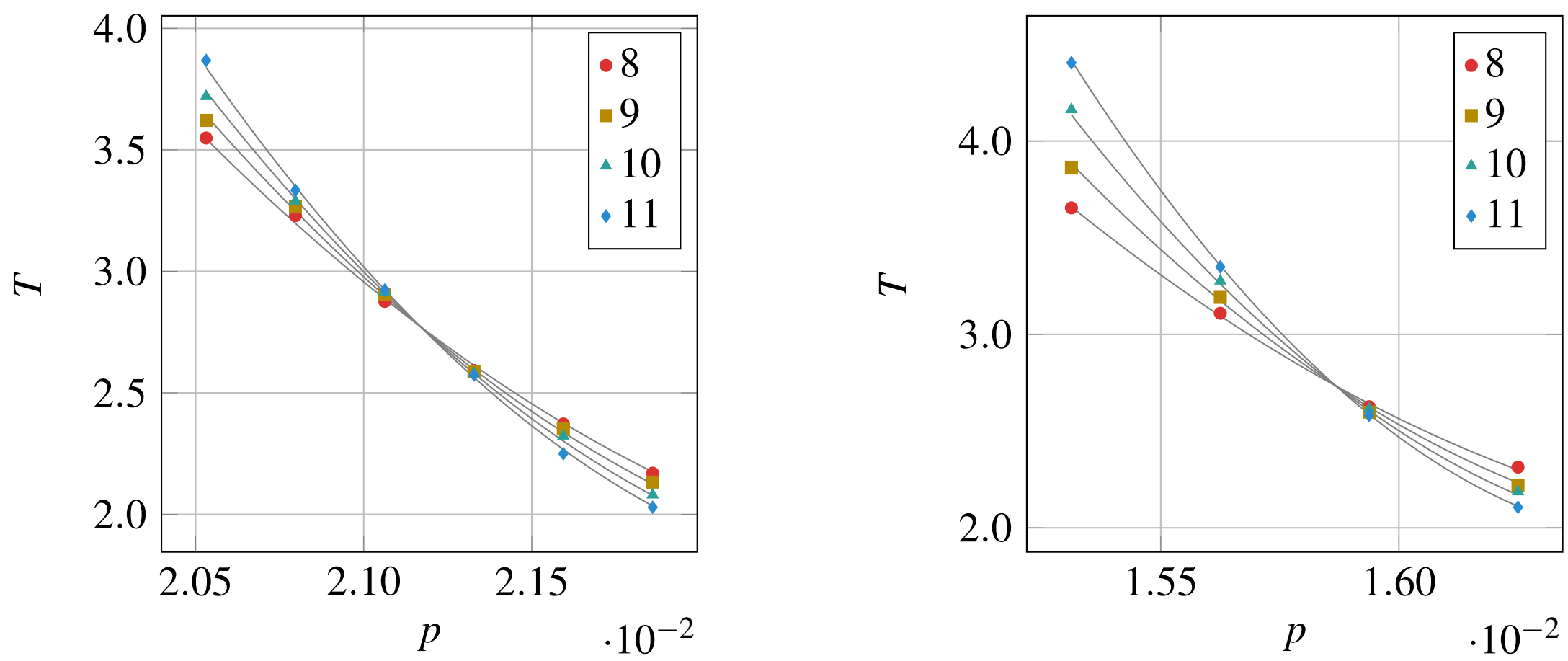

(a) Perfect measurements

(b) Noisy measurements

Figure 6.8: Average memory time $T$ for the 4D toric code depending on the physical error rate $p$ for the Hastings decoder with (a) perfect syndrome measurement and (b) noisy syndrome measurement. The fit was obtained via Equation 6.3.

For perfect syndrome measurements we find

$$\begin{aligned} p_c &= 2.117\% \pm 0.006\%, \\ \nu &= 1.14 \pm 0.18. \end{aligned} \tag{6.4}$$

The data and the fitted function are shown in Figure 6.8(a). The data was obtained by running the simulation $5 \times 10^4$ times for each data point.

For the simulations where we do model syndrome errors with probability $q = p$ we find

$$\begin{aligned} p_c &= 1.587\% \pm 0.002\%, \\ \nu &= 0.65 \pm 0.03. \end{aligned} \tag{6.5}$$

The data and fitted function are plotted in Figure 6.8(b). The data was obtained by running the simulation $5 \times 10^4$ times for each data point.

The value of the threshold only decreases by a factor of roughly 1.3 as opposed to the 2D toric code where it decreases from around 11% to about 3% [45, 64] which is more than a factor of 3.

Note that the memory times in Figure 6.8 are extremely small since we are looking at data which are close to threshold. In order to see the trend for much lower error rates, one can look at Figure 6.9 where the number of Monte Carlo trials is relatively low.

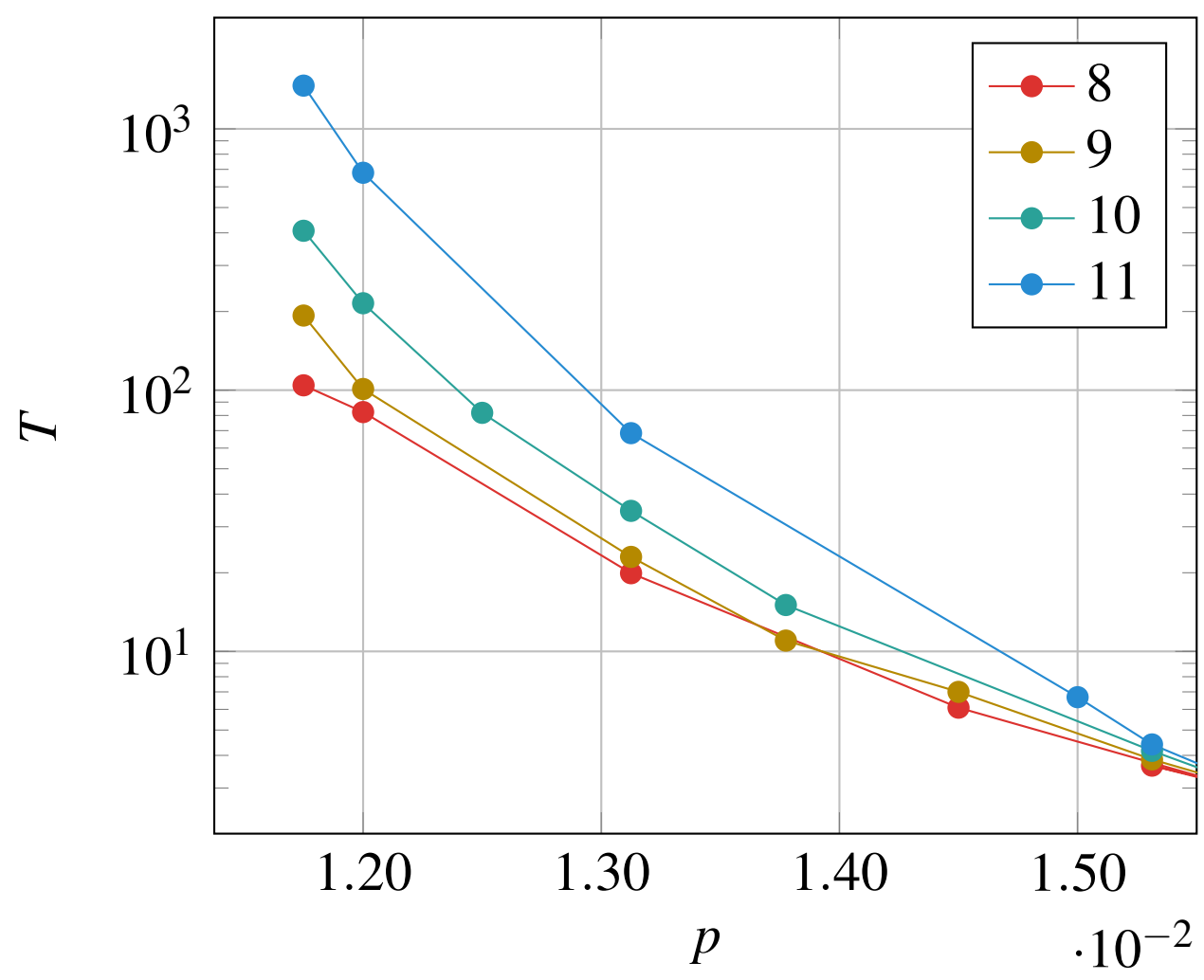

Figure 6.9: Results for the Hastings decoder for smaller values of $p = q$ as those in Figure 6.8(b). The average memory time increases up to around $T = 1500$ for the $L = 11$ tessellation for $p = 1.15\%$. Due to increasing computational demands we were only able to run around 100 trials for each data point.

We observe that the number $N_{res}$ of failures of the decoder due to residual syndromes, is higher than the number $N_{log}$ of failures where the decoder did manage to correct back into a code state but applied a logical operator. This may be explained due to the fact that errors for which the local decoder gets stuck require an error of size $O(L)$ to occur in contrast to errors which lead to a logical error which require at least $O(L^2)$ errors, showing that the decoder is energy-barrier limited. This could explain the fact that the ratio $N_{res}/N_{log}$ increases with system size $L$ as can be seen in Figure 6.10.

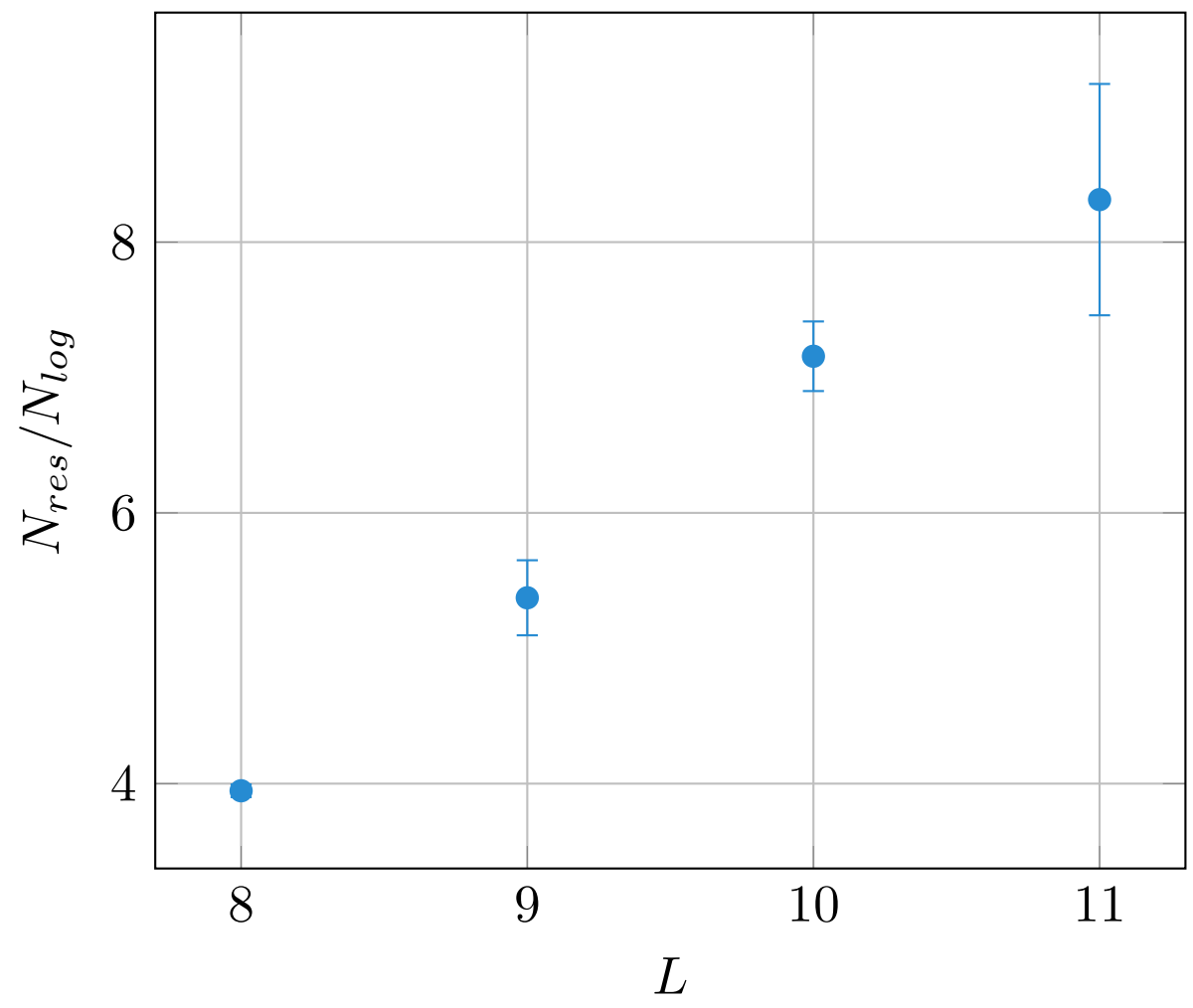

Figure 6.10: Dependence of the ratio $N_{res}/N_{log}$ on size $L$ at $p = 2.1062\%$ assuming perfect syndrome measurement.

## 6.3 Cellular automaton decoders

An alternative to using a full computer to determine the recovery operation could be *cellular automata*. Given that cellular automata can be implemented using very simple and fast electrical components, this decreases the effective rate of errors which can occur during an error correction cycle. Using cellular automata also simplifies the implementation in hardware: To reduce noise the physical qubits are kept at low temperatures. Having to connect them to an external classical computer means having to transmit the information of the measurement outcome out of the fridge which is a difficult engineering challenge. Cellular automata on the other hand are such primitive devices that they can be manufactured small enough to be put within the fridge and close to the qubits. In this section we will introduce two different cellular automaton rules. We will also perform numerical simulation to analyze their respective performance.

While we only consider cellular automaton decoder for 4D codes, there have been proposals to decode the toric code with cellular automata where the update rules move the positions of violated stabilizer checks. The first decoder of this kind was introduced in [64] and analyzed further in [65]. A different approach was taken in [66, 67] where the cellular automata effectively mediate an attractive force between violated stabilizer checks by mimicking a Coulomb interaction.

### 6.3.1 Toom's rule

We first discuss Toom's rule which was introduced in [68]. Consider a 2D square tessellation where each face is associated with a degree of freedom which can take the values $+1$ and $-1$ (the Ising model). Each face is surrounded by four edges which we will label 'north', 'east', 'south' and 'west'. Two neighboring faces share exactly one edge which is either in the east of one face and the west of the other or in the north of one and in the south of the other. Every edge is associated with the parity between the two faces incident to it. An edge-check is non-trivial if and only if the two faces incident have different values. Toom's rule states that for each face the value is flipped if and only if the parity checks of the north and the east edges are non-trivial. If we view Toom's rule as a cellular automaton, it means that the cell of the automaton resides on faces, above the qubit, and the automaton processes the syndrome of its NE edges (see Figure 6.11).

To turn Toom's rule into a decoder for the 4D toric code we can for example apply the update rule to every 2D plane of the 4D hypercubic tessellation. We thus partition the set of all planes into 6 groups, each of a set of parallel planes. We then apply the rule on the first group, say, on all $x$-$y$-planes, then the second group, i.e. all $x$-$z$-planes, etc. Within each group the CA rule is applied on all qubit faces in parallel. One needs to fix an orientation in the tessellation so that the vertex between the 'North' and 'East' edges is the vertex with the largest coordinates (modulo L). After this round of applications a new syndrome record can be obtained. It may also be possible to apply Toom's rule in parallel on all qubits, but we have not implemented this.

### 6.3.2 DKLP rule

The DKLP rule counts the number of non-trivial edge-checks and does a majority vote. In case that exactly half of the edge-checks surrounding a face are violated the face is flipped with probability 1/2. The idea behind the rule is that the weight of the non-trivial syndrome is never increased but it may be decreased (see Figure 6.11). According to [40] the update rule can be applied to a set of faces which do not share a common edge. Thus we partition the faces in these 6 planar groups as for Toom's rule, but then we further subdivide each plane into sets of non-overlapping faces by dividing them into a checkerboard pattern and apply the rule only in parallel on each non-overlapping set.

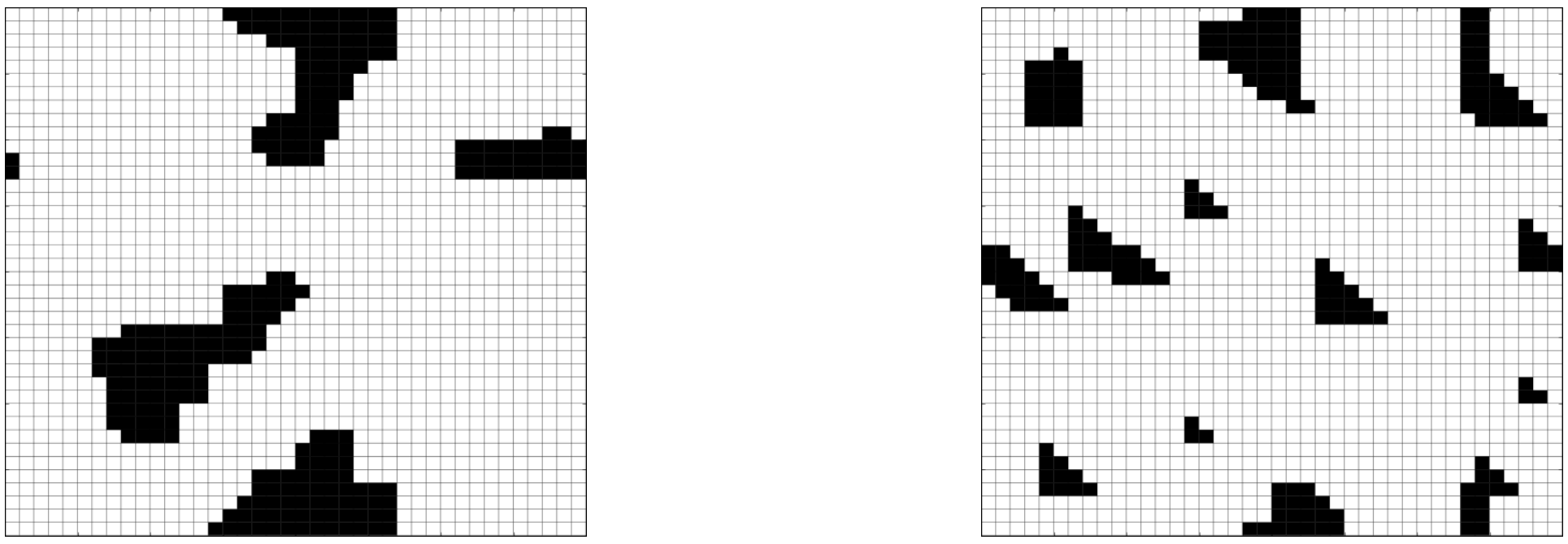

(a) Typical configuration: DKLP rule. (b) Typical configuration: Toom's rule.

Figure 6.11: Illustration of the action of the cellular automaton rules on a 2D classical Ising model (spins on faces, periodic boundaries) where spins have been flipped once (black), each with probability $p = 0.4$ and either (a) the DKLP rule or (b) Toom's rule has been applied several times. In (a) we see that after a few applications there are only large islands left which slowly shrink at their boundary. (b) The dynamics of Toom's rule is slightly different due to its anisotropy. The lower-left boundary of the triangular-shaped islands is left invariant, while the upper right boundary shrinks diagonally until the island is removed.

### 6.3.3 Numerical simulation

To simulate the performance of the cellular automaton decoders, we proceed similarly as for the Hastings decoder. After applying the decoder we check whether we could correct back to the original code state by running the same decoder indefinitely under perfect syndrome measurements. If no logical operator was applied and the decoder did not get stuck we move on to the next time step.

The results for the DKLP decoder for different sizes can be seen in Figure 6.12. Every data point was obtained by $4 \times 10^3$ trials. We observe that for increasing size the crossing point is receding in the direction of lower physical error rate.

In contrast to the results on the Hastings decoder we see that the average memory time is much higher when we are close to the cross-over point: $O(10^2)$ compared to $O(10^0)$ for the Hastings decoder. For the codes that we consider the cross-over point occurs between 0.5% and 0.6% without syndrome errors and between 0.4% and 0.5% with syndrome errors.

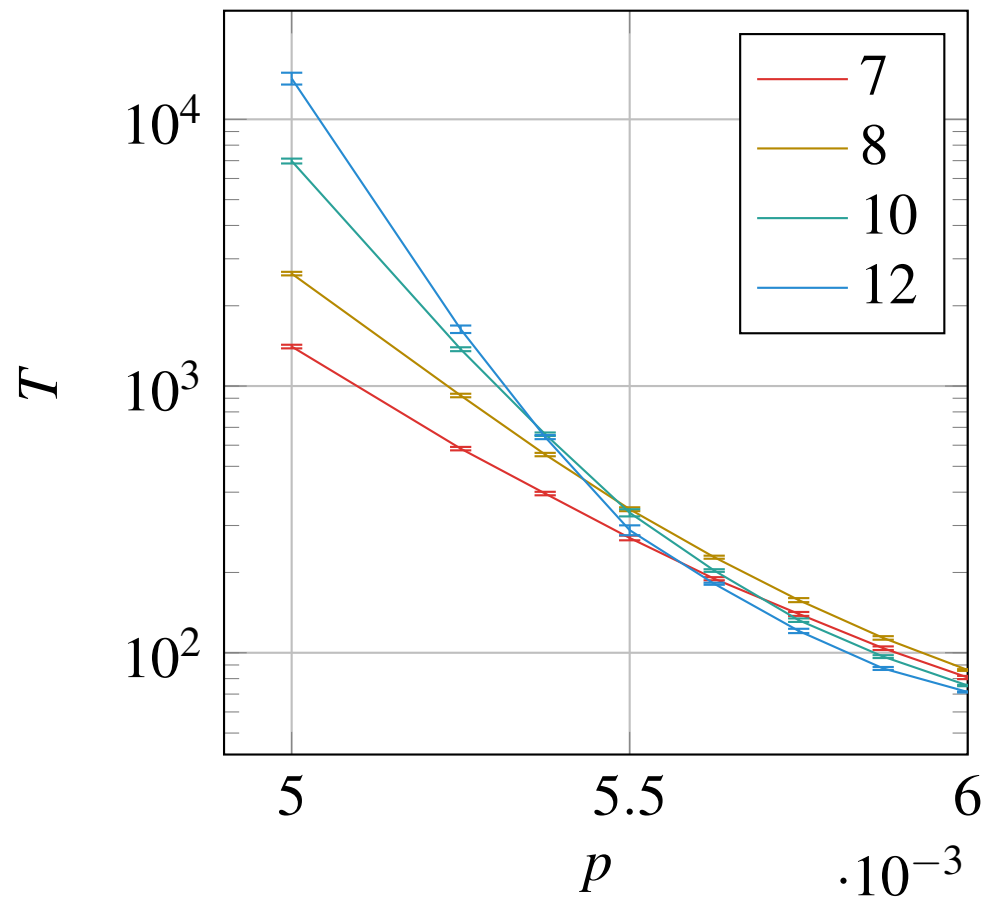

(a) Perfect measurements

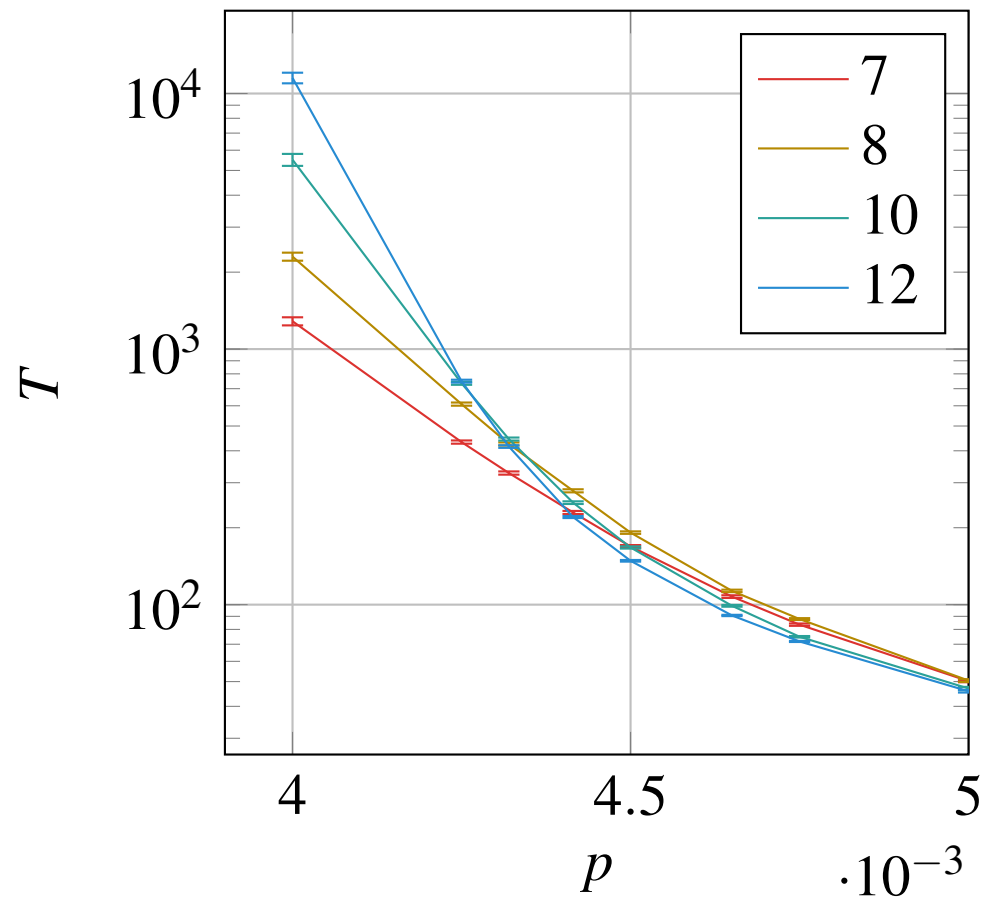

(b) Noisy measurements

Figure 6.12: Results for the DKLP decoder for the 4D toric code. At each time-step the update rule is applied to every plane once. (a) Without syndrome errors (b) With syndrome errors

In [40] the authors anticipated that a decoder based on Toom's rule will perform better than the DKLP decoder. Our results seem to confirm their intuition. In Figure 6.13 we can see that the cross-over points now lie in the region between 0.9% and 1% assuming perfect syndrome measurement and between 0.7% and 0.8% when including syndrome errors. Every data point in Figure 6.13 was obtained by $4 \times 10^3$ trials.

For these cellular automaton decoders we have also observed that in the regime where the physical error rate is low the number of failures due to non-correctable syndromes is increasing with the size of the code (hence energy-barrier limited).

In the regime where classical computation is fast compared to the syndrome acquisition rate it is possible to apply the update rule multiple times. In the Hastings decoder we have similarly allowed to partition the space in neighborhoods and apply correction multiple times. In Figure 6.14 we show the results when decoding using Toom's rule with noisy syndrome where we repeat the process of applying the update rule to every plane 30 times. We see that the performance of the decoder improves and that the crossover region is now between 1.25% and 1.45% which is close to the estimated value of the threshold for the Hastings decoder. Applying the update rule 100 times did not result in a further shift of the crossover region. Similarly, the memory time of the Toom's rule decoder is in the same regime as for the Hastings decoder (cf. Figure 6.9).

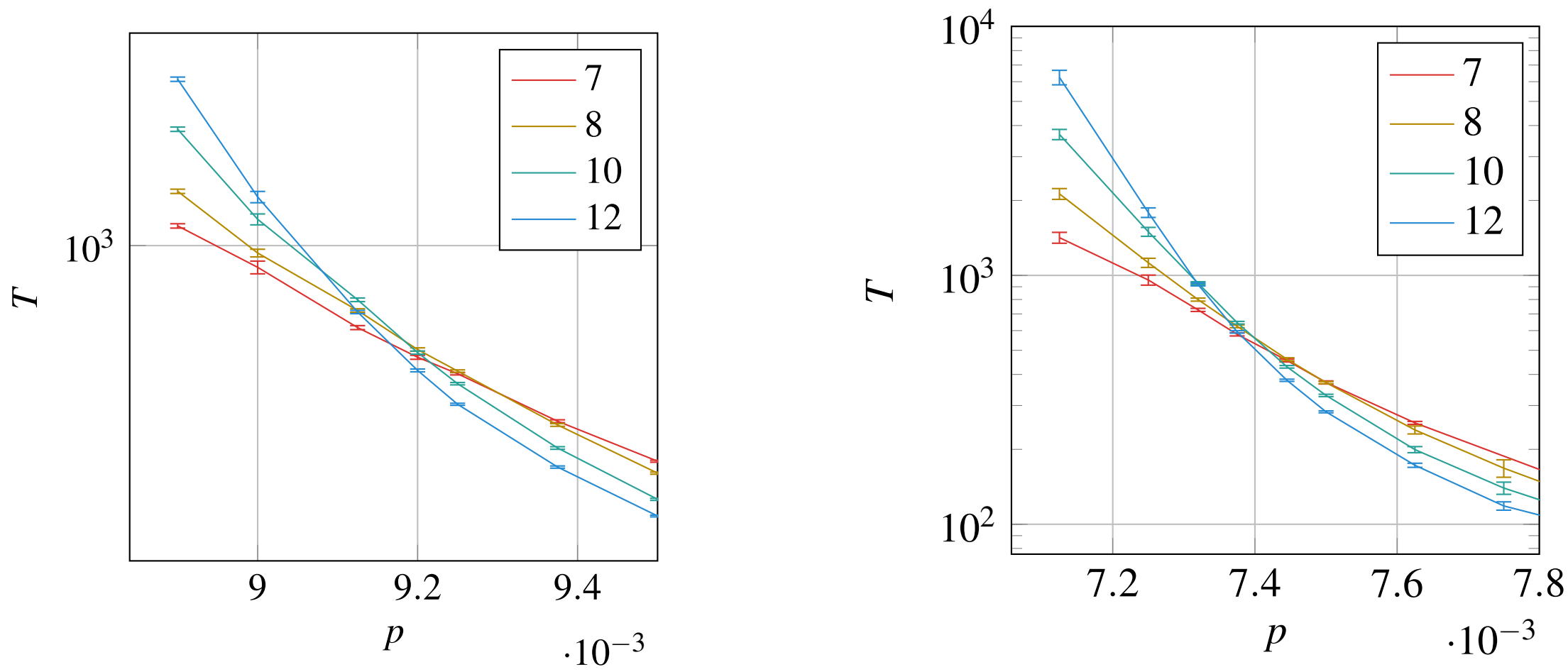

(a) Perfect measurements

(b) Noisy measurements

Figure 6.13: Result for the Toom's rule decoder for the 4D toric code. At each time-step the update rule is applied to every plane once. (a) Without syndrome errors (b) With syndrome errors.

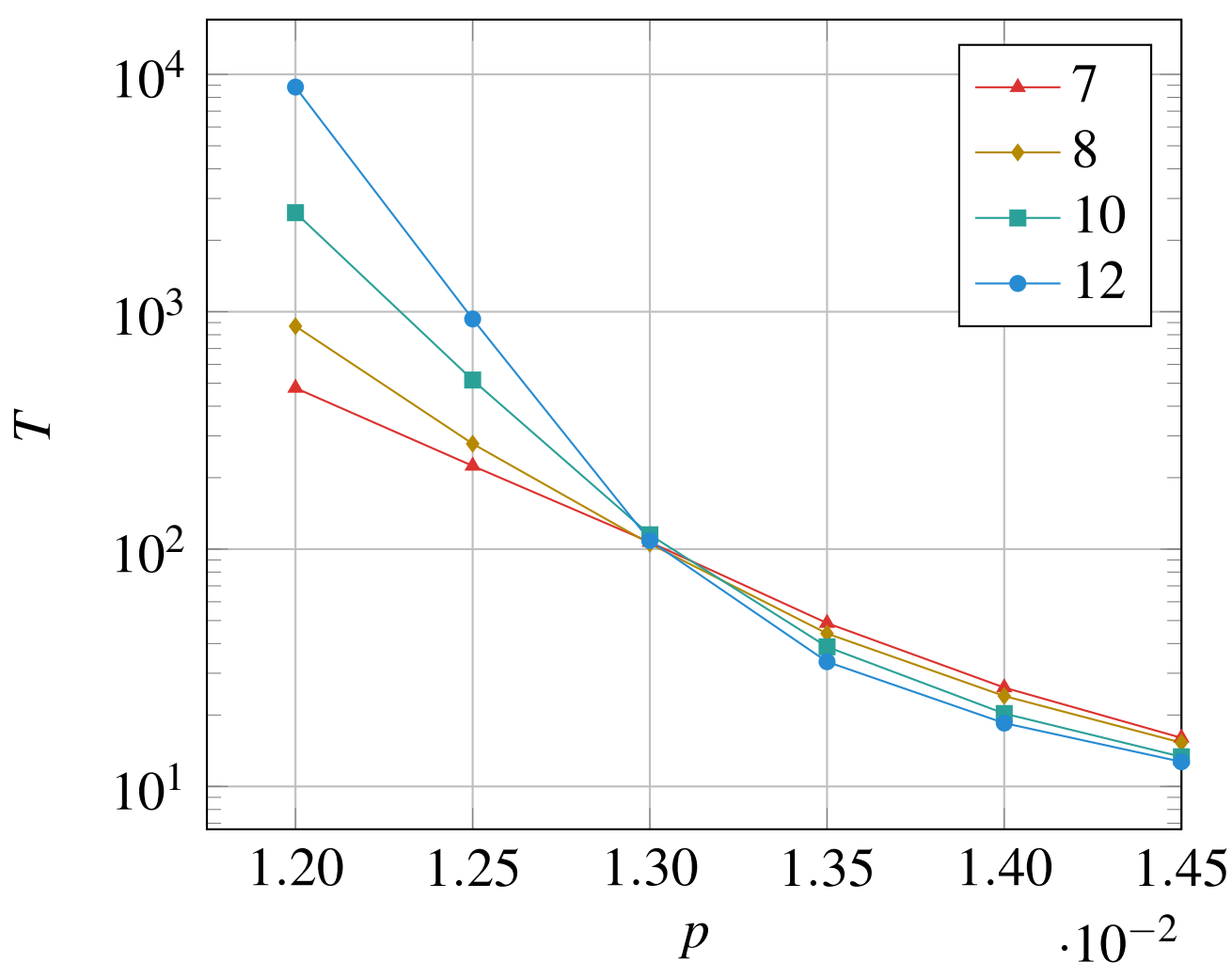

Figure 6.14: Result for the Toom's rule decoder for the 4D toric code with syndrome errors. At each time-step the update rule is applied to every plane 30 times.

Overall, we thus find that the Toom's rule decoder performs quite well given the locality of the correction rules, in particular if one allows the rules to be applied multiple times in a QEC cycle. The Hastings decoder is more computationally intensive and applies more non-local error correction but its benefits as compared to Toom's rule are not clear given the current data.

### 6.3.4 DKLP decoding the 4D hyperbolic code

The DKLP rule can be immediately adapted to work on the 4D hyperbolic code example that we discussed in Section 5.4. The only difference as compared to the 4D toric code is that we did not partition the set of faces to be non-overlapping. The results can be seen in Figure 6.15. As we only have one example of a 4D hyperbolic code we cannot determine any crossing points. However, it is very clear that the decoder performs much better as compared to the 4D toric code.

## 6.4 Decoding with neural networks

In this section we explain how to use neural networks as a means to determine a recovery procedure. Feed-forward neural networks have similar advantages as cellular automata in that they can be implemented with very primitive, low-power electrical circuits [69]. For example, in [70] the authors report on an integrated circuit implementing a network of 1 million neurons on a $240\mu\mathrm{m} \times 390\mu\mathrm{m}$ CMOS chip. The chip draws $20\mathrm{mW}/\mathrm{cm}^2$ as compared to $50\mathrm{W}/\mathrm{cm}^2 - 100\mathrm{W}/\mathrm{cm}^2$ for a modern CPU or $30\mathrm{W}/\mathrm{cm}^2$ for an FPGA [71].

Neural networks have been considered to decode the surface code, see [11, 12, 13, 14]. However, in all these approaches the decoder is tailored to the specific system size. This could be problematic as training the network usually takes an exponential amount of time which could render this approach infeasible for large system sizes. Here we present a neural network decoder that uses *convolutional networks* which are usually used for image recognition. The advantage is that we only need to train once on a fixed system size. After the training the network can then be scaled up arbitrarily.

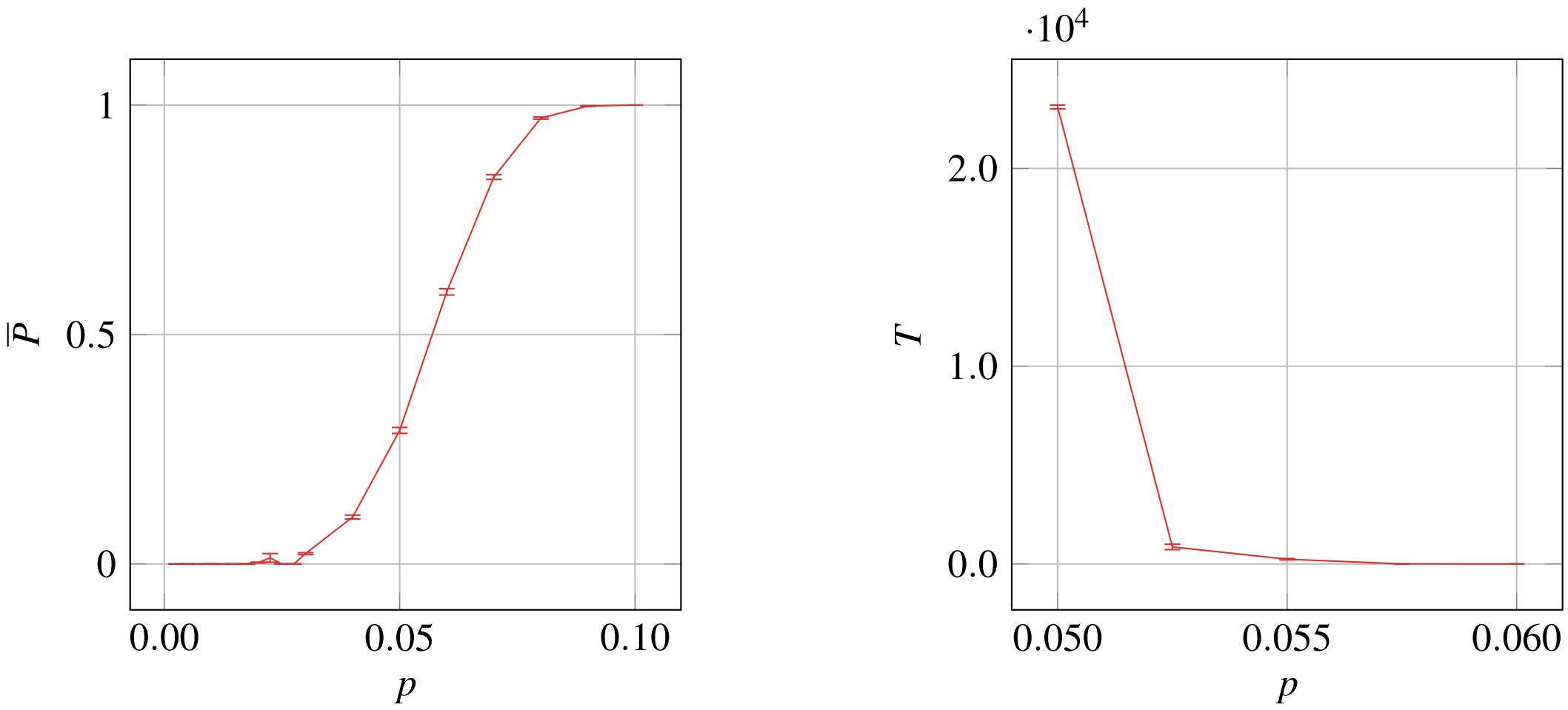

(a) Perfect measurements

(b) Noisy measurements

Figure 6.15: Applying the DKLP decoder to the 4D hyperbolic code example.

Before going into the details of the implementation we will discuss the basics of machine learning with artificial neural networks.

### 6.4.1 The principles of neural networks

A *neural network* is a directed, multipartite graph consisting of *layers* $l = 0, \dots, L$. The vertices in layer $l$ are connected to vertices in the following layer $l+1$.

The vertices of the network are called *neurons*. The main idea behind the neural network is the following: Each neuron computes a primitive function $f_{w,b} : \mathbb{R}^q \to [0,1]$ which takes $q$ real values as input and returns a single value in the interval $[0,1]$. Each input value is given by a neuron of the previous layer connected to it. The subscripts $w \in \mathbb{R}^q$ and $b \in \mathbb{R}$ are parameters which can differ for each neuron. Before we discuss the function $f_{w,b}$ in more detail, let us first understand how the network performs a computation. The neurons in the first layer do not have any predecessors and hence do not take any input. Their output is simply set to be the input of the network, which is a bit string $x \in \{0,1\}^m$. The output values of the neurons in the last layer are interpreted as the output of the network. We see that the network describes a function $F : \{0,1\}^m \to [0,1]^n$. The first layer $l = 0$ is called the *input layer* and the last layer $l = L$ is called the *output layer*. All other layers $l = 1,\ldots,L-1$ are called *hidden layers* since they are considered to be internal to the network.[1]

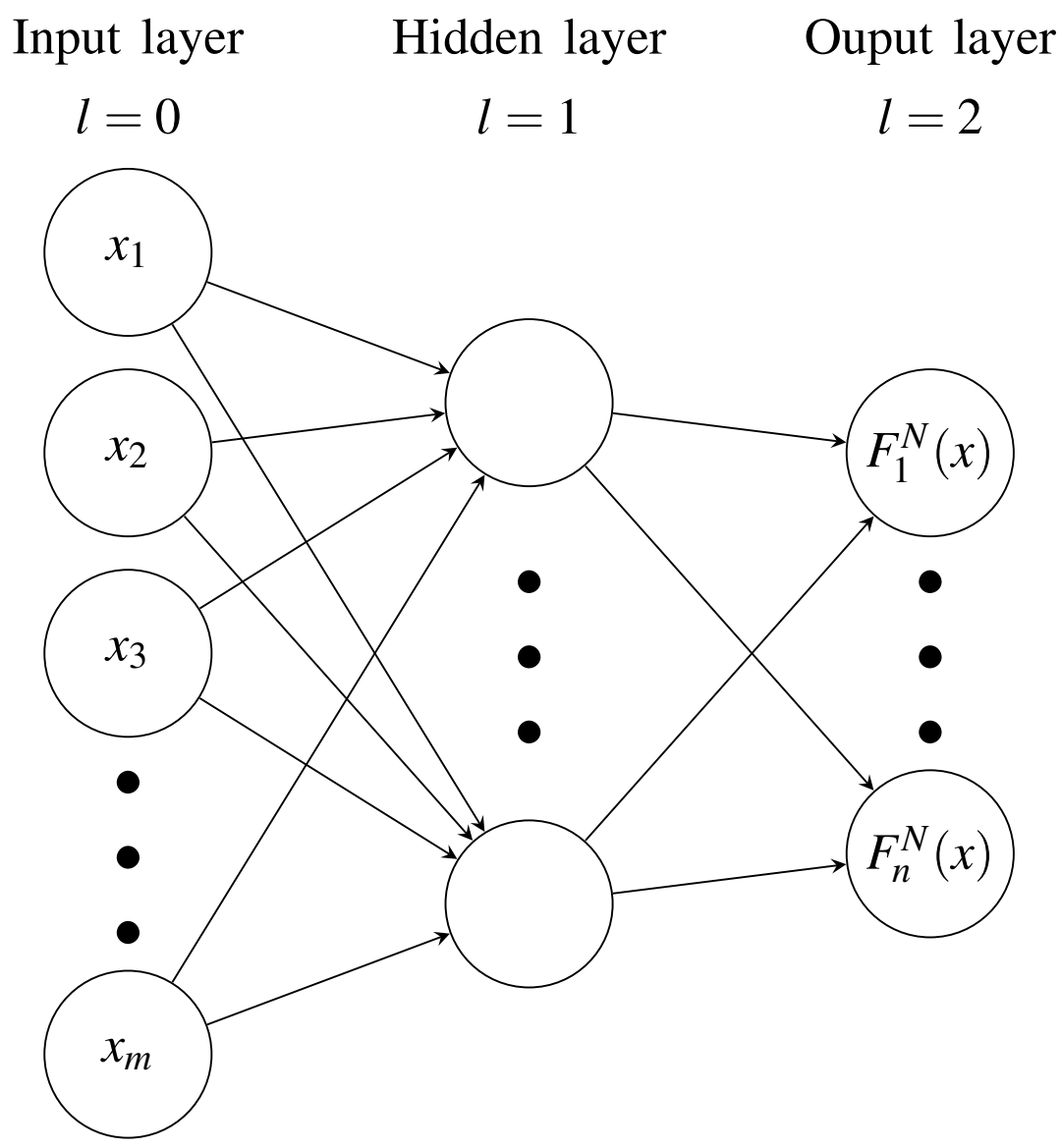

Figure 6.16: A neural network consisting of 3 layers. The network takes input $x \in \{0,1\}^m$ which is represented as the first layer of neurons. The values of neurons in the hidden layer and the output layer are given by the function $f_{w,b} : \mathbb{R}^q \to [0,1]$ evaluated on their input (indicated by $q$ incoming arrows). The parameters $w \in \mathbb{R}^q$ and $b \in \mathbb{R}$, called weights and bias, can be different for each neuron. The values of the neurons in the last layer are the output of the network $F^N(x) \in \mathbb{R}^n$ (the $N$ stands for "network").

The parameters $w$ and $b$ are called *weights* and *biases*. They define a linear map $w \cdot y + b$ where $y$ is the input of the neuron. The function that each neuron computes has the form:

$$f_{w,b} : \mathbb{R}^q \to [0,1], \quad y \mapsto \sigma(w \cdot y + b) \tag{6.6}$$

[1]Alternatively, one could remove the input layer and consider the first hidden layer as taking the input from outside.

where $\sigma$ is the is the non-linear *sigmoid function*

$$\sigma(z) = \frac{1}{1+\exp(-z)}, \tag{6.7}$$

which is plotted in Figure 6.17. This function is chosen to mimic biological neurons: A biological neuron has branches called *dendrites* and a single special branch called an *axon*. The dendrites react to the electrical potential along their membrane. They are analogous to the input of our artificial neurons. Once the electrical potential along the dendrites exceeds some threshold the neuron fires an electrochemical pulse which is created along the axon which in turn stimulates other neurons. The locus where the biological neurons connect are called *synapses* and the weights $w$ model the strength with which they connect. The bias $b$ enables us to set the threshold differently for each neuron. The sigmoid function $\sigma$ can be thought of as a smooth version of the Heaviside step function $\Theta$ (see Figure 6.17). The smoothness of $\sigma$ will become important when we discuss the training of neural networks.

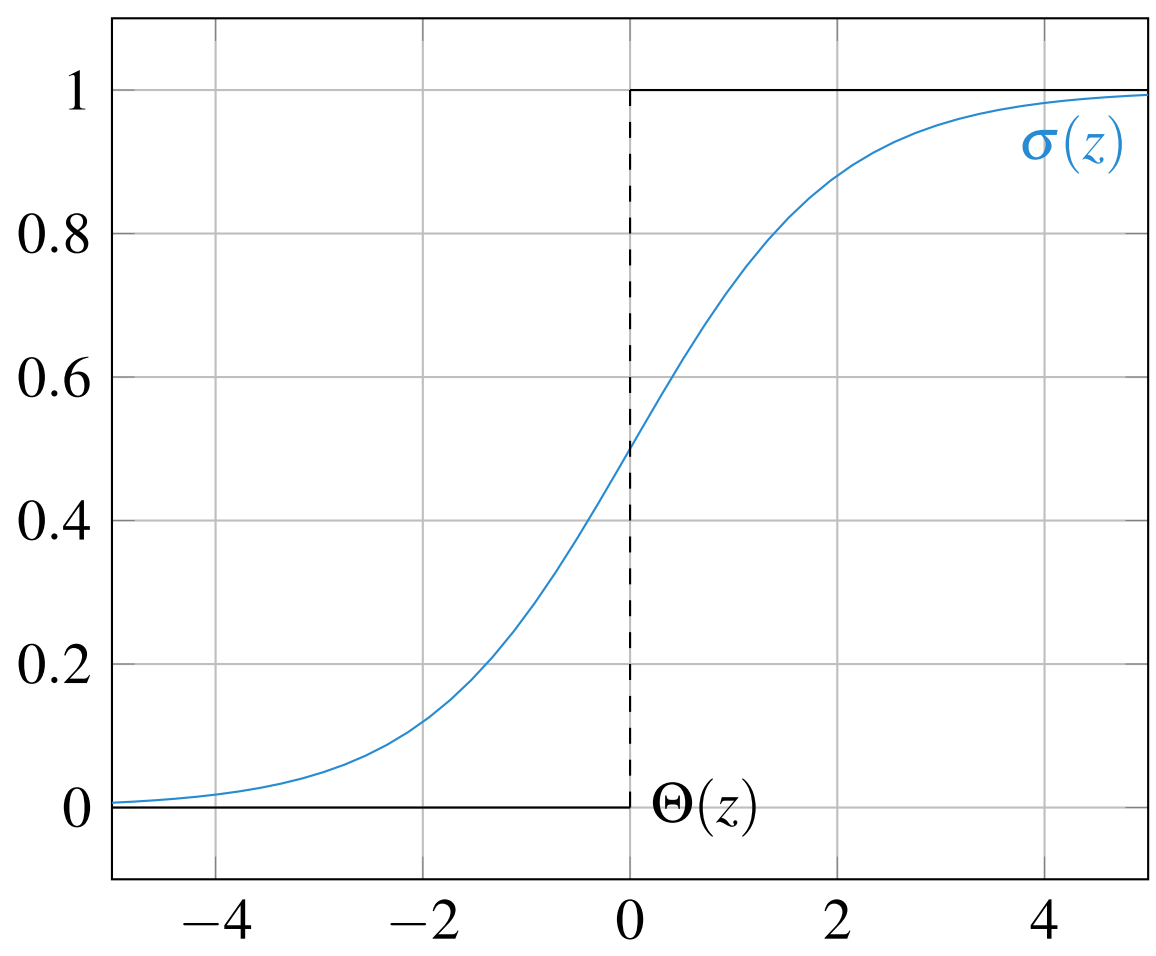

Figure 6.17: The sigmoid function $\sigma$ is a smooth version of the Heaviside step function $\Theta$.

Let us have a look at a small example which gives some intuition why neural networks are able to perform interesting computations: Consider a single neuron which takes $q = 2$ inputs and has weights $w = (-12, -12)$ and bias $b = 17$. The neuron computes the following values for each

input:

$$\begin{aligned} f(0,0) &= \sigma(17) \approx 1, \\ f(1,0) &= \sigma(5) \approx 1, \\ f(0,1) &= \sigma(5) \approx 1, \\ f(1,1) &= \sigma(-7) \approx 0. \end{aligned} \tag{6.8}$$

We observe that these are approximately the input/output relations of the NAND gate. The approximation can be made arbitrarily precise by increasing the absolute values of the weights $w$ and the bias $b$. Since any Boolean function $F : \{0,1\}^m \to \{0,1\}^n$ can be computed by a network of NAND gates, there consequently also exists a representation of $F$ as a neural network. However, in practice it is more efficient to adapt the connectivity of the network to the problem at hand. We will discuss the choice of network connectivity in more detail later.

In this section we considered inputs to be bit strings. However, it can be advantageous to allow the input to be real-valued. In classic applications of neural networks such as image classification and speech recognition the inputs are most naturally represented as elements of $\mathbb{R}^m$.

### 6.4.2 Training

In the previous section we have seen that neural networks are a powerful ansatz to model functions $F : \{0,1\}^m \to [0,1]^n$. The question is how to choose the individual weights and biases of the neurons to make the network compute $F$, or at least give a good approximation. This task can be formulated in terms of an optimization problem where pairs of input and desired output $(x, F(x))$ are used to find the right weights and biases. In our setup we assume that the inputs of $F$ are weighed by some probability distribution $P : \{0,1\}^m \to [0,1]$. The distribution $P$ prioritizes certain inputs over others and effectively reduces the dimensionality of the input space. In principle we would want to optimize over all possible pairs of inputs and outputs of $F$ (while taking $P$ into account). However, this is generally not practicable so that we restrict ourselves to optimizing over some subset $D \subset \{(x, F(x)) \mid x \in \{0,1\}^m\}$. The set $D$ is sampled according to the distribution $P$. The optimization of the network is called *training* and the sample $D$ is called the *training data* or *training set*.

There is no developed theory on how much training data is needed to guarantee that a neural network will be able to approximate a given function to some desired measure of precision. There are so far only a large variety of heuristics on how to choose the network and tweak details of the training process to obtain better results [72]. The appeal of neural networks partially stems from the following fact: In order to train a neural network to model some function $F$ we do not need to be able to compute $F$ ourselves. We only need to have a sufficiently large training set $D$. This is what makes neural networks a powerful tool: Even when we have no algorithm to compute $F$, where $F$ could be as complicated as the optimal move in a game of Go or whether a picture shows a dog, we can still train a neural network and hope that it will pick up on some structure which underlies the function $F$.

We will now discuss the training of neural networks. We denote the weight vector of the $i$th neuron in layer $l$ by $w_i^l$ and the $j$th entry of this vector by $w_{i,j}^l$. Similarly, the bias of the $i$th neuron in the $l$th layer is labeled $b_i^l$. These are the parameters that we need to optimize. An essential ingredient for the optimization is a measure of how good a neural network performs on the training data $D$. This measure is called the *cost function* $C_D(w_{i,j}^l, b_i^l)$ which maps the values of the weights $w_{i,j}^l$ and biases $b_i$ of the neural network into $[0,1]$. If the value of the cost function is small then this is an indicator that the network performs well. For reasons that will become apparent in the following discussion, we demand $C_D$ to be differentiable. An obvious choice for the cost function is the average squared $L^2$ norm $\|\cdot\|^2$ of the difference of the networks output $F_N(x, w_{i,j}^l, b_i^l)$, which depends on the choice of the weights $w_{i,j}^l$ and biases $b_i^l$, and the desired value $F(x)$ over all elements of the training set $D$:

$$C_D(w_{i,j}^l, b_i^l) = \frac{1}{2|D|} \sum_{(x,F(x))\in D} \|F^N(x, w_{i,j}^l, b_i^l) - F(x)\|^2 \tag{6.9}$$

To optimize the weights and biases we perform an iterative procedure called *gradient descent* which we will discuss now. Generally, gradient descent is a tool to find a local minimum of a differentiable function $f : \mathbb{R}^n \to \mathbb{R}$ which is close to some initial point $x_0 \in \mathbb{R}^n$. In the first step we evaluate the negative gradient $-\nabla f$ at $x_0$. By following the negative gradient for a small enough distance we will obtain a point $x_1 := x_0 - \eta_0 \nabla f(x_0)$ such that $f(x_1) \leq f(x_0)$. Iterating this process gives a sequence $x_0, x_1, x_2, x_3, \ldots$, where

$$x_{i+1} := x_i - \eta_i \nabla f(x_i). \tag{6.10}$$

If the parameters $\eta_i$ where chosen small enough we have that $f(x_{i+1}) \leq f(x_i)$ so that the sequence $x_i$ will converge towards the location of a local minimum. Clearly, we do not want to choose the $\eta_i$ too small or otherwise the rate of convergence of the $x_i$ will be slow. However, if we are choosing $\eta_i$ too large it will make us overshoot the location of the local minimum. There are situations in which $f$ satisfies conditions, i.e. if $f$ is convex and smooth, in which there exists an explicit choice of $\eta_i$ for which the convergence can be guaranteed. In the context of training neural networks the parameters $\eta_i$ are collectively referred to as the *learning rate*. At the time of writing there is no developed theory on how to choose the learning rates optimally and we have to consider heuristics. An overview of several heuristics can be found in [73].

Let us now apply gradient descent to optimize neural networks. The setup is the following: We have some set of training data $D$, a neural network with some initial choice of weights and biases and a cost function $C_D$. The task is to find weights and biases which (locally) minimize the cost function $C_D$. This confronts us with the problem of how to compute the gradient $\nabla C_D$. This is solved by the second major ingredient of the training of neural networks: The *backpropagation algorithm*. The backpropagation algorithm consists of two steps: In the first step we compute the cost function $C_D$ of the neural network (see Equation 6.9). To evaluate the output of the network $F^N$ we evaluate the input $x$ on the first hidden layer and then feed the output of the first hidden layer into the second hidden layer and so forth until we obtain the output of the network at the last layer. In the second step of the backpropagation algorithm we compute the derivative of the cost function with respect to all weights and biases. The derivatives can be computed in linear time in the number of neurons. Obtaining the derivatives is a matter of applying the chain rule several times. To simplify notation we introduce the variable $s_i^l = \sum_{k \in pred(i,l)} w_{i,k}^l f_k^{l-1} + b_i^l$, where $pred(i,l)$ are the predecessors of the $i$th neuron in the $l$th layer. Remember that the value of a neuron is $f_i^l = \sigma(s_i^l)$. The derivatives of the cost function $C_D$ with respect to the weight $w_{i,j}^l$ can be expanded as

$$\frac{\partial C_D}{\partial w_{i,j}^l} = \frac{\partial C_D}{\partial s_i^l} \frac{\partial s_i^l}{\partial w_{i,j}^l}. \tag{6.11}$$

The second factor of Equation 6.11 is simply

$$\frac{\partial s_i^l}{\partial w_{i,j}^l} = f_i^{l-1}. \tag{6.12}$$

The form of the first term of Equation 6.11 depends on whether $l$ is a hidden layer or the output layer. For $l = L$ we expand over the values of the neurons in the output layer $F_k^N = f_k^L$

$$\frac{\partial C_D}{\partial s_i^L} = \sum_{k=1}^{n} \frac{\partial C_D}{\partial f_k^L} \frac{\partial f_k^L}{\partial s_i^L} = \frac{\partial C_D}{\partial f_i^L} \sigma'(s_i^L). \tag{6.13}$$

For all hidden layers $l < L$ we expand over the sums $s_k^{l+1}$ of neurons which are in the next layer and connected to the $i$th neuron in layer $l$

$$\frac{\partial C_D}{\partial s_i^l} = \sum_{k \in succ(i,l)} \frac{\partial C_D}{\partial s_k^{l+1}} \frac{\partial s_k^{l+1}}{\partial s_i^l} = \sum_{k \in succ(i,l)} \frac{\partial C_D}{\partial s_k^{l+1}} \, w_{i,k}^{l+1} \sigma'(s_i^l), \tag{6.14}$$

where $succ(i,l)$ indicates the set of all neurons in the next layer connected to the $i$th neuron in layer $l$. The derivatives with respect to the biases $b_i^l$ proceeds completely analogously, the only difference being that Equation 6.12 evaluates to 1.

Note that in order to compute Equation 6.14 for some layer $l$ we need to have the result of layer $l+1$. Hence we first evaluate Equation 6.13 for the output layer and then go backwards through the layers of the network (hence the name backpropagation). Finally, having obtained all derivatives with respect to the weights and biases allows us to perform a single step in the gradient descent (see Equation 6.10).

In the discussion above we made two simplifications which we are going to address now: The first simplification was the choice of the cost function. The $L^2$ norm is very intuitive but it leads to a very slow convergence of the training. The reason for this can be seen in Equation 6.13 and Equation 6.14. The derivatives are proportional to the derivative of the sigmoid function $\sigma'(z)$ which is close to 0 when $|z|$ is sufficiently large. This is avoided by choosing the *cross-entropy* as cost-function:

$$C_D = -\frac{1}{|D|} \sum_{x \in D} \left( \sum_{k=1}^{n} \left[ F_k(x) \log\left(F_k^N(x)\right) + (1 - F_k(x)) \log\left(1 - F_k^N(x)\right) \right] \right) \tag{6.15}$$

The cross-entropy has a less intuitive form. However, since $F(x), F^N(x) \in [0,1]$ for all inputs $x$ one can see that the cross-entropy is (a) positive and (b) small when the output of the network is close to the desired output. The cross-entropy has the advantage that in Equation 6.11 the derivatives of the sigmoid function cancel, see [74, 75, 76] for a derivation.

The second simplification was taking the average over the whole training set $D$. In practice, the training set is usually subdivided into several subsets called *batches* to speed up the computation.

Furthermore, part of the available training data is kept aside and not used for the training of the network. This data set is the called the *validation set* $V$ and it is only used to evaluate the cost function after every step of the training. The reason to keep the validation set separate is to check whether the neural network performs well on data outside of the training data. To summarize the training procedure:

1. Initialize the weights and biases.
2. Pick a batch $B \subset D$ and learning rate $\eta$.
3. Perform a single step of the gradient descent, using backpropagation to compute the gradient.
4. Compute the cost function $C_V$ on the validation set.

As long as $C_V$ keeps descending we repeat steps 2 - 4. The initial values in step 1 are usually chosen to be random.

### 6.4.3 Decoding

To use a neural network for decoding we first train it on pairs of random errors and their syndrome. After the training we have a network which takes a syndrome as input and returns a probability distribution over the faces, indicating how likely it is that an error occurred on each face (see Figure 6.19). To decode back to a code state we apply a flip to the $m$ faces which have the highest probability of error. We then update the syndrome and feed it back into the network.

As already mentioned at the beginning of this section, we consider neural networks with a certain structure. If we were to use a neural network with several layer as the one in Figure 6.16 then the number of variables would grow with the size of the code. This is indeed the case with all previous proposals which decode with neural networks [11, 12, 13, 14]. The networks we consider instead are called *convolutional networks* which are commonly used for image recognition. The name convolution itself comes from image processing and it is used to blur, sharpen or highlight edges. The convolution is a map which is being applied to each pixel. The map simply takes a weighted sum of the pixel value and all its nearest neighbors and assigns it to that pixel. The weights are always the same, so a convolution is determined by a $3 \times 3$ matrix of weights for the pixel and its 8 neighbors. The convolution in the neural network works in essentially the same way, except that our regions are not 2D but 4D. The weights of the image convolution correspond to the weights and biases of neurons which we apply.

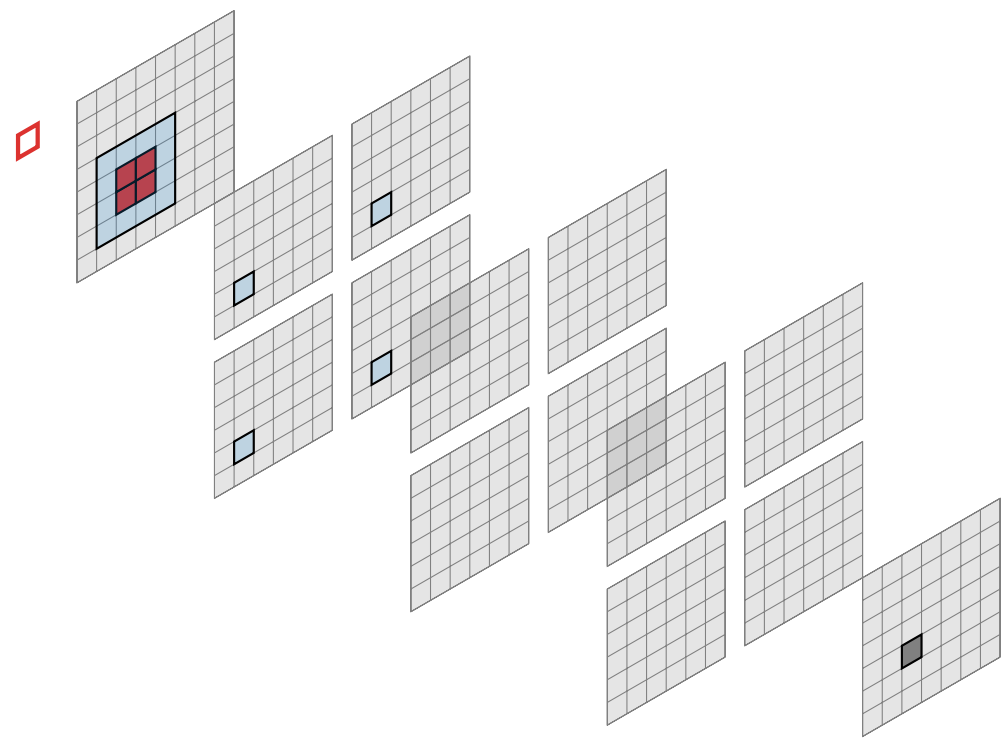

Figure 6.18: Illustration of the neural network. Each array of neurons is shown as a 2D square grid where each square is a neuron. However, for our purposes the array has the same dimension as the code (3D or 4D). The network consists of a single input layer (leftmost array) which receives the measurement results (red boundary of a square face). The input layer is followed by three hidden layers. The number of channels in each hidden layer is 4 in this illustration. The final layer returns the probability distribution as output. A single convolution is highlighted: The blue neurons in the first hidden layer are connected to all blue neurons in the input layer.

The networks we consider here consist of an input layer which receives the result of the parity check measurements, one convolution of size $3 \times 3 \times 3 \times 3$ followed by two convolutional layers with kernel size $1 \times 1 \times 1 \times 1$, i.e. a hypercube. See Figure 6.18 for an illustration. We chose the number of channels in the hidden layer to be 15 as this number gave good performance. However, we did not optimize this.

After the linear map of the final convolutional layer, a softmax function is applied:

$$x_i \to \frac{e^{x_i}}{\sum_i e^{x_i}} \tag{6.16}$$

The softmax function is usually applied when the output has to be a probability distribution, i.e. non-negative numbers that sum to 1. However, in the neural network decoder this is not a necessity, as during the decoding process we only care about the relative order of the output numbers, which the softmax layer preserves. Curiously, among the limited number of neural networks that we trained, we saw that the ones with the softmax layer performed slightly better. The networks were trained on systems of size $L = 5$ with errors sampled with an error rate close to where we believed the threshold to be.

In our simulations we saw that the neural network decoder also suffers from the energy-barrier limit that we have seen for the local decoders. In the neural network decoder we implemented a function that removes the parallel lines by flipping all qubits between them.

For the 3D toric code we can visualize the neural network decoder while it is operating, see Figure 6.19, where the process of decoding from a state to which an error has been applied back to the code state is shown. The syndrome is indicated by red loops. Note that sometimes loops end in the figure which is due to the periodicity of the boundaries. The opaqueness of the faces indicates the output of the neural network. It can be interpreted as how certain the network is that a face contains an error. We can see in Figure 6.19(c) that when the syndrome forms long straight lines the network assigns an equal probability to all incident faces. However, when there are many violated syndromes adjacent to a face, the network assigns a high probability to this face supporting an error. This is quite similar to the considerations taken in the construction of the Hastings decoder and the cellular automaton decoders.

### 6.4.4 Numerical simulation

Her we present numerical simulations of the performance of the neural network decoder. We only consider prefect syndrome measurement since, as of now, the results for noisy measurement have been inconclusive.

#### Monte-Carlo trials

A single Monte-Carlo trial consists of the following steps. We apply a single random $Z$-error independently on each qubit with probability $p$. The syndrome information is given to the neural network which returns the probability distribution. The qubit with the highest probability is flipped and the syndrome updated. This is repeated until the syndrome is either completely removed or only parallel lines are left. In the latter case we remove those lines by performing the following steps:

1. Make a list of current edges in the support of the syndrome. Order the violated edges in the

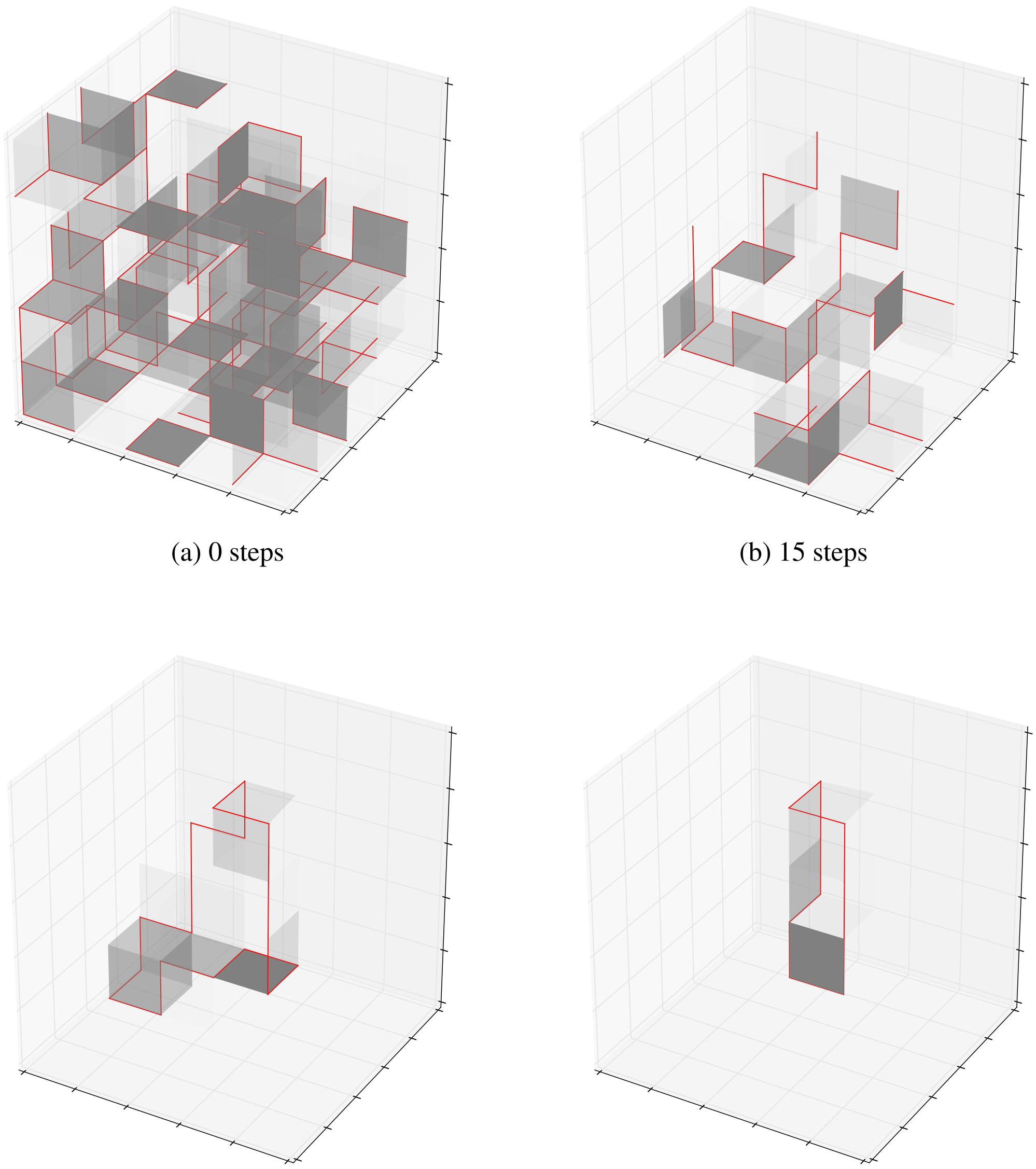

(a) 0 steps (b) 15 steps (c) 30 steps (d) 40 steps

Figure 6.19: Applying the neural network decoder to the 3D toric code with $L = 5$. The syndrome is highlighted in red. The neural network outputs a probability distribution over the faces indicating where it believes an error to be present. The probability of each face is indicated by its opaqueness. Each figure shows the current syndrome and the output of the network during the decoding. In each step the decoder flips the face with the highest probability.

list by their direction.

2. For each edge $e$ in the list: Assume $e$ has coordinates $(x_0, x_1, x_2, x_3)$ and $(x_0, x_1, x_2, x_3) + \vec{e}_i$. We look for the closest edge $e'$ with coordinates $(y_0, y_1, y_2, y_3)$ and $(y_0, y_1, y_2, y_3) + \vec{e}_j$ in the support of the syndrome such that $i = j$. We then change one of the $x_i$ in $e$ so that $e$ and $e'$ get closer by flipping the corresponding qubit.

3. Update the syndrome.

4. Repeat the steps above until no parity check is violated or a time limit is exceeded.

In the latter case we declare failure. Otherwise we can check whether a logical operator has been applied.

### Results of the simulation

In Figure 6.20 we see the results for decoding $Z$-errors on the 3D toric code for sizes $L = 6, 8, 10, 12$. We observe a crossing at 17.5%. We compare this to the minimum-weight decoder (which is implemented using an integer program solver and hence not computationally efficient). It has a slightly better performance with a crossing at around 23%. Note that the 3D toric code is not self-dual and the protection against $X$-errors is much lower.

The results for the 4D toric code are shown in Figure 6.21. We perform the same analysis as for the Hastings decoder by assuming a scaling behavior in the variable

$$x = (p - p_c)L^{1/\nu}, \tag{6.17}$$

to determine the critical error probability $p_c$. We expand the logical error probability $\overline{P}$ for small $x$ (around $p = p_c$ where the dependence on the system size $L$ is small):

$$\overline{P}(p, L) = A + Bx + Cx^2. \tag{6.18}$$

The fit in Figure 6.21 was obtained by fitting Equation 6.18 to the data for $p \in [0.065, 0.075]$. The fitting parameters $p_c$ and $\nu$ are:

$$\begin{aligned} p_c &= 7.1\% \pm 0.3\% \\ \nu &= 0.65 \pm 0.02 \end{aligned} \tag{6.19}$$

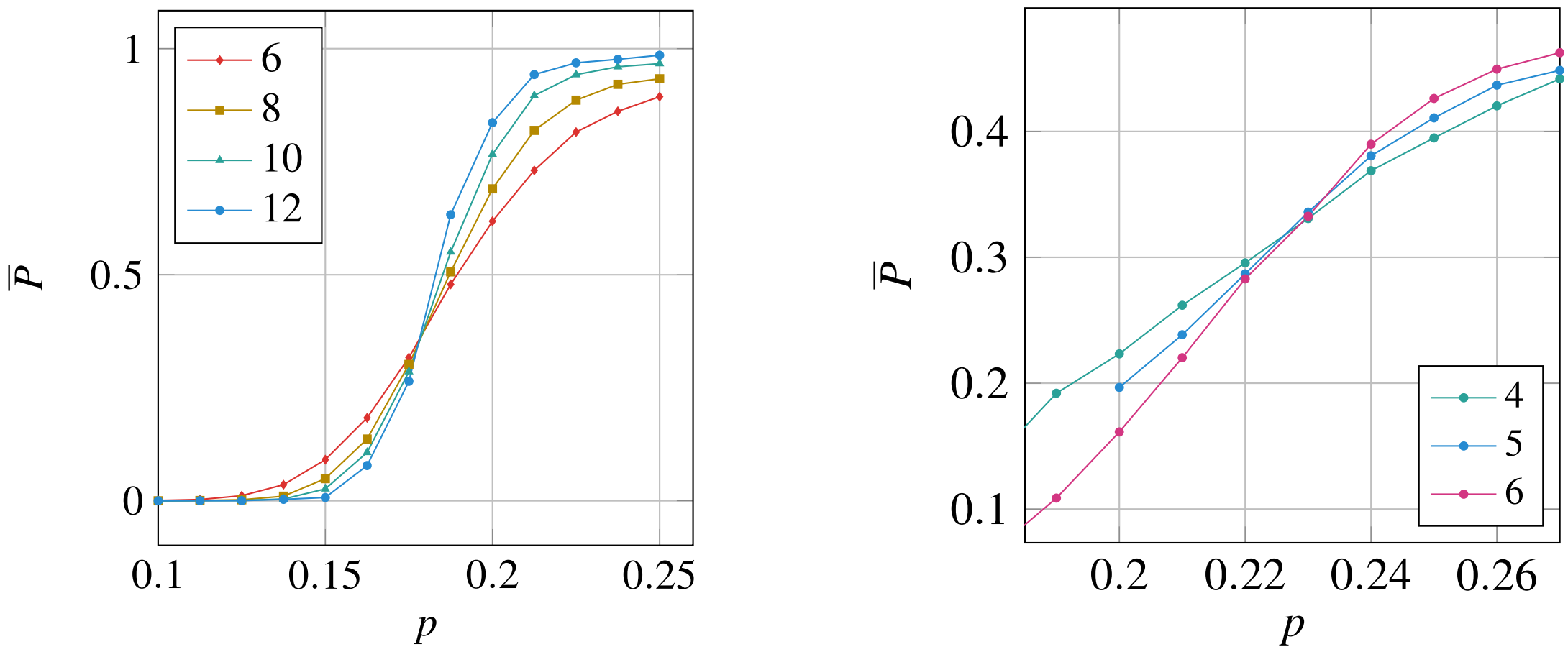

(a) Neural network decoder

(b) Minimum-weight decoder

Figure 6.20: (a) The results of the numerical simulation for the 3D toric code for $Z$-errors only, assuming perfect measurements. We considered system sizes $L = 6, 8, 10, 12$. The lines cross at around 17.5%. (b) Results for the minimum-weight decoder which has exponential run-time. The lines cross at around 23%.

This result has to be compared to the threshold of the minimum-weight matching decoder on the 4D toric code which is around 11%. The neural network decoder performs slightly worse, but it is computationally efficient. Although, as for the Hastings decoder it is not entirely clear whether these results will persist when going to much larger system sizes.

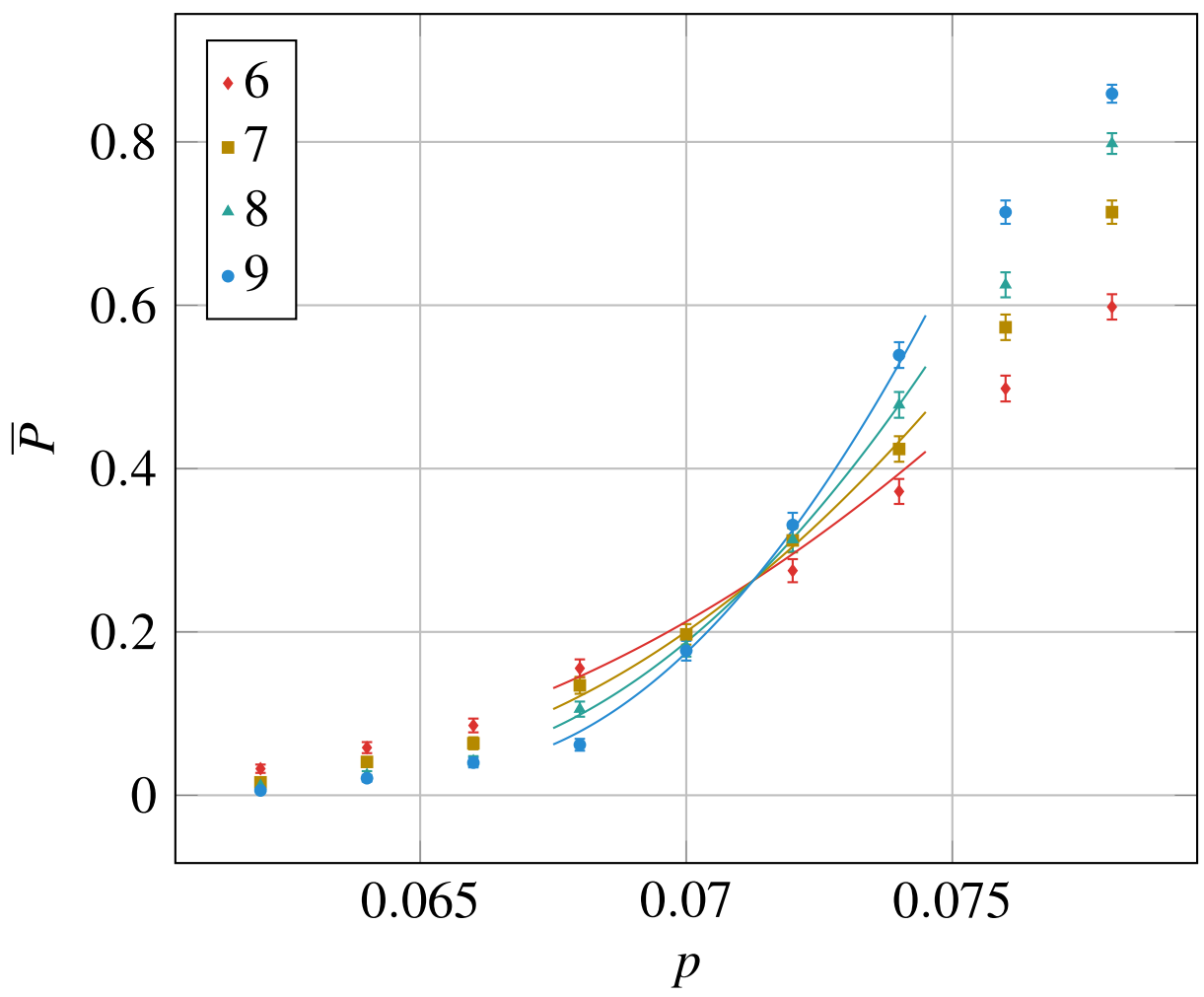

Figure 6.21: The results of the numerical simulation for the 4D toric code assuming perfect measurements. We considered system sizes $L = 5, 6, 7, 8$. The lines indicate the values of Equation 6.18.

# Chapter 7

# Conclusion and outlook

In this thesis we have dealt with quantum codes which are defined in curved and higher-dimensional geometries. Such codes allow for overhead savings and simplified decoding schemes.

**Chapters 3 and 4** dealt with quantum codes derived from tessellations of closed hyperbolic surfaces. We have discussed how such tessellated surfaces can be constructed using reflection groups. The tessellations we considered were regular, meaning that all faces are identical $r$-gons and the same number $s$ of faces meets at every edge. Codes derived from such tessellations have their number of encoded qubits scaling with the number of physical qubits. As the stabilizer measurements on these codes have to be performed on $r$ and $s$ qubits simultaneously we consider the codes with the lowest $r$ and $s$ as the most promising candidates. These codes are the $\{5,4\}$-codes which offer an encoding rate of $1/10$ and have hence almost the same weight measurements as the surface code. We have enumerated all codes of this and some other hyperbolic code families and defined extremal codes which have the property that they have their distance to physical qubit ratio is maximal among all codes of this family.

We gave numerical evidence that the decoding threshold of the $\{5,4\}$-codes under minimum weight decoding against independent $X$- and $Z$-noise is $p_c = 2.5\%$ when measurements are perfect and $p_c = 1.3\%$ when measurements are subject to noise. It was also shown numerically that in the latter case the number of times that the measurements need to be repeated only has to be logarithmic in system size. We also showed that using hyperbolic codes over the toric code or the surface code we can potentially save physical qubits by orders of magnitude while obtaining the same rate of error suppression. In the numerical analysis we saw that the performance of the hyperbolic codes is not only determined by the distance but also by the number of lowest weight logical operators. Hence, codes which are not extremal may be interesting as well and give good performance.

Regarding the construction of the tessellated surfaces themselves we remark that those may be of independent interest for investigating other statistical physics models. For example, it is known that percolation exhibits various new phenomena in hyperbolic spaces, e.g. several definitions of the percolation threshold which are identical for Euclidean tessellations obtain different values in hyperbolic spaces.[1]

By making them planar we saw that we obtain small examples of codes with small savings in physical qubits as compared to the surface code. However, we saw that such planar constructions can not have the same asymptotic properties as the ones obtained from closed surfaces. Another way of changing the hyperbolic surfaces is putting Euclidean tessellations on the faces of a hyperbolic tessellation, thereby diluting the effects of curvature. These semi-hyperbolic codes make it possible to interpolate between hyperbolic and Euclidean codes.

There may be ways of generalizing the hyperbolic code construction in a different direction. In our discussion we have reduced the geometry of a curved surface to properties of a group. The derived quantum code was completely determined by the structure of this group and some distinguished subgroups which gave rise to the cells of the tessellation. Being tied to geometry is helpful to gain intuition but it also leads to constraints as the result by Delfosse shows. A construction based on group theory alone may be able to circumvent such constraints.

In **Chapter 5** we discussed 4-dimensional quantum codes. We explained why such codes are interesting for implementations by discussing the local linear dependencies between the stabilizer checks which can be thought of as encoded by a classical code. Due to a result shown in [55] this implies that measurements do not have to be repeated to be able to decode. The noisy syndrome can simply be decoded using the classical code. The local linear dependency also provides a way to decode the code locally.

We introduced a 4D version of the surface code and show that this code encodes a single qubit. Furthermore we showed that any closed loop is also a boundary so that any way in which a noisy syndrome is fixed the result can be used for decoding. We mentioned the 4D hyperbolic codes and provide a single example.

An interesting future direction of research regarding the single-shot property is whether codes

[1] Private communication with Leonid Pryadko. To be published.

can be optimized to exhibit greater protection against syndrome noise.

In **Chapter 6** we implemented decoders for the 4D codes. We discussed the minimum-weight decoding problem for 4D codes. We gave some evidence that the tesseract code has the same threshold as the 4D toric code under minimum-weight decoding and assuming perfect measurements. Performing simulations with single-shot noisy measurements turned out to be inconclusive due to even-odd effects.

We analyzed a local decoding scheme proposed in [61] for 4D hyperbolic codes. We made some adaptations to the 4D toric code and extended the decoder to use single-shot fault-tolerance. However, it turned out that decoders based on cellular automata give comparable, if not better performance, while being much simpler. For the 4D hyperbolic code example we have observed a much better performance compared to the 4D toric code.

We proposed a decoder for 4D codes based on neural networks which utilizes convolutional networks which allow us to decode arbitrary large codes by only having to train the network once. We performed numerical simulations for this decoder as well and saw that it has a better performance compared to the purely local decoders.

The neural network decoder may be improved in several ways. The most obvious improvement is performing a grid search to optimize various parameters such as the number of filters, the number of hidden layers etc. It could also help to not only train the network on the syndrome of random errors, but also on partially decoded syndromes which appear during the decoding process. A more ambitious way to improve this scheme is using techniques like the ones used in the Go-program ALPHAGO. We can view the (simulation of the) decoding procedure as a single player strategy game where only after a large number of moves at the end it is revealed whether the decisions we made were good. Such a situation is quite common and is solved in ALPHAGO and other programs by a technique called *reinforcement learning*.

# Appendix A

# Families of hyperbolic codes with constant distance

The result of Section 3.3.2 guarantees that there exists an infinite family of hyperbolic codes with distance lower bounded by the number of physical qubits $n$. This raises the following question: Does the distance of a hyperbolic surface code necessarily increase with $n$? This would imply that we could pick any tessellated hyperbolic surface and obtain a code with a guaranteed lower bound on its distance. In this section we will show that this is generally not the case. There exist families of tessellated hyperbolic surfaces of increasing size which have essential cycles of constant length. They are quotient surfaces $\mathbb{H}^2/\Gamma_L$ indexed by a parameter $L > 2$ where each $\Gamma_L$ contains some short and some long translations. It is not obvious that such subgroups $\Gamma_L$ exists as they have to be (a) consistent with the structure of the tessellation, meaning that each $\Gamma_L$ is a subgroup of $G_{r,s}$, and (b) normal in $G_{r,s}$.

The construction works for tessellations with Schläfli symbols $\{r,4\}$ with $r \geq 6$ and $r$ even. Let us consider a group generated by $x, y, g_1, \dots, g_r$. The generators $g_i$ fulfill the relations

$$g_i^2 = e, \tag{A.1}$$

$$(g_i g_{i+r/2})^L = e, \tag{A.2}$$

$$(g_i g_j)^2 = e, \quad |i-j| \neq r/2. \tag{A.3}$$

We identify the generators $g_i$ with the edges of a $r$-gon, so that there is a natural action of the dihedral group $D_r = \langle x, y \mid x^2, y^2, (xy)^r \rangle$ on the generators $g_i$. For $x$ we have

$$xg_ix = g_{r-i+1} \tag{A.4}$$

and for $y$ we have

$$yg_1y = g_1, \tag{A.5}$$

$$yg_{r/2+1}y = g_{r/2+1}, \tag{A.6}$$

$$yg_iy = g_{r-i+2} \tag{A.7}$$

for all $i = 2 \ldots r/2$.

From $g_i^2 = (g_i g_{i+r/2})^L = 1$ follows that the parallel reflections $g_i$ and $g_{i+r/2}$ generate groups isomorphic to the dihedral group $D_L$. The full symmetry group of the finite tessellated surface is generated by $g_i, x, y$ and is therefore isomorphic to a semi-direct product of $r/2$ copies of $D_L$ with $D_r$

$$G_L = (D_L \times \cdots \times D_L) \rtimes D_r. \tag{A.8}$$

Defining $a = g_1$, $b = x$, $c = y$ we obtain the same relations as in Equation 3.9:

$$a^2 = b^2 = c^2 = (ab)^4 = (bc)^r = (ca)^2 = e \tag{A.9}$$

The groups $G_L$ can be used to construct a family of hyperbolic surfaces. We identify faces, edges and vertices with cosets with respect to the subgroups $\langle ab \rangle$, $\langle ac \rangle$ and $\langle bc \rangle$ as discussed in Section 3.1.5. The rotation around the center of a face is $\rho = ab = g_1x$ and the rotation around a vertex is $\sigma = bc = xy$. All group elements which are not conjugates of the above act fixed-point free.

Let us consider an example to give a more intuitive understanding of the group $G_L$. Choosing $r = 4$ we obtain the symmetry group of the Euclidean square grid on the torus. The group of symmetries of the torus $T^2$ is the group $\mathrm{Isom}(T^2) = (O(1) \times O(1)) \rtimes O(2)$, which is a semi-direct product of the translation group $O(1) \times O(1)$ with the rotation group $O(2)$. We see that the group $(\mathbb{Z}_L \times \mathbb{Z}_L) \rtimes D_4$ forms a discrete subgroup.

From the size of the group we can deduce the number of edges in the tessellation of the surface. Since $|G_{4,q}|$ is equal to the number of fundamental triangles (see Figure 3.5) and since there are 4 different such triangles for every edge (there are two faces and two vertices incident to every edge) we obtain

$$|E| = \frac{|G^L_{r,4}|}{4} = \frac{(2L)^{r/2}\cdot 2r}{4} = \frac{r}{2}(2L)^{r/2}. \tag{A.10}$$

Let us now consider the length of the essential cycles of these tessellated surfaces. First of all we observe that we will always have $r/2$ essential cycles of length $L$ which correspond to translations $(g_i g_{i+r/2})^L$. However, we have the translations $(g_i g_j)$ where $|i-j|>1$ and $|i-j|\neq r/2$. These translations are of length 4 and do not grow with $L$.

# Appendix B

# The codespace of the Tesseract Code

In Section 5.3 we have defined the Tesseract Code which is a $[[n = 6L^4 - 12L^3 + 10L^2 - 4L + 1, k = 1, d = L^2]]$-code. We will now show how removing $B$ gives us a stabilizer code which encodes a single qubit by showing that $\dim H_2(A,B) = 1$. Additionally, we will show that, unlike for the 4D toric code, there are no essential 1-cycles or essential 3-cocycles. This means that when we fix a noisy syndrome, as described in Section 5.1.2, we are guaranteed to obtain a valid syndrome which can be decoded.

Let us first describe how we can compute the dimensions of the relative homology groups before we go into the details. For continuous spaces it can be shown that one can 'shrink' them to a lower dimensional space while leaving their homology the same. We will use this fact by first identifying the space $A$ with a continuous $L \times L \times L \times L$-box in $\mathbb{R}^4$, which we will also denote by $A$, and then shrink the quotient space $A/B$ (in which all points in the subspace $B \subset A$ are identified). How are the relative homology groups $H_i(A,B)$ related to the homology of $A/B$? The relative homology groups $H_i(A,B)$ stay invariant when factoring out $B$ on both sides. This is called the *excision theorem* (see [25] for more background). Hence we have that the dimensions of the relative homology groups of $A$ with $B$ factored out are given by the dimensions of the relative homology groups of the quotient space $A/B$ with a single point factored out

$$\dim H_i(A,B) = \dim H_i(A/B, B/B) = \dim H_i(A/B, \text{point}). \tag{B.1}$$

Let us now go into the details of the calculation. Since $A$ is a hypercubic box tiled by hypercubes, it can be identified in the obvious way with $[0,L]^4 \subset \mathbb{R}^4$. The set of cells $B$ forms a subspace of $A$ under this identification:

$$B = \{(x,y,z,w) \in A \mid x \in \{0,L\} \text{ or } y \in \{0,L\}\}. \tag{B.2}$$

In fact, $B$ is a connected subspace of the boundary of $A$.

The process of continuously shrinking a topological space $A$ (within itself) onto a subspace $M \subset A$ is called a *deformation retraction*. Formally a deformation retraction is described by a continuous map $f : [0,1] \times A \to A$ such that $f(t,\cdot) = id_M$ for all $t$, $f(0,\cdot) = id_A$ and $f(1,A) = M$.

Corollary 2.11 in [25] states that if there exists a deformation retraction of $A$ onto $M$ (more generally: a homotopy equivalence) then $\dim H_i(M) = \dim H_i(A)$ for all $i$.

We define a deformation retraction of $A/B$

$$f_t^w : A/B \to A/B,\ (x,y,z,w) \mapsto (x,y,z,(1-t)\,w), \quad t \in [0,1] \tag{B.3}$$

onto the subspace

$$M = \{(x,y,z,0) \in A/B\}. \tag{B.4}$$

Similarly, there is a deformation retraction

$$f_t^z : M \to M,\ (x,y,z,0) \mapsto (x,y,(1-t)\,z,0), \quad t \in [0,1] \tag{B.5}$$

onto the space

$$N = \{(x,y,0,0) \in A/B\}. \tag{B.6}$$

The space $N$ is a $L \times L$ square in the $x$-$y$-plane with all points with $x \in \{0,L\}$ or $y \in \{0,L\}$ identified. Hence we have that $N$ is homeomorphic to the sphere $\mathbb{S}^2$. With Proposition 2.22 from [25] we have

$$\dim H_i(A,B) = \dim H_i(A/B, \text{point}) = \dim \tilde{H}_i(\mathbb{S}^2) = \begin{cases} 1, & i = 2 \\ 0, & i \neq 2 \end{cases} \tag{B.7}$$

where $\tilde{H}_i$ is the reduced homology group defined as $\tilde{H}_i(X) = H_i(X)$ for $i \neq 0$ and $\tilde{H}_0(X) = H_0(X) \oplus \mathbb{Z}_2$. The last equality follows from the fact that homology groups $H_i(X)$ with $i > 0$ are not affected by removing vertices. Furthermore, the 0th homology group $H_0(X)$ counts the number of connected components of $X$, since for any vertex in $X$ we factor out all other vertices which are connected

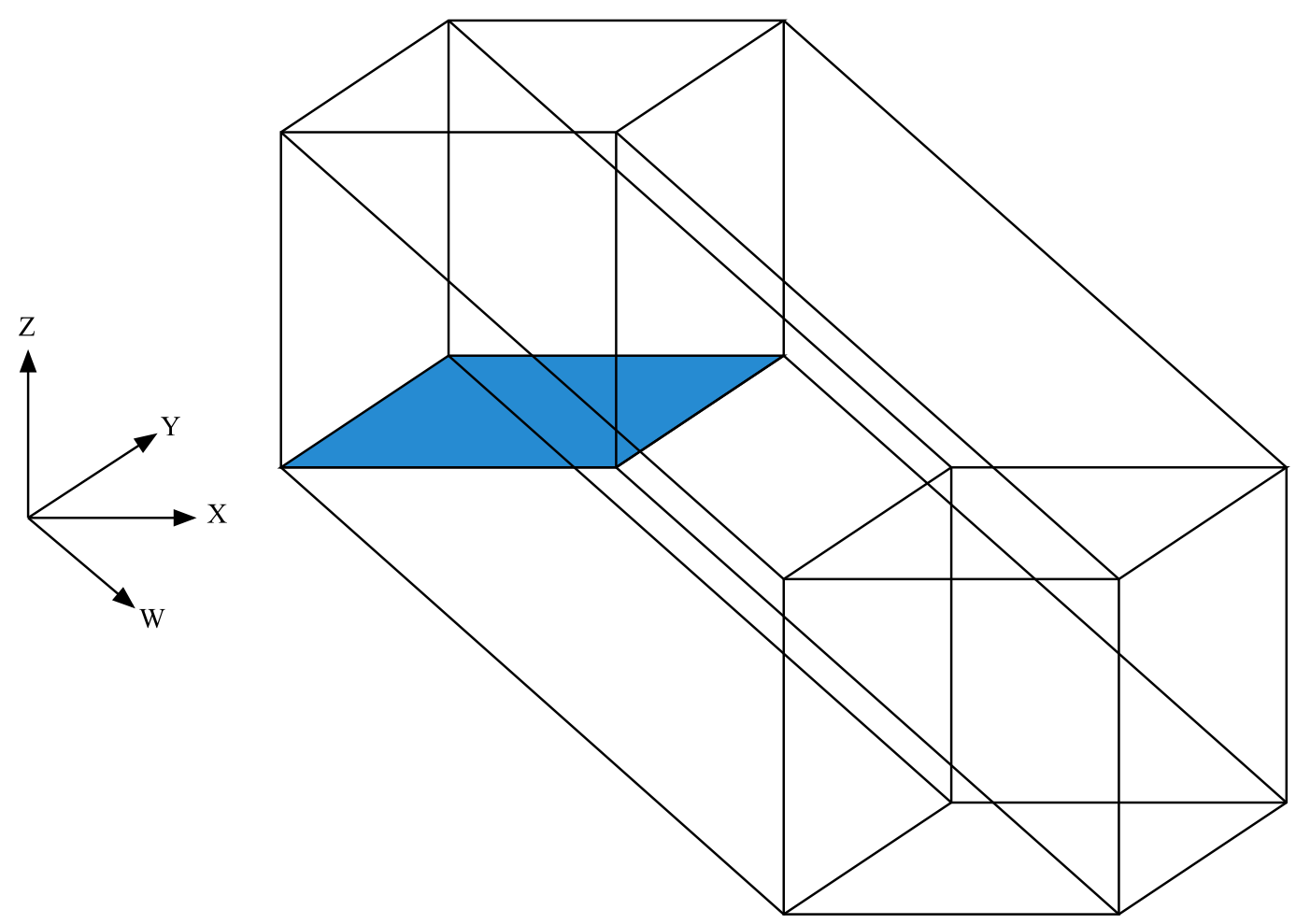

(a) The manifold $A/B$. All points with coordinates $(0,y,z,w)$, $(L,y,z,w)$, $(x,0,z,w)$ or $(x,L,z,w)$ are identified. The subspace $N$ is highlighted in blue.

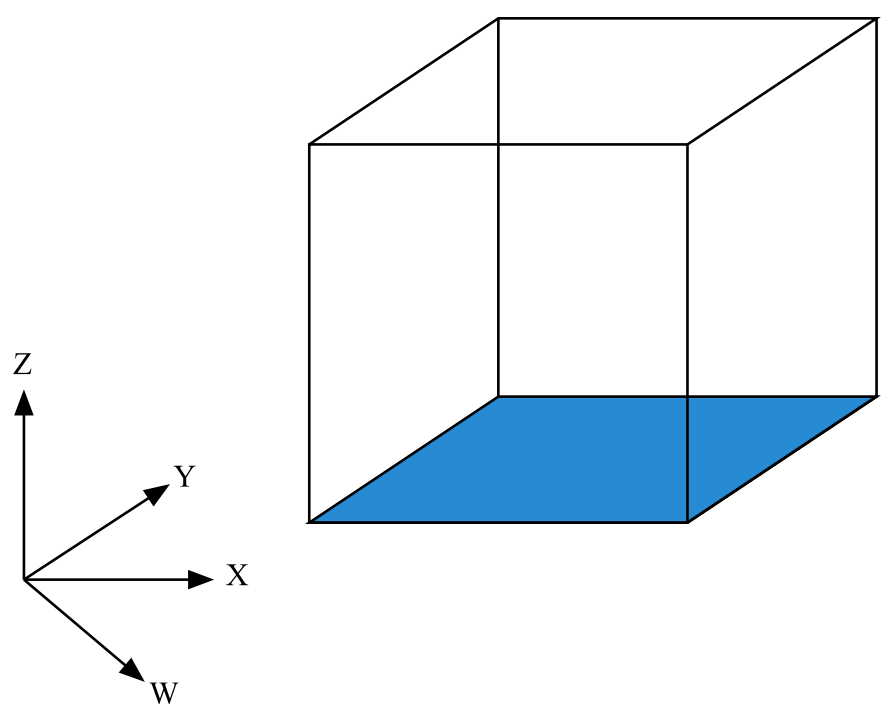

(b) The manifold $M \subset A/B$. The subspace $N$ is highlighted in blue.

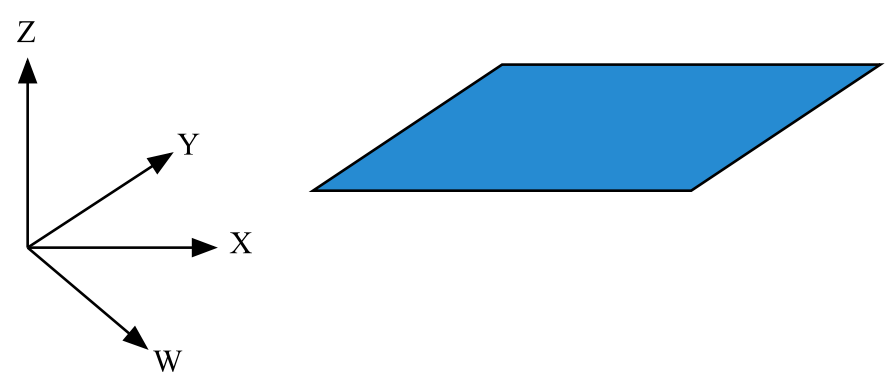

(c) The manifold $N$ which is homeomorphic to a sphere $\mathbb{S}^2$.

Figure B.1: Illustration of the deformation retraction. The map $f_t^w$ retracts the space $A/B$ into $M$ along the $w$-direction. Similarly, $f_t^z$ retracts the space $M$ into $N$ along the $z$-direction. The spaces $A/B$ and $N$ must hence have isomorphic homology groups.

by a string of edges to it. When $X$ has only a single connected component $\dim H_0(X) = 1$ and thus $\dim H_0(X, \text{point}) = \dim \tilde{H}_0(X) = 0$.

Equation B.7 shows that there are no non-trivial 1-cycles since $\dim H_1(A,B) = 0$. To show that there are no non-trivial 3-cocycles we need the generalization of Poincaré duality to relative homology, which is known as *Poincaré-Lefschetz duality*. It states that for a compact manifold $A$ and $B, C \subset \partial A$ with $\partial B \cap \partial C = B \cap C$ we have that $\dim H^i(A,C) = \dim H_{n-i}(A,B)$. By taking $C = \{(x,y,z,w) \in A \mid z \in \{0,L\} \text{ or } w \in \{0,L\}\}$ we thus have

$$\dim H^3(A,B) = \dim H_1(A,C) = \dim H_1(A,B) = 0. \tag{B.8}$$

The second equality follows from the fact that $B$ is equal to $C$ up to relabeling coordinates.

# List of publications

- N.P. Breuckmann and B.M. Terhal, "Space-Time Circuit-to-Hamiltonian Construction and its Applications," *IOP Journal of Physics A: Mathematical and Theoretical*, vol. 47, no. 19, p. 195304, 2014.

- N.P. Breuckmann and B.M. Terhal, "Constructions and Noise Threshold of Hyperbolic Surface Codes," *IEEE Transactions on Information Theory*, vol. 62, no. 6, pp. 3731-3744, 2016.

- N.P. Breuckmann, K. Duivenvoorden, D. Michels, and B.M. Terhal, "Local Decoders for the 2D and 4D Toric Code," *Quantum Information and Computation*, vol. 17, no. 3-4, 2017.

- N.P. Breuckmann, C. Vuillot, E. Campbell, A. Krishna and B.M. Terhal "Hyperbolic and Semi-hyperbolic Codes for Quantum Storage," *IOP Quantum Science and Technology*, vol. 2, no. 3, 2017.

- K. Duivenvoorden, N.P. Breuckmann and B.M. Terhal, "A Renormalization-Group Decoder for 4D Toric Codes," in preparation

- N.P. Breuckmann and X. Ni, "Decoding Higher-Dimensional Quantum Codes with Neural Networks," in preparation

# Curriculum vitae

**Personal Data**

| | |
|---|---|
| Name: | Nikolas Peter Breuckmann |
| Date of Birth: | 1988-04-13 (Duisburg, Germany) |
| Nationality: | German |

**Education**

| | |
|---|---|
| 2004 - 2007 | *Abitur*<br>Hittorf-Gymnasium Recklinghausen |
| 2008 - 2010 | *Bachelor of Science in Physics (Minor in Computer Science)*<br>RWTH Aachen University<br>Thesis - Quantum Subsystem Codes: Their Theory and Use<br>(under supervision of Prof. Barbara Terhal) |
| 2009 - 2011 | *Bachelor of Science in Mathematics (Minor in Physics)*<br>RWTH Aachen University<br>Thesis - Logical and Algorithmic Aspects of Rank Notions over Rings<br>(under supervision of Prof. Erich Grädel) |
| 2010 - 2013 | *Master of Science in Physics (Track: Quantum Field Theory and Gauge Theories)*<br>RWTH Aachen University<br>Thesis - From Quantum Circuits to Hamiltonians:<br>Analysis of a Multi-Time Construction for QMA<br>(under supervision of Prof. Barbara Terhal) |